\documentclass[natbib]{svjour3}

\usepackage{amssymb,graphicx,latexsym,multirow,rotating,url}

\journalname{Space Science Reviews}
\newcommand{\ao}{Applied Optics}
\newcommand\apj{ApJ}%
\newcommand\apjl{ApJ}%
\newcommand\apjs{ApJS}%
\newcommand\aap{A\&A}%
\newcommand\jgr{J.~Geophys.~Res.}%
\newcommand\mnras{MNRAS}%
\newcommand\nat{Nature}%
\newcommand\pasj{PASJ}%
\newcommand\solphys{Sol.~Phys.}%
\newcommand{\aapr}{{A\&AR}}
\newcommand{\aaps}{{Astron. Astrophys. Suppl. Ser.}}

\newcommand{\azh}{Astronomicheskii Zhurnal}
\newcommand{\Ha}{{H$\alpha$}}
\newcommand{\Hb}{{H$\beta$}}
\newcommand{\La}{{L$\alpha$}}
\newcommand{\Lb}{{L$\beta$}}
\newcommand{\Lg}{{L$\gamma$}}
\newcommand{\Le}{{L$\epsilon$}}
\newcommand{\kmps}{km~s\ensuremath{^{-1}}}
\newcommand{\arcsec}{\hbox{$^{\prime\prime}$}}
\newcommand{\logT}{\ensuremath{\log T}\mbox{(K)}}
\newcommand{\cc}{\mbox{cm}\ensuremath{^{-3}}}
\newcommand{\eg}{{\it e.g.,}\/}

\newcommand{\ie}{{\it i.e.,}\ }

\newcommand{\ga}{\lower.4ex\hbox{$\;\buildrel >\over{\scriptstyle\sim}\;$}}
\newcommand{\la}{\lower.4ex\hbox{$\;\buildrel <\over{\scriptstyle\sim}\;$}}

\begin{document}

\title{Physics of Solar Prominences: I - Spectral Diagnostics and Non-LTE Modelling}


\author{N.~Labrosse \and P.~Heinzel \and J.-C.~Vial \and T.~Kucera \and S.~Parenti \and S.~Gun\'ar \and B.~Schmieder \and G.~Kilper}

\authorrunning{Labrosse et al.} 

\institute{N. Labrosse \at
              Department of Physics and Astronomy,
		University of Glasgow,
		Glasgow G12 8QQ,
		United Kingdom \\
              \email{n.labrosse@physics.gla.ac.uk}           
           \and
           P. Heinzel \and S. Gun\'ar \at
	   Astronomical Institute, 
		Academy of Sciences of the Czech
		Republic,
		Dr. Fri\v{c}e 298/1,
		CZ-25165 Ond\v{r}ejov,
		Czech Republic
	   \and 
	   J.-C. Vial \at
	   Institut d'Astrophysique Spatiale,
		Universit\'e Paris XI/CNRS,
		91405 Orsay Cedex,
		France
	   \and 
	   T. Kucera \at
	   NASA/GSFC, Code 671,
		Greenbelt, MD 20771,
		USA
	   \and 
	   S. Parenti \at
	   Observatoire Royal de Belgique, 3 Av. Circulaire,
		1180  Bruxelles,
		Belgique\\
	   Now at Institut d'Astrophysique Spatiale,
		Universit\'e Paris XI/CNRS,
		91405 Orsay Cedex,
		France
	   \and 
	   B. Schmieder \at
	   Observatoire de Paris, LESIA, 5 place Jules Janssen, F-92190 Meudon cedex, France
	   \and
	   G. Kilper \at
	   Rice University, Dept. of Physics and Astronomy 6100 Main St., Houston, TX 77005, USA\\
	   Now at NASA/GSFC, Code 671, Greenbelt, MD 20771, USA
}

\date{Received: date / Accepted: date}

\maketitle

\begin{abstract}

This review paper outlines background information and covers recent advances made via the analysis of spectra and images of prominence plasma and the increased sophistication of non-LTE (\ie\ when there is a departure from Local Thermodynamic Equilibrium) radiative transfer models. We first describe the spectral inversion techniques that have been used to infer the plasma parameters important for the general properties of the prominence plasma in both its cool core and the hotter prominence-corona transition region. We also review studies devoted to the observation of bulk motions of the prominence plasma and to the determination of prominence mass. However, a simple inversion of spectroscopic data usually fails when the lines become optically thick at certain wavelengths. Therefore, complex non-LTE models become necessary.  We thus present the basics of non-LTE radiative transfer theory and the associated multi-level radiative transfer problems. The main results of one- and two-dimensional models of the prominences and their fine-structures are presented. We then discuss the energy balance in various prominence models. Finally, we outline the outstanding observational and theoretical questions, and the directions for future progress in our understanding of solar prominences.
	
\end{abstract}

\tableofcontents

\section{Introduction}
\label{sec:1}

Solar prominences are large magnetic structures confining a cool (temperature $T \lesssim {10^4}$~K) and dense (electron density $10^9-10^{11}$~cm$^{-3}$) plasma in the hot solar corona.  Typically, the prominence plasma is 100 times cooler and denser than its coronal surroundings. This situation raises important questions about the origin of the prominence plasma, and the energy and force equilibria which allow it to remain in the corona for a relatively long time.

Many important characteristics of solar prominences, along with a presentation of some of the mathematical and physical tools necessary to understand the observations, are given in the monograph by Einar Tandberg-Hanssen \citep{1995nsp..book.....T}. Because of the impressive amount of exquisite observations obtained since that book was published some fifteen years ago, and the ever increasing power of numerical models and simulations developed in order to explain these observations, we felt that it was time to present an updated vision of the current status of our knowledge in the field of solar prominence physics.

\begin{figure}
	\center
	  \includegraphics[width=\textwidth]{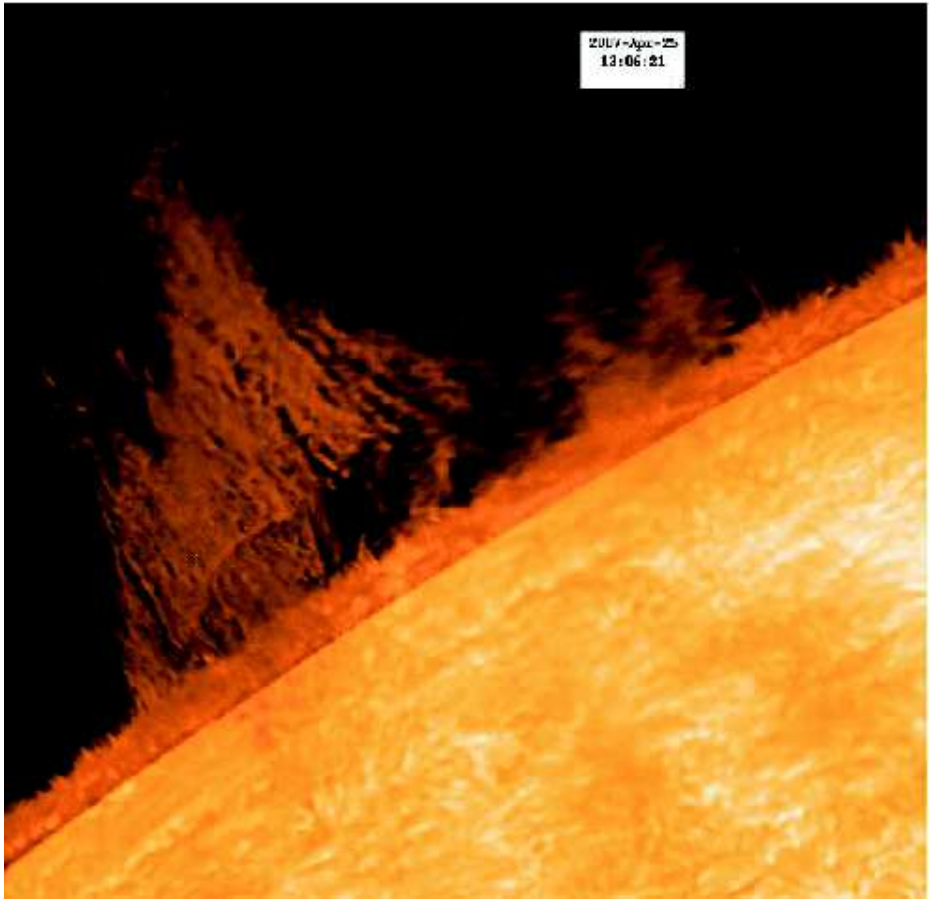}
	  \caption{A quiescent solar prominence observed above the limb in H$\alpha$ {with the Hinode/SOT NFI filter \citep{2008ApJ...686.1383H}}.}
	  \label{fig:0-sot}
\end{figure}

Early ground-based  observations (\eg\ in the H$\alpha$ line of hydrogen at 6564~\AA) have led to a distinction being made between a \textit{prominence}, which appears in emission when it is seen above the solar limb, and a \textit{filament} when it is seen against the solar disk in absorption. This fundamental difference in the appearance of a prominence is easily explained in terms of opacity {and emissivity} of the plasma. In this review, we will often use both terms interchangeably as they refer to the same object.
The prominence radiation in H$\alpha$  allows us to view the densest parts of the structure in detail. The H$\alpha$ line is routinely observed from the ground as well as from space (Fig.~\ref{fig:0-sot}), as it is currently done by the SOT instrument \citep[Solar Optical Telescope,][]{2008SoPh..249..167T} on {the} Hinode {satellite} \citep{2007SoPh..243....3K}.

\begin{figure}
	\center
	\includegraphics[width=\textwidth]{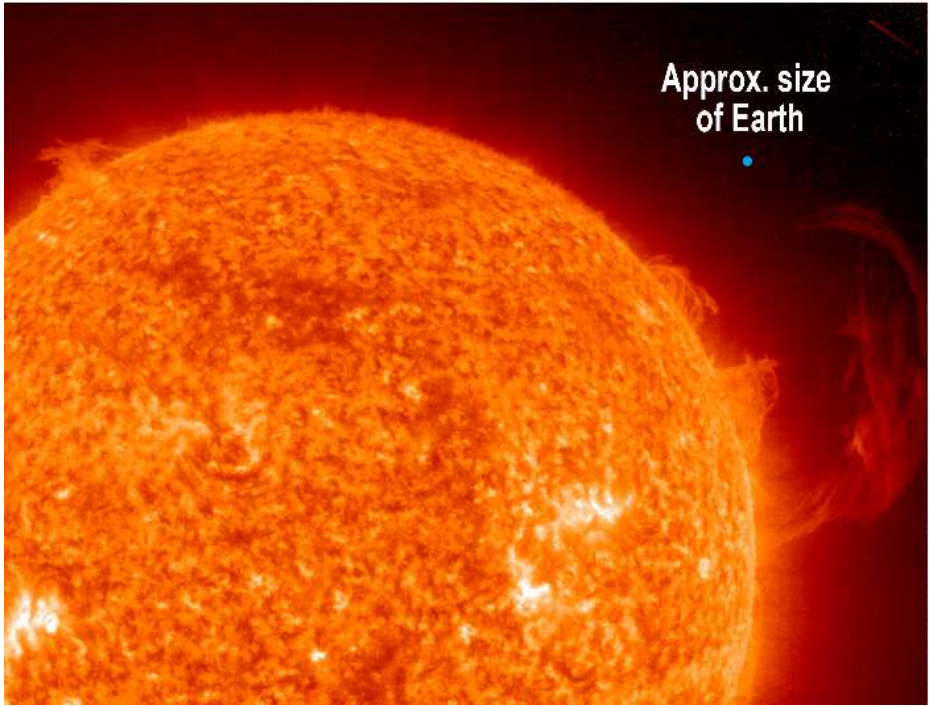}
	\caption{Quiescent and eruptive prominences observed by SOHO/EIT on 21 April 2005 in the 304~\AA\ channel where emission comes predominantly from He~II.}
	\label{fig:promearth}
\end{figure}

A further distinction is usually made between \textit{quiescent} and \textit{active} prominences.
{For a discussion on the classification between different types of prominences and filaments, we refer the reader to Sects.~2.2.1 and 5 of the companion review paper \citep[hereafter Paper~II]{2010SSRv..tmp...32M}.}
Quiescent prominences are relatively stable features lying mostly outside active regions. Their lifetimes range from a few days up to several months (several solar rotations).  Their dimensions are in the range of a few $10^4-10^5$~km in length, and a few $10^3-10^4$~km thick (with fine structures of the order of $10^2-10^3$~km). Their heights are of the order of $10^4-10^5$~km. Figure~\ref{fig:promearth} shows several prominences observed with the Extreme ultraviolet Imaging Telescope \citep[EIT,][]{1995SoPh..162..291D} on board the Solar Heliospheric Observatory \citep[SOHO,][]{1995SoPh..162....1D} on April 21, 2005 in the 304~\AA\ channel (where emission is predominantly from He~II). 
The size of the Earth is indicated for comparison.
While they are globally stable, high-resolution movies from H$\alpha$ ground-based and space-based observations do show that even quiescent prominences exhibit rapid variations in their fine structure on time-scales of a few minutes \citep[\eg ][]{2007SoPh..246...65L,2008ApJ...676L..89B}.

Active prominences are dynamical features typically occurring in the vicinity of active regions and are usually short-lived (their lifetime is smaller than the lifetime of the associated active region).  Their altitude is often smaller than that of quiescent prominences.
Both quiescent and active prominences form an integral part of the solar corona.  The conditions for their stability are inherently related to the associated magnetic configuration (see Paper~II{, Sect.~2}).
Prominences and filaments may undergo large-scale instabilities which will disrupt their equilibria and lead to eruptions. These eruptions are often associated with flares and Coronal Mass Ejections (CMEs). 

In this review, we are primarily interested in the determination of the prominence plasma parameters which play a role in the force and energy equilibria, while Paper~II describes recent progress in four areas of prominence research: their magnetic structure, the dynamics of prominence plasmas (formation and flows),  the dynamics of magneto-hydrodynamic (MHD) waves in prominences, and  the formation {and large-scale patterns} of the filament channels in which prominences are located.
Both papers focus primarily on non-eruptive prominences. 

We outline general goals of the prominence spectral diagnostics and non-LTE modelling.
Non-LTE stands for departures from Local Thermodynamic Equilibrium, or LTE.
Particular approaches and techniques, together with the most important results, {are} detailed in 
subsequent sections. The ultimate goal of the prominence spectral diagnostics is the determination
of all important physical (thermodynamic) quantities and their variations in
space and time. Some of these quantities {(as known in the eighties)} are listed in Table~\ref{tab:hvar}, while others will be mentioned in following sections. 
\begin{table}
\caption{Main prominence plasma parameters in the cool core and in the PCTR (prominence-to-corona transition region): Electron temperature, microturbulent velocity, electron density, gas pressure, hydrogen ionization ratio, flow velocities. Adapted from \cite{1990LNP...363..294E}.}
\label{tab:hvar}
\begin{tabular}{lll}
\hline\noalign{\smallskip}
Physical Parameter & Core & PCTR  \\
\noalign{\smallskip}\hline\noalign{\smallskip}
$T$ (K) & $4300 - {10000}$ & $10^4 - 10^6$ \\
$\xi_t$ (km s$^{-1})$ & $3 - 20$ & 30 \\
$n_\mathrm{e}$ (cm$^{-3}$) & $10^9 - 10^{11}$ & $10^6 - 10^8$\\
$p$ (dyn cm$^{-2}$) & $\sim 0.02 - 1$ & $\sim 0.2$\\
$N(\mathrm{H^+})/N(\mathrm{H^0})$ & $0.2 - 0.9$ &\\
$V$ (km s$^{-1})$ & $\sim 5$ & $\sim 10$\\
\noalign{\smallskip}\hline
\end{tabular}
\end{table}
Primarily we are interested in those quantities which play a role
in establishing the prominence equilibria, namely the momentum and energy equilibrium.
The other main physical quantity, the magnetic field, is presented in Paper~II.

Two basic approaches are used to obtain the information on the plasma parameters. The first one is {\em a direct inversion of spectral data}, which is possible under some specific conditions. Namely, in the case of both cool as well as hot, \textit{optically thin} plasmas (plasmas from which all {emitted} photons leave freely without interaction), one can obtain spatially averaged values of some parameters, as we discuss later. For hotter optically thin plasmas, no radiative transfer is needed, but in some cases one has to solve the statistical equilibrium equations for multilevel atoms and ions in order to compute the line emissivities. 
Sect.~\ref{sec:3} presents spectral inversion techniques for prominence plasma parameters in the cool part. The inferred values of the {electron temperature, non-thermal velocities, electron density, ionization degree, pressure, and abundances,} are given.
Sect.~\ref{sec:4} follows the same approach and focuses on the plasma of the PCTR (prominence-to-corona transition region). The mechanisms of line emission from an optically thin plasma are first described, and then the inferred values for {the electron temperature, electron density, gas pressure, and non-thermal velocities} are presented. {The small-scale structure of the PCTR is also discussed.}
Sects.~\ref{sec:5} and \ref{sec:6} review measurements of velocities and observations of various types of bulk motions in prominences, and the determination of the mass of the prominence plasma.

The second approach requires the solution of the radiative transfer problem, using the coupled equations of radiative transfer, statistical equilibrium, and other constraint equations. This is generally called non-LTE radiative transfer,  or non-LTE modelling. It is applied to multilevel atoms or ions, and is necessary in the case of optically thick plasmas. We devote a substantial part of this paper to prominence non-LTE modelling. Contrary to direct inversion techniques, the non-LTE modelling represents \textit{a forward method}. This means that starting from a given prominence model (spatial distribution of temperature, pressure, gas density), one evaluates the excitation and ionization balance for given species, determines the opacities and emissivities, and finally solves the transfer equation along the line-of-sight to get the emergent {\em synthetic spectrum}. The latter is then compared to the observed spectrum. In this way, one can adjust the initial model in order to get an optimum agreement with the observations. This procedure is iterative and the final models are called {\em semi-empirical models}, {provided} that they are at least partially data-driven.
In Sects.~\ref{sec:7} and \ref{sec:8}, we present the basics of the radiative transfer theory and of multilevel non-LTE problems {relevant to prominences}. Then, Sects.~\ref{sec:9} and \ref{sec:10} present the basic results from one-dimensional {(1D)} and two-dimensional {(2D)} non-LTE radiative transfer computations, with particular emphasis on the hydrogen and helium spectra of model prominences.

The radiation which prominences or filaments emit, and which we observe in the form of monochromatic images or in the form of spectra, has actually a two-fold importance for the prominence physics. On one hand, it provides us with the diagnostics of the prominence structure as we explained above.
On the other hand, the radiation field inside the prominence plays a crucial role in the global energy budget of prominence structures via the radiative losses or gains, and has also to be considered when evaluating self-consistently the prominence internal structure (momentum equilibria). The prominence densities are generally low enough so that the radiation field determines to a large extent the plasma ionization state which enters the {equation of state of the plasma} -- the relation between the gas pressure and density depends on the temperature and on the degree of ionization of hydrogen and helium. Therefore, the non-LTE physics must be integrated into the global radiation magneto-hydrodynamical (RMHD) modelling in order to obtain consistent models. Such theoretical models which are not data-driven are usually called {\em ab initio models}. They should play a crucial role in our understanding of the prominence physics.
Considerations on energy balance issues are given in Sect.~\ref{sec:11}.

Finally, Sect.~\ref{sec:12} summarises open issues related to the material reviewed here, and we make some suggestions on possible future directions to progress in our understanding of solar prominences, both observationally and theoretically.

In the following,  we describe various aspects of the prominence spectral diagnostics and non-LTE modelling in detail, and summarize the most important results. Our discussion is focused mainly on optical, ultraviolet (UV), and extreme ultraviolet (EUV) diagnostics, with some extension to selected infrared (IR) lines. However, we don't include the complex issue of the prominence polarimetry. The Stokes polarimetry is currently used to diagnose the prominence/filament magnetic fields via the Hanle effect,  and the results are briefly discussed in Paper~II {{(Sect.~2.1.4)}}.

\section{Spectral Inversion for the Cool Prominence Plasma}
\label{sec:3}
\label{s:DenPress}

In this section, the focus is on relatively direct measurements of temperature, density, pressure, and other quantities (such as ionization degree and abundances) from the direct inversion of Gaussian spectral lines. Methods involving extensive modelling {for more complex cases} are described in Sect.~\ref{sec:9}. Previous reviews can be found in \eg\ \citet{1990LNP...363..187H} and \citet{1995nsp..book.....T}.

\subsection{Temperature and Non-Thermal Velocities}
\label{s:Temperature}
The simplest way of deriving the temperature is to analyse the profile of the line.
{In this section, we limit ourselves to optically thin plasmas emitting spectral lines which are Gaussian-shaped.}
For a Gaussian-shaped line, the Doppler width is given by:
\begin{equation}\label{eq_th}
\Delta \lambda_D = \frac{\lambda_0}{c} \left(\frac{2kT}{m} \right)^{1/2} \ ,
\end{equation}
where $\lambda_0$ is the rest wavelength position of the line, $T$ is
the ion temperature, and $m$ is the mass of the ion. In the literature, we may find different values for
$T$. In some cases the temperature of maximum ionization fraction is
taken. In other cases, $T$ corresponds to the temperature of 
formation of the line.

{Gaussian-shaped spectral lines from optically thin plasmas have a width that generally exceeds the thermal broadening. Unresolved motions can be responsible for additional broadening.}
If a line is broader than expected from the thermal component, an
additional non-thermal term $\xi$ is included in (\ref{eq_th}):
\begin{equation}\label{eq_th2}
\Delta \lambda_D = \frac{\lambda_0}{c} \left(\frac{2kT}{m}+ \xi^2 \right )^{1/2} \ .
\end{equation}
The parameter $\xi$ is the non-thermal velocity (NTV) component and may have different
interpretations. Generally, it describes an averaged value along the
line-of-sight (LOS) of the unresolved motions,  due to waves or turbulence. Equation~(\ref{eq_th2}) also assumes that the non-thermal motions have a Maxwellian distribution. Uncertainties associated with the line width measurement and the determination of $T$ affect the result for $\xi$.  This is particularly true for cool lines produced by neutrals and singly ionized ions \citep{1998ApJ...505..957C, 2007SoPh..241...39B}, for which it is difficult to establish the formation temperature. 
Estimating this quantity as a function of temperature is important for understanding the distribution of energy in the different layers of the prominence-corona interface. Important implications in the energy budget may be deduced (see Sect.~\ref{sec:5}).

\begin{figure}
	\center
\includegraphics[width=\textwidth]{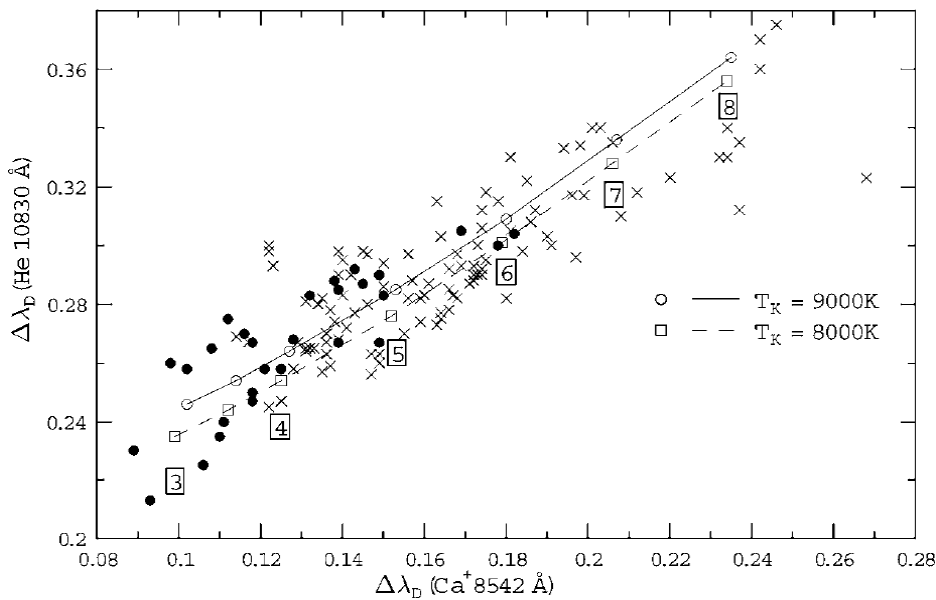}
\caption{Relation between observed Doppler widths of He~I 10830~\AA\ and Ca~II 8542~\AA\ for two prominences (dots and crosses). The calculated curves for kinetic temperatures of  8000~K and 9000~K are shown along with the  values of the NTVs (values in the squares in \kmps). From \cite{2003SoPh..217..133S}.}
\label{fig:sw03}
\end{figure}

The thermal velocity and the NTV may be deduced by a line fitting technique and by using the measured Full Width at Half Maximum (FWHM$_\mathrm{meas}$). Under the assumption that the instrumental profile is Gaussian (FWHM$_\mathrm{inst}$), the true solar line width (FWHM$_\mathrm{sun}$) is given by:
\begin{equation}
	\mathrm{FWHM_{sun}}^2=\mathrm{FWHM_{meas}}^2-\mathrm{FWHM_{inst}}^2 \ .
\end{equation}
An immediate technique for separating the temperature from $\xi$ consists
in measuring as many lines from different atoms and ions of different
atomic masses as possible \citep[Fig.~1 of][]{1990LNP...363..187H}.
Results {obtained by \cite{2003SoPh..217..133S}} with Ca~II 8542~\AA\ and He~I 10830~\AA\ are shown in Fig.~\ref{fig:sw03}: they derived a temperature in the range $8000-9000$~K and $\xi$ between 3 and 8~\kmps\ {in two quiescent prominences}. 
{It is difficult to assess whether the dispersion of observational values results from uncertainties or intrinsic variations inside each prominence.}
It should be
pointed out that the technique assumes that the lines are optically
thin. Otherwise, temperature and $\xi$ are overestimated.

\subsection{Electron Density}
\label{s:Density}

As is clear in the Hvar model \citep[see Table~\ref{tab:hvar}, and][]{1990LNP...363..294E}, and {as we will see} below, measured values of the electron density  vary greatly. Some of these variations are, no doubt, due to {differences between} the various techniques {that are used}, but {there are also variations among} individual prominences, and between parts of the same prominence.

A primary ground-based technique for measuring the electron density is to use the
Stark effect. The Stark effect broadens lines in such a way that the density can be
estimated, \eg\ from the number of the highest resolvable Balmer line \citep[see][]{1939ApJ....90..439I}.
However, the broadening can also be due to electric field and
instrumental broadening, so this technique essentially provides an upper
limit on $n_\mathrm{e}$.  The method was applied on numerous
occasions by Hirayama \citep[\eg ][] {1971SoPh...17...50H,1972SoPh...24..310H,1985SoPh..100..415H}, who obtained upper limits on the electron density  between 
$10^{10}$ and $10^{11}$~\cc\ in quiescent prominences.

Lines in the mid-infrared have also been used as a tool for electron density diagnostics. 
Based on the observations of the H~I line at 808.3~cm$^{-1}$ (12.37~$\mu$m) in a quiescent prominence, \citet{1985SoPh..102...33Z} obtained an upper limit to $n_\mathrm{e}$ of $1.9\times10^{10}$~\cc. \citet{1998SoPh..179...89C} measured Stark broadening  in the  spectra of a quiescent and two time-frames of an active prominence. They found upper limits to $n_\mathrm{e}$ of $2\times10^{10}$~\cc\ in the former, and $7.4\times10^{10}$ and $3.5\times10^{10}$~\cc\ from the two spectra in the active region prominence.

The Hanle effect, which is a depolarization due to the magnetic field, can also be used to calculate $n_\mathrm{e}$ by utilizing the fact that there is additional {collisional} depolarization that increases with density in some lines and not in others. \cite{1986A&A...156...79B,1986A&A...156...90B,1994SoPh..154..231B} used the Hanle effect to measure $n_\mathrm{e}$ in a set of combined 24 prominences, finding values between $1\times10^9$ and $6.4\times10^{10}$~\cc.

Landman developed a technique for estimating $n_\mathrm{e}$ from ratios of resonance lines of Na, Mg, Sr$^+$, and Ba$^+$, which draws on some assumption on the intensity of the \La\ line. His revised average value for a combination of quiescent and active region prominences is $9\times10^{10}$~\cc\ \citep{1986ApJ...305..546L}, with a range of about a factor of two.

\citet{1983A&A...119..261K} determined an average electron density limit of $\ge5.7\times10^{9}$~\cc\ for a prominence observed during an eclipse, using a diagnostic based on the ratio of the hydrogen \Hb\ intensity to that of the white-light continuum.
\citet{1986A&A...162..307S} used Thompson scattering measurements during the same eclipse to estimate $n_\mathrm{e}$ in a highly activated/erupting prominence as $\ge2\times10^9$~\cc.  These calculations provide only lower limits because they assume a homogeneous source; a filling factor less than one would lead to a
higher $n_\mathrm{e}$.
A more complex diagnostics of this type was recently developed by \cite{2009SoPh..254...89J}{ who derived similar density values}.

There have been a number of electron density estimates based on emission measure methods. 
{The emission measure yields the amount of plasma emitting the observed radiation along the LOS, and is described in Sect.~\ref{sec_em}.}
Such methods require {the} modelling of the source volume, especially
along the LOS, and of the filling factor. For instance,
\citet{1993ApJ...418..510B} determined the emission measure of a
quiescent prominence from microwave data. By estimating the extent of
the source along the LOS and assuming a filling factor
between 0.03 and 0.3, they derived an electron density in the range $1
\times10^{10} - 3\times 10^{10}$~\cc. 
Similarly, \citet{1993A&A...274L...9H} used microwave observations of an
erupting prominence to calculate $n_\mathrm{e}=10^{10}$~\cc, assuming a filling
factor of 0.33 and $T=6000$~K. 
\citet{1998SoPh..183..323L} provided an emission measure estimate using
\Ha\ line fits combined with a cloud model. Using an estimate of the
prominence thickness, they calculated $n_\mathrm{e}=1.8\times10^{10}$~\cc.

\citet{1993ApJ...412..853K} observed what they thought to be second order
plasma radiation emitted at 1.4 GHz from an erupting prominence. From this
assumption they calculated a lower limit to the density of $6\times10^9$~\cc.

{Thus we see that we have measurements of prominence electron density from $10^9$ to $10^{11}$~cm$^{-3}$. Other (indirect) methods, such as non-LTE modelling (described in Sect.~\ref{sec:9}), or prominence seismology (Paper~II, Sect.~4.4), yield similar electron density values.}
{The aim of on-going work is to shorten the two-orders-of-magnitude range of values, inasmuch as the large diversity of prominences allows for such an achievement.} 

\subsection{Ionization Degree}
\label{s:Ionization Degree}

The electron density is not the only density to be determined. If one wants to derive \eg\ the total mass of the prominence or the {gas} pressure, one has to measure the density of neutral hydrogen (along with minor atoms and ions such as He). This raises the issue of the ionization degree of the hydrogen plasma, that we define  as the ratio of the proton density (roughly equated here to the electron density $n_\mathrm{e}$) to the neutral hydrogen density (roughly equated here to the population density of the ground level, $n_1$).  Actually, there are not many studies devoted to this issue. They require a simultaneous derivation of the $n_\mathrm{e}$ and $n_1$ densities. We have discussed electron densities in the section above. Neutral hydrogen densities are addressed in Sect.~\ref{sec:6}.  

The earlier measurements {from} Skylab/ATM \citep[Apollo Telescope Mount,][]{1977ApOpt..16..825T} provided $n_\mathrm{e}/ n_1 $ values between 1 and 2 \citep{1980ApJ...240..908O,1981SoPh...69..313K}. From the ratio of \La\ to Ca~II K intensities {measured by the LPSP (Laboratoire de Physique Stellaire et Plan\'etaire) instrument \citep{1978ApJ...221.1032B} on OSO-8 (Orbiting Solar Observatory),} \citet{1982ApJ...253..330V} derived a value of 3.4 (within a factor 3), definitely higher than the 0.09 value derived by \citet{1984ApJ...279..438L} which led to suspiciously high pressures of the order of 6~dyn cm$^{-2}$.  
The Hvar reference atmosphere gives $N(\mathrm{H^+})/N(\mathrm{H^0})$ between 0.2 and 0.9 (Table~\ref{tab:hvar}). 

{The ionization degree of the plasma plays an important role in the momentum balance.}
Its importance with respect to the damping of MHD waves in prominences is emphasized in Sect.~4.3.3 of Paper~II \citep[see also][]{2009SSRv..tmp...39O}.

At this stage, we should emphasize that only a comparison between the full set of profiles (including hydrogen) and the predictions of non-LTE modelling can provide an adequate answer for this important parameter.

\subsection{Gas Pressure}
\label{s:Pressure}

Early ground-based efforts to measure prominence gas pressure are discussed extensively by \citet{1990LNP...363..187H}.  A number of early values measured with ATM, OSO-4, and OSO-6, yielded values between 0.01 and 0.08~dyn cm$^{-2}$ for  quiescent prominences, although \citet{1979SoPh...61..319M} found pressure values between 0.07 and 0.22~dyn cm$^{-2}$ in a more active prominence.

One often used pressure diagnostic is the ratio of H$\beta$ to Ca~II
8542~\AA, originally suggested by \citet{1978ApJ...221..677H}. The method has
yielded a range of values from low $0.01 - 0.04$~dyn cm$^{-2}$
\citep{1978ApJ...221..677H,1997A&A...319..669S,2000SoPh..196..357S} to
high values of $0.3 - 0.5$~dyn cm$^{-2}$ \citep{1988A&A...197..274B}.  \citet{2002A&A...385..273G}
revised models of the relevant line emissions.  They applied the results to the data of \cite{2000SoPh..196..357S} and found  the pressure to be above 0.1~dyn cm$^{-2}$.
Although the values vary greatly from prominence to prominence, the
ratio tends to be fairly uniform for a given prominence
\citep[\eg ][]{1998A&A...334..280D,2003AN....324..338D}. For instance
\citet{2000SoPh..196..357S} report variations in the pressure of about
30\% over a given prominence.

As discussed in Sect.~\ref{s:Density}, \citet{1986ApJ...305..546L}  used ground-based
spectra to calculate the electron density. He then derived a pressure of 1.4 dyn cm$^{-2}$, a value which, like the
density values (whether electron or neutral hydrogen), is higher than that calculated using the H$\beta$ to Ca~II
ratio discussed above.

\subsection{Abundances}
\label{s:Abundances}

The precise measurement of abundances could be a precious source of information on the origin of prominence material, and hence on the prominence formation mechanisms.
Establishing whether the prominence plasma {more closely} resembles the photospheric {or chromospheric plasma} tells us something about its formation. The different models for the formation of prominences invoke \textit{plasma condensation} (where {chromospheric} abundances are expected) or \textit{siphon} and \textit{emerging flux} models, where prominence material originating from the photosphere is transported into the corona (leading to photospheric abundances). We refer the reader to Sect.~3 of Paper~II for an in-depth discussion of the different models of prominence formation.

However, very few studies aimed at measuring quiescent prominence abundances exist in the literature, probably because of the difficulty in treating this {partially}-ionized environment. 

\cite{1980ApJ...235..268M} derived absolute element abundances for C, N, O, and Si in different solar regions, including a prominence, taking as reference a Si~IV line from ATM measurements (which was then used to scale the abundances to hydrogen). His averaged values in the prominence indicated {the} N abundance to be less, C as abundant as, and O more abundant than in the quiet Sun. However, there are important uncertainties in these measurements,  primarily arising from  the atomic physics calculations (uncertainties of at least 20\%).

The Ne/Mg abundance ratio in quiescent prominences was obtained from Skylab by \cite{1998ApJ...494..450S}. Their results, obtained under the assumption that {the two} lines used are formed at the same temperature, supported a photospheric origin for the prominence plasma. \cite{1995ApJ...440..884S}  arrived to a similar conclusion by finding Ne-enriched plasma (that is, close to the photospheric value) in emerging flux regions and close to the footpoints of a filament.
 \cite{1986ApJ...308..982W} derived the same ratio for an erupting prominence and found a value intermediate between that for the photosphere and corona. This was probably due to the heating and mixing of the prominence plasma with the surrounding coronal environment. 

These analyses are difficult {to perform, because of} the uncertainties in the absolute UV line photometry coupled with the uncertainties in many atomic parameters.

Let us note that non-LTE modelling techniques can also be used to study abundances. For instance, \cite{1978ApJ...221..677H} inferred a helium abundance of $0.100\pm0.025$ by number with respect to hydrogen.

\section{Spectral Inversion for the Prominence-Corona Transition Region}
\label{sec:4}

In this section, we describe plasma diagnostic techniques that apply to most of the  EUV ($100-1000$~\AA) and UV  ($1000-4000$~\AA) plasma emission, \ie\ coming from the chromosphere (temperature of $6 \times 10^3-10^4$~K) to the corona ($T > 10^6$~K), {with particular emphasis on} the PCTR ($T> 10^4$~K). 

\begin{figure}
	\center
	\includegraphics[width=\textwidth]{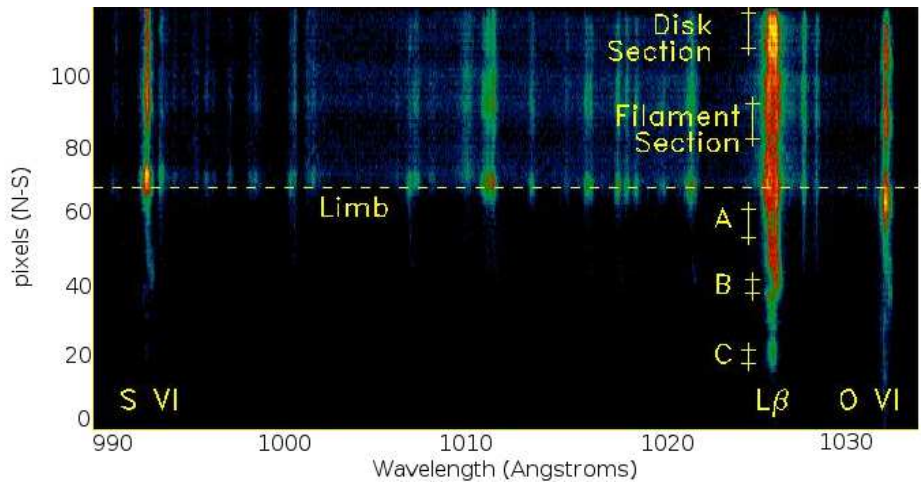}
	\caption{A spectrum obtained by SOHO/SUMER in S~VI, \Lb, and O~VI. {A portion of the disk, a filament, and the filament extension above the limb as a prominence are visible. A, B, and C point to three different parts of the prominence. {Adapted from \cite{1999SoPh..189..109S}.}}
	\label{fig:sumerspec}}
\end{figure}

Since 1995, the two most important instruments for UV-EUV spectroscopy of prominences have been SUMER (Solar Ultraviolet Measurement of Emitted Radiation) and CDS (Coronal Diagnostic Spectrometer) aboard SOHO. Each instrument has its own advantages, and both are therefore complementary and extremely useful to study the PCTR. SUMER wavebands include a large number of chromospheric and transition region (TR) lines. In addition, it has a high spectral resolution. Figure~\ref{fig:sumerspec} shows an example of a spectrum obtained by SUMER in S~VI, \Lb, and O~VI. The solar disk, the filament on the disk, and the prominence {above the limb}, are visible on the spectrum.
CDS consists of two types of spectrometers: the Normal Incidence Spectrometer (NIS) and the Grazing Incidence Spectrometer (GIS). CDS wavebands are richer in density sensitive lines at coronal temperatures than those of SUMER. 
\cite{2002SoPh..208..253P} reviewed the great contribution of these instruments to solar prominence physics, along with other SOHO instruments.
These two spectrometers were joined in 2007 by EIS (Extreme-ultraviolet Imaging Spectrometer) on Hinode. The possibility {of obtaining} combined observations with these three instruments {provides} a rare opportunity for detailed spectroscopy of the solar atmosphere. This has led to a number of {joint} observation campaigns dedicated to prominences \citep[\eg][]{2008ApJ...686.1383H} {as well as other targets}.

We summarise the main characteristics (spatial and spectral resolution, wavelength coverage) of current space-based spectrometers which significantly contribute to solar prominence science in Table~\ref{tab:instr}.
\begin{table}
	\begin{minipage}{\textwidth}
	\caption{Main characteristics of current space-based spectrometers. References: 1 -- \cite{1995SoPh..162..189W}; 2 -- \cite{1995SoPh..162..233H}; 3 -- \cite{2007SoPh..243...19C}. \label{tab:instr}}
	\begin{tabular}{p{2.65cm}p{2.30cm}p{2.15cm}p{2.30cm}c}
		\hline\noalign{\smallskip}
		Instrument / Observatory & Wavelength range (\AA, 1st order) & Spectral pixel size & Spatial pixel size (arcsec/pixel) & Ref.\\
		\noalign{\smallskip}\hline\noalign{\smallskip}
		SOHO/SUMER & $660-1610$ & $42-45$~m\AA\ pixel$^{-1}$ & 1 & 1\\
		SOHO/CDS & $150-800$ & 70~m\AA\ pixel$^{-1}$ (NIS1), 117~m\AA\ pixel$^{-1}$ (NIS2), 400~m\AA\ for 12 pixels (GIS) & 1.68 & 2\\
		Hinode/EIS & $170-210$ and $250-290$ & 22~m\AA\ pixel$^{-1}$ & 1 & 3\\
		\noalign{\smallskip}\hline
	\end{tabular}
\end{minipage}
\end{table}
The CDS {spatial and spectral resolutions were affected by} the temporary loss of the SOHO satellite in 1998. Post-recovery values are different (in some cases the performance was degraded), and these can be found on the CDS website\footnote{\url{http://solar.bnsc.rl.ac.uk/software/uguide/uguide.shtml}}.
Post-recovery CDS data show distorted line profiles, but standard routines have been made available in the CDS software package to recover the original line profiles.

Let us also mention the Ultraviolet Coronagraph Spectrometer \citep[UVCS,][]{1995SoPh..162..313K}, the other SOHO spectrometer. UVCS can be used to study erupted prominences high in the corona ($R>1.5R_\odot$). Even though the spatial and spectral resolutions are lower than those of the low corona instruments, UVCS data are unique in their kind. 

More details on the diagnostic techniques presented in this section can be found in \cite{1977ApJ...215..652F, 1992str..book.....M, 1994A&ARv...6..123M}; and \cite{2008Phillips}.
These techniques apply to optically thin plasmas with electron densities $n_\mathrm{e} < 10^{13}~\mathrm{cm^{-3}}$. In such environments, the radiation which is emitted by the source completely escapes the source itself without being absorbed. This is true for most of the lines emitted in the PCTR, although some exceptions apply.  
There are many atomic processes involved in the formation of the observed radiation. These include excitation and de-excitation of atomic levels,  and ionization and recombination for each ion. However, not all these processes act at the same rate, and in  most cases of interest we can draw  a simplified picture. 
Electron temperature, electron density, plasma velocities and other plasma parameters can be derived by measuring the observed spectral lines and applying the appropriate inversion {technique}. 
Table \ref{tab_ion} lists the main optically thin lines used in recent works to diagnose the PCTR. 
\begin{table} 
\caption{Main lines detected by SOHO instruments that can be used to diagnose the plasma conditions in the PCTR: emitting ion, wavelength, formation temperature, and spectrometers detecting each line.} \label{tab_ion}
 \begin{tabular}{lccc}
\hline\noalign{\smallskip}
Ion & $\lambda$~(\AA) & \logT & Instrument\\
\noalign{\smallskip}\hline\noalign{\smallskip}
Si II       & 1259        &  4.1           &  SUMER \\
C II        & 1037         &  4.4            &  SUMER, UVCS\\
Si III      & 1206         &  4.7           & SUMER\\
C III       & 977, 1175    &  4.8            & SUMER\\
Si IV       & 1402       &  4.9              & SUMER\\
O III       & 525         &  4.9            &  SUMER, CDS\\
N V         & 1242       &  5.3           & SUMER, UVCS\\
O IV        & 554        &  5.3            & CDS\\
O V         & 1218       &  5.4	     &  SUMER\\
O VI        & 1037       &  5.6            & SUMER, UVCS\\
\noalign{\smallskip}\hline
\end{tabular}
\end{table}
This table also lists  the line wavelength, the  formation temperature of the line and the SOHO spectrometers that detect the line. 

The UV-EUV prominence emission also contains various continua emitted from the cool part of the structure (mainly hydrogen). Specific diagnostic techniques can be applied to derive, for example, the plasma temperature from the Lyman continuum (see Sect.~\ref{sec:9.1}).

\subsection{Line Emission from an Optically Thin Plasma}
\label{sec:thinplasma}

The optically thin emission discussed in this section is assumed to come from a plasma in ionization equilibrium: the population of a given ionization state of an element is constant in time. All the lines listed in Table \ref{tab_ion} are optically thin in prominences. Optically thick lines are discussed in Sects.~\ref{sec:9} and \ref{sec:10}.

The  emissivity of the transition (power per unit volume per unit solid angle) due to the spontaneous decay of electrons from the ion's excited level $j$ to lower level $i$ (\textit{bound-bound} transition) is given by:
\begin{equation}\label{eq_P}
	P(\lambda_{ji})= \frac{hc}{4\pi\lambda_{ji}} n_j A_{ji} ~~ \mathrm{[erg ~cm^{-3}~ s^{-1}~ sr^{-1}]} \ ,
\end{equation}
where  $n_j$ represents the number density of the level $j$, and $A_{ji}$ is the spontaneous radiative decay rate from level $j$ to level $i$.

Under the PCTR density conditions, the ionization and recombination processes are slower than the collisional excitation in populating excited levels. In this case the ionization state of an element can be treated independently from the excitation balance. The number density $n_j$ is obtained by solving the statistical equilibrium equations for {a number of} low-lying levels and including all the relevant excitation and de-excitation mechanisms.

Further simplification arises under the {\it coronal  approximation}. This assumes that only the ground level ($g$) and the excited level ($j$) are responsible for the emitted radiation,  and it is generally valid for resonance lines. In this case the statistical  equilibrium equation is reduced to: 
\begin{equation}\label{eq_stat}
n_\mathrm{e} n_g C_{gj} = n_j A_{jg} \ ,
\end{equation}
where $C_{gj}$ is the collisional excitation rate and $n_\mathrm{e}$ the electron number density.

In this case the problem of calculating the emission is reduced to considering excitation from the ground level to the  excited one,
 balanced by spontaneous radiative decay back to the ground level. The line  intensity is then given by the integration of (\ref{eq_P}) in the emitting volume $V$ of cross-sectional area~$A$, along the LOS:
\begin{equation}\label{eq2}
	I(\lambda_{jg}) = \frac{hc}{4\pi\lambda_{jg}A} \int_V {n_\mathrm{e} n_{g} C_{gj} \mathrm{d}V}  ~~\mathrm{[erg~cm^{-2} s^{-1} sr^{-1}]} \ ,
\end{equation}
where (\ref{eq_stat}) was also used. 
The population of the $g$ level can be expressed in the form of:
\begin{equation}\label{eq4}
	n_{g} = \frac{n_{g}}{N_\mathrm{ion}} \frac{N_\mathrm{ion}}{N_\mathrm{el}} \frac{N_\mathrm{el}}{N_\mathrm{H}} \frac{N_\mathrm{H}}{n_\mathrm{e}} n_\mathrm{e} \ ,
\end{equation}
where $n_{g}/N_\mathrm{ion} \sim 1$ under the assumed conditions, $N_\mathrm{ion}/N_\mathrm{el}$ is the ion {density} relative to the total number density of the element;  $N_\mathrm{el}/N_\mathrm{H}$ is the abundance $Ab$ of the element with respect to hydrogen, and $N_\mathrm{H}/n_\mathrm{e} = 0.8$ {in the case of a fully ionised hydrogen-helium plasma}. 

The coefficient $C_{gj}$ is given by:
\begin{equation}\label{eq5}
	C_{gj}=\frac{8.63\times 10^{-6} \Upsilon_{gj}(T)}{\omega_g} T^{-1/2} \exp(-\frac{hc}{\lambda_{jg}kT}) \ ,
\end{equation}
where $\Upsilon_{gj}(T)$ is the thermally averaged collision strength (which is related to the electron excitation cross-section), $\omega_{g}$ is the statistical weight of the ground level, and a Maxwellian velocity distribution for the {colliding} electrons is assumed \citep{1997A&AS..125..149D}. 

{Inserting} (\ref{eq4}) and (\ref{eq5}) into  (\ref{eq2}), we obtain:
\begin{equation}\label{eq6}
	I(\lambda_{jg})= \frac{1}{4\pi A} \int_V{Ab~ G(T)~ n^2_\mathrm{e} \mathrm{d}V} \ ,
\end{equation}
where
the element abundance is assumed constant in the emitting volume. We notice that the line intensity is proportional to $n_\mathrm{e}^2$.
In (\ref{eq6}), we have introduced the function $G(T)$, called the \textit{contribution function}, which contains all the atomic physics parameters and is peaked in temperature (see Fig.~\ref{fig_gt} for an example):
\begin{figure}
	\center
 \includegraphics[width=\textwidth]{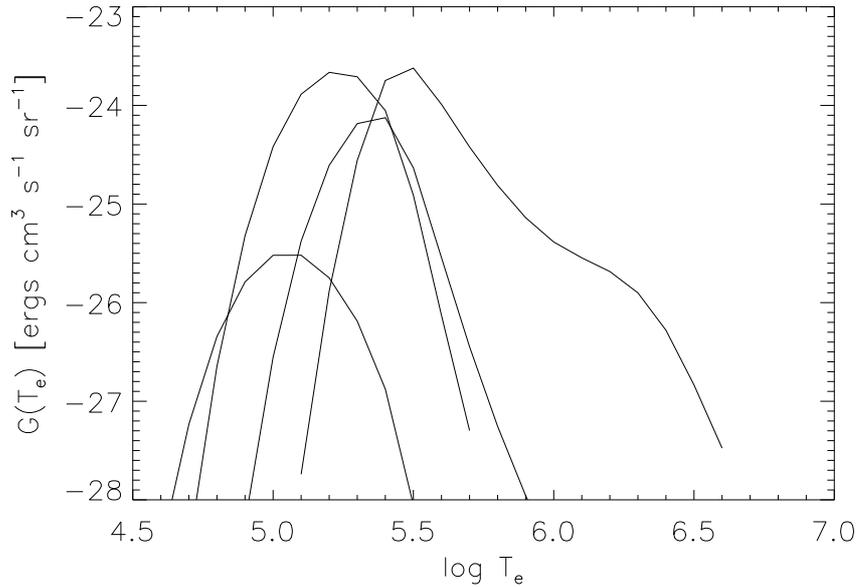}
\caption{Contribution functions for the lines belonging to the O~III-VI ions listed in Table~\ref{tab_ion} (CHIANTI v. 5.1)}\label{fig_gt}
\end{figure}
\begin{equation}
	G(T)= \frac{8.63\times 10^{-6} \Upsilon_{gj}(T)}{\omega_g} \frac{N_\mathrm{ion}}{N_\mathrm{el}} \frac{N_\mathrm{H}}{n_\mathrm{e}} T^{-1/2} \exp(-\frac{hc}{\lambda_{jg}kT}) \frac{hc}{\lambda_{jg}} \ .
\end{equation}

Equation~(\ref{eq6}) was derived under simple conditions. The coronal approximation does not always apply, so the statistical equilibrium equations should be solved  for all low-lying level  populations, including all the relevant excitation and de-excitation processes in a multi-level atom. For example, the proton collision excitation and de-excitation are important for those transitions where the energy between the levels involved is much smaller than $kT$.

In addition, 
the calculation of $C_{gj}$ is quite complex because it requires the knowledge of the electron excitation cross-section. The accuracy of this calculation depends mainly on the approximations used \citep[see][for a review]{1994A&ARv...6..123M}.

If the radiative decay transition probability is small enough, other mechanisms  may start to compete with it in de-exciting the level. This is the case for \textit{forbidden} or \textit{intersystem} transitions,  which de-excite the level by electron collisions (for electron densities large {enough to allow collisional de-excitations to play an important role}). The levels which are sensitive to this  are called \textit{metastables}, and their populations may become as important as that of the ground level. The intensities of their lines depend on the electron density, making them good candidates for electron density diagnostics (see Sec. \ref{sec_ne}).

The inversion {technique} aimed at deriving the plasma parameters from the spectral line intensities requires the knowledge of all the terms in (\ref{eq2}). This includes the ionization {degree of the element under consideration} $N_\mathrm{ion}/N_\mathrm{el}$ in (\ref{eq4}). This term is calculated by solving  a set of equations invoking the  ionization balance among the ionization stages from 0 to $z{+1}$, where $z$ is the stage of ionization of interest. For each {stage $k\le z$}, all the relevant ionization and recombination processes are taken into account:
\begin{equation}
N_k q_k = N_{(k+1)}\alpha_{(k+1)} \ ,
\end{equation}
where $q$ and $\alpha$ are the ionization and recombination rate coefficients, respectively.
 For an optically thin  plasma, the dominant processes are: collisional ionization, excitation followed by auto-ionization, and radiative and dielectronic recombinations \citep[see][for more details on these processes]{1992str..book.....M}.

 When dealing with spectroscopic diagnostic techniques to derive physical parameters of the emitting plasma, it is often necessary to compare the measured emission with that expected under assumed conditions. There exist various atomic databases and spectral codes that allow the calculation of the line intensity from (\ref{eq6}). Some are freely available, \eg\ CHIANTI\footnote{\url{http://www.chianti.rl.ac.uk/}}  \citep{1997A&AS..125..149D} and APEC/APED \citep{2001ApJ...556L..91S}.
In the  physics of the solar UV-EUV emission the most commonly used is CHIANTI.
This database, built by an international collaboration (USA/Italy/UK), contains atomic data (energy levels, wavelengths, radiative transition probabilities and excitation data) for a large number of ions. It  provides software for diagnostics analysis using the Interactive Data Language (IDL).  This code is kept up-to-date with  results of new atomic physics calculations {and laboratory measurements}.
CHIANTI also includes several ionization equilibrium calculations and elemental abundances. This is an important aspect that allows comparisons of results under different assumptions. The diagnostics analysis, in fact,  depends on these assumptions. 
Note that CHIANTI does not take into account the effect of velocity fields.
The investigator  should also be aware that uncertainties arising from atomic physics calculation are usually at least 10\%.

\subsection{Electron Temperature}
\label{pctrdiag:t}

Knowing the thermal structure of prominences is an essential element to understand their energy balance. It is  still not understood how a prominence can stay stable for days with a core plasma at about $8000$~K surrounded by the 1~MK corona. 
There exist several spectroscopic techniques to derive the electron temperature from the observed optically thin emission. Here we present those which are most often used in the analysis of prominence observations: the \textit{emission measure} and \textit{differential emission measure} techniques, which both involve the electron density at a given temperature.

\subsubsection{Emission Measure}\label{sec_em}

The Emission Measure (EM), first introduced by \cite{1961AZh....38...45I} and \cite{1963ApJ...137..945P}, is a quantity which yields the amount of plasma {emissivity} along the LOS. 
\cite{1963ApJ...137..945P} used a simple approach by noticing that for most lines, the contribution function $G(T)$ is a narrow, peaked function of temperature. This author assumed that $G(T)$ can be approximated by a constant function, equal to $0.7\ G(T_\mathrm{max})$, in an interval $\Delta T$ around the temperature $T_\mathrm{{max}}$ where the function reaches its maximum, and zero elsewhere. The temperature interval is such that $\Delta T/T_\mathrm{max} \sim 1-2$. Under this condition, the optically thin line intensity can be rewritten as:
\begin{equation}
	I(\lambda_{jg})= \frac{1}{4\pi} Ab~\langle \mathrm{EM} \rangle \langle G(T)\rangle \ ,
	\label{eq:em}
\end{equation}
where $\langle \mathrm{EM} \rangle$ is the \textit{column emission measure} of the plasma at a temperature $T$ in the interval $\Delta T$, {averaged along the LOS $h$}:
\begin{equation}
	\langle \mathrm{EM} \rangle= \int_h n_\mathrm{e}^2 \mathrm{d}h~~~\mathrm{[cm^{-5}]} \ .
\end{equation}
In (\ref{eq:em}), $\langle G(T)\rangle$ is the contribution function as defined by \cite{1963ApJ...137..945P}. The emission measure so defined can be estimated directly from the observed intensity of a single line. It should be remembered, however, that this is a crude estimation of the real emission measure which  takes into account the emission of the spectral line over the total temperature interval.

Emission measure analyses on prominences have been performed since  the time of Skylab. \cite{1979ApJ...232..929M} derived the EM under the assumption of \cite{1964SSRv....3..816P} and found a steeper decrease of EM with height for hot material ($T>10^5$) than for cooler areas. Further indications led the authors to conclude that their observations better fit a picture with cool threads having a thin TR (see Sect.~\ref{s:FillingFactor}). However, while the core material remains constant with height in the prominence, the TR of the thread gets thinner.

\subsubsection{Differential Emission Measure}\label{sec_dem}

When several lines formed at different temperatures are available, a more sophisticated technique may be used to derive the distribution in temperature of the plasma along the LOS, that is the \textit{differential  emission  measure}, or DEM:
\begin{equation}
	\mathrm{DEM}(T) = n_\mathrm{e}^2 \frac{\mathrm{d}h}{\mathrm{d}T} ~~~\mathrm{[cm^{-5}~K^{-1}]} \ .
\end{equation}
In this case the total line intensity can be expressed by:
\begin{equation}\label{eq_dem}
	I(\lambda_{jg})= \frac{1}{4\pi} Ab \int_{T} G(T) \mathrm{DEM}(T) \mathrm{d}T \ .
\end{equation}
The DEM contains information on the multi-temperature nature and the fine structure of the prominence along the LOS, and the dominant physical processes at work.
To derive the DEM, (\ref{eq_dem}) needs to be inverted using the observed line intensities, through the calculation of $G(T)$ and assuming a constant elemental abundance. 
The wider the temperature range of the lines, the more the inversion procedure is constrained.
However, this inversion is not an easy task and requires some assumptions concerning the solution. In addition, the solutions are sensitive to the atomic physics calculations and their uncertainties.
There exist different codes \citep[\eg ][]{1991AdSpR..11..281M,1997A&A...327.1230L}  available to solve the DEM problem. 
Note that one can find slightly different DEM definitions in the literature. For example, the DEM may be defined {for a} unit volume ($\sim n_\mathrm{e}^2 {\mathrm{d}V}/{\mathrm{d}T}$). This shows that some precaution is needed when comparing results from various sources. 

In the case of prominences, the DEM provides insight into the temperature gradient in the PCTR, which is one of the key parameters for the prominence modelling.
A strong temperature gradient in the PCTR was identified decades ago in Skylab data. Skylab results indicated a {lower} pressure and thinner TR for prominences than for the chromosphere-corona TR \citep[\eg][]{1990LNP...363..106V}.
The radiative losses in the PCTR seem to be too high to be compensated by either waves or conduction \citep[see \eg][]{2008A&A...480..537A}. However, these conclusions often rely on models in which, due to the lack of detailed information, important assumptions concerning the prominence fine structure and plasma conditions are made.

Deriving the DEM from observations is not a {trivial} task. Furthermore, given a DEM, there is not a unique way to obtain the density and temperature distributions as functions of the path length $h$. It should be noted that {the} emission from prominences spans the whole UV-EUV temperature range, from $10^4$ to $10^6$~K. The DEM calculation does not {provide any information on} the filamentary origin of the prominence emission, so that a comparison with expected DEMs from different geometric models is needed.
Based on observational evidence, there are two main models that could satisfy the observed thermal properties and relate them to the unresolved fine structure {(see Fig.~\ref{f:Pojoga})}. The first model considers threads having a cold core and a thin TR interfacing with the coronal environment \citep[\eg][]{1991SoPh..132...81C}. The observed EUV emission originates mostly from the thread's TR, and the DEM gives a direct information on the gradient of PCTR of the threads. The second model pictures prominences as made of isothermal threads, each of them having a different temperature in the observed UV-EUV range \citep[\eg][]{1994scs..conf..357P}. In this case, most of the EUV emission comes from the threads, and the DEM relates to the amount of threads at a given temperature.

\begin{figure}
	\center
  \includegraphics[height=4.6cm]{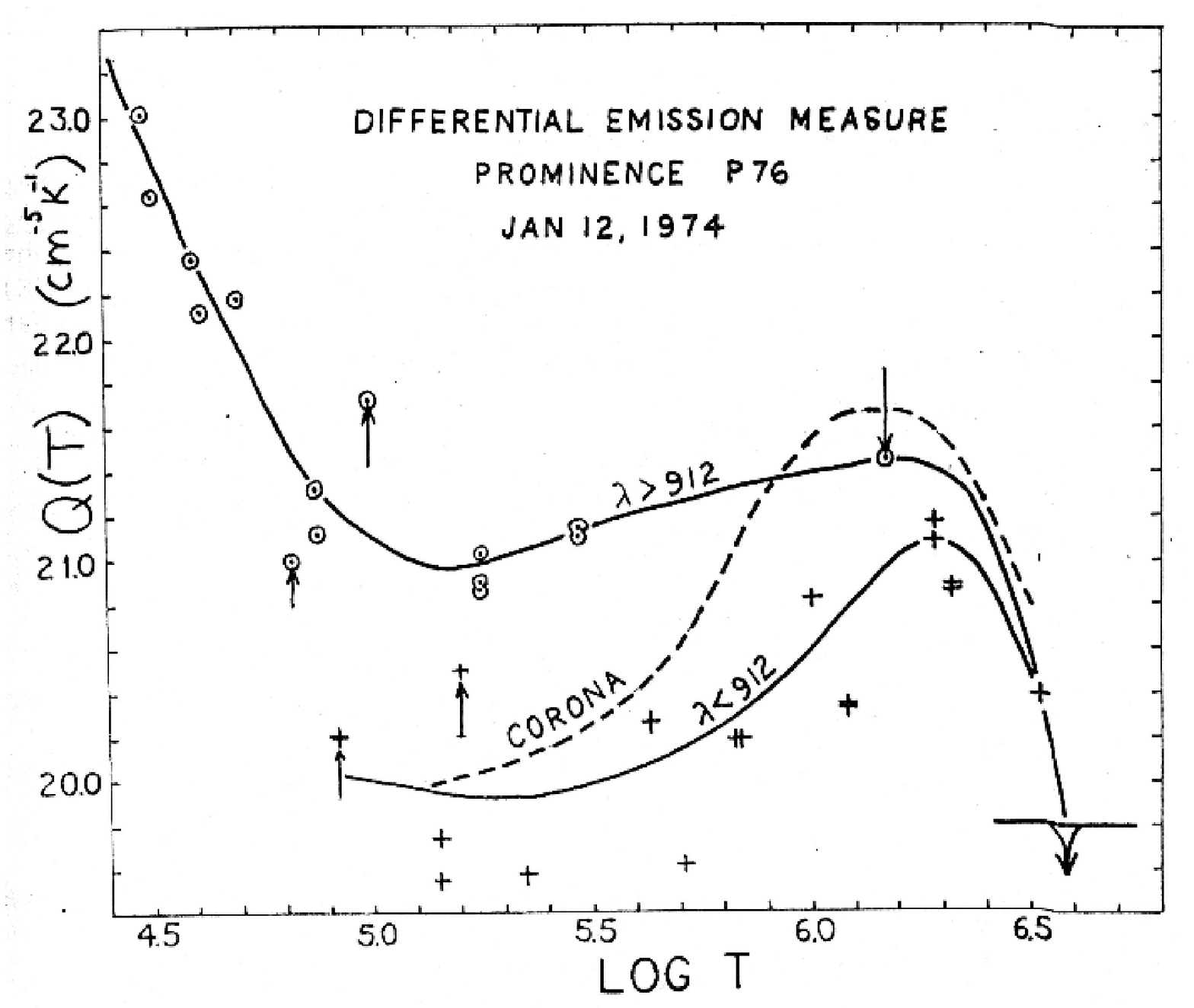}
 \includegraphics[width=4.7cm]{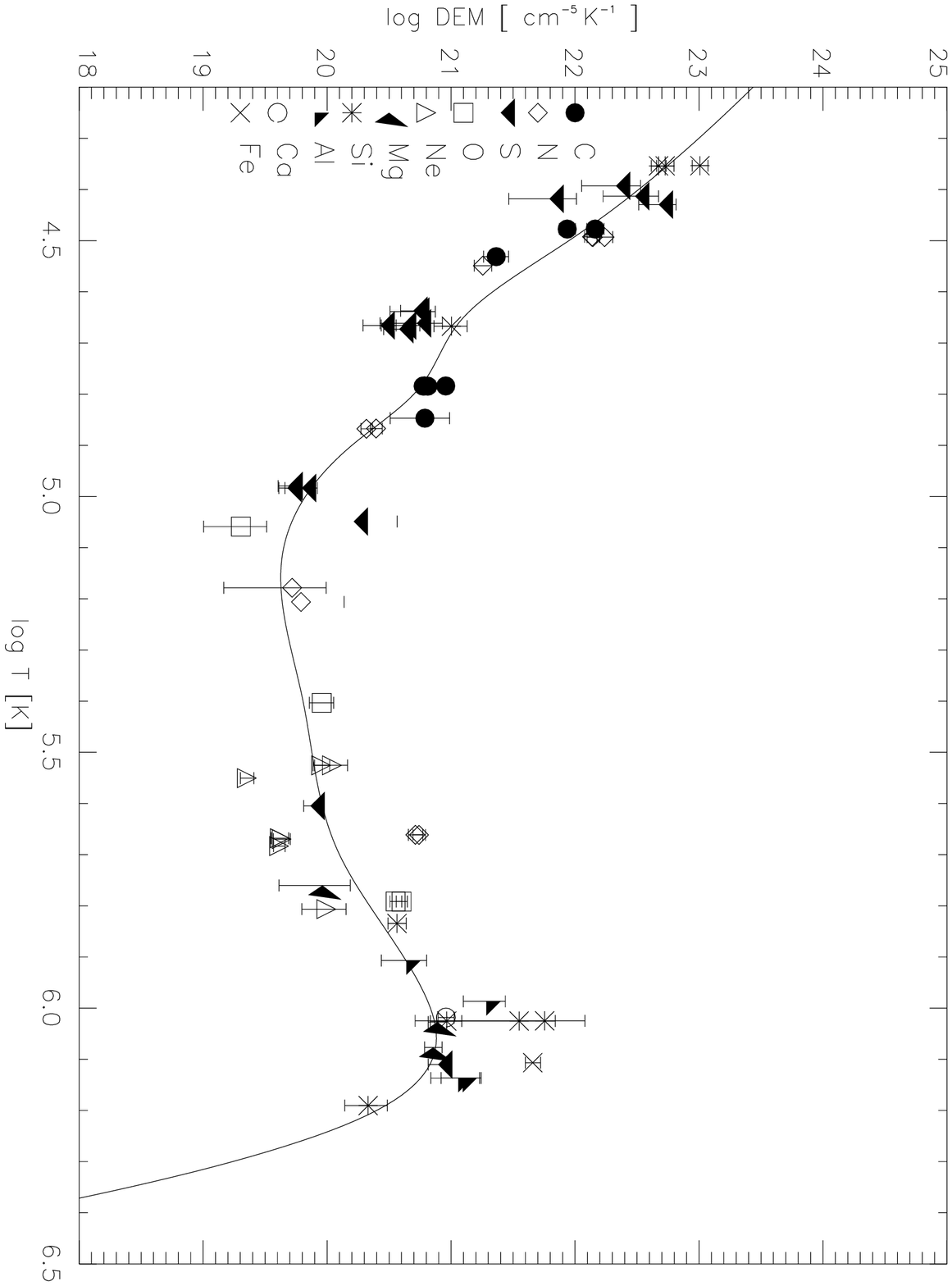}
\caption{DEMs for a quiescent prominence from \cite{1986NASCP2442..127S} (left) and  \cite{2007A&A...469.1109P} (right). }
\label{fig_dem}  
\end{figure}

\paragraph{Quiescent prominences.}
The DEM derived from \cite{1986NASCP2442..127S} was obtained including the effect of H Lyman continuum absorption (Fig.~\ref{fig_dem}). They tested three geometric models (threads with hot sheaths and cool core; isothermal threads; threads with temperature gradient along the longitudinal magnetic field) and did not  arrive at a conclusive result.
\cite{1988sscd.conf..151E} showed a steeper gradient of the DEM below $10^5$~K than at higher temperatures. He concluded that extra, unknown, heating mechanisms should play a role, in addition to thermal conduction, in order to balance the observed radiative losses. 

It is important to constrain the DEM calculation with as many lines as possible, distributed over a wide temperature interval. Since the early measurements mentioned above, the development of multi-wavelength instruments at high resolution has allowed us to obtain more reliable  results. 

Examples of these results are those of \cite{1993A&A...273..267W}, who used lines from the rocket instrument High Resolution Telescope and Spectrograph {\citep[HRTS,][]{1975JOSA...65...13B}} over the range $4.3< \log T <  6.1$; \cite{2004SoPh..223...95C} who combined SUMER and CDS data to better constrain the results; and \citet[][see our Fig.~\ref{fig_dem}]{2007A&A...469.1109P} who used more than 60 SUMER lines (lying above the H Lyman continuum head) from their prominence spectral atlas \citep{2004SoPh..220...61P,2005A&A...443..679P}.
These results show that the DEMs in prominences are very similar to those of the quiet Sun. The main differences are to be found in the amplitude of the gradients at low and high temperatures and on the position of the minimum between them.

The minimum temperature at which the DEM can be derived is limited by the optical thickness of the plasma, because the total intensity of the line is modified. 
The DEM is correctly derived for optically thin emission, while lines at low stages of ionization or neutrals may be affected by opacity and by the hypothesis behind the DEM derivation {\citep{2007ASPC..368...81A}}. Even if the effect is less strong than in the quiet Sun, some absorption is still present \citep{1986NASCP2442..127S}. For example, \cite{2007A&A...469.1109P} put a limit to about 4.2 in $\log T$.

For temperatures smaller than $\log T \approx 5-5.5$, the DEM decreases roughly as $T^{-2}-T^{-3}$ \citep[see][for discussion]{2006ApJ...645.1525K}. \cite{1988sscd.conf..151E, 2004SoPh..223...95C}; and \cite{2007A&A...469.1109P} found the minimum of the DEM between $\log T = 5.1-5.2$, which is a smaller temperature than what was found for the quiet Sun  and for activated prominences  \citep[$\log T=5.4-5.6$,][]{2006ApJ...645.1525K,2007A&A...469.1109P,2008ApJ...673..611K}.

Discrepancies among the results stimulate further investigation. Beside the uncertainties arising from the different methods used, the lines selected and the assumed theoretical atomic physics, such differences may have a physical origin.
It may be that the differences are linked to the different levels of activity in the prominence. Also, \cite{1991SoPh..132...81C} showed that the DEM is proportional to $\cos\theta$ (the angle between the  direction of the magnetic field and the gradient of temperature, which enters in the thermal conduction term in the energy equation), so that the derived thickness of the {PCTR} changes with the angle of view of the observed prominence. 
{The modelling of \cite{2005A&A...442..331H} is consistent with this picture.}

Another aspect worth mentioning is that the derived DEM generally includes that of the background and foreground emissions. {Studies} addressing this are mentioned in the next section.

\paragraph{Activated and erupting prominences}
We report on a few examples of temperature measurements in activated and erupting prominences. 
The intensification of activity in  prominences involves higher velocity fields and temperatures.
\cite{2006ApJ...645.1525K} studied a jet in an activated prominence on  17 April 2003 using TRACE \citep{1999SoPh..187..229H}, SUMER, and \Ha\ from the Kanzelh\"ohe Solar Observatory. Their  DEM analysis on different parts of the jet shows curves with similar profiles below $\log T =5.4$ (the local minimum), but with varying gradients and coronal components for higher $T$. The low temperature part of the DEMs is following a power law function with index equal to $-3$, although this may be artificially steepened by a gap in the temperature coverage in the DEM. {This value is steeper than that found for quiescent prominences and active regions.}
These DEMs are one of the few existing in the literature where a background emission has been removed from the data. This is a very important point (see the authors' discussion). It is difficult, generally, to have  good data for the instrumental stray light and real background emission. The authors are more uncertain on the background subtraction at coronal temperatures than in the low TR part. 

\cite{2008ApJ...673..611K} also studied an erupting prominence with SUMER. Applying again the background subtraction, they calculated the DEM in various parts of the prominence. 
They found the TR DEM to be similar to that of the non-erupting prominence, also with a minimum at $\log T=5.4$. In addition, they found significant hydrogen Lyman continuum absorption (at wavelengths below 911~\AA) which varies inside the structure. In particular, the effect  was less important
near the base of the prominence and in a kink region, which they estimated to be areas of mild heating. 

Interesting results come from attempts to measure erupted filaments in the high corona obtained with UVCS data. Applying the EM technique,  \cite{2000ApJ...529..575C}
 found a flat EM distribution in the range $4.6<\log T <5.5$ at 1.7 $R_\odot$. In one case \cite{2003ApJ...597.1118C} found temperatures up to $1.6 \times 10^6 \mathrm{K}$ in the filament upper edge, through detecting Si~XII and Mg~X emission. This could be the result of prominence plasma being heated and ionized as it merges with the ambient corona.

\subsection{Electron Density and Gas Pressure}

Here we focus on relatively direct measurements of density and pressure in the PCTR. More extensive reviews of {previous} prominence density measurements can be found in reviews by \citet{1990LNP...363..187H} and \citet{1995nsp..book.....T}.

\subsubsection{Electron Density from Line Ratios}
\label{sec_ne}

Lines formed through allowed transitions, such as those we have described in the previous section, have an intensity which depends on the electron density squared (\ref{eq6}).  

\begin{figure}
	\center
 \includegraphics[width=\textwidth]{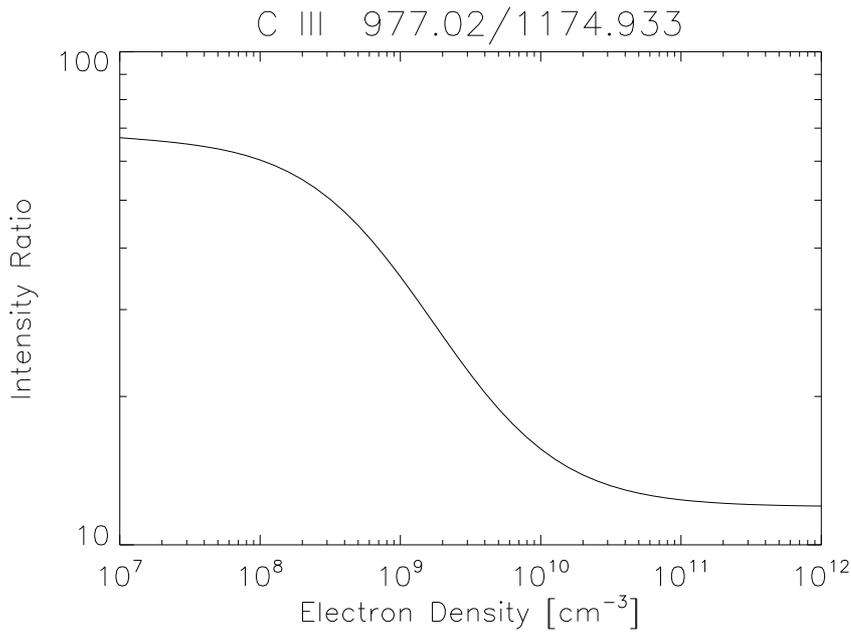}
\caption{The C III density-sensitive ratio (CHIANTI v. 5.1).}\label{fig_c3}
\end{figure}

In forbidden and intersystem transitions, the upper level is a metastable ($m$) and its radiative decay rate is very low ($1-100~\mathrm{s}^{-1}$). {This means that if the electron density is high enough, the collisional de-excitation can compete with spontaneous emission in depopulating the level. This happens at different densities, depending on the actual lines used in the analysis (see an example in Fig.~\ref{fig_c3}). Note that a high rate of collisions could also result in a level population of $m$ comparable to that of the ground level. } 

The properties of the metastable level populations can be used to diagnose electron density.
Let us consider the general case of two lines from the same ion formed through the transitions $j \rightarrow g$ and $m \rightarrow k$ .
Assuming the same emitting volume, from (\ref{eq2}) their ratio may be written as:
\begin{equation}\label{eq14}
\frac{I(\lambda_{jg})}{I(\lambda_{mk})}= \frac{n_g}{n_m} \frac{C_{gj}}{C_{mk}} \frac{\lambda_{mk}}{\lambda_{jg}} \ .
\end{equation}

At low electron densities, only lines arising from the levels excited from the ground state $g$ are visible. On the contrary, at higher electron densities, the lines originating from the de-population of the metastable level will be also visible. 
Looking at the term $n_g/N_\mathrm{ion}$ in (\ref{eq4}), we see that for allowed transitions $n_g \propto n_\mathrm{e}$, while $n_{m} \propto F(T,n_\mathrm{e})$, because of the density dependence of the population of this level. This means that the intensity ratio of (\ref{eq14}) is a function of density:

\begin{equation}
	\frac{I(\lambda_{jg})}{I(\lambda_{mk})} \propto \frac{n_\mathrm{e}}{F(T,n_\mathrm{e})} \ .
\end{equation}

This variation in density  dependence  of lines formed from the same ion  becomes a tool
to estimate the averaged electron density of the emitting region: the ratio of  line {intensities} from an allowed and forbidden transition is a function only of density (assuming a constant temperature in the emitting volume). 
It should be pointed out, however, that this ratio is an averaged value based on the emitting plasma along the LOS. 

{Since 1995, electron density values have been calculated using SUMER. They have an advantage over some earlier measurements, in that the ratios consist of intensities of lines from a common ion, reducing possible errors related to abundances and temperature variations.}
In prominence studies, for example, a commonly used line ratio is the one between the 977 \AA\ and the 1174.9 \AA\ lines produced by C~III (the line formation temperature is $6.3 \times 10^4$~K). The dependence of the theoretical line ratio on the electron density is shown in Fig.~\ref{fig_c3}. 
The comparison of the measured ratio with the values shown in Fig.~\ref{fig_c3} gives a unique solution for the electron density in the density range $10^8 - 10^{11}~\mathrm{cm^{-3}}$.

\subsubsection{{Results of Electron Density Determinations}}

As is clear below, measured densities  vary greatly. {Some of these variations are likely due to differences between the various techniques, but there are also true variations between prominences, and within prominences themselves.}

Space-based measurements of $n_\mathrm{e}$ have been done using EUV observations. As a result, they are chiefly focused on emission from lines in the hotter portions of the prominence, \ie\ the PCTR. Thus it would not be surprising if the density values were somewhat different than those measured using emission from \Ha. The chief method has been to use line ratios, which have the advantage that they are independent of {the} filling factor, although they are based on a number of assumptions which are described above. One such assumption is that, as pointed out by \cite{1993A&A...273..267W}, the calculations assume ionization balance, which may not be correct in activated or erupting prominences.

{The electron density can also be determined via an emission measure method.} 
\citet{2003ApJ...584.1084C} derived density values from TRACE observations of a jet  in an active region prominence. The emission measure was combined with different models of the source volume incorporating a number of threads with Gaussian density profiles about their central axes. With a Gaussian e-folding width of {1000~km} and a slab of depth {3450~km}, the on-axis electron density was about $10^{10}$~\cc.

Electron density values for a number of space-based observations are listed in Table~\ref{t:DensPress}, along with the temperatures of formation of the lines used for the diagnostics. These temperatures are taken from the original papers, and different assumptions based on different ionization equilibrium calculations result in different values.
\begin{table}
\begin{minipage}{\textwidth}
\caption{{Electron} density and {gas} pressure measurements from space-based EUV observations. The temperatures shown in the table correspond to the formation temperatures of the lines used for the diagnostics in the corresponding papers. References: 1 -- \citet{1972ApJ...178..515N}; 2 -- \citet{1976SoPh...50..365O}; 3 -- \citet{1979SoPh...61..319M}; 4 -- \citet{1979ApJ...232..929M}; 5 -- \citet{1983SoPh...84...63P}; 6 -- \citet{1986ApJ...308..982W}; 7 -- \citet{1993A&A...273..267W}; 8 -- \citet{1997SoPh..175..411W}; 9 -- \citet{1999ESASP.446..467M}; 10 -- \citet{2003ApJ...584.1084C}; 11 -- \citet{2004SoPh..223...95C}; 12 -- \citet{2006ApJ...645.1525K}; 13 -- \citet{2007A&A...469.1109P}. Pressure values marked with * are based on $n_\mathrm{e}T$ values quoted in the paper assuming $p=2n_\mathrm{e}kT$, with $k$ {being} the Boltzmann constant.}
\label{t:DensPress}
\begin{tabular}{p{1.44cm}p{1.59cm}p{1.44cm}p{1.30cm}p{1.25cm}p{1.60cm}c}
\hline\noalign{\smallskip}
Instrument & Ions & Prominence & $n_\mathrm{e}$          & $T$  & $p$                      & Ref.\\
           &      &          & ($10^9$~\cc)    & (K)  & (dyn~cm$^{-2}$)          &        \\
\noalign{\smallskip}\hline\noalign{\smallskip}
OSO-4 and OSO-6 & C III & quiescent & 1 & 80000 & 0.01 & 1\\
\\
ATM & C III & 9 quiescent & $0.7-2.7$ & 90000 & $0.17-0.59$*  & 2\\
\\
ATM & multi-species line ratios using & 3 quiescent & & $56000 - 67000$ & $0.04-0.08$ (quieter) & 3\\
& Si~III, C~III, Si~IV, O~III &&&& $0.07-0.22$ (activated)\\
\\
ATM  & C III/S IV & quiescent& $9.4-40$ &&& 4\\
 & C III / EM &                    &$8.4-29$    &&&\\
\\
SMM/UVSP      & Si IV/O IV   & quiescent      &  500, 300 and below & 100000; 180000 && 5\\
\\
ATM & O IV, Ne V, Mg VII   & erupting       & $8-300$ &&& 6\\
\\
HRTS & multi-species line ratios using & activated & 100 & 100000 & $0.28 - 2.8^*$ & 7\\
& O~IV, N~IV, Si~IV, N~V& quiescent &&&\\
\\
SUMER         & O IV line ratio & erupting       &  $3 - 300$ &&& 8\\
\\
SUMER         & O IV              & quiescent       & $1.3-14$    &  $170000-200000$ &  $0.03-0.31$ & 9\\
\\
TRACE         & EM + model      & active region & $7-19$   & 250000  && 10\\
\\
SUMER & C III & quiescent & 2 & 70000 & 0.04 (0.02 edge, 0.06 center) & 11\\
\\
SUMER  & O V & activation in plage & 3 & 250000 & & 12\\
\\
SUMER & C III & quiescent & $0.6,3.6$ & 70000 & $0.012,0.07$ & 13\\
\noalign{\smallskip}\hline
\end{tabular}
\end{minipage}
\end{table}
Like the values derived from optical data, the values from EUV line
ratios vary over a significant range.  Most, but not all,  values derived from EUV line ratios are
in the range $10^9-10^{11}$~\cc\ {(the lowest values being mostly related to quiescent prominences)}. The density values also vary over the prominence. In quiescent prominences, \citet{2004SoPh..223...95C} measured values from $1\times10^9$ to $3.2\times10^9$~\cc, and \citet{2007A&A...469.1109P} from $5.75\times10^8$ to $3.6\times10^9$~\cc.
\citet{1997SoPh..175..411W} measured density values from $3\times10^9$
to $3\times10^{11}$~\cc\ in an erupting prominence.
{This large range of values partly reflects the invalidity of most hypotheses made in the respective derivations (techniques of line ratios, EM, etc).}

\subsubsection{Gas Pressure}\label{s:PressurePCTR}

Pressure values for the PCTR are based on the density sensitive line ratios discussed above, and are also listed in Table~\ref{t:DensPress}.  

More recently, measurements have varied to a greater extent.  As with the density measurements, these include variations within the prominences.
For instance,  \citet{2004SoPh..223...95C} found pressure values of 0.02 at the edge (0.06~dyn cm$^{-2}$ at prominence centre), and \citet{1999ESASP.446..467M} found values in the range $0.03-0.31$~dyn cm$^{-2}$. 
{The aim of future observations will be to reduce the one-order-of-magnitude range of values within each type of prominence.}

\subsection{Small-Scale Prominence Motions and Non-Thermal Velocities}

Spectroscopic measurements of line position (Doppler shift) and width (Doppler width) are commonly used to study prominence motions. 
The PCTR lines are generally {Gaussian, and are} narrower than those of the solar TR; however, excess  broadening has also been observed \citep{1995nsp..book.....T}. The method to derive thermal and non-thermal broadening {is} described in Sect.~\ref{s:Temperature}.

For quiescent plasma, additional broadening may have different origins: the ion and electron temperatures may not be in equilibrium, the ionization equilibrium may fail, and absorption effects may be present \citep{1992str..book.....M}. 
For a quiescent prominence, it is generally assumed that the first two effects do not occur, or can be neglected, mostly because of a high enough rate of collisions. The effect of plasma absorption may be overcome by a careful analysis of the data, with the identification of modified line profiles and intensities \citep[\eg][]{1998ApJ...505..957C, 2007A&A...469.1109P}. 
Once those lines are excluded, measurements of widths larger than the thermal ones give information on the unresolved, non-thermal motions inside the prominences. 
However, careful attention should be given to the possible under-estimation of the instrumental profiles. 

A correlation of NTVs with temperature on the quiet Sun has been observed by several authors
\citep[\eg][]{1998ApJ...505..957C, 2007A&A...469.1109P}. A similar (but not identical) behaviour is also observed in prominences. The line width seems to increase, increasing the total brightness of the line \citep{1993A&A...273..267W, 1998A&A...334..280D, 1999SoPh..187..405W}, even though this is not always the case \citep{1998A&A...334..280D}.
All these observational aspects imply that the non-thermal widths may have an origin in the unresolved fine structure of the prominence (each thread having different plasma conditions and dynamics), and/or the presence of waves {(Paper~II, Sect.~4)} and micro-turbulence.
The dynamics of a prominence, including the non-thermal motion component, increases in activated and erupting prominences. 

\begin{figure}
	\center
  \includegraphics[width=0.49\textwidth]{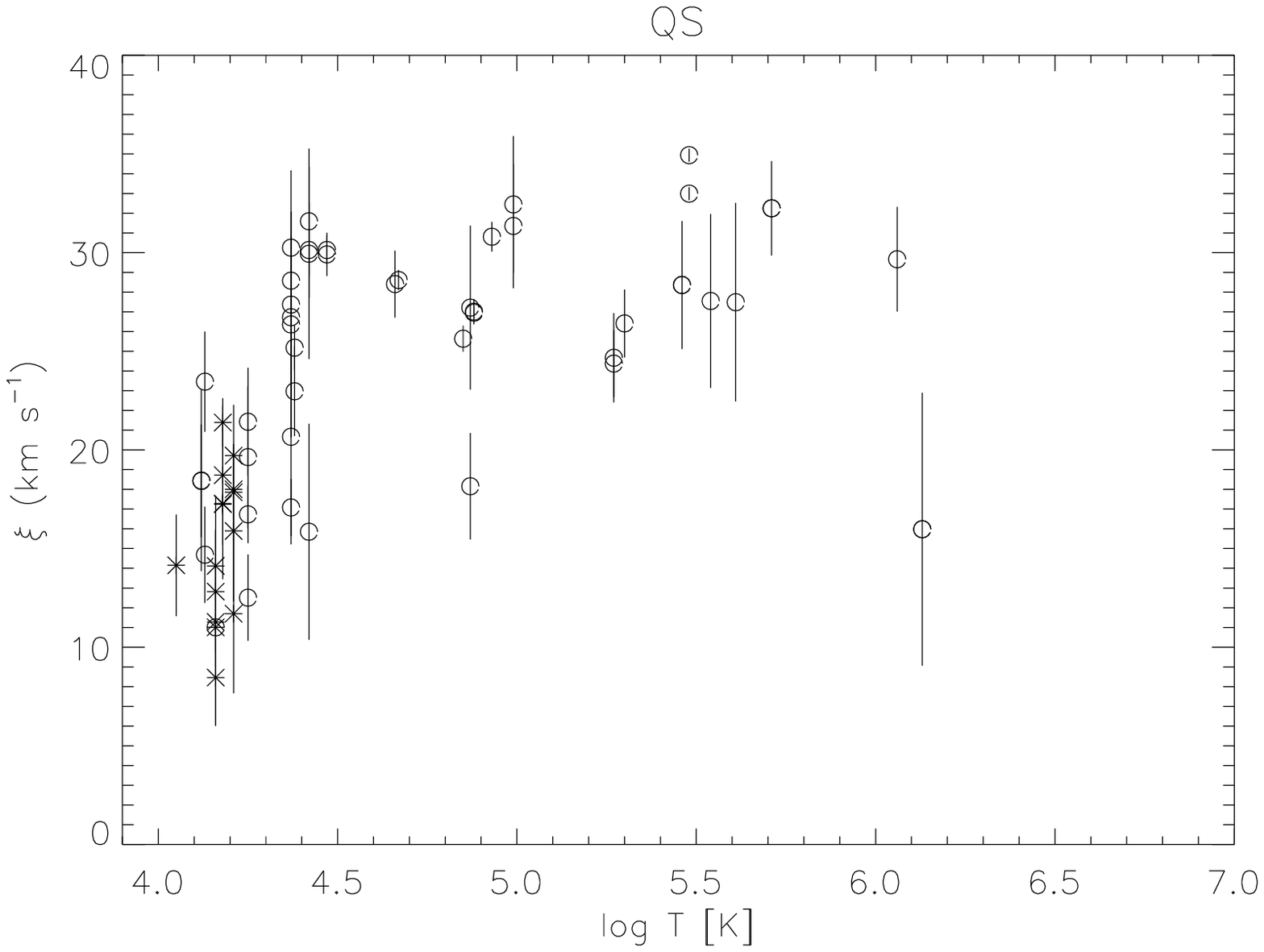}
 \includegraphics[width=0.49\textwidth]{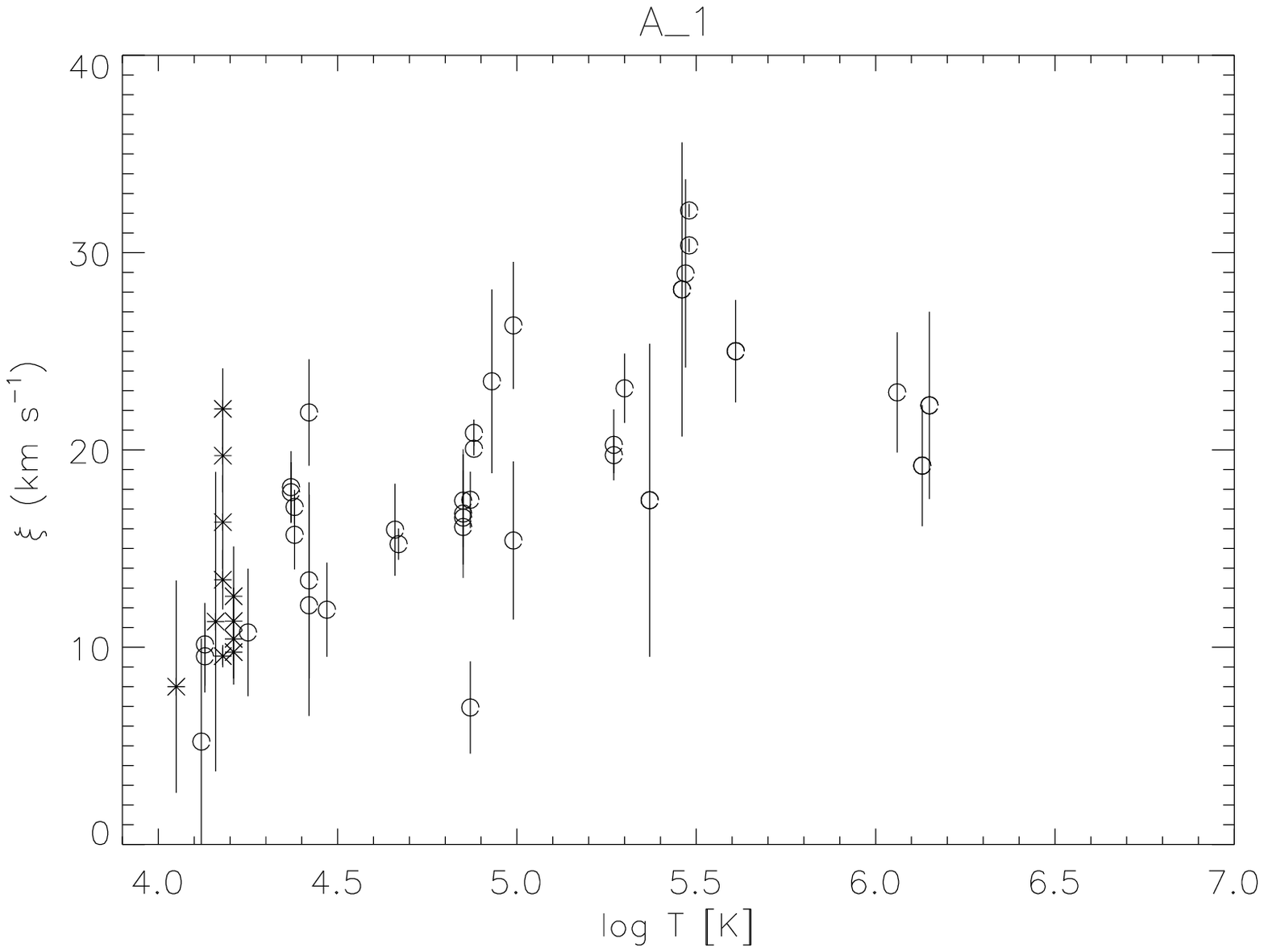}
\caption{Quiet Sun (left) and quiescent prominence NTV (right) as function of temperature \citep{2007A&A...469.1109P}.}
\label{qs_pr_sp}
\end{figure}

Figure~\ref{qs_pr_sp} shows the NTV as function of temperature measured by \cite{2007A&A...469.1109P} using SUMER data for the quiet Sun (left) and a quiescent prominence (right) in 1999. 
These plots show a similar correlation of the NTV with temperature:  an increase of velocity from chromospheric to TR temperatures, and a decrease towards coronal values. In both cases, the maximum NTV is found at $\log T=5.5$.
The main differences between the two distributions are the smaller velocities in the prominence, and a different gradient in temperature.
These two aspects have been reported by several authors. However, the amplitude of the gradient difference and the absolute values of the velocities differ \citep{1993A&A...273..267W, 2004SoPh..223...95C}.

Somewhat different behaviours of the NTV-$T$ distribution can be found in the literature. For example, \cite{2003SoPh..217..133S} report the variation with temperature of the UV line widths in the form $\Delta\lambda /\lambda$, without separating the thermal and non-thermal components. They noticed a discontinuity in this distribution at around $\log T =4.8$ with a sudden increase of the velocity. They attributed this to the different structures of the ``cooler'' and ``hotter'' plasma.  However, the impossibility of resolving the prominence finest structure leaves ambiguity in this kind of interpretation.  These authors also studied the widths of hydrogen Lyman lines. They estimated an NTV $> 30$~\kmps. However, these lines are optically thick, and this result should be considered with some precaution.

\begin{figure}
	\center
 \includegraphics[width=\textwidth]{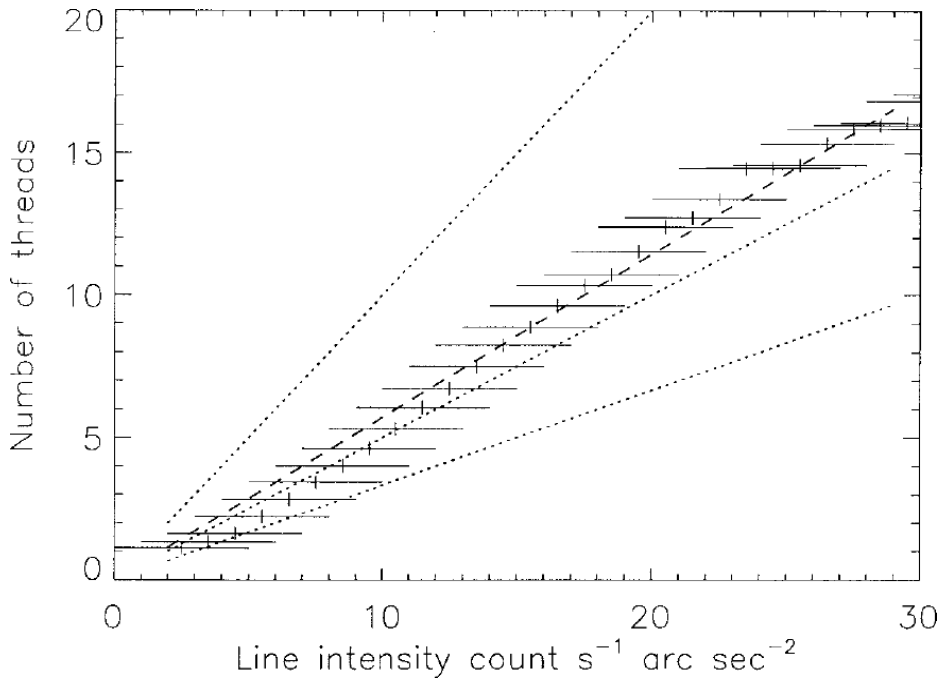}
\caption{The number of threads in the LOS (crosses) as a function of the line intensity. From \cite{1999SoPh..187..405W}}
\label{fig_wiik99}
\end{figure}

It is clear that high-resolution measurements of the  unresolved motions, as well as Doppler shifts, may bring important information on the prominence fine structures and their dynamics. Using SUMER, \cite{1999SoPh..187..405W} performed a statistical analysis of the moments of the N~V 1238~\AA\ line ($\log T = 5.3$) in a prominence. They then used the results from the multi-thread model of \cite{1993A&A...273..267W} to infer the properties of the structure's fine geometry. They estimated the  NTVs as a function of the line intensities, and found an averaged value of 23~\kmps. This value, together with the measured line shift, was used to infer the number of threads along the LOS (Fig.~\ref{fig_wiik99}). 
Their results were consistent with the multi-thread model, where the intensity along the LOS is proportional to the number of threads, and the velocity inside the thread has a Gaussian distribution given by the value found for the averaged NTV.

A variation of the NTV within the prominence (central region versus periphery) has been noticed in some cases \citep{1979ApJ...232..929M,1998A&A...334..280D,2003SoPh..217..133S}, but not in others \citep{1980SoPh...68..187V}. This piece of information may bring insight on the heating deposition in prominences.
The velocities seem to increase towards the peripheral regions and to be more marked for hotter lines.

\subsection{Small-Scale Structure}
\label{s:FillingFactor}

It is clear from numerous observations that prominences are made up of many fine structures, very likely smaller than the best resolution currently achievable ($\approx150$~km). In addition to images showing that prominences are made up of numerous small-scale structures, measurements of the filling factor in prominences indicate that these small-scale structures inhabit only a small percentage of the prominence volume, while efforts to model small-scale prominence structures indicate that there are numerous threads along the LOS. 

\begin{figure}
	\center
\includegraphics[width=\textwidth]{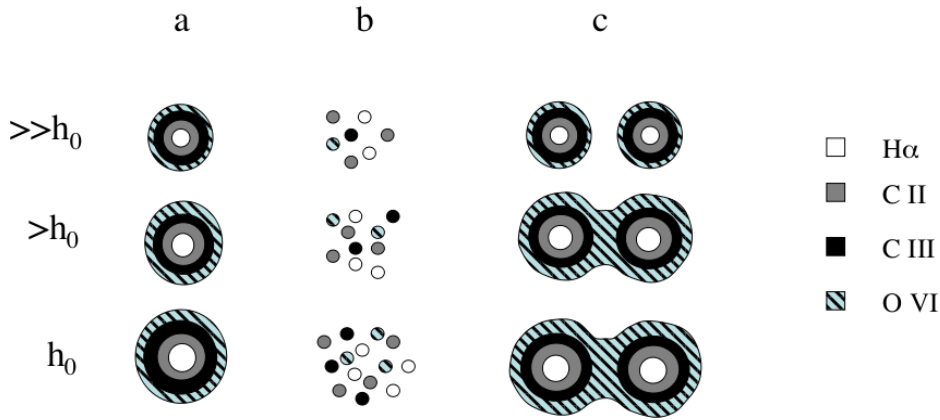}
\caption{Possible models of  the PCTR. a) Each cold core ($\sim 8000$~K) is surrounded by a PCTR. b) Individual threads at different temperatures. c) Cold cores in a larger enveloping PCTR. Here the properties vary with height ($h_0$). The length of the threads  is not specified -- in some cases the prominence is modelled as a series of short knots {or} blobs. {Adapted} from \citet{1994scs..conf..357P}.
\label{f:Pojoga}}
\end{figure}

In addition to being relevant to the understanding of the structure of prominences in general, this question also relates to the question of the structure of the PCTR. As depicted in Fig.~\ref{f:Pojoga}, there {are} a number of models describing the possible spatial relationships between cold and TR temperature prominence material.

Filling factors can be formulated in various ways. Here we will discuss measurements of the volume filling factor which measures the fraction of the volume filled by material, $f_V$. Another filling factor discussed in the literature is the filling factor perpendicular to the LOS, $f_s$, which describes the degree to which the entrance slit of the instrument is filled by the fine structure elements \citep[see][]{1996SoPh..164..211H}. This factor depends on the instrumental resolution as well as the physical properties of the object.
The filling factor  calculations discussed here assume that the prominence consists of either material of uniform density or of empty space, which is, of course, a simplification {which could affect spectral diagnostics} \citep[see][]{2000ApJ...531..585J}.

For the PCTR, the basic method for measuring the filling factor is to divide the {volume} emission measure $\mathrm{EM}= \int_Vn_\mathrm{e}^2\mathrm{d}V$ derived from optically thin emission, by the square of the density derived from a line ratio diagnostic  (see Sect.~\ref{sec_em}). This method has been used, for instance, by \citet{1979ApJ...232..929M}, who calculated {volume} filling factors of $0.018-0.024$, and in an erupting prominence by  \citet{1986ApJ...308..982W} who found  a {volume} filling factor of $\le0.023$.

Lyman continuum absorption measurements (see Sect.~\ref{sec:EUVAbsorptionMass}) have been used to calculate the filling factor of neutral hydrogen and helium in the prominence.
\citet{1998SoPh..183..107K} used data from CDS to calculate $f_s\ga 0.3$. Using an estimate of the prominence dimension along the LOS and standard values for the density, this yields a volume filling factor $f_V$ of  $0.001-0.1$. 
\citet{2001SoPh..199..115C} used observations of a filament by CDS and common estimates for the prominence density to   find a volume filling factor of cold material of $\la 0.2$.

Given that the fraction of space taken up by prominence material is relatively small, how is this material structured? Clearly the structures must be on a small-scale.  \cite{1998A&A...334..280D} cite a lack of small-scale (2") spatial coherence in lines formed at temperatures from $10^4-10^5$~K as evidence of multi-thermal threads on that scale or smaller.

A number of researchers have approached this problem by modelling the distributions of intensity, shift, and width of spectral lines. Using such an analysis on HRTS EUV data, \citet{1993A&A...273..267W}  calculated that there were 15 threads contributing to the emission from each pixel at 20000~K and about  30 threads at TR temperatures. \cite{1999SoPh..187..405W} found a similar result of $15-20$ threads for N V emission observed with SUMER. Both works note spatial scales on the order of a few 100~km or less.   

\citet{2004SoPh..223...95C} based a multi-thread model on multi-line measurements of intensity and Doppler velocity. They find that the number of threads increases with temperature, with two to eight 50~km threads needed at $\log T \approx 4.8-5.6$, and  35 threads at $\log T\approx 5.8$.  Over the entire prominence, this would mean about  20 low temperature and 800 high temperature threads.
These thread estimates are consistent with line moment analyses using \Ha\ data performed by 
\citet{1991SoPh..136..317M} and \cite{1991SoPh..131..107Z} who found $\approx10$ threads in a resolution element.
There are non-LTE models  which also use multiple threads to model prominence hydrogen emission (see Sects.~\ref{sec:9} and \ref{sec:10}). \citet{1991SoPh..136..317M} note, however, that there seems to be a discrepancy between velocity-based and density-based thread models, with the density-based models indicating that there are many more threads. They suggest that the velocity threads may in fact be bundles of smaller density structures.

\citet{2001SoPh..199..115C} studied a filament using both EUV (including Lyman {continuum} absorbing coronal lines, as mentioned above) and radio data. They found that the data support a model in which cool threads surrounded by TRs are suspended in the hot coronal gas.

\citet{2008ApJ...673..611K} compared times and locations in an erupting prominence  in which there was substantial Lyman {continuum} absorption and in which the absorption has disappeared. The change in TR emission between  the two regions was about a factor of five. This would require at least 2 or 3 alternating layers of 
cool and TR temperature material along the LOS. 

Although the concepts of filling factor and characteristics of small-scale structures are connected, the details of how are still not clear.

\section{Bulk Motions}
\label{sec:5}

Prominences have long been known to be dynamic structures, displaying internal
motions of various kinds even when globally at rest. A number of
reviews have been written on this topic, \eg\ \citet{1988dsqs.work...15S} and \cite{1995nsp..book.....T}. In this section, we discuss internal
bulk motions of non-erupting prominences. Oscillatory motions are
described and discussed in Paper II, Sect.~4.

A good understanding of flows (their trajectories, velocities, and thermal properties) is highly important as a test for models of prominence formation and stability, {as they may help us to distinguish between different mechanisms of mass supply, and for the closely related issue of the prominence magnetic field structure.  These are discussed in more detail in Paper~II (Sects.~2 and 3).} For this, we need to know the detailed trajectories of prominence plasma blobs, including the plasma origin, and any change in temperature and velocity.  How do their motions compare to various magnetic field models -- especially in the prominence barbs {(Paper~II, Sect.~2.1.3)}? What connections, if any, are there between flows at different temperatures?

In the last decade, new instrumentation has yielded more information concerning flows. In the visible, the combination of high temporal and
spatial resolutions, along with \Ha\ spectral information, has allowed new investigations: counter-streaming flows have been seen in filaments on the disk, using high-resolution instruments which can provide images with a spatial resolution of $\approx 0.2\arcsec$, and a time resolution of $\approx1$~minute or higher, along with some Doppler information. 
The SOT instrument on board Hinode allows for equally good temporal and spatial resolution with excellent image stability over long periods of time, offering another probe of prominence motions in \Ha\ and Ca~II H {(3968~\AA)} lines.

In the EUV range, progress has been made using spectrographs (SUMER, CDS{, and EIS}) which can give information in a range of lines formed at chromospheric, TR, and coronal temperatures. Higher cadence UV and EUV imaging information from TRACE and, at times, from EIT, has been important as well.

\subsection{Measurements of Velocities}

Measurements of velocities in filaments and prominences are not trivial. There are two main methods that have been used to infer motions. Motions in the plane of the sky are measured by tracking actual features, while LOS motions are detected using Doppler shifts {or, in some cases, line profile distortions}.  Sometimes the plane-of-sky and line position methods can be combined to good effect, as with the ground-based \Ha\ observations {reported by, \eg} \citet{1998Natur.396..440Z,2003SoPh..216..109L}; and \citet{2006SoPh..234..115C}, but this can only be done to a very limited extent in EUV \cite[\eg ][]{2006ApJ...645.1525K}. 
With either method alone, we have no  direct information on the 3D structure. However, in some cases we can make estimates based on the knowledge of the orientation of prominence features as viewed over several days, or from two points of view using the STEREO spacecrafts.

Observations of motions in the plane of the sky have the advantage that it is possible to pick out actual moving features, although there are cases, especially in optically thin plasmas, in which multiple layers of plasma can make the data hard to interpret. Feature tracking is also insensitive to evenly moving flows.  A common variation of feature tracking is to use the time slice method \citep[][and Fig.~\ref{f:Lin03}]{2003SoPh..216..109L}. 
The intensity or velocity along a  slice of the  image of the filament is followed versus time, and the slope of the brightening or darkening in
this 2D diagram gives a measurement of the velocity of the feature in
the plane of the sky. This method can be easily applied to high spatial
resolution images, \eg\ with the Swedish Solar Telescope (SST) or the
Dutch Open Telescope (DOT) at La Palma, and has also been used with EUV images.
Another technique for tracking flows in the plane of the sky is  local
correlation tracking (LCT), although this method is not good at
isolating fine features.
In the case of the relatively low resolution EUV observations, the
traceable moving features often make up the minority of the
emission. There can also be problems in correctly tracing features
in low cadence ($\sim10$~min) data sets.

Doppler measurements more reliably give access to steady flows.  However, especially in optically thin plasmas, they represent an integration of the LOS velocity over many different features so that the measurement provides an average value along the LOS. 
{We should also note that current space-based EUV spectrometers do not have an absolute wavelength calibration, and large uncertainties may therefore arise in establishing the reference line position for Doppler shift measurements.}
For observations of filaments on the disk,  the signal coming from the prominence must be disentangled from the chromospheric  background. Various techniques have been developed to do this, principally based on cloud model methods \citep{1964PhDT........83B,1996A&A...309..275M,2007ASPC..368..217T}. In such models, four parameters are derived for the cloud:  the LOS velocity $V$,  the source function $S$, the optical thickness $\tau$, and  the line width $\Delta \lambda _{D}$.  $S$ and $\tau$ are strongly coupled but $V$ can be easily computed (see also Sect.~\ref{sec:cloud}).

\subsection{Quiescent Prominences}

Even the most stable quiescent filaments exhibit flows of $5-20$~\kmps. Recent research has focused on variations and  motions in thread-like structures that seem to make up filaments.

\citet{1998Natur.396..440Z} reported fine-scale counter-streaming in \Ha\ both along filament spines  {(Paper~II, Sect.~2.1.3)} and vertically in prominence barbs.
Moving features were traced over 10000 to 100000~km at speeds of 5 to 20~\kmps\ perpendicular to the LOS. The results were obtained by using movies  of the filament taken in  the red and blue wings of the \Ha\ line, which revealed oppositely directed flows. This is consistent with an earlier study of an active region filament by \citet{1991A&A...252..353S} in which the authors  report Doppler measurements of possible intermittent counter-streaming in filament footpoints with LOS velocities   of  $\pm 5$~\kmps. They used a cloud model to calculate velocities of $\pm15$~\kmps $\pm15$\%.

\begin{figure}
	\center
	\includegraphics[angle=90,width=0.45\textwidth]{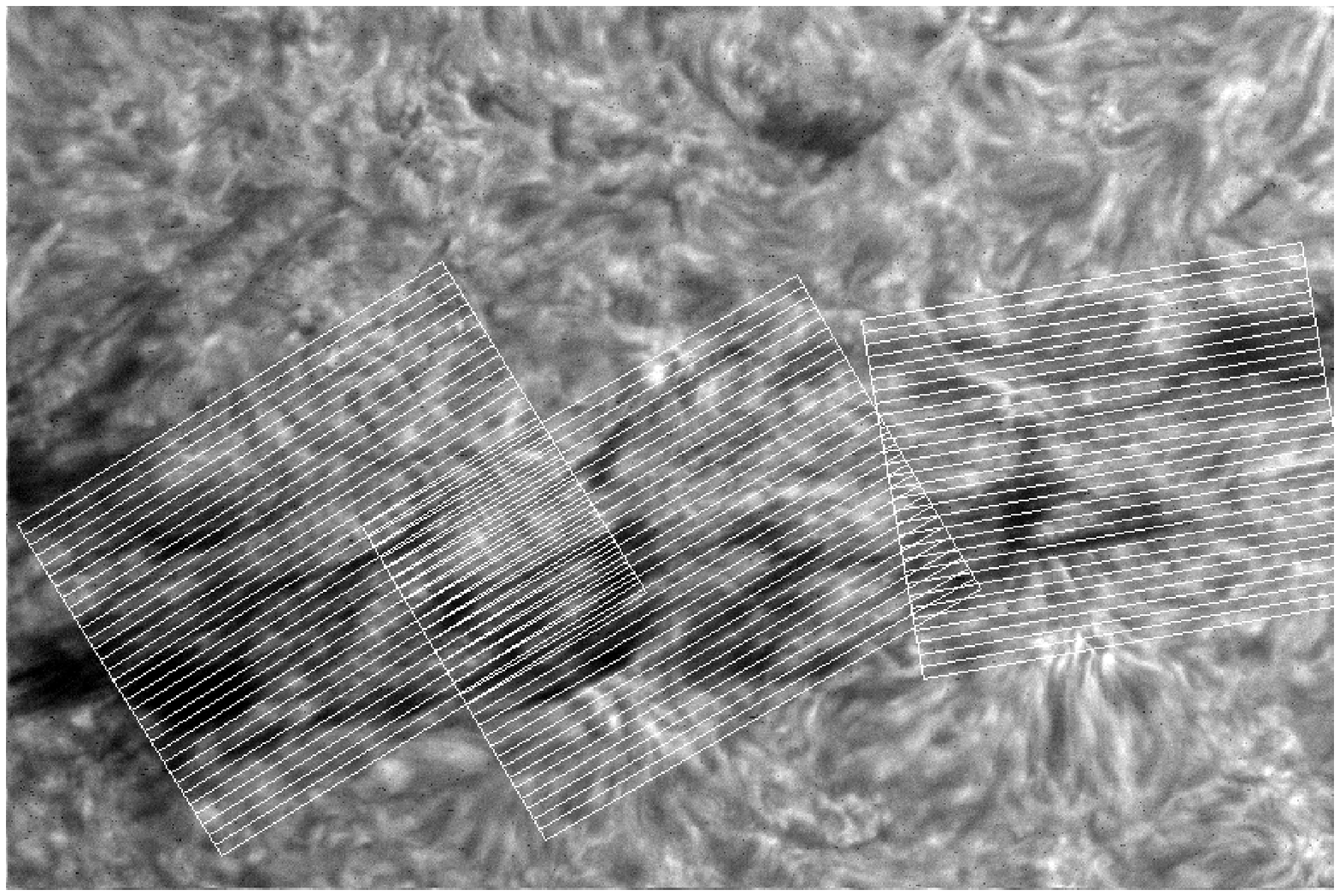}
	\includegraphics[width=0.45\textwidth]{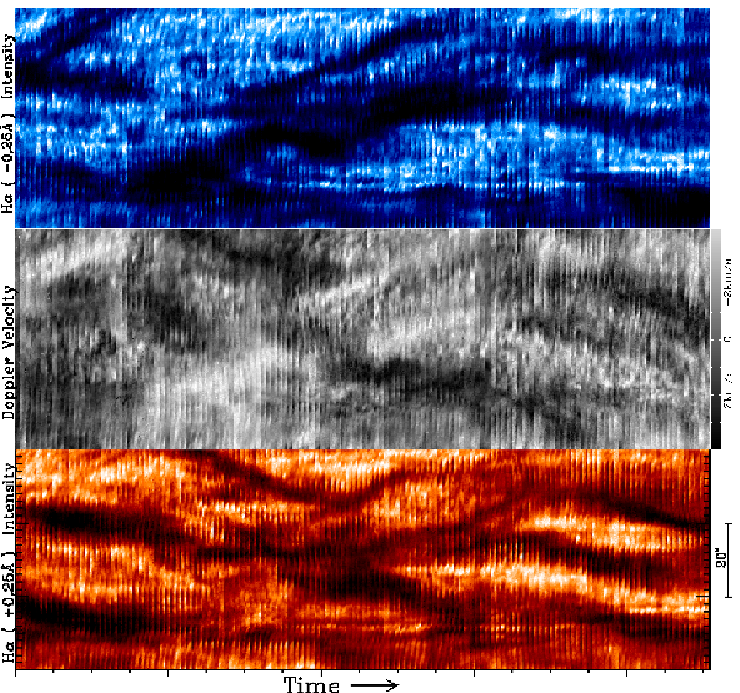}
\caption{Images illustrating the use of time slice diagrams to analyze flows in a \Ha\ filament observed  with the SST {(19 June 1998)}. The left panel shows the  locations of parallel slices used for analysing velocities in \Ha\ fine structures. At right from top to bottom are  the blue wing intensities, Doppler shift, and red wing intensities along a single time slice. Diagonal features show motions along the slice. From  \citet{2003SoPh..216..109L}. \label{f:Lin03}}
\end{figure}

Later studies by \citet{2003SoPh..216..109L, 2005SoPh..226..239L} and \citet{2008SoPh..247..321S} found counter-streaming in threads in \Ha\ with average transverse velocities in the range $7-15$~\kmps\ and speeds in one instance as high as 30~\kmps.
{They also report motions of the threads relative to one another with typical velocities of $2-3$~\kmps, and report a case of sideways motion in an isolated bundle of threads in a barb} which, they suggest, is related to photospheric motions. Fig.~\ref{f:Lin03} shows the time-slice techniques used by \citet{2003SoPh..216..109L} to analyse the data.

\citet{2006SoPh..234..115C,2007JKAS...40...67C} used a cloud model to
analyse quiescent filament observations along the spine and in a
barb at five points in the \Ha\ line. They found the data
consistent with a filamentary structure in which some filaments show
streaming in different directions with LOS velocities of $15\pm3$~\kmps,
while others contain plasmas at rest or moving only very slowly
($0\pm3$~\kmps).

{These observations and others listed in Table~\ref{t:velocities-ha} are crucially important in understanding the magnetic structure of prominences, especially in the barbs. The material seems to be moving both up and down the barb structures in fairly close proximity, in such a way that these opposite flows seem to be crossing when seen in the plane of the sky. The downward motions are slower than what would be expected if the material were falling unimpeded under the force of gravity. This indicates that some process is at work, which provides an upwards force to the cool material {\citep[\eg][]{2001SoPh..198..289M}}, while still allowing it to move up- and downwards fairly readily. These observations are a challenge to most common models of the magnetic field in prominence barbs (see discussion in Paper~II{, Sect.~2.1.3}).}

\begin{figure}
	\center
\includegraphics[width=\textwidth]{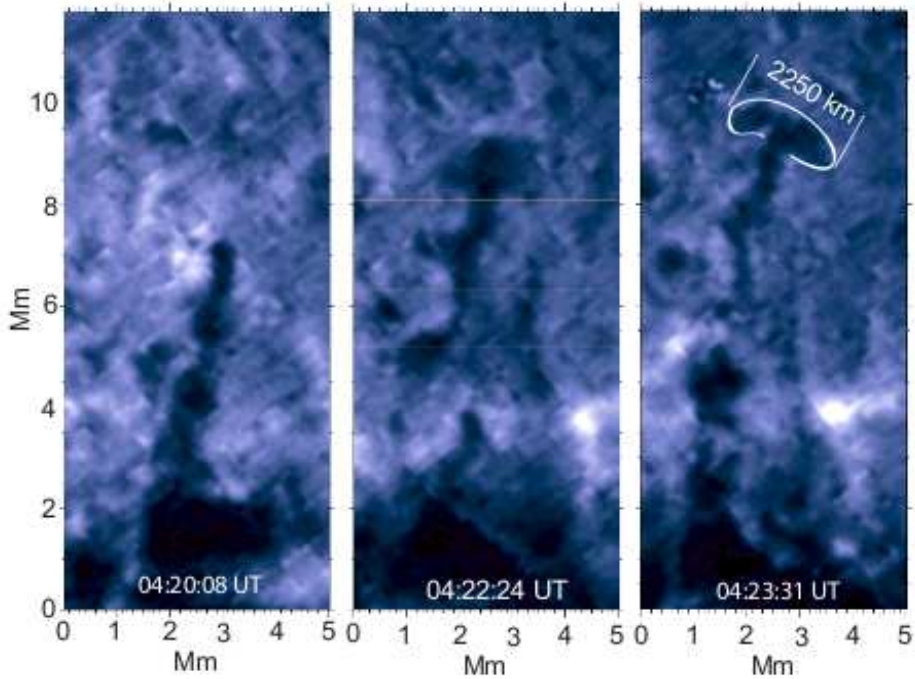}
\caption{Dark upward moving features observed {on the limb} with Hinode/SOT {(30 November 2006)} {in the Ca~II H (3968~\AA) line} by \citet{2008ApJ...676L..89B}.}
\label{f:Berger08}
\end{figure}

{The Hinode SOT instrument has made possible images of prominences and prominence flows on the limb with unprecedented resolution and pointing stability.}
\citet{2008ApJ...676L..89B} discuss very different types of motions observed in prominences on the limb in \Ha\  and Ca~II H (3968~\AA) with SOT. In observations of large, quiescent hedgerow prominences, they report two types of vertically moving features. One of these consists of bright downflow streams with velocities of about 10~\kmps\ and lifetimes of about 10 minutes. These bright downflows are sometimes seen to interact with what appears to be large scale vortex rotations.  They also report turbulent-looking, dark upflows which ascend with velocities of 20~\kmps\ in the plane of the sky, and have a large-headed ``mushroom cap'' shape (Fig.~\ref{f:Berger08}). 

\begin{figure}
	\center
\includegraphics[width=\textwidth]{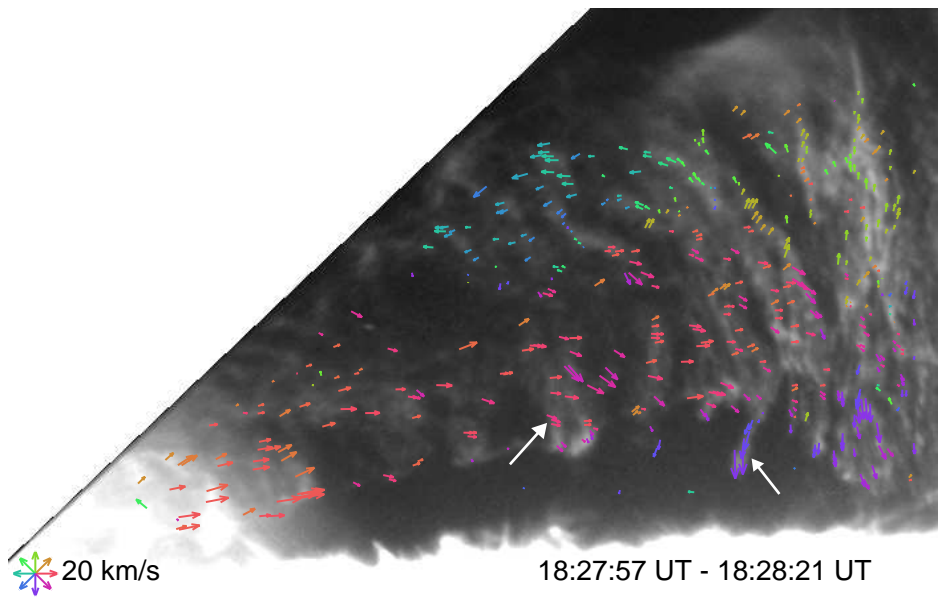}
\caption{Vertically oriented features seen in a hedgerow prominence {(17 August 2007)} observed with 
Hinode/SOT {in the \Ha\ line} by \citet{2008ApJ...689L..73C}. 
They appear to move both horizontally and vertically in the plane of the sky. White 
arrows point out a newly formed feature and one which is falling downwards.}
\label{f:Chae08}
\end{figure}

\citet{2008ApJ...689L..73C} analysed another hedgerow prominence observed with SOT in \Ha.
This prominence showed numerous vertically oriented structures. They were seen to move 
horizontally across the field of view at speeds of $10-30$~\kmps\ until a point at which they 
appear to shed downwards moving blobs (see Fig.~\ref{f:Chae08}). 
The bright blobs move downwards with acceleration 
of 0.015 to 0.083~km s$^{-2}$, far lower than the gravitational acceleration of 0.27~km s$^{-2}$. 
The change in direction from horizontal to vertical motion results in a vortical-appearing  motion.

These observations of prominences at the limb present a somewhat different picture than that presented by \Ha\ observations on the disk, which seem to describe relatively direct flows along straight thread-like structures with {inclined} up- and downflows in the barbs, and more or less horizontal motions along the spine. How do these disk observations correspond to the complex motions observed in hedgerow-type prominences on the limb? Perhaps the vertical prominence motions observed with SOT are providing a side view of the flows in barbs seen in filaments on the disk.

\subsubsection{Flows Observed in UV and EUV}
\label{s:EUVPromFlows}

In many ways, observations in the UV and EUV are quite limited compared to \Ha\ observations  since they have not yet reached the combination of high temporal, spatial, and often spectral resolutions available in \Ha. However, the insight they provide with regard to temperature information makes them important to our understanding of basic prominence properties.

Although fine  counter-streaming threads cannot be observed directly in the EUV, larger-scale counter-streaming flows in prominences have been observed in {H, He~I, and He~II} Lyman continuum absorption features in coronal lines by TRACE \citep{2008SoPh..247..321S}. Also, Doppler observations combined with modelling indicate that prominences do show such behaviour in EUV.
\cite{1998A&A...334..280D} observed a quiescent hedgerow prominence in a number of chromospheric and TR lines observed by SUMER.
Doppler shifts were generally correlated between many, although not all,  of the EUV  lines. 
In general, however, they did not see detailed spatial coherence in the intensities of the different lines, leading them to suggest that the emission may come from threads formed at different temperatures.

{\citet{1999SoPh..187..405W} observed a polar crown prominence in the N~V line at 1238~\AA\ with SUMER. They found the prominence to be quite dynamic, with structures changing on time scales of a few hours and  modelled the 
prominence as a collection of tiny threads with a Gaussian velocity distribution with a width of 23~\kmps.}

A number of studies of EUV observations of prominences have described relatively fast motions. \citet{1999ApJ...520L..71W} tracked individual knots and brightening in quiescent prominences seen by EIT at 304~\AA\ with a 20~min cadence.  He measured speeds in the plane of the sky of $10 - 70$~\kmps\ which increased as the moving features ``surged'' from one end of the  prominence as seen above the limb  to the other.
These motions were most clear during the period of prominence formation.  Motions continued later in prominence development, with the prominence becoming increasingly complex and filamentary. In a later paper, \citet{2001ApJ...560..456W} suggested that such jets are related to flux cancellation occurring near the footpoints of the prominence barbs. Such jet-like motions have also been observed in active region prominences (Sect.~\ref{s:ARPromFlows}).

\citet{2003SoPh..212...81K} analysed  prominences in quiet and plage regions
using CDS wide band movies taken in He~I, O~V, and Mg~IX.
They found multi-thermal features with velocities across the line of
sight in the $5-70$~\kmps\ range. Some of the motions appeared to be horizontal 
in nature. For others, the direction in three dimensions was not as clear. \citet{2006ApJ...645.1525K} {measured a velocity of about 40~\kmps in a jet like feature in an activated prominence.}

\citet{1993A&A...273..267W} observed two prominences using the HRTS
instrument. One prominence, described as hotter and more active
(including a footpoint eruption during the observations), was oriented
so that the LOS was along its axis. It showed Doppler shifts of $\sim
30$~\kmps\ along the edge of the prominence and LOS velocities between
$\pm10$~\kmps\ in the rest of the prominence.  
The second prominence in a plage
region had a more north-south orientation. It appeared cooler and
quieter, and showed much lower bulk velocities of $\sim 3$~\kmps.  One
footpoint, however, exhibited a LOS velocity
of 50~\kmps.

One question of interest is whether there is a variation in velocities of plasmas formed at different temperatures. A general trend of higher velocities {detected} in EUV observations as opposed to \Ha\ observations can be seen in Tables~\ref{t:velocities-ha} and \ref{t:velocities-euv}, which summarise many of the results discussed in this section. 
\begin{table}
	\caption{Velocity flows in quiescent filaments and prominences observed in \Ha. Acronyms: BBSO -- Big Bear Solar Observatory; SVST -- Swedish Vacuum Solar Telescope{, now SST}.
	\label{t:velocities-ha}}
	\begin{tabular}{p{1.65cm}p{1.65cm}p{2.05cm}p{1.65cm}p{1.05cm}p{1.65cm}}
		\hline\noalign{\smallskip}
		Observation & Speed (\kmps) & Notes & Method & Telescope & Source\\
		\noalign{\smallskip}\hline\noalign{\smallskip}
		{\bf \underline{On Disk}}&&&&&\\
		
		Spine, barbs &  $5-20\perp$ LOS & Counter-streaming, vertical in & Doppler time slice  & BBSO & \citet{1998Natur.396..440Z} \\
		& &  barbs, horizontal in spine\\
		\\
		Polar crown, spine & 8 $\perp$ LOS & Counter-streaming horizontal in & Doppler time slice & SVST & \citet{2003SoPh..216..109L}\\
		& & spine\\
		\\
		Spine, barbs& $15\pm10 \perp$ LOS & & Doppler time slice & SST & \citet{2005SoPh..226..239L}\\
		Barb & $5-13\perp$ LOS &$\perp$ to thread orientation&&&\\
		\\
		Spine, barbs & 10 $\perp$ LOS & Intermittent counter-streaming & Doppler time slice & Hida & \citet{2008SoPh..247..321S}\\
		\\
		Spine, barbs & 15 $\pm$ 3 LOS & Counter-streaming + stationary & Doppler shifts, cloud model & BBSO & \citet{2006SoPh..234..115C,2007JKAS...40...67C}\\
		\\
		Barb & 10 $\perp$ LOS & Upwards & Feature tracking & SVST & \citet{2004ApJ...612..519V}\\
		
		\underline{\bf Limb}&&&&&\\
		
		Hedgerow & $\le35\perp$ LOS & Vertical & Feature tracking & SOT & \citet{2008ApJ...689L..73C}\\
		& $10-30\perp$ LOS & Horizontal &&&\\
		\\
		Hedgerow & 10 $\perp$ LOS & Bright down- & Feature & SOT & \citeauthor{2008ApJ...676L..89B}\\
		& & flow & tracking & & (2008)\\
		& 20 $\perp$ LOS & Dark {upward moving feature} &&& \\
		\\
		{Hedgerow} & {$2-10$ LOS} & {Horizontal} & {Doppler shifts} & {MSDP} & {\cite{2009arXiv0911.5091S} }\\
		& {$2-25$ $\perp$ LOS} & {Up- and Down-flows} & {Feature tracking} & {SOT} &\\
		\noalign{\smallskip}\hline
	\end{tabular}
\end{table}
\begin{table}
	\caption{Velocity flows in quiescent filaments and prominences observed in EUV. LCT stands for Local Correlation Tracking.
	\label{t:velocities-euv}}
	\centering
	\begin{tabular}{p{1.65cm}p{1.65cm}p{2.05cm}p{1.65cm}p{1.05cm}p{1.65cm}}
		\hline\noalign{\smallskip}
		Observation & Speed (\kmps) & Notes & Method & Telescope & Source\\
		\noalign{\smallskip}\hline\noalign{\smallskip}
		{\bf \underline{On Disk}}&&&&&\\
		
		Barb & $<6$ LOS & Vertical & Doppler shift & SUMER & \citet{1999SoPh..186..259K}\\
		
		\underline{\bf Limb}&&&&&\\
		
		Polar crown & $\pm$ 10 LOS & Prominence oriented $\parallel$ LOS & Doppler shift & HRTS & \protect{\citet{1993A&A...273..267W}} \\
		Prominence  & $30\pm3$ LOS & Prominence oriented $\perp$ LOS &&&\\
		Barb & 50 LOS&&&&\\
		\\
		Polar crown & 23 LOS & Width of velocity distribution & Doppler + threads & SUMER & \citet{1999SoPh..187..405W} \\
		\\
		Hedgerow & $\pm8$, max$\approx$25 LOS & & Doppler shift & SUMER & \protect{\cite{1998A&A...334..280D}}\\
		\\
		Prominence & $10-40\perp$ LOS & vertical & LCT & ATM &\citet{1999SoPh..185..113P}\\
		\\
		Three prominences & $10-70\perp$ LOS & & Feature tracking & EIT & \citet{1999ApJ...520L..71W} \\
		\\
		& $5-70\perp$ LOS & Mostly horizontal & Feature tracking & CDS & \citet{2003SoPh..212...81K}\\
		\\
		Prominence & $22\pm3 - 28\pm2$ & $\log T = 4.2 - 4.6$ & Doppler & SUMER & \cite{2004SoPh..223...95C}\\
		&	$35\pm2$ LOS & $\log T = 4.8 - 5.6$ & & &\\
		\\
		Prominence jet & $38\pm4$ & Feature seen in \Ha\ and PCTR & Feature tracking, Doppler & TRACE, SUMER & \citet{2006ApJ...645.1525K}\\
		\noalign{\smallskip}\hline
	\end{tabular}
\end{table}
However, it is not clear if this trend is real, or if it is the result of differences in observing capabilities and target selection. There have been some efforts to combine observations of \Ha\ and EUV motions, but they have been somewhat limited by the lower resolution of the EUV data.  

As to comparisons of plasmas observed at different temperatures in the EUV, \citet{1993A&A...273..267W} and \cite{2006ApJ...645.1525K, 2008ApJ...673..611K} found motions at temperatures below about $2.5\times10^5$~K to be quite similar, and \cite{1999SoPh..185..113P} found similar motions in the range $10^4$ to $10^6$~K. On the other hand,  \citet{1998A&A...334..280D} report variations. Using Doppler measurements, \citet{2004SoPh..223...95C} found an increase in velocities with temperature:  from 16000 to 40000~K, velocities were $22 - 28$~\kmps, while for 60000 to  400000~K, the mean velocity was 35~\kmps.

\subsubsection{Cross-Field Diffusion}

Most of the motions discussed up to this point are assumed to be along or with magnetic field lines. Another possible type of bulk flow is cross-field diffusion of neutral atoms. \citet{2002ApJ...577..464G} calculated the diffusion times for neutral atoms in a simple prominence model to be 22 days for hydrogen and 1 day for helium.  This difference is used to explain a decrease of the ratio of He to H as a function of height inside stable quiescent filaments derived from He I 10830~\AA\ and \Ha\ observations by \cite{2007ApJ...671..978G}, and thus an overabundance of neutral helium in the lower part of the filaments.
However, for more active or erupting filaments, \citet{2009ApJ...704..522K} report that the variation in the H/He ratio disappears and is accompanied by  a general increase in absorption in both lines in filaments prior to eruption,  suggesting that this injection of mass prior to eruption happens quickly enough or mixes the material such that  cross-field diffusion cannot cause the separation of neutral H and He.

{\cite{2009arXiv0911.5091S} measured the velocity vector in prominence flows observed by SOT (prominence shown in Fig.~\ref{fig:0-sot}) by combining  measurements of transverse velocities from SOT images and Dopplershifts derived from \Ha\ profiles observed with the MSDP spectrograph operating on the solar tower in Meudon. Although the motions appear nearly vertical in the plane of the sky, the velocity vectors suggest that the flows are inclined by an angle of 30 to 90 degrees. Perhaps the vertical prominence  motions (down flows) observed by SOT provide a side view of the flows in barbs seen in filaments on the disk.  \cite{2009arXiv0911.5091S}  suggest that the dark bubbles rising up inside the hedgerow prominences may be due to higher magnetized regions which could correspond to parasitic polarity emergence in the filament channel. }
\citet{2008ApJ...676L..89B} mention cross-field diffusion as a possible cause for the motions they observe with SOT (Fig.~\ref{f:Berger08}).

\subsection{Flows in Active Region Prominences}
\label{s:ARPromFlows}

One of the signatures of active region prominences is their high level
of activity, which includes easily seen flows along the prominence axis
\citep{1995nsp..book.....T}. A number of studies report the formation of active region prominences via abrupt jet-like flows, often associated with observed activity in the magnetic field.

\citet{2003ApJ...584.1084C} found fast ($80-250$~\kmps), jet-like and
eruptive behaviour in a forming active region filament observed by TRACE
 in the 1600 and 171~\AA\ channels. The material was often visible in
absorption at 171~\AA, but appeared to be heated with time so that it
eventually showed coronal temperature emission in the EUV. They found
that the jets were associated with cancelling magnetic flux.
\citet{2005ApJ...631L..93L} reported \Ha\ observations of two active
region filaments formed by surges with measured velocities of 150 and
180~\kmps\ perpendicular to the LOS. The filaments appeared relatively
stable, and one lasted at least two days before going over the limb.
{Observations of the formation of two active region filaments by \citet{2004SoPh..223..119S} 
revealed a dynamic process with downward velocities of 20~\kmps, but less frequent \Ha\ up-flows.}
Other studies have reported shorter-lived filaments also formed by
surges \citep{1976SoPh...50..399Z,1999SoPh..190...45L}.

{\cite{2000SoPh..195..333C} report motions in an active region prominence seen in \Ha\ including a flow system connected to
neighbouring loops that injected material with speeds of $10-40$~\kmps.
\citet{2002SoPh..209..153D} and \citet{2008AdSpR..42..803L}  described bulk flows in \Ha, including counter-streaming, in existing active region filaments. Velocities were generally in the range 5 to 15 \kmps, but as high as 25 \kmps.
\citet{2002SoPh..209..153D} report especially active periods characterised by twist motions and downflows which they  associate  with episodes of nearby magnetic activity in parasitic polarity regions seen in magnetograms. }
\citet{2007Sci...318.1577O} observed an active region prominence in Ca~II with SOT. They found ubiquitous horizontal flows. Some exhibited steady velocities of about 40~\kmps, while others showed more complex acceleration.

In addition to the quiescent prominences mentioned in Sect.~\ref{s:EUVPromFlows}, \citet{2003SoPh..212...81K} also analysed a  prominence in an active region observed in \Ha\ and with TRACE at 1216 and 1600~\AA.
In one case, a horizontally moving  feature with a plane of the sky velocity between 5 and 40~\kmps\ was observed in both \Ha\ and  the 1600~\AA\ band (thought to represent C~IV emission), suggesting a source with a cool core and a TR. 

{
\subsection{Summary}
As seen in Tables~\ref{t:velocities-ha} and \ref{t:velocities-euv}, there is a significant range of flow velocities that have been measured. While comparing these values, one should keep in mind obvious differences such as disk \textit{vs} limb observations, or cool part \textit{vs} PCTR. The true orientation of the flows clearly needs to be carefully appreciated before attempting a sensible discussion of their possible causes and consequences (\eg\ enthalpy flux). We come back on these issues in Section~\ref{sec:12}. Observations from the forthcoming Interface Region Imaging Spectrograph (IRIS\footnote{\url{http://iris.lmsal.com/}}) may help in addressing some of the questions outlined above, as this instrument will be able to measure both LOS and transverse velocity components of neutral and ionized plasma by combining spectroscopy and imaging at high spatial and temporal resolutions.
}

\section{Mass Determination}
\label{sec:6}

Prominence density, total mass loading, and  mass composition are critical values for {theoretical models of the origin of the prominence plasma (Paper~II, Sect.~3)}. An accurate determination of these quantities is not a trivial task and, typically, only order of magnitude estimates can be made.  Efforts over the past decade have significantly improved these measurements to start testing models of prominence support, mass draining, mass loading, dynamics within the material, and eruption mechanisms.

\subsection{Mass Estimates From Neutral Hydrogen Density and Geometrical Considerations}
\label{sec:6-coldens}

A rough estimate of the mass $M$ of a \textit{cool} prominence (low ionisation ratio) can be made using the following relation: 
\begin{equation}
M \approx N_\mathrm{H} m_\mathrm{H} V \ ,
\label{eq:6-estimate}
\end{equation}
where $N_\mathrm{H}$ and $m_\mathrm{H}$ are the mean neutral hydrogen number density and the mass of the hydrogen atom, respectively, and $V$ is the volume occupied by the prominence plasma.
{The mass obtained in this way is only a lower limit to the total mass of the prominence.}
Using typical values of $N_H\approx 3\times10^{10}$~cm$^{-3}$ and $V\approx 10^{26}-10^{29}$~cm$^{3}$ {(using lower and upper limits for the typical dimensions given in Section~\ref{sec:1})}, one obtains $M\approx5\times 10^{12}-10^{15}$~g.

{Note that the precise determination of the dimensions of prominences and filaments, and in particular their geometrical thickness, is far from trivial.}
The volume $V$ of the prominence in (\ref{eq:6-estimate}) can be estimated by making assumptions about the morphology. For instance, the geometrical thickness can be determined by tracking an object over several days and comparing its appearance when on the disk and above the limb.  If this is not possible, then the thickness can be set equal to the apparent width of the structure.

More accurate techniques are based on the determination of the hydrogen column density at each spatial position within the prominence or filament.
One solution is to use non-LTE calculations consistently solving the radiative transfer problem in a prominence slab as described in {Sect.~\ref{sec:7}}. { Let us mention here the work of \cite{1996SoPh..164..211H} who, by combining polarimetric data in the \Ha\ and He~I D3 lines on one hand, and observed \Ha\ intensities (compared to theoretical ones) on the other hand, obtained values for the geometrical thickness ranging between a few hundreds~km up to a few $10^4$~km for 18 measurements in different prominences. They then derived a corresponding total column mass of $10^{-5}$~g cm$^{-2}$.} 
A different kind of non-LTE calculations can also be made using the so-called cloud model \citep[][{and also Sect.~\ref{sec:cloud}}]{1996A&A...309..275M, 1999A&A...346..322H}.  Using the cloud model with the optical thickness, Doppler width, source function, and velocity as input parameters, the absorption in H$\alpha$ can be calculated and compared with observations. However, uncertainties in the ionisation fraction of hydrogen result in a large range of values for the total hydrogen number density.

\subsection{Determinations from EUV Continuum Absorption}
\label{sec:EUVAbsorptionMass}

\begin {figure}
\center
	\includegraphics[width=\textwidth] {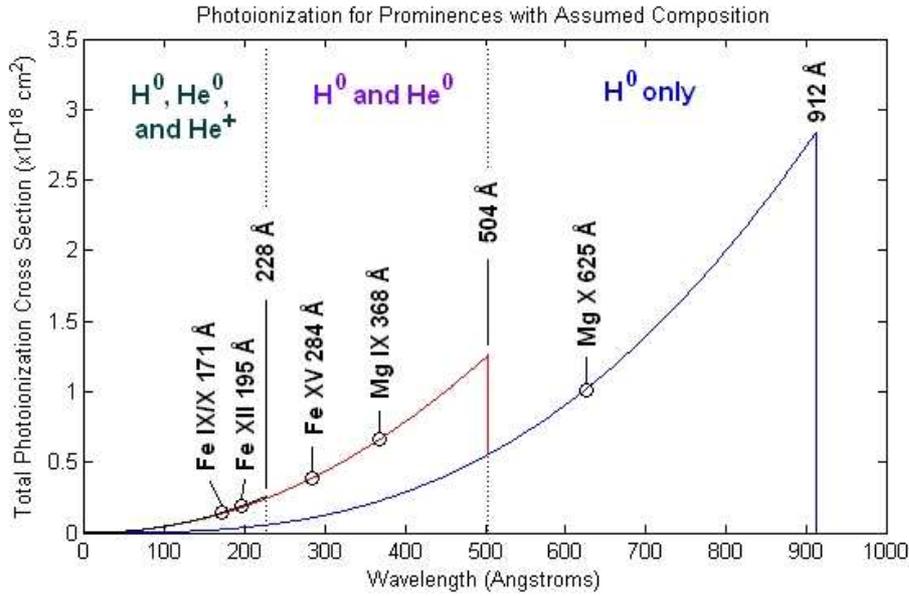}
	\caption{Plot of the average photo-ionization cross-section per atom/ion in a prominence with an assumed composition (45\%~H$^0$, 45\%~H$^+$, 9\%~He$^0$, and 1\%~He$^+$). In this plot, the contributions to the total photo-ionization cross-section are separated by the absorbing species: those due to H$^0$ (below the blue line, wavelength shorter than 912~\AA), He$^0$ (between the blue and the red line, which starts at 504~\AA), and He$^+$ (between the red and the black line, which starts at 228~\AA).}
	\label{fig:photo-ionization_cross-section}
\end {figure}

Prominences are observed in EUV coronal lines as dark structures embedded in the brighter background of coronal radiation (see Fig.~\ref{fig:6-xrt-195}, right).  Since the prominence material is much cooler than the corona (where such lines are formed), the usual complicated combination of atomic processes is reduced to the relatively simple process of absorption, which is the absorption by the photo-ionization continuum. For prominences in EUV, the main absorbers are neutral hydrogen, neutral helium, and singly-ionized helium, for which the cut-off wavelengths are 912~\AA, 504~\AA, and 228~\AA, respectively.  The photo-ionization cross-section is inversely proportional to some power of the photon frequency \citep{Keady2000,2005ApJ...622..714A}; a plot of the average cross-section per atom/ion for an assumed prominence composition is provided in Fig.~\ref{fig:photo-ionization_cross-section}.  
The used composition is based on assumed helium abundance and on the model-dependent ionization fraction for hydrogen and helium \citep{1978ApJ...221..677H, 1999A&A...349..974A, 2004ApJ...617..614L}. Since the vast majority of electrons are in the ground state in H and He \citep{1973ApJ...186.1043M,1987A&A...183..351H,2001A&A...380..323L}, only the ground state photo-ionization cross-sections need to be used in an analysis of the continuum absorption (in general, much weaker subordinate continua overlap the resonance ones).  Much of the research described below has been conducted to calculate the number density and total mass, and to reduce errors related to the foreground radiation, unknown depth of the material, and emissivity blocking due to the cavity.

\begin {figure}
	\centering
	\includegraphics [scale=.3] {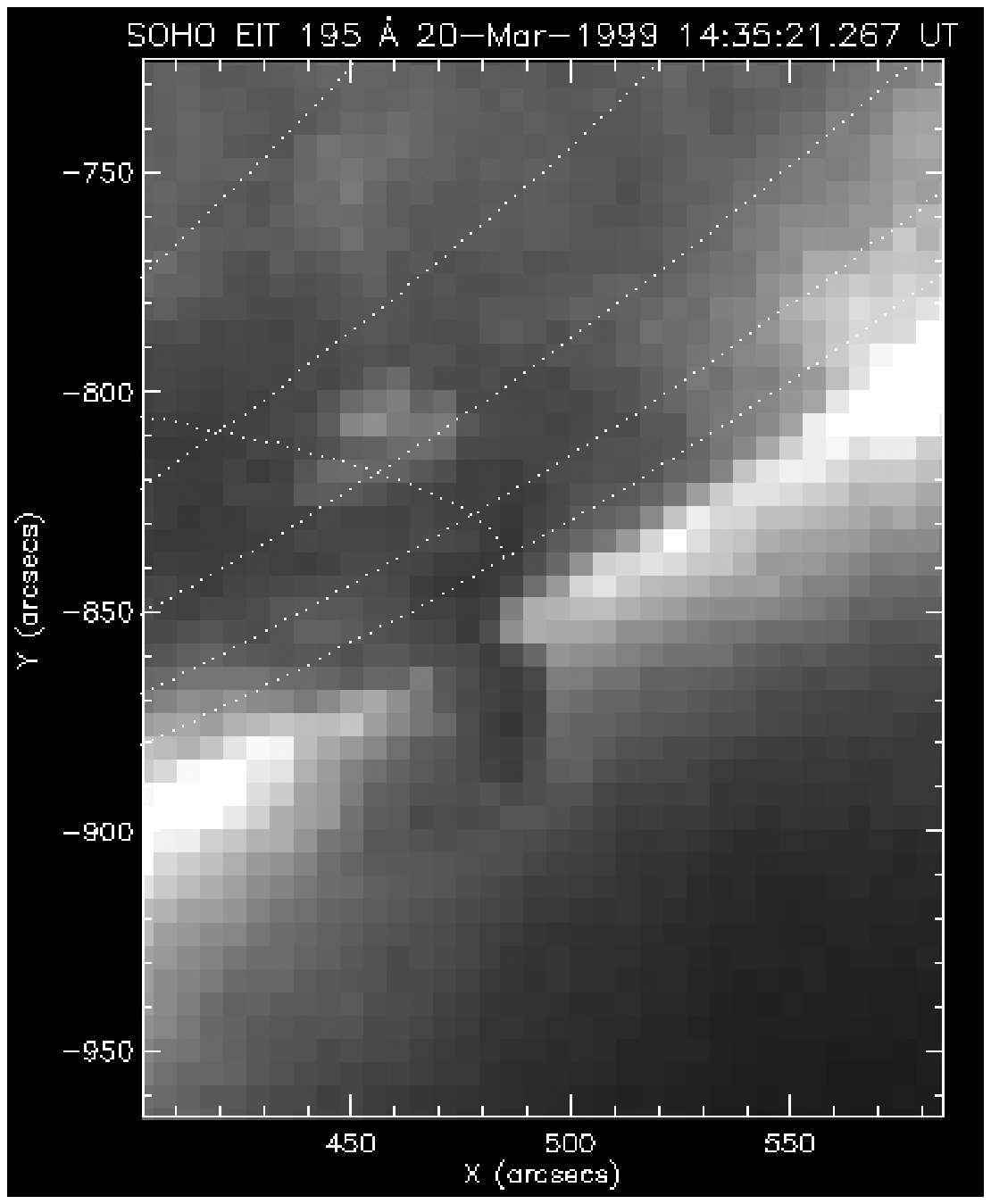}
	\hspace {.1 in}
	\includegraphics [scale=.3] {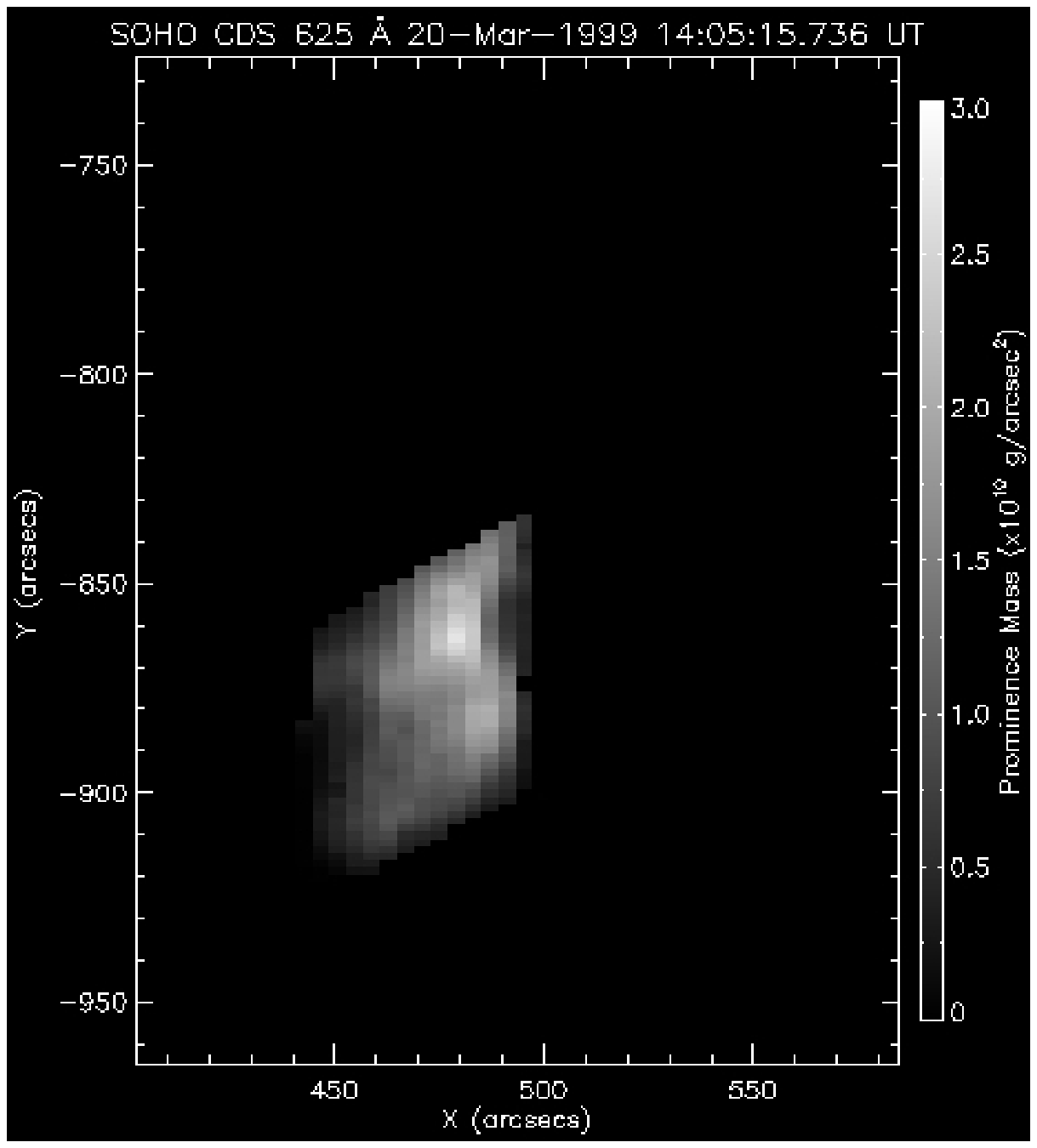}
	\hspace {.1 in}
	\includegraphics [scale=.3] {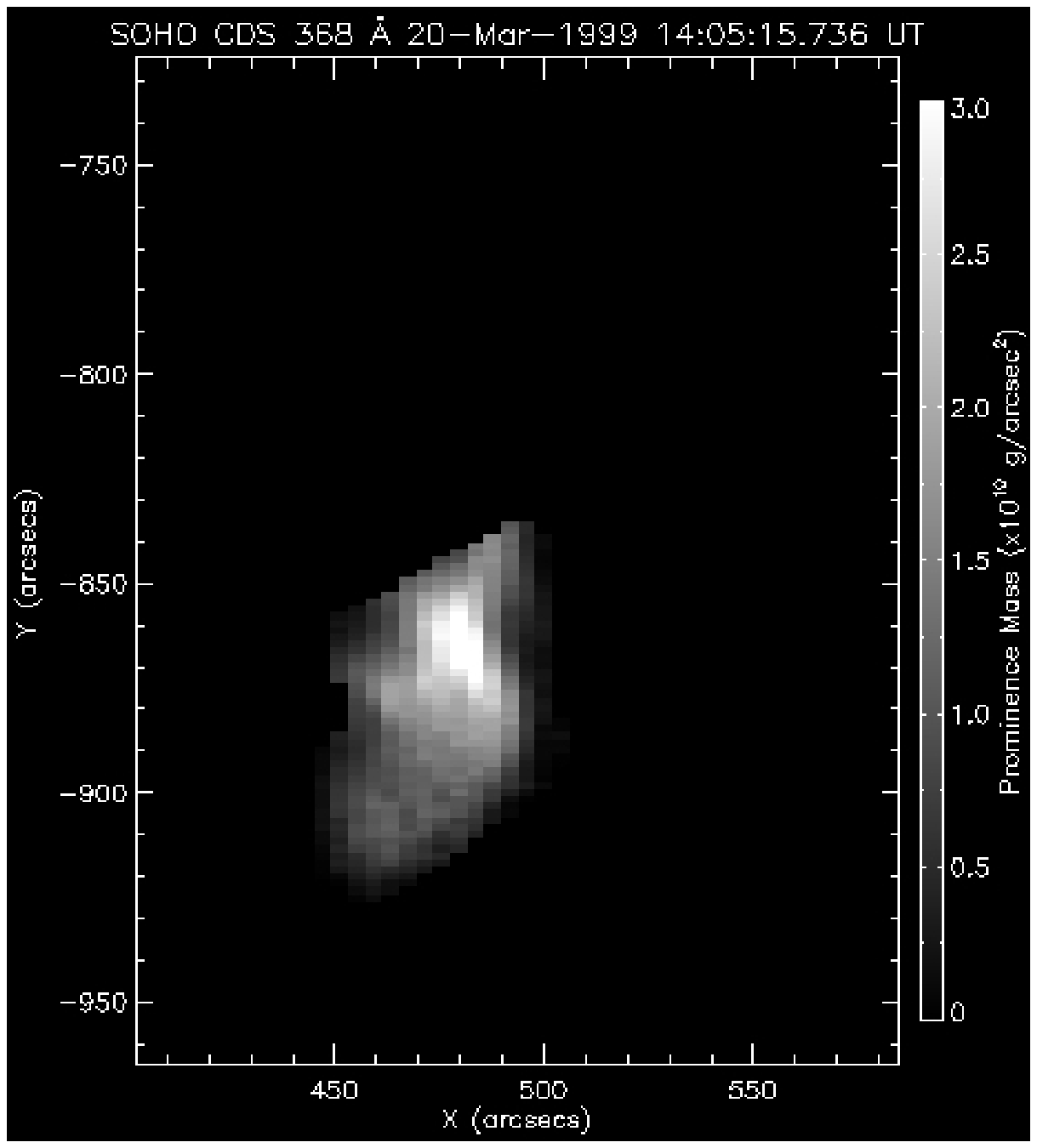}
	\caption{Figures from \citet{2009Gilbert} of a prominence on 20 March 1999 at around 14:30 UT in EIT 195~\AA\ (left), the mass map calculated from CDS 625~\AA\ absorption due only to H$^{0}$ (center), and the mass map calculated from CDS 368~\AA\ absorption (right), which is due to a combination of H$^{0}$ and He$^{0}$.  In the mass maps, a brighter pixel corresponds to greater mass, and both wavelengths have the same scaling in brightness, ranging from zero to $3.0 \times 10^{10}$~g arcsec$^{-2}$.  Note the additional mass detected in the lower part of the prominence at 368~\AA, presumably due to absorption from He$^{0}$.}
	\label{fig:MassMaps}
\end {figure}

\begin{figure}
  \includegraphics[width=\textwidth]{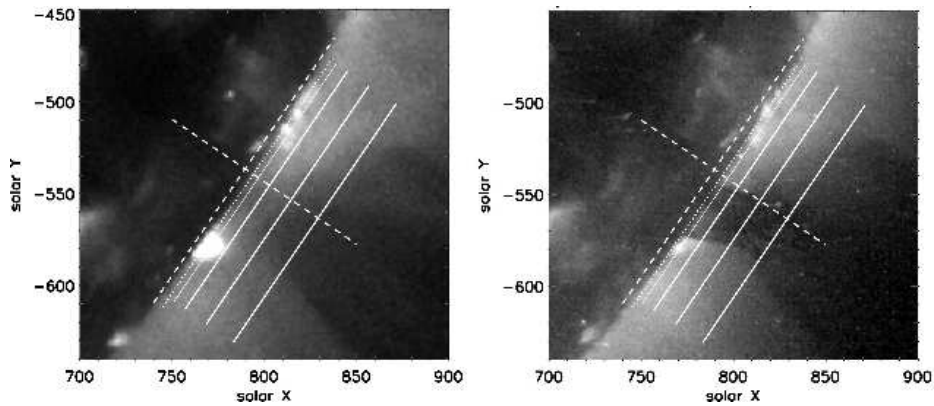}
	\caption{Multi-wavelength images obtained with XRT (left) and with TRACE at 195~\AA\ (right) of a prominence observed on 2007 April 25 at 13:19 UT.  The prominence is visible in absorption at 195~\AA\ but is transparent to X-rays, allowing a determination of the emissivity blocking due to the cavity and a more accurate calculation of the continuum absorption.  Adapted from \cite{2008ApJ...686.1383H}.}
	\label{fig:6-xrt-195}
\end{figure}

The earliest calculations of the continuum absorption in EUV were done by \citet{1976SoPh...50..365O} who used the filament observations taken by the Harvard EUV spectrometer on ATM. Since then, a large number of cases has been observed by SOHO, TRACE, STEREO, and Hinode.  Mass measurements were first attempted by \citet{1998SoPh..183..107K} using several spectral lines observed with CDS to estimate a column density of neutral hydrogen of the order of $10^{18}$~cm$^{-2}$.  Follow-up mass calculations have been done by \citet{1999PhPl....6.2205G} using TRACE observations, by \citet{2000SoPh..197..313P} for an active region filament eruption, and by \citet{2005ApJ...618..524G, 2006ApJ...641..606G} in a large study of 23 prominences that took into account the coronal radiation in front of the prominence material, correcting an underestimation of the prominence mass, to yield total mass values ranging from $1 \times 10^{14}$~g to $2 \times 10^{15}$~g.  
Similar values have been obtained for a quiescent filament by \cite{2003ESASP.535..447H}.
\citet{2009Gilbert} extend the method by interpolating the coronal background emission for the entire prominence area, yielding a mass calculation at each pixel to form a 2D map (Fig.~\ref{fig:MassMaps}). 
Total mass estimates of the relatively small prominences analyzed are on the order of $10^{14}$~g, with margins of error $\sim20\%$.

A related line of study has used multi-wavelength EUV observations to determine the overall 3D structure of prominences, with an emphasis placed on the geometry and extended area of prominences and their related cavities in the EUV.  The early work on non-LTE radiative transfer diagnostics found a much larger opacity (sometimes $1-2$ orders of magnitude) for the hydrogen Lyman continuum below 912~\AA\ as compared to the \Ha\ line opacity, which can explain observations of extended filament structures in EUV co-aligned to much narrower \Ha\ counterparts \citep{2001ApJ...561L.223H, 2003SoPh..216..159H, 2003A&A...401..361S, 2004SoPh..221..297S}.  This indicates the presence of more filament material than can be visible in \Ha. Further research utilized these findings to determine the vertical extent of material in filaments, their 3D structure, and the amount of emissivity blocking, which is due either to the lack of coronal emission in the cavity or to the presence of the absorbing material \citep{2004A&A...421..323S,2005ApJ...622..714A,2006A&A...459..651S,2007SoPh..242...43A}.  \citet{2008ApJ...686.1383H} used observations from the X-Ray Telescope on Hinode  \citep[XRT,][]{2007SoPh..243...63G} {of the prominence shown in Fig.~\ref{fig:0-sot}} to determine the cavity blocking and isolate the continuum absorption (Fig.~\ref{fig:6-xrt-195}). 
{The column density can be related to electron density and effective thickness using non-LTE models (see, \eg\ Sect.~\ref{sec91:heasley}).}
They calculated the optical thickness at various wavelengths and a neutral hydrogen column density of $1-5 \times 10^{19}$~cm$^{-2}$.

Previous determinations of the prominence mass using radio measurements by \citet{1995SoPh..156..363I} were expanded upon and compared with EUV measurements by \citet{2001SoPh..199..115C}, who came to similar conclusions about the emissivity blocking as the studies described above.  \citet{2004A&A...420..307D} went further and attempted to distinguish between a prominence with a transition region and corona below it, or a thread-like PCTR without any quiet-Sun transition region below \citep[which is discussed more in][]{2008ApJ...673..611K}.  They and \citet{2005A&A...443.1055C} also calculated the neutral helium-to-hydrogen ratio to be $N(\mathrm{He^0})/N(\mathrm{H^0}) = 0.1-0.2$.

{Although the various methods do not lead to large variations in the determination of the mass, one should be aware of the uncertainties due to the adopted hypothesis and also be aware of the unknown nature of the \textit{extended} filaments (possibly leading to some blocking or absorption) which could play a role in the  total mass determination.}

\section{Basics of Radiative Transfer}
\label{sec:7}

Observations show that there are central cool parts of prominences (or fine prominence structures) surrounded by the gradually hotter PCTR {(which is optically thin in most transitions)}. Densities are generally higher in the central parts. In these regions, several spectral lines and also some continua (\eg\ the hydrogen Lyman continuum) are optically thick and this requires a proper treatment of the radiative transfer. 
{In this section, we introduce the basics of the radiative transfer relevant to prominence physics and outline its most important aspects to understand how prominence diagnostics can be done from the analysis of optically thick lines.}

We will start with the relatively simple 1D slab models. These models have proven to be very useful for the basic understanding of radiation processes in prominences and, moreover, they describe relatively well the global prominence radiation properties as observed at lower spatial resolution. Indeed, the prominences look like vertical plasma slabs when observed above the limb with low resolution. On the other hand, high-resolution images show that prominences are highly heterogeneous, containing many fine-structure threads, blobs or fibrils. This is discussed in Sect.~\ref{sec:10}.

The {\it radiative transfer equation} determines the modification of the {\em specific intensity} of radiation $I(\nu)$ along an elementary geometrical path $ds$ due to absorption and emission processes ($\nu$ is the frequency):
\begin{equation}
	\frac{\mathrm{d}I_{\nu}}{\mathrm{d}s} = -\chi_{\nu} I_{\nu} + \eta_{\nu}.
\end{equation}
The absorption and emission coefficients are denoted as $\chi_{\nu}$ and $\eta_{\nu}$, respectively.
Using $x$ as the reference coordinate in a simple 1D prominence slab (see Fig.~\ref{fig:1d_rt}):
\begin{figure}
	\center
	\includegraphics[width=\textwidth]{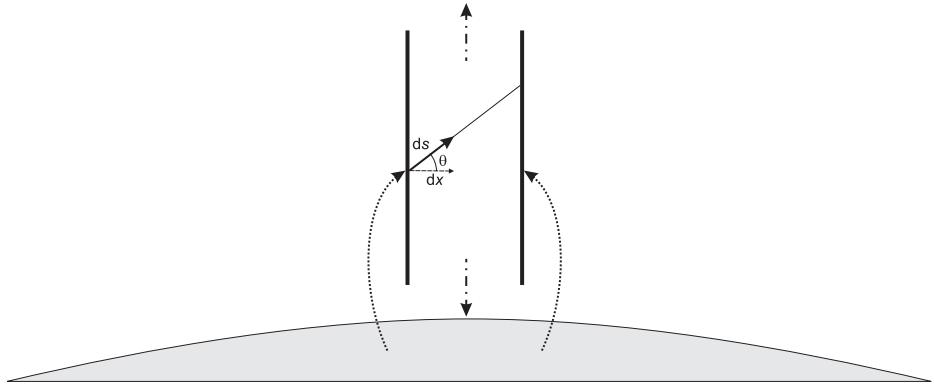}
	\caption{Sketch of the 1D geometry of the radiative transfer problem in the case of a vertical slab (prominence) above the limb. {Dotted arrows indicate the incident radiation.}\label{fig:1d_rt}}
\end{figure}
\begin{equation}
	\frac{\mathrm{d}x}{\mathrm{d}s} = \cos \theta \equiv \mu \ .
\end{equation}
Defining $\tau_{\nu}$ as the {\em optical depth} at frequency $\nu$ {with}:
\begin{equation}
	\mathrm{d}\tau_{\nu} = - \chi_{\nu} \mathrm{d}x \ ,
\end{equation}
we can express the transfer equation in its standard 1D plane-parallel form:
\begin{equation}
	\mu \frac{\mathrm{d}I_{\nu}}{\mathrm{d}\tau_{\nu}} = I_{\nu} - S_{\nu} \, .
\end{equation}
The {\em source function} is then:
\begin{equation}
S_{\nu} \equiv \frac{\eta_{\nu}}{\chi_{\nu}} \, .
\end{equation}

The radiative transfer equation must be solved
numerically because the source function usually strongly
depends on the radiation field. Only after the source function
is specified, the integration of the radiative transfer equation is
relatively simple (the so-called formal solution). In Sect.~\ref{sec:8}, we discuss various modern approaches used for solving the non-LTE transfer problem under prominence conditions.

A critical issue in all prominence non-LTE modelling is a realistic determination of the boundary conditions for the radiative transfer equation. Since prominences are
typically rather low-density objects, the scattering of the incident
radiation plays a dominant role in determining the source functions.
The boundary conditions at the surfaces of modelled structures are
specified by the radiation coming from the surrounding solar
atmosphere (or possibly from other structures around) for all
directions and considered frequencies. The most important irradiation
comes from the underlying photosphere and chromosphere, and also in UV or EUV
lines and continua  from the chromosphere-corona TR and the
corona. The actual conditions depend on the geometry of the problem and
on the line or continuum transitions under consideration. However, once
the source function is determined and we want to compute the {\em
synthetic spectrum}, the formal solution of the radiative transfer equation has to be performed
along a prescribed LOS, typically the LOS along which
we observed the structure. In this case, only the incident radiation
in the direction of the LOS is considered in the boundary conditions of
the radiative transfer equation. Observing the prominence on the limb, the LOS directed towards the
observer contains the coronal radiation {emitted by material located} behind {and in front of} the prominence.
This may be relevant for some UV or EUV transitions, but not  for
'cool' lines {such as the hydrogen \La\ line. In fact, the \La\ line shows emission from the corona, but this is quite negligible in comparison to values of the prominence source function determined by the incident chromospheric and TR radiation.} On the disk, when
observing the filaments, the situation is different. In such case one
has to consider the radiation passing along the LOS from the solar surface below
the filament. This poses a great challenge because we don't see
directly this {region}, and thus it is difficult to estimate this incident radiation. This is a typical problem of the so-called cloud model. 
For Lyman lines, a possible solution is discussed by \citet{2006A&A...459..651S}.

Depending on the atomic model considered, the incident radiation for
various line and continuum transitions were specified by various
authors. In most cases, the observed specific radiation intensities are
used. Compiled data from various sources have been given \eg\ by
\citet{1993A&AS...99..513G} and \citet{2005A&A...442..331H} for
hydrogen, \citet{1992SoPh..138..123R} for hydrogen subordinate
continua, \citet{2001A&A...380..323L} and \citet{2007A&A...463.1171L}
for helium. These and complementary data files can be obtained from the
authors.

\subsection{Opacity and Emissivity in Spectral Lines}

We introduce the notion of the line absorption profile for a transition between the lower and upper atomic levels $i$ and $j$, respectively:
\begin{equation}
\phi_{\nu} = \frac{1}{\sqrt\pi  \Delta\nu_{\rm D} } H(a,x) \ .
\end{equation}
$H(a,x)$ is the Voigt function, $a$ is the damping parameter $a=a_i+a_j$ pertinent to the respective atomic levels $i$ and $j$, and $x=\Delta\nu / \Delta\nu_{\rm D}$ is the frequency displacement from line centre expressed in units of the Doppler width.
The Voigt function is a convolution between a Gaussian and a Lorentzian profile. This reflects the two dominant line broadening mechanisms: Doppler broadening in the line core, and collisional broadening in the wings of the line.
The absorption profile is normalised:
\begin{equation}
\int_0^\infty \phi_{\nu} \, d\nu =1 \, .
\end{equation}

The energy absorbed in the line is $E_j - E_i = h \nu_{ij}$, where $E_{i,j}$ is the excitation energy of a given atomic level, and $\nu_{ij}$ is the line centre frequency.

Using the Einstein coefficients for absorption ($B_{ij}$), spontaneous emission ($A_{ji}$), and stimulated emission ($B_{ji}$), the absorption coefficient corrected for stimulated emission is written as:
\begin{equation}
\chi_{\nu} = n_i B_{ij} \frac{h \nu_{ij}}{4 \pi} \phi_{\nu} - n_j B_{ji} \frac{h \nu_{ij}}{4 \pi} \psi_{\nu} \ ,
\end{equation}
and the emission coefficient as:
\begin{equation}
\eta_{\nu} = n_j A_{ji} \frac{h \nu_{ij}}{4 \pi} \psi_{\nu} \, .
\end{equation}
In these relations, $n_i$ and $n_j$ are the atomic level populations. The frequency dependence of the absorption and emission processes is given by the profiles $\phi_{\nu}$ and $\psi_{\nu}$ , respectively. In most practical cases one assumes that $\psi_{\nu} \equiv \phi_{\nu}$, and this approximation is called \textit{complete redistribution} (see Sect.~\ref{sec:scat}).

\subsection{Line Source Function}

As we have seen, the line source function is defined as the ratio of emission and absorption coefficients, \ie\
\begin{eqnarray}
	S_{\nu} & \equiv & \frac{\eta_\nu}{\chi_\nu} \label{eq:snu}\\
	& = & \frac{n_j A_{ji} \psi_{\nu}} {n_i B_{ij} \phi_{\nu} - n_j B_{ji} \psi_{\nu}} \\
	& \simeq & \frac{n_j A_{ji}}{n_i B_{ij} - n_j B_{ji}} \rho_{ij}(\nu)
\end{eqnarray}
with $\rho_{ij}(\nu) \equiv \psi_{\nu}/\phi_{\nu}$.                                                                              The term of
stimulated emission in the denominator has been simplified because of its small contribution.
This corresponds
to the so-called non-LTE situation typical for prominences
and filaments. The atomic level populations and $\rho_{ij}(\nu)$
depend on the radiation intensity and thus are coupled to the
radiative transfer equation.
Contrary to that, in LTE the level populations are
given by the Boltzmann distribution and the source function is simply
equal to the Planck function. Unfortunately, LTE is of no use in the physics of
prominences because of rather low plasma densities. Basic
concepts of the non-LTE physics are well explained in the textbook of
\citet{1978stat.book.....M}.

\subsection{Partially-Coherent Scattering in Prominence Plasmas}
\label{sec:scat}

In the general case of  photon scattering in spectral lines, the emission profile differs from the absorption one, and $\rho(\nu) \neq 1$. The emission profile $\psi_{\nu}$ is expressed in terms of the so-called {\em scattering integral} $\bar{J}$:
\begin{equation}
\label{eq:jbar}
\bar{J} = \int_{0}^{\infty} J_{\nu'} \phi_{\nu'} d\nu' \ ,
\end{equation}
\begin{equation}
	\psi_{\nu} = \frac{1}{\bar{J}} \int_0^\infty R_{\nu',\nu} J_{\nu'} \mathrm{d}\nu' \ ,
\label{eq:scatint}
\end{equation}
where $J_{\nu'}$ is the mean intensity at frequency $\nu'$, and $R_{\nu',\nu}$ is the {\em redistribution function}, i.e. the probability that the radiation absorbed at frequency $\nu'$ will be reemitted at frequency $\nu$. As we have mentioned
above, the complete redistribution (CRD) assumes that $\rho=1$.
In such a case, the redistribution function has the simple form $R_{\nu',\nu}=\phi_{\nu'}\phi_{\nu}$ and, when inserted into (\ref{eq:scatint}),
one gets $\psi_{\nu} \equiv \phi_{\nu}$. The photon frequencies are
completely uncorrelated in this case. The other extreme is purely
coherent scattering in which the scattered photon has the same frequency as the absorbed one. 

A realistic situation is partially-coherent scattering, somewhere between purely coherent scattering and CRD, and we thus call it \textit{partial redistribution} (PRD). For resonance lines (transitions between the ground state $g$ and an upper level $j$), one has in PRD:
\begin{eqnarray}
R_{\nu',\nu} & = & \gamma R_{II} +  (1 - \gamma) R_{III}\\
\gamma & = & \frac{A_{jg}}{A_{jg} + Q_{\rm E}} \, .
\end{eqnarray}
The function $R_{II}$ follows from purely coherent scattering in the
atom's frame, while $R_{III}$ reflects the complete redistribution in
the atom's frame due to elastic collisions having the rate $Q_{\rm E}$.
$\gamma$ is the branching ratio, i.e. the probability that the
coherence in the atom's frame is destroyed by collisional perturbation
of the upper atomic state. $R_{\nu',\nu}$ is the velocity-averaged
(i.e. in the observer's or laboratory frame) redistribution function,
here also averaged over all directions. The critical importance of PRD
for resonance lines like hydrogen \La\ emitted by quiescent
prominences was first clearly demonstrated by
\citet{1987A&A...183..351H}, although the necessity to consider PRD for
prominences was mentioned by several authors before
\citep[see][]{1978SoPh...57...27C,1983BAICz..34....1H}.
However, the subordinate lines like the hydrogen H$\alpha$ line which arise between two excited atomic levels can be well described by the CRD approximation -- the coherence is partially destroyed by the lower-level broadening \citep[{this is further discussed in}][]{1983BAICz..34....1H}.

1D and 2D non-LTE models of prominences and filaments, including their
fine structures (threads),  have been constructed using the
angle-averaged PRD for the first two Lyman lines (for higher members of the
Lyman series the coherence effects become less important and one can
use CRD). This is described in  Sects.~\ref{sec:9} and \ref{sec:10}. Within 1D models, PRD
was also used for strong resonance lines of helium
\citep{2001A&A...380..323L} and of Ca~II \citep{2002A&A...385..273G}. The emergent profiles computed with PRD
may significantly differ from those computed assuming CRD and this
substantially affects the resonance-line diagnostics. However, the resulting line profiles do reflect both the PRD scattering physics and the actual shape of the {incident line profiles} to be scattered. This can be
easily understood by inspecting the scattering integral in (\ref{eq:scatint}).
If the frequency distribution of the radiation field is flat enough
over the line absorption profile, then $\psi_{\nu} = \phi_{\nu}$ and we
get the CRD case (we call this special case 'natural excitation'). This
shows how critical is the real shape of $J_{\nu}$ in combination with
the redistribution function. Finally, let us mention that the line
scattering in real heterogeneous prominences is highly anisotropic, and
thus the angle-dependent redistribution functions should be considered
for even more realistic diagnostics. To our knowledge, no realistic
optically thick line transfer was performed for prominences using the
angle-dependent PRD.

\subsection{Formal Solution of the Radiative Transfer Equation in a Finite 1D Slab}

Here we consider two kinds of schematic 1D slabs of a finite geometrical thickness, oriented either vertically above the solar surface and irradiated symmetrically on both sides (the case of prominences seen on the limb, Fig.~\ref{fig:1d_rt}), or oriented horizontally to the solar surface and irradiated mainly from below (the case of filaments). More realistic geometries will be discussed in Sect.~\ref{sec:10}.

The formal solution of the radiative transfer equation gives us the outgoing radiation
intensity at the slab surface (i.e. for $\tau=0$) and in direction $\mu$:
\begin{equation}
	I(0,\mu) = I_0(\tau,\mu) \exp(-\tau/\mu) + \int_0^\tau S(t) \exp(-t/\mu) \mathrm{d}t/\mu,
\end{equation}
where $I_0(\tau,\mu)$ is the {\em incident radiation} on the opposite
side of the slab.

Assuming a constant source function, we get analytically:
\begin{equation}
I(0,\mu) = I_0(\tau,\mu) \exp(-\tau/\mu) +
S [1 - \exp(-\tau/\mu)].
\end{equation}
Two limiting cases are important:
\begin{eqnarray}
\tau \ll 1 & \Rightarrow & S [1 - \exp(-\tau/\mu)] \simeq S \tau/\mu\\
\tau \gg 1 & \Rightarrow & S [1 - \exp(-\tau/\mu)] \simeq S \, .
\end{eqnarray}
Now we can discuss the distinction between prominences and filaments.

For prominences on the limb, the spectral line is in emission ($I_0 = 0$, no background radiation considered), $\mu =1$, and so:
\begin{equation}
I(0) = S [1 - \exp(-\tau)] \, .
\end{equation}
In a special case of an {\em optically thin slab} we get:
\begin{equation}
I(0) \simeq S \tau = \eta D,
\end{equation}
where $\tau \ll 1$ and $D$ is the geometrical thickness of the slab. We use the fact that $S$ and $\eta$ are related by (\ref{eq:snu}), and $\tau = \chi D$.

However, we have to remember that $I_0$ from other directions drive the source function or, equivalently, $\eta$. This is the case in central cool parts, where the radiation scattering is the dominant process determining the source function.
On the other hand, inside the PCTR the temperature
is steeply increasing, emission lines become optically thin
and the
collisional excitation starts to dominate over the scattering.

For filaments on the disk, the spectral lines are typically in absorption, and for $\mu =1$ we get:
\begin{equation}
I(0) = I_0 \exp(-\tau) + S [1 - \exp(-\tau)] \, .
\end{equation}
Since the line source function in central cooler parts is generally controlled
by the photon scattering, we can approximately write:
\begin{equation}
S \simeq \frac{1}{2} I_0,
\end{equation}
where 1/2 is the {\em dilution factor} by which the incident
solar-disk radiation $I_0$ has to be multiplied because there is
roughly only one half of the prominence or filament surrounding from which the
incident radiation illuminates it (assuming  no radiation from the corona).
For filaments seen  against the disk we compute their
contrast as:
\begin{equation}
\frac{I(0)}{I_0} = \frac{1}{2} [1 + \exp(-\tau)]
\end{equation}
and in two limiting situations we get:
\begin{eqnarray}
\tau \ll 1 & \Rightarrow & \frac{I(0)}{I_0} \simeq 1\\
\tau \gg 1 & \Rightarrow & \frac{I(0)}{I_0} \simeq \frac{1}{2} \, .
\end{eqnarray}
For an optically thin filament, the line centre contrast approaches unity, and in the case of a large optical thickness, it becomes 1/2. This is why we can see the filaments as dark structures relative to the background chromosphere. Since $\tau$ decreases from the line centre towards the line wings, we don't see the filaments when shifting the narrow-band filter out of the line centre.

{To summarize}, the reason why we observe prominences on the limb in emission and
filaments on the disk in absorption is the following. The cool prominence
plasma absorbs the radiation coming from the solar disk and scatters
it in all directions. Because there is no coronal
background in 'cool' optical lines, we see on the limb only the
scattered radiation and the line is thus in emission. On the
other hand, the chromospheric background of the filament is the
absorption line which becomes even darker due to filament absorption.
The radiation scattered in the direction toward the observer represents
only a small fraction of the absorbed one and thus cannot compensate
for the absorption. We thus see filaments darker than the background
solar surface.

There is still another specific aspect of the radiation absorption by
prominences or filaments, which is now attracting more and more
attention with respect to prominence mass loading: this is  the absorption
of EUV line radiation by
the resonance continua of H~I, He~I and He~II (see Sect.~\ref{sec:6}).

\subsection{Statistical Equilibrium Equations}

The non-LTE source function is not known in advance and must be
computed by solving the transfer problem. The atomic level populations,
on which the absorption and emission coefficients depend, must be
computed using the {\em equations of statistical equilibrium}
which replace the Boltzmann equation valid in LTE. A general form
of the equations of statistical equilibrium is:
\begin{equation}
\frac{d n_i}{dt} = \sum n_j (R_{ji} + C_{ji}) - n_i \sum
(R_{ij} + C_{ij})
\end{equation}
\begin{equation}
\frac{d n_i}{dt} = \frac{\partial n_i}{\partial t} +
\frac{\partial n_i V}{\partial x} \, .
\label{eq:dni_dt}
\end{equation}
$R_{ij}$ are the radiative rates, those for absorption and stimulated
emission depend on the line and continuum radiation field.
$C_{ij} = n_{\rm e} \Omega_{ij}(T)$ are the collisional rates proportional to the electron density $n_{\rm e}$ and dependent on temperature $T$ through the function $\Omega_{ij}(T)$. {Note that the coefficients $C_{ij}$ are defined here in such a way that they include $n_\mathrm{e}$, unlike the definition (\ref{eq5}) in Sect.~\ref{sec:thinplasma}.}
The time-derivative on the left hand side of (\ref{eq:dni_dt}) splits into the local temporal variations of $n_i$ (\eg\ due to time-dependent heating processes) and the divergence of the flux of atoms in state $i$ ($V$ is the macroscopic flow velocity of the prominence plasma). Other equations to be used are the charge-conservation equation $\sum N_{k} Z_k = n_{\rm e}$, and the total particle number ($N$) evaluation which comes from the  equation of state for the gas pressure $p$:
\begin{equation}
p=NkT \ .
\end{equation}
Here $Z_k$ is the ionization degree of $k$-th species and $N=\sum N_k + n_{\rm e}$ ($N_k$ is the total density of atoms in a given ionization state $k$). Finally, knowing $N$, the electron density, and the atomic abundances together with the atomic masses, one can compute the gas density $\rho$ .

To be more specific, we write the line radiative rates in the form
$R_{ij} = B_{ij} \bar{J}_{ij}$ for absorption, $R_{ji}({\rm spont})
= A_{ji}$ for spontaneous emission and $R_{ji}({\rm stim}) = B_{ji}
\bar{J}_{ij}$ for stimulated emission. 
Then $R_{ji} = R_{ji}({\rm spont})+R_{ji}({\rm stim})${, and $\bar{J}_{ij}$ is the integrated spatially averaged {\em mean intensity} weighted by the absorption profile, defined by (\ref{eq:jbar}).}
This quantity tells us how many line photons
are actually absorbed from the mean radiation field, owing to the
frequency dependence of the absorption coefficient represented by the
line profile function $\phi_{\nu}$.

In the special case of the so-called 'two-level atom model', the equations of statistical equilibrium can be
written simply as:
\begin{equation}
n_1 B_{12} \bar{J}_{12} + n_1 C_{12} =
n_2 A_{21} + n_2 B_{21} \bar{J}_{12} + n_2 C_{21} \ ,
\end{equation}
where 1 and 2 refer to the lower and upper levels of the transition, respectively.
Combined with the expression for the line source function, we get, after some algebra and dropping the subscripts relative to the energy levels, the well-known formula:
\begin{equation}\label{LSouF}
    S=(1-\epsilon)\bar{J}+\epsilon B_{\nu_{0}}\ .
\end{equation}
In the typical case of a UV resonance line ($h\nu/kT\gg
1$), such as the \La\ line of hydrogen, $\epsilon$ can be expressed as:
\begin{equation}\label{destr}
    \epsilon\approx\frac{C_{21}}{(C_{21}+A_{21})}\, .
\end{equation}
It represents the probability of a photon destruction. At high densities, $\epsilon$ reaches unity, and the LTE conditions are achieved. On the other hand, at low densities, $\epsilon$ is very small, and thus large departures from LTE take place. This is the case for solar prominences. Taking the typical temperature and electron density values as $T=8000$~K and $n_{\rm e}=10^{10}$~cm$^{-3}$, we get $\epsilon\simeq 10^{-6}$ for the hydrogen \La\ line. This also means that roughly $10^{6}$ scatterings are needed before the photon is destroyed (thermalized).
This example demonstrates the necessity of using the non-LTE approach for prominence radiative transfer modelling.
A general formulation for both lines and continua within a multilevel atom
can be found in \citet{1978stat.book.....M}.

{The above formulation assumes a stationary state. Non-stationary radiative state of prominences was first considered by \cite{1980SoPh...67..351E}, who estimated the relaxation time needed to reach the statistical equilibrium for Lyman continuum photo-ionization/recombination. For typical prominence densities, this time is of the order of $1-2$ minutes or even longer for low-pressure structures. He finally concludes that these radiative relaxation times are comparable with the time scales of the observed variations of the fine structure in quiescent prominences \citep[\eg][]{1976SoPh...49..283E}. However, detailed time-dependent non-LTE modelling will be required to determine departures of the prominence plasma from stationary state. In this respect, the temperature relaxation towards the radiative-equilibrium state was recently considered by \cite{2007A&A...465.1041G} who also found rather long time scales (larger than $10^2$~s). }

\section{Multilevel Non-LTE Problems}
\label{sec:8}

The solution of the mutually coupled radiative transfer and statistical equilibrium equations (non-LTE problem) for multilevel atoms can be achieved only numerically, and various techniques have been developed for this task. {They are briefly reviewed in this subsection.}

For prominence non-LTE modelling, two types of methods have been extensively used in the seventies and eighties, namely the {\em complete linearization} method -- CL of \citet{1969ApJ...156..681A}, and the {\em equivalent two-level-atom} method -- ETLA described by \citet{1987nrt..book..135A}. The solution of the radiative transfer equation is achieved by the Feautrier method (differential form of the radiative transfer equation), or using its integral form. All these and auxiliary techniques are thoroughly described and discussed in the textbook of \citet{1978stat.book.....M}. 

CL means that all equations which have to be solved simultaneously (namely the radiative transfer and statistical equilibrium equations, but also others like \eg\ momentum balance and energy balance equations) are linearized and then solved as a system of linear algebraic equations.
Because of this, a Jacobi iterative solution is required which converges quadratically. The great advantage of this approach is that all equations are solved simultaneously, and the change in one variable automatically affects other relevant variables. On the other hand, ETLA uses a generalized form of the two-level-atom source function and solves the radiative transfer equation for it, transition by transition. Then the radiation fields thus obtained are used to iterate the statistical equilibrium equations. In this case, the radiative transfer and statistical equilibrium equations are not solved simultaneously for all transitions, but only for one at a time.
In the case of a strong coupling of atomic transitions, this method can
fail or may have convergence difficulties.

A CL-based code, developed for stellar atmospheres, was modified for
prominences in the form of 1D vertical slabs, and extensively used in a
series of papers by \citet[][hydrogen plus helium plasma]{1974ApJ...192..181H}, \citet[][hydrogen plus helium, magneto-hydrostatic equilibrium and radiative equilibrium]{1976ApJ...205..273H},
\citet[][other species included]{1978ApJ...221..677H},
\citet[][hydrogen Lyman continuum modelling]{1983ApJ...268..398H}. CL was
also used by \citet{1983SoPh...85..141F} who modelled the hydrogen
\La\ line. All these studies used the complete redistribution
approach for hydrogen Lyman lines, although \citet{1979phsp.coll...53M}
noticed that PRD may play a role for L$\alpha$.
Later, \citet{1987A&A...183..351H} used their version of CL to study
the hydrogen line formation in an isothermal isobaric 1D slab. In their work, PRD in Lyman lines was first used in a consistent way (with the help of intermediate ETLA iterations), which led to an excellent agreement with OSO-8 observations of {the} \La\ line.

Another prominence code, based on ETLA approach, was developed by Gouttebroze \citep[][and references therein]{2000SoPh..196..349G}, and after extensive tests against the results from CL, it was used to compute a large grid of 140 isothermal isobaric 1D slab models which still represents an important benchmark for other prominence modellers or for observers. A subset of these 'GHV' models was published by \citet{1993A&AS...99..513G}, and the whole set is available from the authors upon request. \cite{1994A&A...292..656H} then used all of those models to generate various correlations between prominence radiation and plasma parameters, which have proved to be very useful for prominence diagnostics.

An extensive prominence non-LTE modelling was also done by two groups in Kiev, namely by Morozhenko, Zharkova, and Yakovkin and Zel'dina. They studied the formation of lines of different elements, by using the integral method for multilevel atoms. Two pioneering attempts are worth mentioning. \citet{1968SvA....12...40Y} have first noticed that PRD might be of importance for hydrogen \La\ modelling in disk filaments, but their approach was rather schematic. Then, \citet{1970asas.book..176M} considered a multi-slab model to account for the prominence fine structures. This was later developed by \citet{1989HvaOB..13..331Z}, who took into account the mutual radiative interaction of individual slabs \citep[see also][]{1989HvaOB..13..317H}. Multi-slab models (sometimes called multi-thread models, according to the fact that we frequently observe many fine-structure threads inside the prominence body)  have also been used, without mutual radiative interaction,  by \citet{1996ApJ...466..496F} and by others. Without radiative interaction, one slab is first modelled in detail, and then several (usually identical) slabs are put together. The emergent synthetic spectrum is finally computed by a formal solution along the LOS through all slabs.
This is discussed in more details in Sect.~\ref{sec:9.1}.

A principally different approach to solve the multilevel non-LTE problem is based on the so-called {\em Accelerated Lambda Iteration} or ALI technique,  widely used in stellar atmospheric modelling since the end of eighties and beginning of nineties. ALI and related methods have proven to be extremely efficient, namely for {multi}-level atoms with complex line and continuum transitions. We recommend the review by \citet{2003ASPC..288...17H}. A specific approach which is now used in some prominence non-LTE codes is called the {\em preconditioning} of the statistical equilibrium equations, and is thoroughly discussed by \citet{1991A&A...245..171R,1992A&A...262..209R}. It was first incorporated into prominence codes by \citet{1995A&A...299..563H} and \citet{1995A&A...302..587P}, and later by \citet{2000SoPh..196..349G}. In stellar atmospheric codes, it is used in connection with CL \citep[hybrid code TLUSTY of][]{1995ApJ...439..875H}, or in connection with ETLA \citep[PANDORA of][]{1992ASPC...26..489A}.
The preconditioning of Rybicki and Hummer was named as MALI (Multilevel ALI) method. Combined with the necessary linearization to account for non-linear terms in the statistical equilibrium equations, it represents a robust tool for prominence non-LTE modelling \citep{1995A&A...299..563H,1995A&A...302..587P}. Using purely the MALI technique, no problems have been met with the convergence, even when some transitions like the hydrogen \La\ line are extremely optically thick (say up to $10^6$) and others are optically thin. However, by adding other constraint equations (such as MHD or energy balance), it is possible that one should have to solve the radiative transfer equation in the strongest transitions (which are most coupled to the plasma thermodynamic conditions) by CL or ETLA, and then combine this with MALI for other less critical transitions.
This 'hybrid' approach is now used in some stellar atmospheric codes.
Within MALI, further sophistications like various types of accelerations or multigrids are now being applied \citep[see the recent work by][]{2009A&A...498..869L}.

Several sophisticated non-LTE codes have been developed and are used for
various types of prominence modelling and spectral diagnostics. The most important
ones which are currently used or are of potential
use in prominence physics are the following: IAS-code of Gouttebroze (1D, 2D, ETLA, MALI, PRD),
Ond\v{r}ejov-code of Heinzel (1D, 2D, MALI, linearization, PRD, radiation magneto-hydrostatic),
and the code of Paletou (1D, 2D, MALI, PRD). Of potential use are also
the codes  PANDORA of Avrett and Loeser (1D, ETLA, MALI, PRD, multielement) or
TLUSTY of Hubeny (1D, hybrid, PRD, multielement). These codes are generally available to
potential users\footnote{The IAS prominence code was described in
\citet{2000SoPh..196..349G} and is available at
\texttt{http://www.ias.u-psud.fr/pperso/pgoutteb/RTC/rtc.html}.}.

\section{Results from 1D Non-LTE Modelling}
\label{sec:9}

One can ask oneself why it is necessary to build and use a heavy computing machinery such as non-LTE radiative transfer calculations in order to get some atmospheric information through comparison with radiative observables. The answer lies in the very complex processes of line and continua formation, and in particular when the solar feature (such as a prominence in our case) is optically thick in some lines and continua. In this case, the atomic level populations are non-linearly coupled to the radiation field which has a non-local character extending to the whole medium. A well known example is the \Ha\  line which may be considered as optically thin in some cases, but is inevitably linked to the \La\ and \Lb\ transitions which are definitely optically thick in prominences. Another answer lies in the importance of \eg\ the Lyman lines in the radiative budget of prominences.
Consequently, the direct  inversion of physical parameters from the observed radiation (as described in Sects.~\ref{sec:3} and \ref{sec:4}) is an impossible task, which leads to the necessity of extensive forward modelling through the solution of the non-LTE problem (as described in Sects.~\ref{sec:7} and \ref{sec:8}). 

In this section, we review the main results obtained from 1D non-LTE computations. {In spite of its intrinsic limitations, so obvious from the point of view of fine-structure observations, we put some emphasis on 1D modelling for the following reasons. On one hand, 1D modelling is an historical milestone since it was performed at a time when observations were limited in terms of spatial and spectral resolution. On the other hand, the progress made in the frame of 1D allows to shed some light on the continuous improvement of modelling in relation with the physics of the radiation (\eg\ PRD \textit{vs} CRD) and the availability of new UV line profiles.} We start by presenting the non-LTE modelling of the hydrogen lines and continua (Sect.~\ref{sec:9.1}), followed by the helium lines and continua (Sect.~\ref{sec:9.2}), and then the spectra from ionized calcium (Sect.~\ref{sec:9.3}). We also devote some space to the case of moving prominences (such as active and eruptive prominences), as the radial component of the prominence plasma will cause a Doppler effect which has significant consequences on the  radiation emitted by the prominence (Sect.~\ref{sec:9.4}).
Results from 2D non-LTE modelling are reviewed in Sect.~\ref{sec:10}.

In the following, we will refer to the hydrogen Lyman series as the Lyman series, unless otherwise stated.

\subsection{Hydrogen Lines and Continua}
\label{sec:9.1}

\subsubsection{Early Observations and Modelling}
\label{sec91:early}

\begin{figure}
	\center
\includegraphics[width=\textwidth]{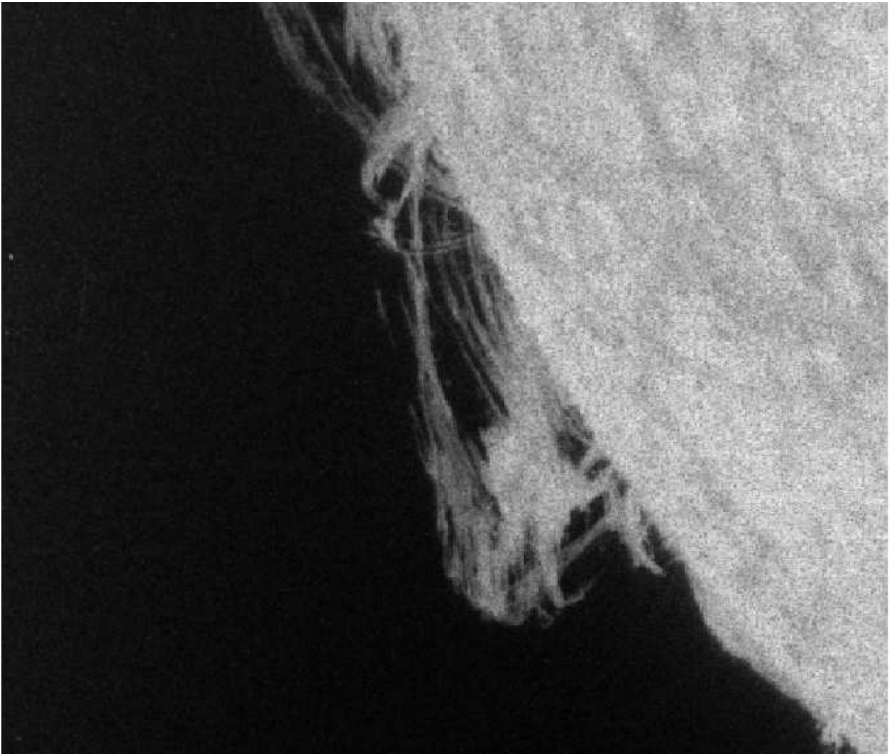}
\caption{\La\ prominence {(3 July 1979)} observed with the Transition Region Camera (TRC). From \cite{1980ApJ...237L..47B}.}
\label{fig:TRC}
\end{figure}

The Lyman series {was} discovered in the solar chromosphere as early as 1960 with the first \La\ spectrum obtained from a rocket by \cite{1960JGR....65..370P} and analyzed by \cite{1961ApJ...133..596M}. The spectrum of the \Lb\ line (more difficult to obtain because located below the 1150 \AA\ cut-off of magnesium fluoride coatings) was recorded by \cite{1965AnAp...28..755T}. 
No prominence spectrum in the Lyman series was obtained. The attempts to measure it with Skylab and HRTS \citep{1979SoPh...61..319M} were hampered by the fact that the \La\ line was too bright for the exposure time used. This is no surprise when one looks at the beautiful \La\ pictures of the Transition Region Camera \citep[][Fig.~\ref{fig:TRC}]{1980ApJ...237L..47B}, and VAULT \citep[][Fig.~\ref{fig:vault}]{2001SoPh..200...63K}, since the prominence \La\ intensity is about the quiet Sun intensity multiplied by a factor of $0.3-0.4$, roughly the value of the dilution factor.

At the end of the 1970s, the non-LTE modelling was ahead of the observations \citep{1976ApJ...205..273H,1976ApJ...210..827H,1978ApJ...221..677H}. Earlier work by \cite{1963PASJ...15..122H}  and \cite{1971SoPh...19..401P}  had shown the importance of the incident (UV and EUV) radiation for ionizing the prominence plasma. The same was valid for the excitation of the atomic levels, especially in the Lyman series. It was realized that with reasonable 1D models, the Lyman series was optically thick, with $\tau$ up to $10^{6}$ for \La. Such an important opacity combined with the strength of the incident radiation as compared to the local (thermal) radiation field in these lines was a natural explanation for the (predicted) bright appearance of prominences in the Lyman series. Moreover, numerical efforts were already performed \citep{1978SoPh...58...47M} to treat the multi-slab radiative transfer within a set of tiny structures (such as threads) in the frame of a two-level atom with application to the \Ha\ line.
As for the Lyman series,  \cite{1964SvA.....8..262Y} and \cite{1968SvA....12...40Y} considered the excitation and ionization from levels 1 and 2, derived electron densities (as proportional to the square root of the population of level 2) and finally produced emergent \La\ profiles for which the frequency variation of the scattering was properly taken into account in the core and the wings of the line.

\begin{figure}
	\center
\includegraphics[width=\textwidth]{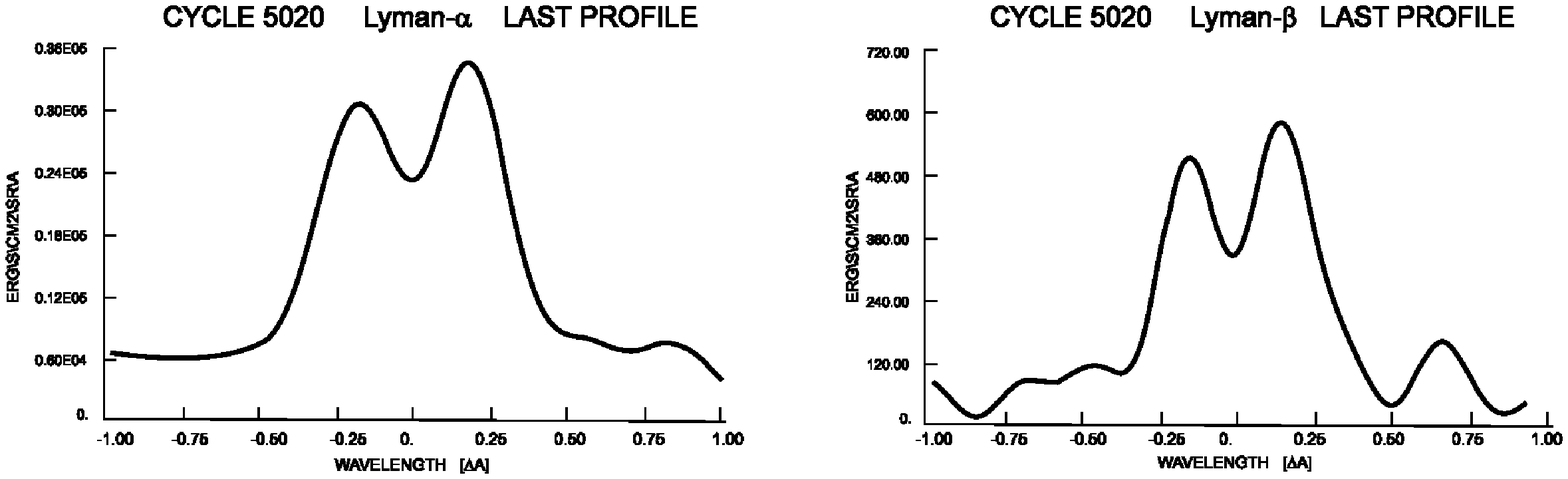}
\caption{Observed  \La\ and \Lb\ prominence profiles. From \cite{1982ApJ...254..780V}.\label{fig:oso8}}
\end{figure}

The first (photoelectric) prominence spectral observations in \La\ along with \Lb, Ca~II, and Mg~II  lines came from OSO-8 \citep[][and Fig.~\ref{fig:oso8}]{1982ApJ...253..330V}. Earlier on, \cite{1978ApJ...220.1001M}  had published a seminal paper on 2D non-LTE transfer in an externally illuminated structure. Their code allowed \cite{1982ApJ...254..780V}  to compare \La, Mg~II h and k, and Ca~II H and K computed profiles to observed ones (see Sect.~\ref{sec:10}).
A two-level atom was assumed and the ionization was treated with the assumption of photo-ionization only but the OSO-8 incident radiation was properly taken into account. This allowed the author to build a reasonable model and evaluate the 2D (boundary) effects. 

\begin{figure}
	\center
\includegraphics[width=\textwidth]{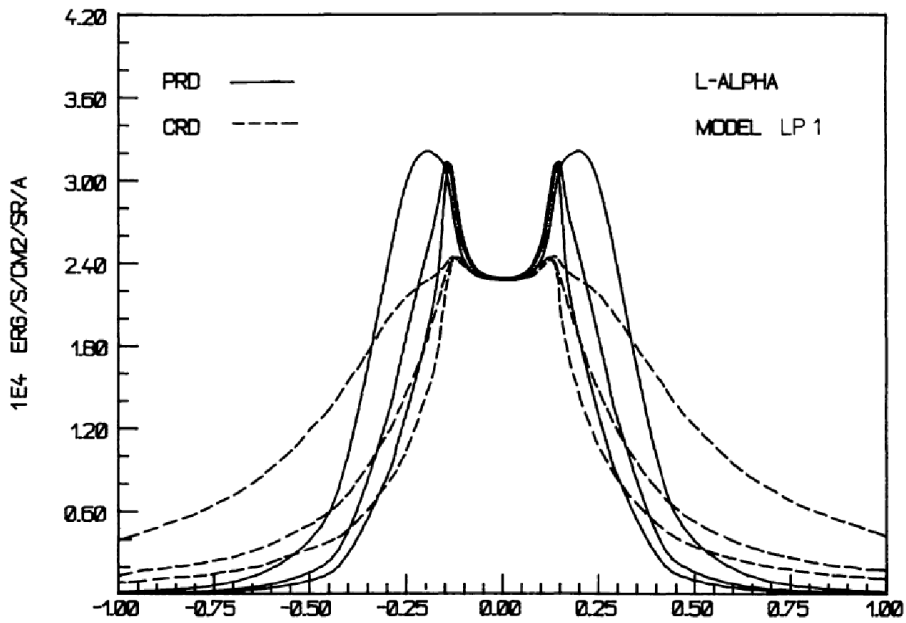}
\caption{\La\ line profiles emergent from a 1D prominence model at three angles. PRD in full line and CRD in dashed line. From HGV1.\label{fig:prdcrd}}
\end{figure}

At this time, the issue of PRD had already been addressed for the case of stellar atmospheres \citep{1962MNRAS.125...21H,1969MNRAS.145...95H}, but no clear way for handling the problem of collisions had been found before \cite{1972ApJ...175..185O} . 
A first treatment in the frame of a 1D prominence modelling was applied by \cite{1979phsp.coll...53M}  who evaluated the influence of PRD on the ratio of \Ha\ to \La\ intensities. \cite{1983A&A...121..155H}  also discussed the influence of PRD on integrated intensities of the \La\ line, and compared with OSO-8 observations. 
Later on, \citet[][HGV1]{1987A&A...183..351H} performed the first realistic modelling with emphasis on the \La\ profile whose detailed shape is very sensitive to PRD (Fig.~\ref{fig:prdcrd}). They showed that PRD not only lowered the far wings of the \La\ line but also had a visible influence on the near wings, namely a reproduction of the incident \La\ profile.

\subsubsection{Computed vs. Observed Line Profiles and Continuum Intensities}
\label{sec91:heasley}

 Since then, a series of modelling efforts was produced by Gouttebroze, Heinzel, and Vial, epitomized by the GHV paper \citep{1993A&AS...99..513G} which, among other observables, provided emergent line profiles. It should be pointed out that the basic work of Heasley, Mihalas and Milkey mentioned above built 1D models on the basis of prominence integrated intensities only (with the assumption of a flat incident spectrum), and did not provide the spectral signatures in terms of UV line profiles of the various models used.
There are a few reasons for that: the incident (chromospheric and coronal) line profiles were not well known, and the proper treatment of the line radiation scattering (see Sect.~\ref{sec:8}) was not easy. Moreover, there were only a few observed prominence spectra to compare model products with.

\begin{figure}
	\center
\includegraphics[width=\textwidth]{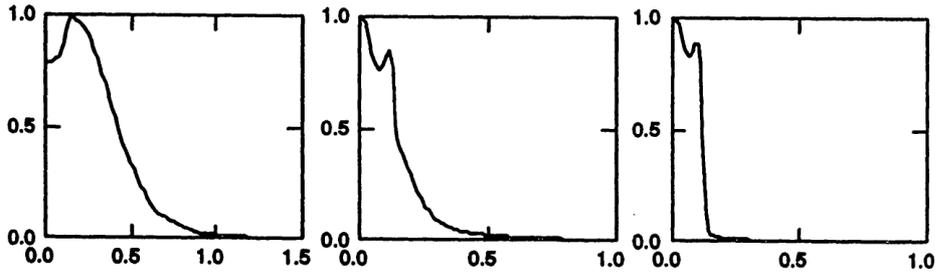}
\caption{\La, \Lb\ and \Lg\ line half-profiles emergent from a 1D prominence model characterised by its temperature (6000 K), its pressure (0.05~dyn cm$^{-2}$), its geometrical thickness (5000~km) and microturbulence (5~\kmps). Abscissae are in \AA. The line profiles are normalised to the value of the maximum intensity, given by GHV in a table reproduced here in Table~\ref{tab:GHVTable}. From GHV.\label{fig:GHVProfiles}}
\end{figure}

\begin{figure}
	\center
\includegraphics[width=\textwidth]{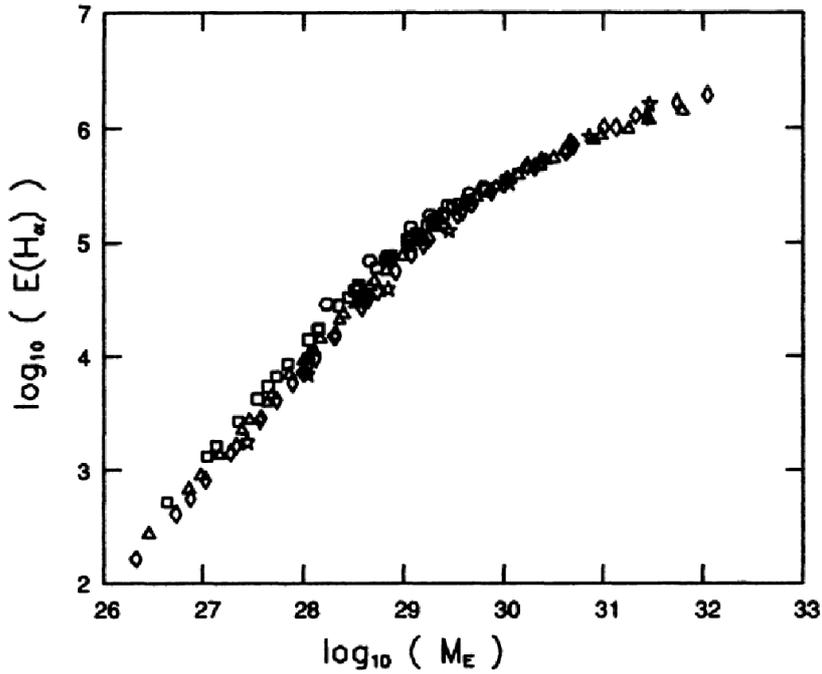}
\caption{Correlation between the \Ha\ intensity and the emission measure (in CGS units). From GHV.\label{fig:GHVHalphaEM}}
\end{figure}

\begin{figure}
	\center
\includegraphics[width=\textwidth]{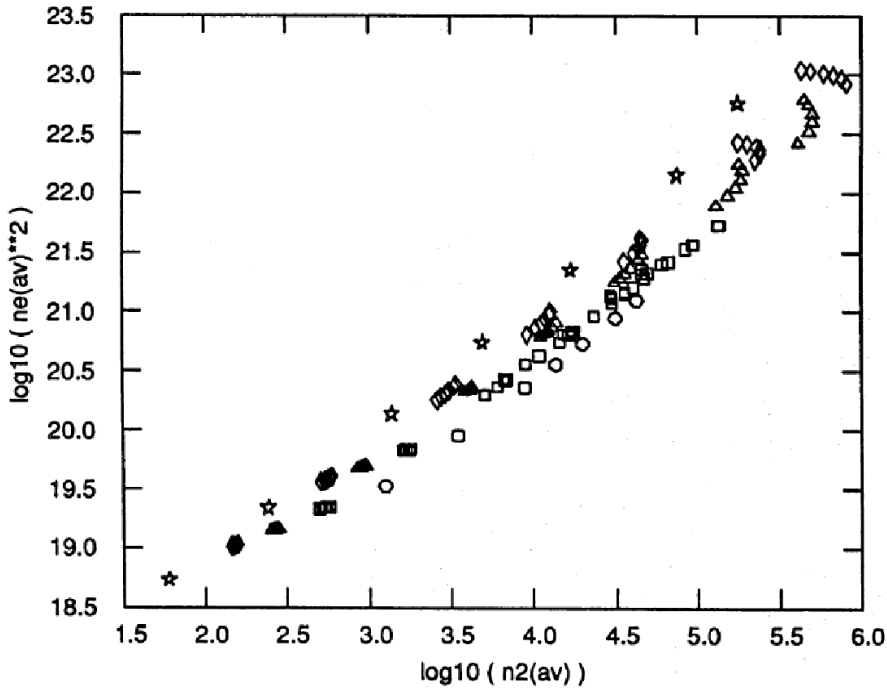}
\caption{Correlation between the square of the electron density and the density of level 2 (in CGS units). {From \cite{1994A&A...292..656H}}.\label{fig:HGV2n2ne2}}
\end{figure}

Working on a set of 140 1D models, GHV were able to produce the corresponding set of observables (especially the Lyman line profiles and continua, see Fig.~\ref{fig:GHVProfiles} and Table~\ref{tab:GHVTable}) comparable with new profiles provided by OSO-8 and by the {Ultraviolet Spectrometer and Polarimeter \citep[UVSP,][]{1980SoPh...65...73W} on the Solar Maximum Mission (SMM) spacecraft}.
\begin{table}
      \caption{Table of radiative quantities computed from the same model as in Fig.~\ref{fig:GHVProfiles}. The six columns give: lower level of the transition; upper level; optical {thickness} of the slab at line centre; maximum intensity of the profile in erg~cm$^{-2}$~s$^{-1}$~sr$^{-1}$~Hz$^{-1}$; integrated intensity in erg~cm$^{-2}$~s$^{-1}$~sr$^{-1}$; full width at half-maximum (in \AA).  From GHV.}\label{tab:GHVTable}
      \begin{tabular}{ccccccc}
      \hline\noalign{\smallskip}
      \multirow{10}{3mm}{\begin{sideways}hydrogen lines\end{sideways}}
      & LL & UL & $\tau$ & $I_{\rm max}$ & $E_{\rm tot}$ & FWHM \\
      \noalign{\smallskip}\hline\noalign{\smallskip}
      & 1 & 2 & 8.60E+05 & 1.56E-08 & 2.74E+04 & 0.84 \\
      & 1 & 3 & 1.38E+05 & 8.19E-11 & 8.07E+01 & 0.29 \\
      & 1 & 4 & 4.79E+04 & 1.63E-11 & 1.17E+01 & 0.24 \\
      & 2 & 3 & 1.72E+00 & 3.00E-06 & 1.14E+05 & 0.53 \\
      & 2 & 4 & 2.37E-01 & 3.53E-07 & 1.49E+04 & 0.32 \\
      & 3 & 4 & 3.09E-02 & 4.17E-07 & 4.43E+03 & 1.17 \\
      & 2 & 5 & 7.94E-02 & 1.16E-07 & 5.39E+03 & 0.27 \\
      & 2 & 6 & 3.71E-02 & 7.02E-08 & 3.43E+03 & 0.26 \\
      & 2 & 7 & 2.07E-02 & 3.50E-08 & 1.77E+03 & 0.25 \\
      \noalign{\smallskip}\hline
      \end{tabular}
\end{table}
This was done with a proper account of the incident radiation profiles \citep[taken from OSO-8 for \La\ and \Lb, and properly scaled for the rest of the Lyman series, see][]{1978ApJ...225..655G}, and a rigorous treatment of the diffusion within the line profiles on the basis of works by \cite{1969MNRAS.145...95H,1973ApJ...185..709M,1985A&A...145..461H}; and \cite{1987A&A...183..351H}. An important feature of this modelling was the use of a 20-level plus continuum atom, which allowed to predict a set of Lyman profiles more complete than what could be actually observed with the then available spectral instrumentation.
They also derived some correlations connecting the observables to the thermodynamic parameters (\eg\ the \Ha\ line intensity to the emission measure, see Fig.~\ref{fig:GHVHalphaEM}).
Another example is the color temperature of the Lyman continuum, which was found representative (at not too high temperatures) of the electron temperature, a confirmation of the result of \cite{1983ApJ...268..398H}. 
Of course, one should also keep in mind the limits and assumptions behind the GHV models (1D, static, isothermal, isobaric). But these correlations \citep{1994A&A...292..656H} provided a unifying thread for combined well-chosen multi-wavelength observations and a valuable  tool for their interpretation. For instance, the above-mentioned correlation between \Ha\ intensity and the emission measure {(see Sect.~\ref{sec_em})} is simply explained by the quasi-constancy of the term $n_\mathrm{e}^2/n_{2}$ (Fig.~\ref{fig:HGV2n2ne2}). 
The two quantities are related to the photo-ionization and recombination terms from and to the second energy level of the hydrogen atom, respectively.

These basic 1D models describe the central cool parts of the prominence or its fine structures reasonably well. A prominence whose physical properties vary with altitude may be simulated by a sequence of individual models. For instance, \citet{2000SoPh..196..349G} computed 20 models to simulate a prominence with a vertical extension of $10^5$~km, a temperature increasing from 5000~K at the bottom to 10000~K at the top, and a pressure decreasing from 0.2 to 0.02~dyn~cm$^{-2}$. 
Other parameters were kept constant. These authors showed how the hydrogen \La\ line profile varies from the bottom to the top of the prominence. Near the bottom, the prominence is very optically thick in \La\ and the emission process is dominated by the scattering of the incident radiation. As the temperature increases and the pressure decreases, the optical thickness decreases. The incident radiation also decreases as the effect of altitude. 
Therefore, the scattering decreases and the thermal emission of the slab becomes more important at the top of the prominence. As a consequence, the line profile is relatively broad and flat at the bottom of the prominence (similar to the incident profile). As altitude increases, the profile is transformed into a thermal emission profile with sharp peaks and low wings. It is worth stressing that as the altitude of the prominence is changed, the boundary conditions of radiative transfer are changed, since the incident radiation is diluted with height.

{Using a simple non-LTE radiative transfer code, \cite{1992A&A...260..419W} derived electron densities $1-5\times 10^{10}$~\cc\   from \Ha\ observations of a prominence observed at Pic du Midi with the MSDP. This corresponds to a gas pressure of $\sim 0.01-0.1$~dyn cm $^{-2}$.
\cite{1996SoPh..164..211H} compared \Ha\ observations with results from GHV and derived electron densities in very good agreement with measurements from \citet[$2.5\times10^9-6.3\times10^{10}$~\cc]{1994SoPh..154..231B}. This enabled them to obtain values for the geometrical thickness ranging between a few hundreds~km up to a few $10^4$~km in different prominences, corresponding to a fairly constant total column mass of $10^{-5}$~g cm$^{-2}$.
\cite{2001A&A...370..281H}, using a more elaborate 1D code  and including a PCTR, derived a constant gas pressure around  0.2~dyn cm$^{-2}$ for reversed profiles and a variation from 0.12 to 0.04~dyn cm$^{-2}$ for non reversed profiles from Lyman lines spectra obtained by SUMER. This corresponds to slightly higher electron densities than in the \cite{1992A&A...260..419W} study.
An other approach  to derive physical parameters of prominences  has been developed recently based on the absorption mechanism of coronal radiation by H, He~I, and He~II continua (see Sect.~\ref{sec:EUVAbsorptionMass}). The opacity derived from observations of EUV lines (TRACE, Hinode/EIS) was consistent with models of prominences having an electron density of the order of $ 10^{11}$~\cc\ with typical parameters for the temperature ($6000-8000$~K) and a thickness of $1-5 \times 10^{3}$~km.  These values yield electron densities in prominences in the range $10^{10}-10^{11}$~\cc\ \citep{2008ApJ...686.1383H}.}

{Let us now discuss the observational material concerning lines and continua.} In the eighties, the only Lyman spectroscopic results came from OSO-8 (\La\ and \Lb) and UVSP \citep[][\La]{1988ApJ...329..464F}. There was a debate about the \La\ reversal, a feature which is about 0.4 \AA\ wide (peak-to-peak) but is perturbed by the geocorona absorption when the observing spacecraft is in low orbit.  Moreover, contrary to the \La\ profile where the geocorona absorption, about 30 m\AA\ wide, is consequently resolved with the OSO-8 20 m\AA\ spectral resolution, the \Lb\ profile, as observed with OSO-8, was obtained with a lower resolution (60 m\AA) which did not resolve the geocorona absorption. This was not of minor importance, since non-LTE modelling always provided a strong disagreement between \La\ and \Lb\ intensities, the \Lb\ computed intensities being much lower than the observed ones (HGV1), a feature also present in the solar chromosphere (see VAL and FAL models)\footnote{The most sophisticated computations of \cite{1995ApJ...455..376H}  led to some (minor) increase in the \Lb\ wings only.}.

\subsubsection{Fine-Structure of Modelled Prominences}

In order to overcome such a discrepancy between \La\ and \Lb\ intensities, several solutions, apart from the graft of a PCTR (see below), were proposed,  most dealing with the concept of fine structuring of prominences and well supported by observations as early as 1960 {\citep{1961PhDT.........1D}}. Nowadays, SST observations show structures down to 0.1\arcsec\ resolution. Various approaches were considered as early as 1978: \cite{1978SoPh...58...47M}  introduced the concept of  \textit{filamentation degree} and considered a 1D multi-slab prominence where the mutual interaction between (isothermal) threads was taken into account. 
On the other hand, \cite{1983SoPh...85..141F,1985SoPh...96...53F},  followed by \cite{1990LNP...363..282V},  considered 1D threads  with a thermodynamic structure determined by a simple energy equation, including conductive energy transport and radiative losses. 
For a discussion of the pros and cons of the two approaches, see \cite{1990LNP...363..279H}  who determined heuristically the effect of the mutual radiative interaction upon the exact shape of the \La, \Lb, and \Ha\ profiles. 
This then led to a systematic modelling of threads where ambipolar diffusion (\ie\ the diffusion of neutral hydrogen atoms with respect to the protons) between the cool core and the ambient corona was taken into account \citep[][FRVG]{1996ApJ...466..496F}. Also note that PRD was taken into account. Agreement between some models and OSO-8 observations was obtained but the high necessary number of threads (about one hundred along any LOS) could be questioned (see Table~\ref{tab:frvg}).
\begin{table}
  \caption{Intensities of \La, \Lb, and \Ha\ (in erg s$^{-1}$ cm$^{-2}$ sr$^{-1}$) for various numbers of threads and different models compared with observed OSO-8 integrated intensities. From FRVG. \label{tab:frvg}}
      \begin{tabular}{cccccc}
      \hline\noalign{\smallskip}
      Number of Threads & Ly$\alpha$(10$^{4}$) & Ly$\beta$ & Ly$\alpha$/Ly$\beta$ & H$\alpha$(10$^{4}$) & H$\alpha$/Ly$\alpha$\\
      \noalign{\smallskip}\hline\noalign{\smallskip}
      \multicolumn{6}{c}{Model A: $p=0.02$ (dyn cm$^{-2}$)} \\
      \noalign{\smallskip}\hline\noalign{\smallskip}
      1 & 1.08 & 587 & 18 & 0.02 & 0.02\\
      10 & 1.47 & 805 & 18 & 0.21 & 0.14\\
      100 & 2.04 & 1100 & 18.5 & 1.99 & 0.97\\
      \noalign{\smallskip}\hline\noalign{\smallskip}
      \multicolumn{6}{c}{Model B: $p=0.05$ (dyn cm$^{-2}$)} \\
      \noalign{\smallskip}\hline\noalign{\smallskip}
      1 & 1.35 & 1780 & 7.5 & 0.06 & 0.04\\
      10 & 1.8 & 2419 & 7.5 & 0.56 & 0.31\\
      100 & 2.47 & 3290 & 7.5 & 4.81 & 1.95\\
      \noalign{\smallskip}\hline\noalign{\smallskip}
      \multicolumn{6}{c}{Model C: $p=0.1$ (dyn cm$^{-2}$)} \\
      \noalign{\smallskip}\hline\noalign{\smallskip}
      1 & 1.76 & 4352 & 4 & 0.13 & 0.08\\
      10 & 2.34 & 5887 & 4 & 1.28 & 0.55\\
      100 & 3.12 & 8003 & 4 & 9.16 & 2.94\\
      \noalign{\smallskip}\hline\noalign{\smallskip}
      \multicolumn{6}{c}{Model D: $p=0.2$ (dyn cm$^{-2}$)} \\
      \noalign{\smallskip}\hline\noalign{\smallskip}
      1 & 2.62 & 10560 & 2.5 & 0.33 & 0.12\\
      10 & 3.43 & 14240 & 2.4 & 2.99 & 0.87\\
      100 & 4.42 & 19350 & 2.3 & 14.9 & 3.4\\
      \noalign{\smallskip}\hline\noalign{\smallskip}
      \multicolumn{6}{c}{Model with $p=0.1$ (dyn cm$^{-2}$) and Cold Isothermal Core (35~km)}\\
      \noalign{\smallskip}\hline\noalign{\smallskip}
      1 & 1.71 & 4127 & 4.1 & 0.33 & 0.19\\
      10 & 2.94 & 5670 & 5.2 & 6.29 & 2.14\\
      100 & 3.86 & 7554 & 5.1 & 18.3 & 4.74\\
      \noalign{\smallskip}\hline\noalign{\smallskip}
      \multicolumn{6}{c}{Model with $p=0.1$ (dyn cm$^{-2}$) and Cold Isothermal Core (45~km)}\\
      \noalign{\smallskip}\hline\noalign{\smallskip}
      1 & 1.97 & 4368 & 4.5 & 1.01 & 0.51\\
      10 & 3.01 & 5563 & 5.4 & 7.5 & 2.49\\
      100 & 3.86 & 7362 & 5.2 & 19.2 & 4.97\\
      \noalign{\smallskip}\hline\noalign{\smallskip}
      \multicolumn{6}{c}{GHV Model with $p=0.1$ (dyn cm$^{-2}$) and $T=6000$ K}\\
      \noalign{\smallskip}\hline\noalign{\smallskip}
      $D=200$~km & 1.63 & 50 & 326 & 1.77 & 1.1\\
      $D=1000$~km & 2.48 & 66.5 & 373 & 7.15 & 2.9\\
      \noalign{\smallskip}\hline\noalign{\smallskip}
      \multicolumn{6}{c}{Observed}\\
      \noalign{\smallskip}\hline\noalign{\smallskip}
      OSO 8 & $2.8-3.6$ & $440-550$ & 65 & $1-10$ & $0.25-2$\\
      \noalign{\smallskip}\hline
      \end{tabular}
\end{table}

Along the classical approach of 1D monolithic (slab) models, another improvement was performed with the graft of a PCTR on top of the isothermal isobaric models \citep{1988dssp.conf...71H}. This was an important step towards the modelling of realistic structures while keeping the full complexity of atomic physics and radiative transfer.
Even though new data in the Lyman series was not expected until SOHO, the modelling (and its Lyman radiative signatures) made impressive progress (see Sect.~\ref{sec:10}).

\begin{figure}
	\center
\includegraphics[width=\textwidth]{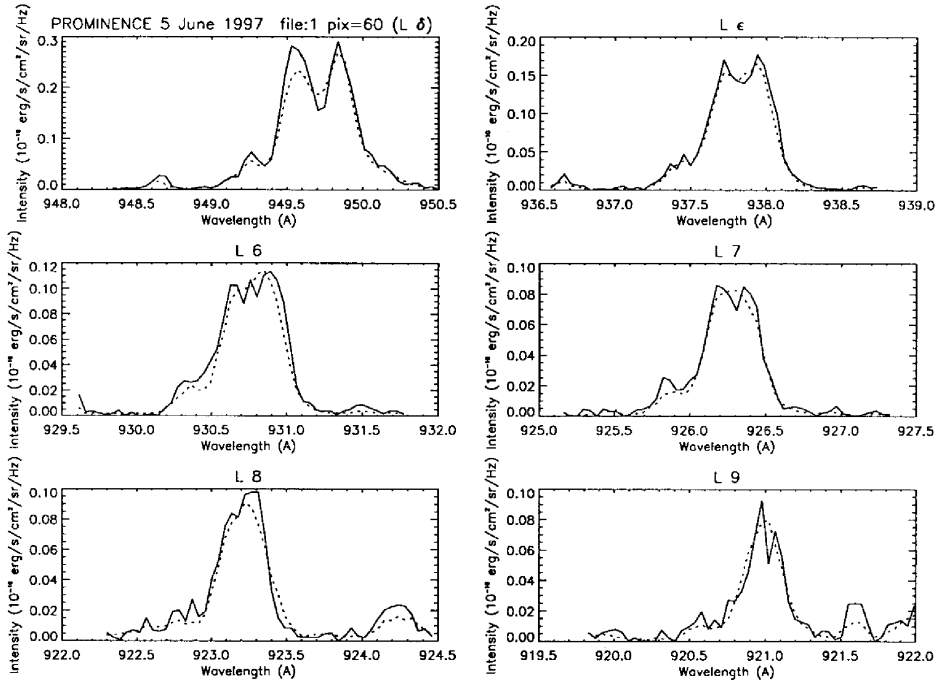}
\caption{L4 to L9 prominence profiles at a given position (solid line) and averaged over the SUMER slit (dashed line). From \cite{1999SoPh..189..109S}.\label{fig:schmieder99}}
\end{figure}

Another path was followed by \cite{1998SoPh..179...75A}  who noticed a strong discrepancy between the physical parameters of the Kippenhahn-Schl\"uter model and the parameters derived from non-LTE analysis. Later on, this led them to develop a 2D multi-component model (Sect.~\ref{sec:10}) whose spectral outputs were compared to incoming SUMER observations of the Lyman series.
Rather rapidly, the Lyman series profiles available in prominences and filaments were confronted to model predictions. From SUMER, \cite{1999SoPh..189..109S}  compared the observed L4 to L9 profiles in a quiescent prominence (Fig.~\ref{fig:schmieder99}) with the profiles provided by three different classes of models: the (isobaric, isothermal) GHV models, the filamentary FRVG models, and finally a superposition of GHV-type models with very small thickness which represent an elementary filamentary structure. 
One can repeat two of the conclusions of the paper: 1) all Lyman lines seem to be formed at the base of the PCTR, and 2) some temperature gradient corresponding to a PCTR is needed to explain the behaviour of several higher Lyman lines. 
It should be pointed out that the modelling used the up-to-date values of incident Lyman intensities provided by SUMER.

\begin{figure}
	\center
\includegraphics[width=\textwidth]{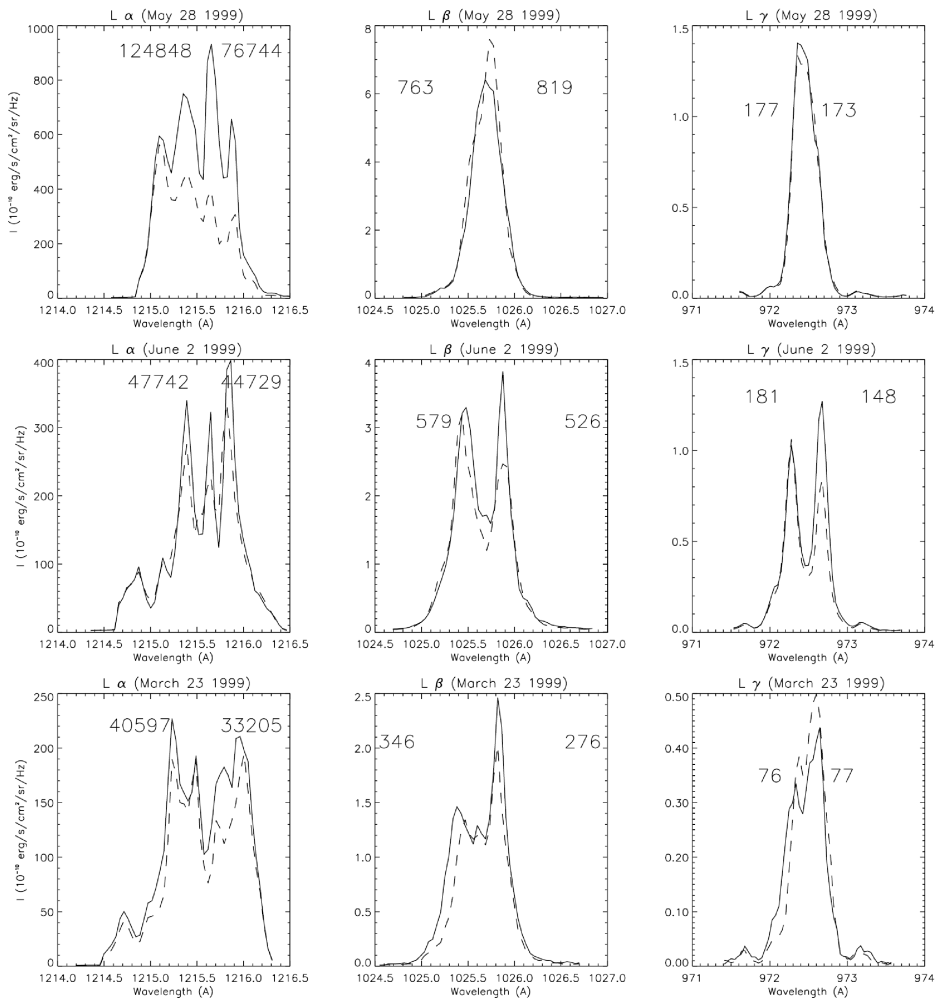}
\caption{\La, \Lb, and \Lg\ profiles {observed by SUMER in} 3 different prominences. Note the large asymmetries in the third case. From \cite{2001A&A...370..281H}.\label{fig:Heinzel01}}
\end{figure}

Similar observations had also been obtained in a filament by \cite{1998SoPh..181..309S}  for L4 to L7 lines. A 1D modelling including various types of PCTR was performed, which validated the concept that a thin PCTR is necessary {in order to obtain Lyman profiles which are not too reversed.} However, the L4 profile was not well reproduced.
L4 to L8 line profiles were observed in the pre-eruption and \textit{disparition brusque} phase of a filament by \cite{2000A&A...358..728S}, and the authors found an agreement between the velocities derived from the Lyman profiles and the velocity of the front of the associated CME bright loop.
In the frame of the Joint Observing Program 107\footnote{\texttt{http://sohowww.nascom.nasa.gov/soc/JOPs/jop107.txt}.}, three quiescent prominences were observed with SUMER from L1 to L9 together with other space and ground-based observatories, including the \Ha\ line \citep{2001A&A...370..281H}. As shown in Fig.~\ref{fig:Heinzel01}, two classes of profiles were obtained: deeply reversed and unreversed ones (a reminder of the differences between OSO-8 and {UVSP} \La\ profiles). 
In both cases, the integrated intensities are similar. Note that because the \La\ profile is obtained on the attenuator part of the SUMER detector, its detailed multiple-peaked profile is not reliable -- but the total intensity is correct. Also note that \cite{2003SoPh..217..133S}  observed up to L19.

The modelling performed along the GHV type of models (isothermal and isobaric) proved to be unsuccessful in reproducing the two classes of observed profiles, especially the unreversed ones. The addition of a PCTR which naturally followed, combined with a magneto-hydrostatic Kippenhahn-Schl\"uter-type model \citep{1999SoPh..184..103H}, proved to be decisive in predicting the two types of profiles, provided the angle between the LOS and the field lines were taken into account. The results from  \cite{2001A&A...370..281H} can be summarized as follows: the PCTR across the magnetic field is very thin because of the strongly reduced perpendicular conduction, while the PCTR along the magnetic field, being governed by strong parallel conduction, is rather extended. Consequently unreversed profiles correspond to observations along the field lines (where the PCTR provides line centre photons) and, on the contrary, deeply reversed profiles correspond to observations across the field lines (where the cool prominence core  \ie the dips, is the only contribution).
The whole piece of work which then followed essentially discussed the various signatures of such models including the possibility of adding some of them in order to reproduce the observed filamentary fine structure.
This led to the development of a full 2D magneto-hydrostatic model of prominence fine structure \citep{2005A&A...442..331H}, which confirmed the sensitivity of the line profile shapes to the direction of the LOS and the magnetic field (see Sect.~\ref{sec:10}).
This idea was tested through the continuous (3 days) observations of a filament/prominence close to the limb \citep{2007SoPh..241...53S} from L3 to L7 (SUMER) and  in \Ha. With the filament being circular in shape, the authors managed to interpret the daily differences of profiles as the result of the changing angle of the LOS with the magnetic field direction.
Another use of the Lyman series (at least from L3 to L7) allowed \cite{2006A&A...459..651S}  to derive various models fitting different portions of a quiescent filament, to {calculate} the optical thickness at the head of the Lyman continuum ($\tau_{912}$), and address the issue of the EUV filament extensions \citep{2001ApJ...561L.223H}. The results raised some questions concerning the central temperature of the \Ha\ filament (found in the range $4000-6000$~K), the ratio of opacities $\tau_{912}/\tau_{\mathrm{H}\alpha}$ (larger than 50, a surprising high value) and the conclusion that EUV filament extensions are, most probably, due to emissivity blocking (i.e. a lack of emitting material at mild or coronal temperatures). 

Note that \cite{1995A&A...299..563H} developed a new code for vertical or horizontal 1D slabs based on the MALI method of \cite{1991A&A...245..171R,1992A&A...262..209R}. This code was subsequently used in all computations made by Heinzel and collaborators. The MALI method has proven to be very efficient and stable for all transitions within multilevel atoms. PRD is routinely used for hydrogen \La\ and \Lb\ lines, following \cite{1995A&A...302..587P} who first applied MALI to 2D models (see Sect.~\ref{sec:10}).

\begin{figure}
\centerline{\includegraphics[width=0.5\textwidth]{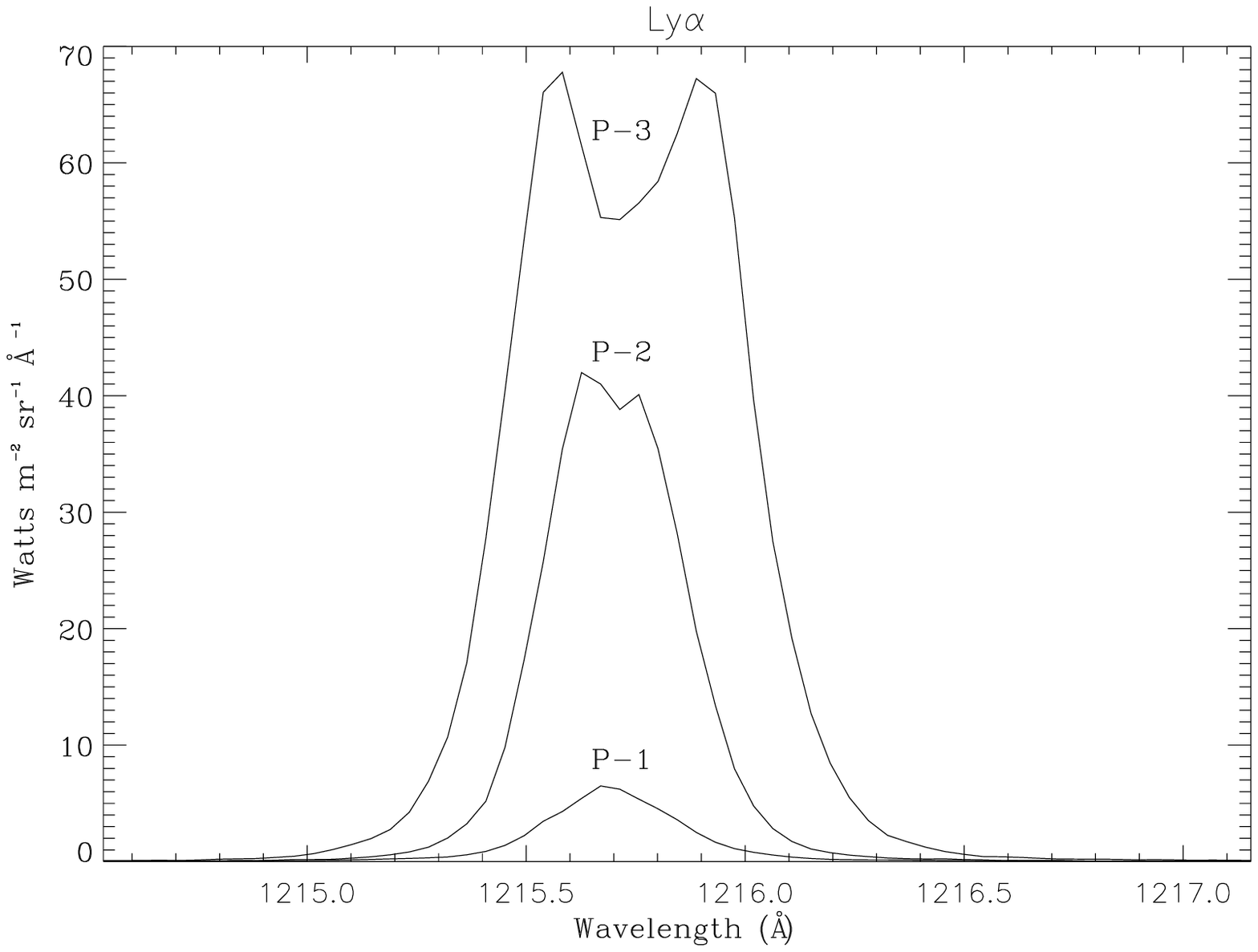}\hspace*{0.5cm}
           \includegraphics[width=0.5\textwidth]{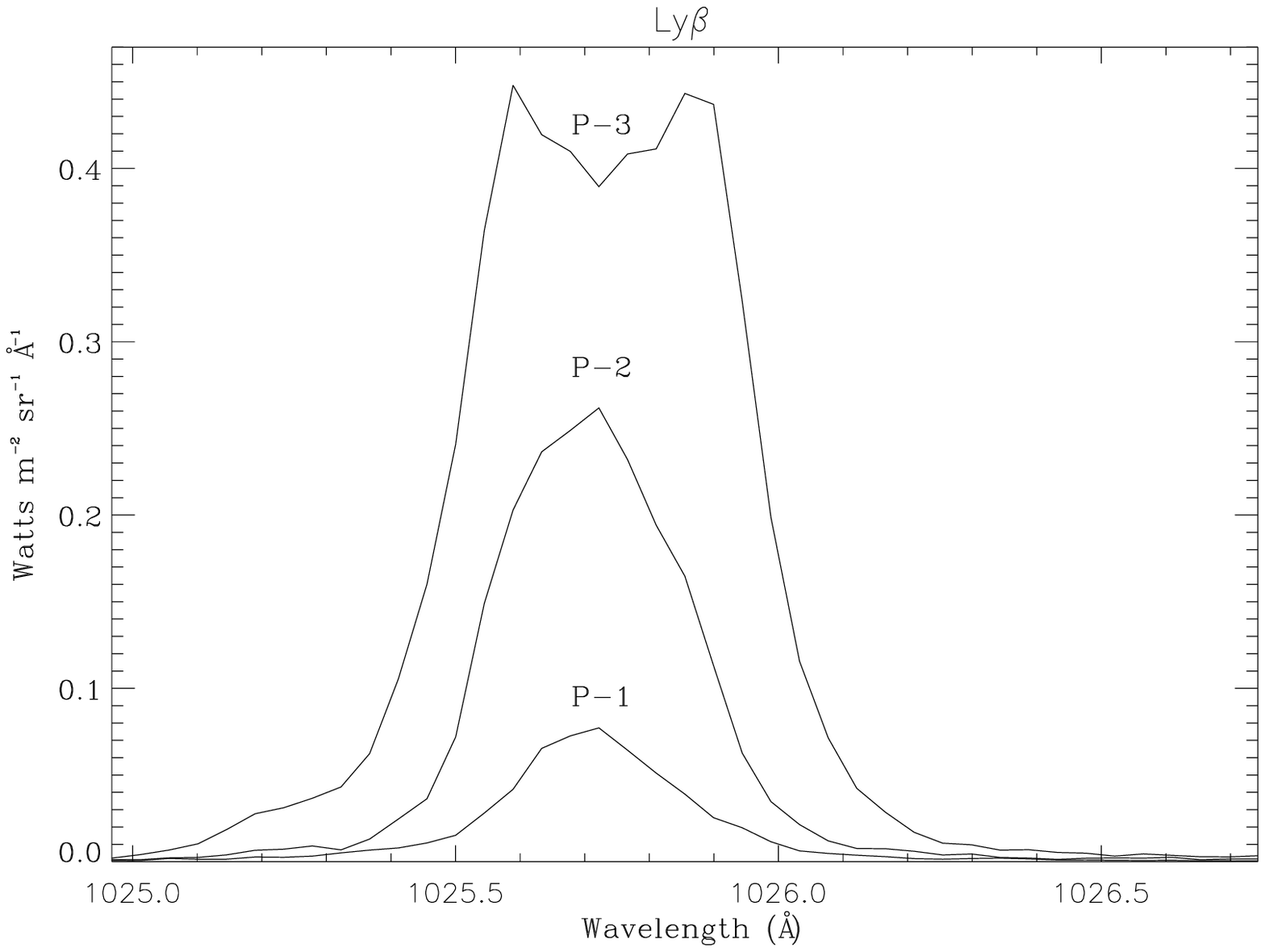}}
	   \caption{Average \La\ and \Lb\ profiles obtained {by SUMER} on 3 different regions of a prominence. From \cite{2007SoPh..246..327V}.\label{fig:LaVial07}}
\end{figure}

Without perturbation by an attenuator on the \La\ channel, \cite{2007SoPh..246..327V}   derived average \La\ and \Lb\ profiles (Fig.~\ref{fig:LaVial07}). 
From the different (spatial) behaviours of \La\ and \Lb\ profiles, these authors worked on a large range of observed \La\ and \Lb\ profiles and intensities, and compared this set of values with predictions of 1D and multi-thread modelling (their Table~3). On one hand, the \Lb\ intensities seem to be better matched by 1D isothermal models, and on the other hand, the observed \La/\Lb\ ratio never goes higher than 180 while the thread modelling leads to much lower values. 
Let us mention that the issue of the number of threads has also been addressed by \cite{1998A&A...332..325P} who used a different technique (namely the dispersion of line intensities) and applied them to \La, \Lb, \Lg, and Lyman continuum Skylab data. They essentially derived a thread number along the LOS of the order of 20 for (Lyman) optically thick lines.

\subsubsection{Electron Temperature From Hydrogen Lyman Continuum}
\label{sec:lycont1D}

\begin{figure}
	\center
	\includegraphics[width=0.49\textwidth]{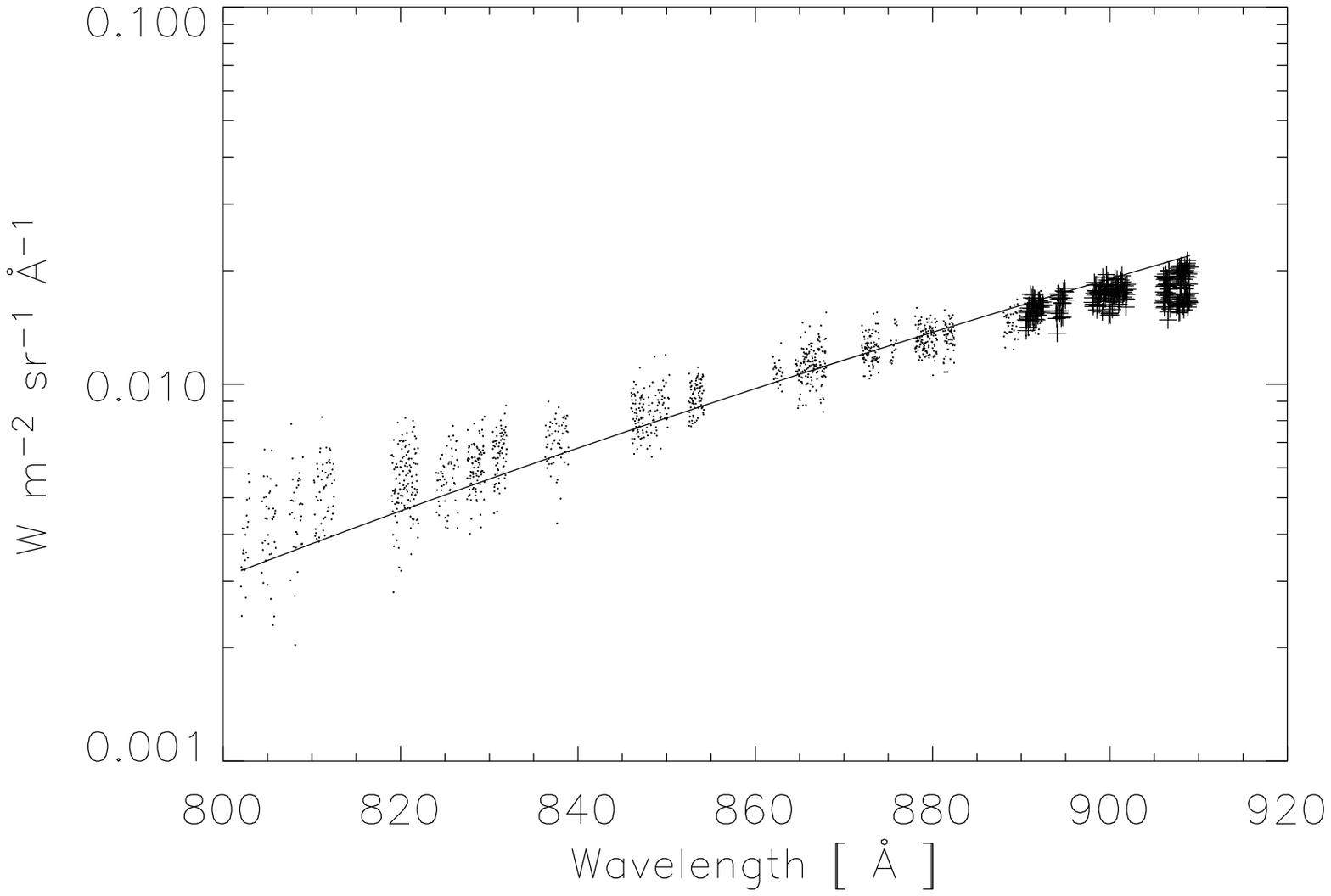}
	\includegraphics[width=0.49\textwidth]{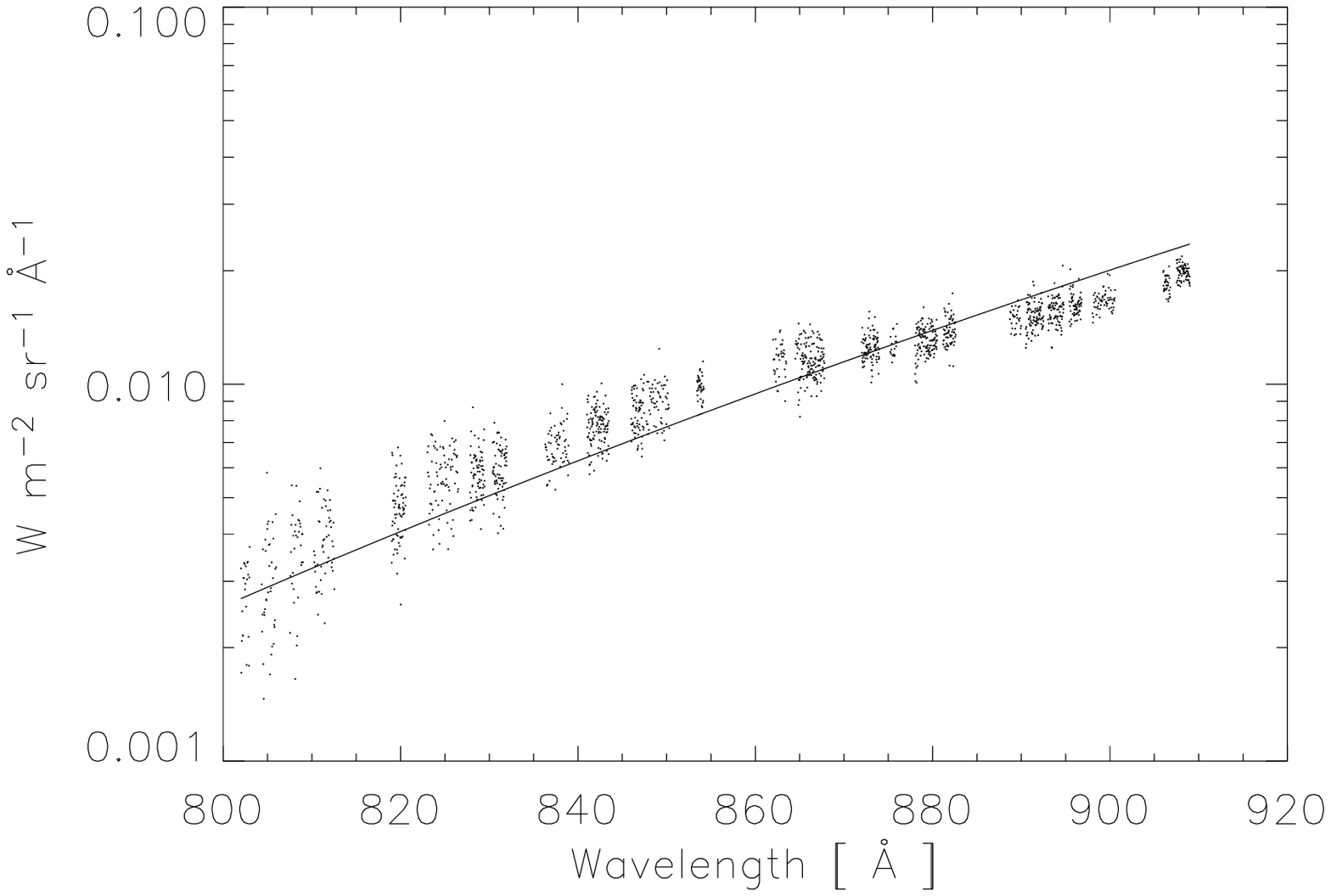}
	\caption{Lyman continuum specific intensity obtained {by SUMER} in two different regions of a prominence. From \cite{2005A&A...443..685P}.\label{fig:par05}}
\end{figure}

The hydrogen Lyman continuum in solar prominences is mostly produced through photo-ionization due to the chromospheric  emission, followed by radiative recombination. This emission is related to regions affected by the penetration of the incident Lyman continuum radiation. Under the assumption that the electron temperature is constant in the emitting region,  the source function ($S_\lambda$) may be approximated by a  black-body radiation {function} ($B_\lambda$), divided by the LTE departure coefficient $b_1$ of the ground state \citep{1970SoPh...15..120N}:
\begin{equation}
S_\lambda = \frac{B_\lambda (T_{\rm c})}{b_1} \ ,
\end{equation}
where $T_{\rm c}$ is the color temperature of the emitting plasma. Under prominence  conditions, $T_{\rm c}$ can represent the electron temperature $T$ \citep{1993A&AS...99..513G}.
In addition,  we may roughly assume that the source function is constant with depth and position, so that the emitted intensity can be written as:  
\begin{equation}\label{eq_cont}
I_\lambda \simeq S_\lambda  = \frac{2 h c^2}{b_1 \lambda^5}~\exp{ (-\frac{hc}{\lambda k T})} \ .
\end{equation}
{Here, }$h$, $c$, and $k$ are the Planck constant, the speed of light, and the Boltzmann constant, respectively.
In this case $T$ and  $b_1$ are obtained by fitting (\ref{eq_cont}) to the observed continuum intensity.
Note that a variation of $T$ with depth can be derived  with the technique used by \cite{1972SoPh...22..358V}  in the chromospheric context.
Only a few measurements by SUMER have been reported. \cite{1998SoPh..183...97O} found a large range of temperatures (from 5000~K to  15000~K), the largest values being probably related to the disappearance (and associated heating) of the prominence. \cite{2005A&A...443..685P} found different temperatures (8280~K and 7560~K) in different parts of the observed prominence. The authors note that the uncertainties are of the order of 250~K, mostly due to the difficult photometric calibration. They also notice some variation of the slope with wavelength (Fig.~\ref{fig:par05}), which can be of solar origin \citep[see][]{1972SoPh...22..358V}. 
It is obvious that such measurements should be repeated on various parts of prominences and on a large set of prominences. It also should be reminded that this technique is only valid well below 15000 K, on one hand, and that it also underestimates $T$ when the layer is too thin (\ie\ optically thin in the Lyman continuum).

\subsubsection{Filament Modelling}
\label{sec:cloud}

There are {various} ways for modelling filaments. Filament modelling in 2D geometry is discussed in Sect.~\ref{sec:10}. In 1D, either one takes the usual geometry of a prominence and  tilts the layer by 90 degrees: its lower boundary receives the whole chromospheric (and part of coronal) radiation. Or, one considers the filament as part of a full 1D atmospheric model starting from the photosphere and ending in the corona. In the latter case, the boundary condition for the incident coronal radiation is more difficult to define.
{Another approach uses} the so-called \textit{cloud model}, proposed by Beckers in his Ph.D thesis \citep{1964PhDT........83B} in order to deal with structures such as spicules, surges and filaments lying above the photosphere and the lower chromosphere, and seen against the solar disk. It has been the object of many developments, the latest ones being summarized in \cite{1996A&A...309..275M} -- see also the review by \cite{2007ASPC..368..217T}. According to \cite{1996A&A...309..275M}, once the source function is known with respect to depth, three important parameters can be derived: the total optical thickness, the Doppler width and the Doppler shift. The source function can be derived from non-LTE modelling such as performed by \cite{1993A&AS...99..513G}, using the geometry of horizontal slabs. \cite{1996A&A...309..275M} successfully applied their method to an Arch Filament System observed in \Ha\ with the Multichannel Subtractive Double Pass (MSDP) of the Vacuum Tower Telescope (VTT). Such a method is especially efficient for observations with a wide field of view and a high temporal cadence.

\subsubsection{Summary}
Both observations and non-LTE modelling of Lyman lines have made a tremendous progress during the last few decades. The synergy between observations and modelling has come from challenging observations in a variety of structures{, the completeness of 1D modelling shedding light on radiative signatures}, and the move from 1D to 2D and multi-thread modelling. 
The proof of a necessary PCTR where the various Lyman lines are (partly) formed at various temperatures is not a minor feature, even if the issue had been raised more than twenty years ago.
So, the picture of a bundle of magnetostatic threads with a low central temperature ($< 7000$~K) and an overall low pressure ($< 0.02$~dyn cm$^{-2}$) seems to be the right one.

\begin{figure}
	\center
\includegraphics[width=\textwidth]{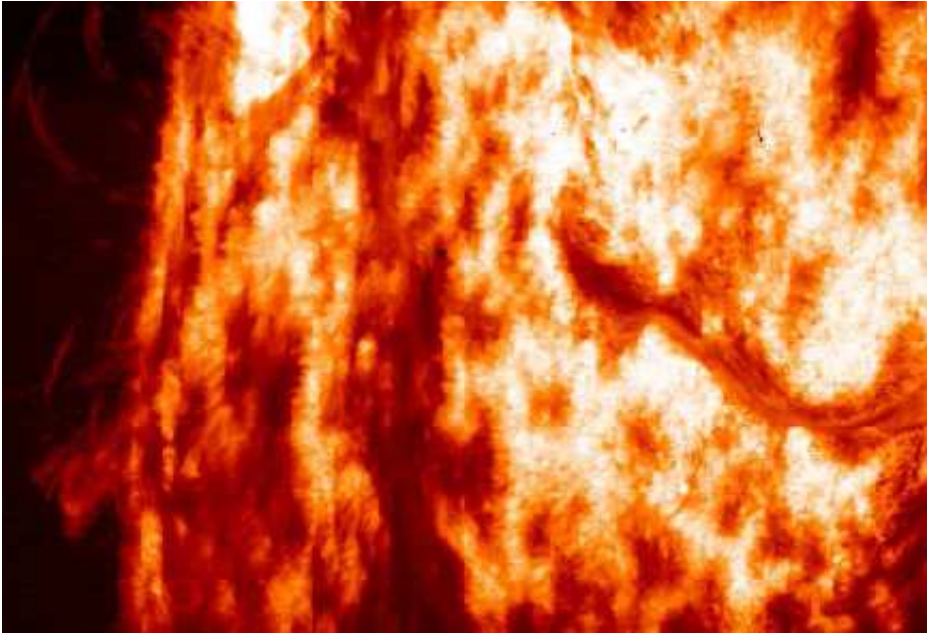}
\caption{Composite picture of prominence and filament observed in \La\ with VAULT. From \cite{2001SoPh..200...63K}. \label{fig:vault}}
\end{figure}

However, one should realize that the range of line profiles and consequently physical parameters remains very large. Moreover, the number of threads assumed in the various modellings (larger than 10 along the LOS) implies that observations are insufficient for resolving them, in spite of progress in spatial resolution, as shown by the VAULT pictures (Fig.~\ref{fig:vault}). 
In the case of filament observations, there could also be a limit due to the very high opacity, especially in the \La\ line.
This means that for a given set of observations performed with the present instrumentation, there might be more than one solution. One way to go around this problem is to combine the analysis of hydrogen lines with lines emitted by other elements (\eg\ He -- see Sect.~\ref{sec:9.2}).
It should also be realized that multithread modelling does not so far take into account the radiative interaction between threads. 

On the observational side, one should be aware that the first ten Lyman lines can only be obtained through, at least,  four consecutive exposure times with SUMER, which means that observations are never strictly simultaneous. The total SUMER sequence for the Lyman lines takes about a couple of minutes, a duration which may appear very small in comparison with the overall lifetime of prominences (at least a few days), but which is comparable to the lifetime of fine structures, as evidenced by the SOT images.
Having in mind the potential offered by high spatial, spectral and temporal resolution of Lyman line profiles, further progress can only be obtained with a dedicated fast UV spectrograph combined with a filament/prominence magnetograph.

On the modelling side, a great step has been made with the use of a non-empirical (magneto-hydrostatic) 2D model of dips (see Sect.~\ref{sec:10}), but fully 3D MHD models are certainly necessary for taking into account the complexity and dynamics of prominences and their fine structures.

\subsection{Helium Lines and Continua}
\label{sec:9.2}

\subsubsection{Early Studies}

After solving the integral diffusion equation for the radiation, \citet{1982SoPh...81..339I} calculated the populations of selected neutral helium levels with quantum number $n \leq 4$ under low-temperature plasma conditions. They found that the excitation and ionization of helium are maximum near the boundaries. Other authors have considered different helium lines in different geometries in order to compare with specific observations. \citet{1984SoPh...92..153M} investigated the excitation of singlet helium levels in the frame of homogeneous and filamentary prominences. \citet{2000MNRAS.313..761L} computed the He~I 5876~\AA\ (D3) emission line profile using the two-cloud model and taking into account the multiplet of He~I.

However, \citet{1974ApJ...192..181H} were the first to deal with the non-LTE hydrogen and helium spectra in model prominences self-consistently. Using simple 1D homogeneous and static models, they were able to make the first satisfactory comparisons with observed intensities of helium lines in quiescent solar prominences.
These authors removed several restrictive assumptions that were made before (see references in their paper). For example, they used a multi-ion helium model atom with 20 energy levels and treated the radiative transfer problems in detail in the resonance lines and continua for  neutral \textit{and} singly ionized helium. Their prominence models consisted of isobaric and isothermal slabs. Then, a series of papers following the same approach was published in the following years \citep{1976ApJ...210..827H,1978ApJ...221..677H, 1983ApJ...268..398H}.
However there were some limitations, such as the use of CRD (see Sect.~\ref{sec:scat}) in the  formation of the resonance lines, or the use of frequency-independent incident intensities for each line. They presented integrated intensities, line ratios and optical depths for various hydrogen, helium, and ionized calcium lines.  However, no  emergent line profiles were shown (see discussion in Sect.~\ref{sec91:heasley}).

\subsubsection{Isothermal Isobaric Models}

{Using a helium system with 34 energy levels (29 for He~I, 4 for He~II, and one level for fully-ionized helium),} \citet{2001A&A...380..323L} computed the neutral and ionized helium spectra emitted by a model quiescent solar prominence. {The geometry was that of a} 1D plane-parallel slab standing vertically above the solar surface. Their computations allowed departure from LTE
for all atomic levels, and the radiative transfer equations were solved
for all optically thick lines and continua. PRD was
included in the calculations of resonance lines. This represented a new
step compared to the above-cited papers towards a more physical
approach. They obtained several radiative quantities related to the
emergent spectrum. An important point is that they also presented the
emergent line profiles. This is particularly useful for the
interpretation of prominence spectra, especially when one observes
optically thick lines such as the resonance lines of neutral (He I
584 and 537~\AA) or ionized helium (He II
304 and 256~\AA) from which the derivation of the
plasma  parameters is not straightforward.

These authors presented detailed comparisons with the pioneering work of Heasley and co-workers. This showed the importance of considering the PRD effects in the formation of the resonance lines of H and He, and of using detailed, frequency-dependent incident line profiles.
In fact, \cite{1987A&A...183..351H} had already shown that PRD is important in  calculating hydrogen line profiles (see also Sect.~\ref{sec91:early}), but the study by \citet{2001A&A...380..323L} also revealed that PRD for the radiative transfer calculations in hydrogen lines and continua is important for the subsequent modelling of helium lines and continua, since the electron density structure inside the prominence slab changes between CRD and PRD \citep{1987A&A...183..351H}.

\begin{figure}
	\center
	  \includegraphics[width=0.9\textwidth]{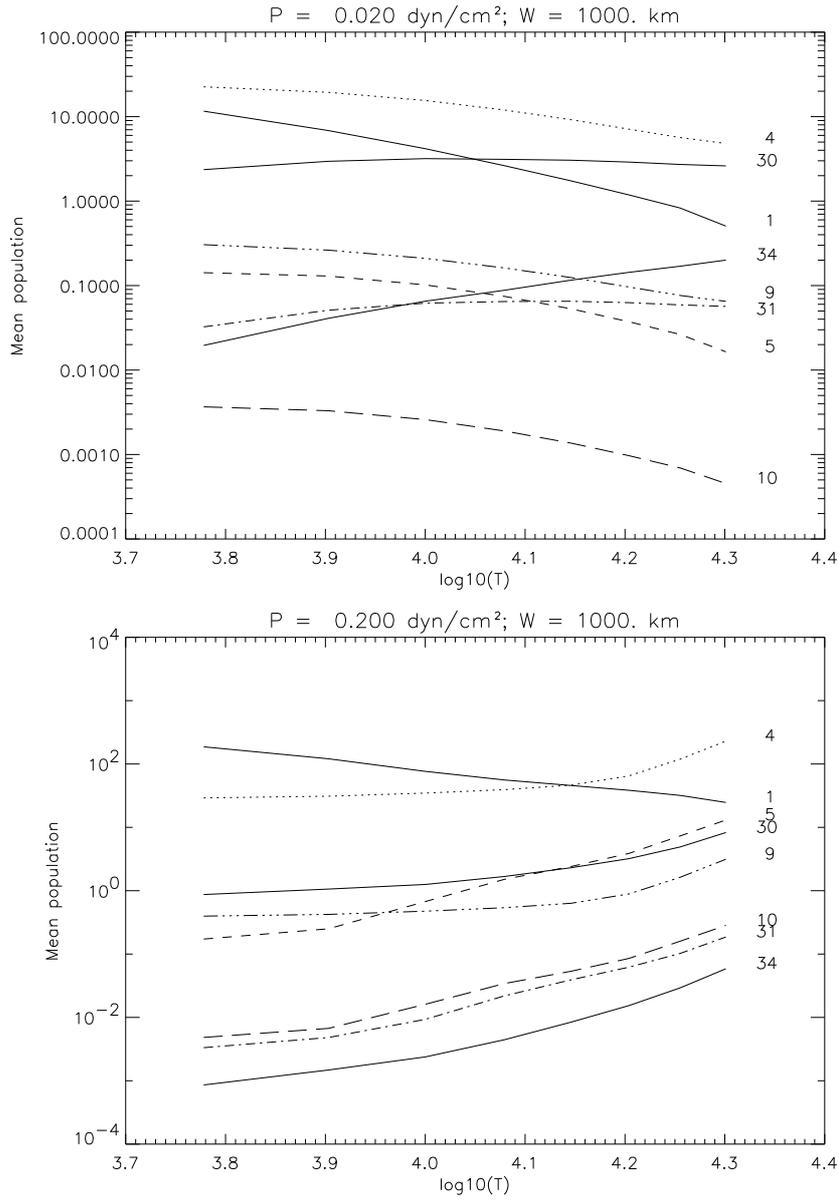}
	  \caption{Mean population densities (in cm$^{-3}$) as a function of temperature for two different pressures (top: 0.02 and bottom: 0.2 dyn cm$^{-2}$) and one slab width (1000~km). Solid lines: ground states of He I (1), of He II (30), and He III continuum (34). The population densities for these levels are divided by 10$^8$. Singlet excited levels represented: $\rm 1s2p~^1P$ (5, dashes) and $\rm 1s3d~^1D$ (10, long dashes). Triplet levels are: $\rm 1s2p~^3P$ (4, dots) and $\rm 1s3d~^3D$ (9, long dashes/dots). Level 31 is the $n=2$ level of He II (short dashes/dots). From \cite{2001A&A...380..323L}.}
	  \label{fig:LG01-fig7}
  \end{figure}

\begin{figure}\sidecaption
	  \includegraphics[width=0.7\textwidth]{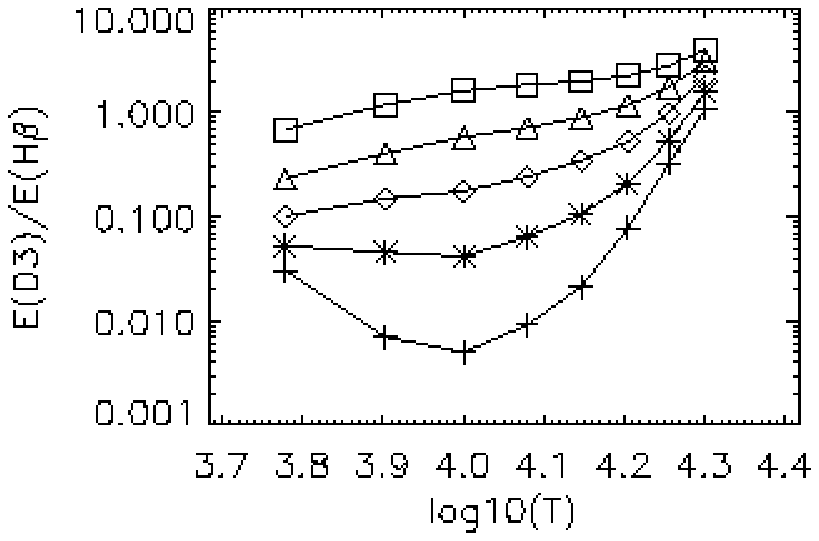}
	  \caption{$E({\rm D}3)/E(\mathrm{H\beta})$ as a function of temperature at 5 pressures. The line ratio is shown for five different pressures: 0.02 ($\Box $), 0.05 ($\triangle $), 0.1 ($\Diamond $), 0.2 ($\ast $), and 0.5~dyn cm$^{-2}$ (+). From \cite{2001A&A...380..323L}.\vspace{1cm}}
	  \label{fig:LG01-fig20}
  \end{figure}

Using a grid of 480 models, \citet{2001A&A...380..323L} investigated the effects of the slab temperature, pressure, and width, as well as of the He abundance, on the mean population densities and on the line profiles and integrated intensities. This investigation illustrated the complexity of the coupling existing between the different line transitions which results from the coupling between the radiative transfer and statistical equilibrium equations. Their results can be briefly summarised as follows: at low temperatures/pressures, the main mechanism of formation of the He resonance lines is the scattering of the incident radiation. A substantial amount of ionized helium can be found at the boundary between the prominence and the corona. This ionized helium is produced via the photo-ionization of neutral helium by the incident radiation, and it can scatter the incident radiation at 304~\AA. Consequently, even \textit{cool} prominences can in principle emit in the He~II line at 304~\AA.  Fig.~\ref{fig:LG01-fig7} shows population densities of several energy levels of He~I and He~II averaged across the prominence slab as a function of temperature. 
It is interesting to note that the population of the He~II ground state (level 30 in this figure) can be larger than the population of the ground state of He~I (level 1) for temperatures just over $10^4$~K and at a low pressure of 0.02~dyn cm$^{-2}$.
At higher temperatures/pressures,
collisional effects become non negligible and contribute to the thermal
component of the neutral helium lines. However the temperatures and
pressures considered in these isothermal and isobaric models were not
high enough to produce a significant collisional contribution in the
formation of the ionized helium lines. The effect of the slab width is
usually negligible on the optically thick resonance lines: when the
line centre is saturated, only the optically thin wings of the line
profile may become brighter for larger slab thickness. 
\citet{2001A&A...380..323L} also examined the correlations between
various line intensities as a function of the prominence plasma
parameters. They notably revisited the ratio between the He~I D3 line and the H$\beta$ line of hydrogen (see  Fig.~\ref{fig:LG01-fig20}).
This
study also showed that the different sensitivities of line intensities
(optically thick vs. optically thin, singlet vs. triplet) could be used
together with hydrogen lines, to improve the diagnostics of the helium
abundance in prominences.

\citet{2002ESASP.505..421G} considered the superposition of several slabs along the LOS (without radiative interaction) and found that it improves the agreement between observed and computed integrated intensity ratios in some cases ($E(\mathrm{H}\alpha)/E(\mathrm{H}\beta)$ and $E(\mathrm{D}3)/E(\mathrm{H}\beta)$), but not for $E(\mathrm{Ca~II}~8542)/E(\mathrm{H}\beta)$.

A comparison between observed and computed  H and He I line
profiles was presented in \citet{2006sf2a.conf..549L}. Profiles of the
\Lb, \Le\ and He I 584~\AA\ lines were obtained by SUMER in a prominence observed in June 2004. The
agreement was satisfactory, but the non-simultaneity of the acquisition
of the profiles of each line prevented the authors to make an in-depth
study. They concluded that the temperature of the prominence central
part was $\sim 8600$~K and the pressure 0.03~dyn cm$^{-2}$.\\

\subsubsection{Models with PCTR}

It is necessary to include temperature and pressure variations in the prominence slab if we want to understand how these cool structures can exist in the hot corona.
Prominence models with a PCTR usually represent some sort of semi-empirical atmospheric model.
This means that the temperature structure is determined empirically in order to achieve agreement between the synthetic spectra and observations. By observations, we mean either the data for a specific prominence or filament or, more generally, some kind of canonical data set. However,  problems arise with the variations of the gas pressure or density. In fact, all non-LTE prominence models so far used are either isobaric, or consider certain magneto-hydrostatic equilibrium (MHS{ -- see also Paper~II, Sect.~2.2)}.

\citet{1999A&A...349..974A} addressed the question of energy balance in quiescent solar prominences using 1D slab models in MHS equilibrium and with a PCTR. Their prominence models included two parts, namely the core of the prominence body with the base (cool part) of the PCTR, where the plasma is still optically thick for some lines, and the outer part of the PCTR where the plasma becomes optically thin. They present pressure and temperature profiles inside the slab. These profiles were then used by \citet{2004ApJ...617..614L} and in subsequent papers for the modelling of He lines (for the inner part of the prominence only).
Later on, \citet{2001A&A...370..281H} used the same kind of models and showed that they were able to reproduce different types of hydrogen Lyman line profiles and integrated intensities that were observed by the SUMER spectrometer. In this respect, the presence of the PCTR in the modelling was found to be critical, especially for the low members of the Lyman series (see Sect.~\ref{sec:9.1}).
\citet{2002ESASP.506..451L} presented a study of the PCTR structure and its signatures on the emergent spectra of hydrogen, helium, and ionized calcium. They computed the spectra emitted by a 1D prominence model in MHS equilibrium as in \citet{1999A&A...349..974A}. The radiative transfer equations were solved for all optically thick transitions (when the optical thickness is $\geq 0.1$) out of LTE. They presented line intensity ratios as well as line profiles. It is worth noting here that it is the first time that predicted spectra are computed simultaneously for H, He and Ca~II in prominence models with a PCTR. They have shown that the presence of the PCTR affects the line profiles in different ways, depending on the optical depth and the region of formation of the spectral lines. For example, the inclusion of the PCTR has a dramatic impact on the resulting emergent profiles of hydrogen and helium resonance lines. However, it is also evident in their study that through the radiative coupling between optically thick and optically thin lines, the intensities of the latter are also affected by the inclusion of the PCTR in the models.

\begin{figure}
	\center
	  \includegraphics[width=0.49\textwidth]{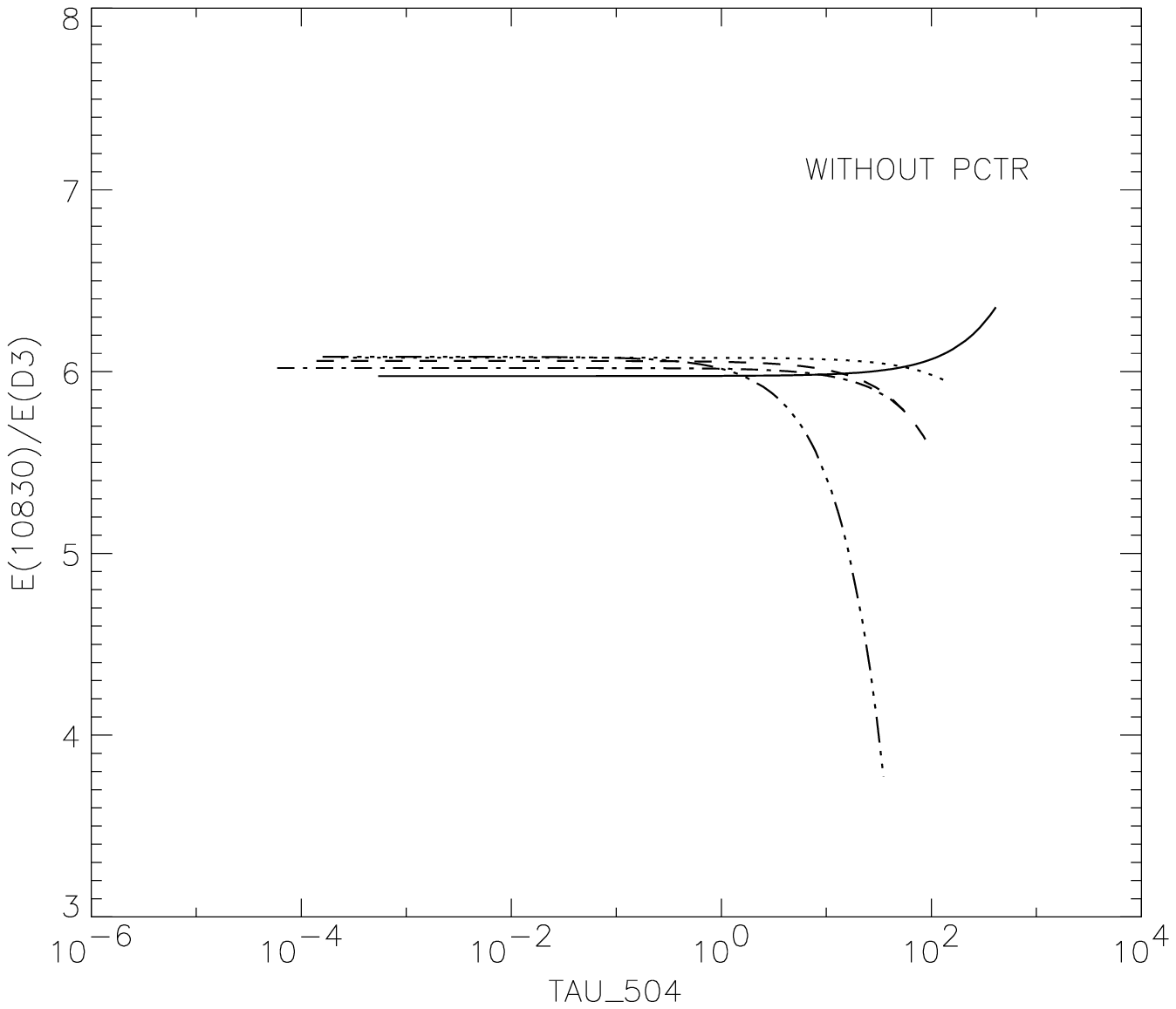}
	  \includegraphics[width=0.49\textwidth]{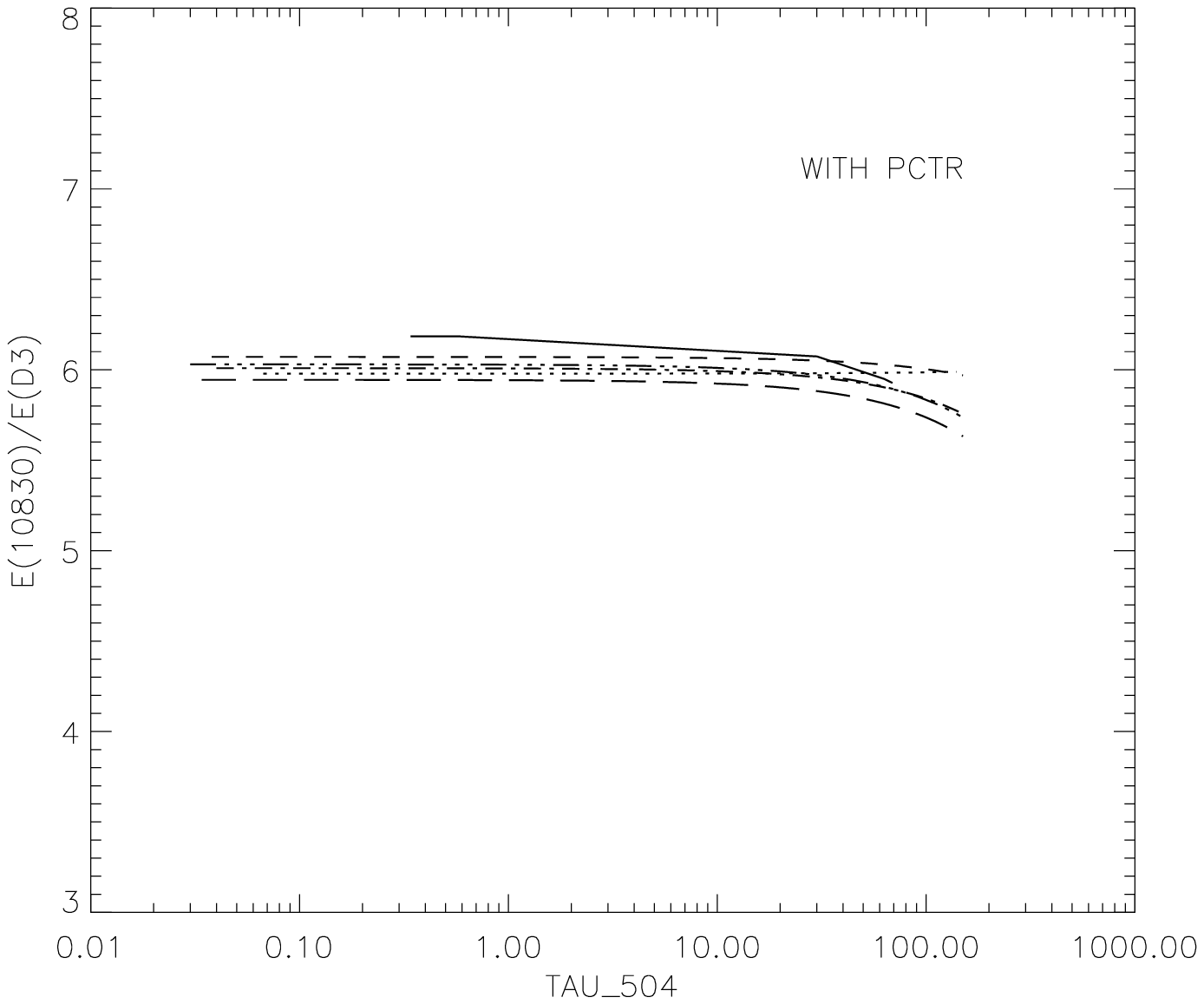}
	  \caption{$E(10830)/E(\mathrm{D}3)$ as a function of the optical thickness at the head of the He~I ionization continuum edge (504~\AA). Left: isothermal isobaric models; right: models with PCTR. The different curves distinguish the models according to their mean temperature. Solid line: mean temperature $T<6000$~K; dotted line: $6000\le T < 9000$~K; dashed line: $9000 \le T < 12000$~K; dash-dotted line: $12000 \le T < 16000$~K; dash-dot-dotted line: $16000 \le T < 20000$~K; long-dashed line: $T \ge 20000$~K. {Note the different opacity scales in the left- and righ-hand figures.} Here, all isothermal models have a mean temperature less than 20000~K. From \cite{2004ApJ...617..614L}.}
	  \label{fig:lg04}
\end{figure}

\citet{2004ApJ...617..614L} investigated the He I triplet line
intensities in prominence models with and without a PCTR.  This study
focused on the emergent intensities of the neutral helium triplet lines
(such as He~I 10830~\AA\ and D3 at 5876~\AA) and the level populations of the
relevant helium states. The authors found that the presence of the PCTR
affects the emitted intensities of the triplet lines 
by reducing
the impact of collisional
excitations at high temperatures in comparison with the isothermal and
isobaric case. A simple study of helium energy level populations
demonstrated how statistical equilibrium is changed when a transition
region is present (Fig.~\ref{fig:lg04}). 
This points to the necessity of including an
interface between the prominence body and the corona to predict all
emergent intensities whatever the region of formation of the radiation
is, due to the non-local nature of the coupling between the radiative transfer and statistical equilibrium equations. 
\citet{2004ApJ...617..614L} also found a correlation between most
of the He I triplet line ratios and the altitude of the model
prominence. This allowed the authors to solve some long-standing
discrepancies in comparisons between predicted triplet line intensity
ratios and observations by extrapolating the computations to higher
altitudes.

\subsubsection{Summary}

After the initial attempt by \cite{2006sf2a.conf..549L}, there should be more attempts to compare simultaneously observed H and He line profiles with computed ones. {In this respect, the He~II Balmer lines in the wings of the H Lyman lines \citep{2009SoPh..257...91E} provide an excellent opportunity to simultaneously observe H and He spectra (representative of cool and hot material, respectively), and put new additional constraints to be taken into account in future modelling efforts.}
The modelling of the helium spectrum should also be extended to disk filaments.
Much progress can be expected from the modelling of the helium spectrum in 2D geometry (discussed in Sect.~\ref{sec:10}), such as presented by \cite{2009A&A...498..869L} and \cite{2009A&A...503..663G}.

\subsection{Ca~II Spectra}
\label{sec:9.3}

Apart from hydrogen and helium lines, the optical and IR spectral region contains, among others, also five lines of ionized calcium which are frequently observed in prominences. These five lines are the UV Ca~II resonance lines H and K plus three IR lines (triplet). The high spatial and temporal resolution images produced in the Ca~II H line by SOT (Fig.~\ref{f:Berger08}) were presented in \citet{2008ApJ...676L..89B}. In SOT movies, highly dynamical fine-structure features are recorded, sometimes showing dark bubbles moving upward. The question thus arises whether these brightness changes are due to the ionization state of calcium or due to other effects.
This may be related to prominence energy and momentum balance.

{Early on, }\citet{1978ApJ...221..677H} computed the ratio $r$ of integrated intensities of the Ca~II IR line at 8542~\AA\ to the hydrogen H$\beta$ line. They found that in the temperature range $6500 - 9000$~K, this ratio is relatively insensitive to temperature, practically independent of column mass, and decreasing with gas pressure. This theoretical relation was then used to derive the gas pressure from observations. 
However, the model curve was restricted only to  pressures below 0.2~dyn cm$^{-2}$. Calcium spectrum formation was discussed also by \citet{1982AAfz...47...34M}. Hydrogen and calcium semi-empirical models were then produced by \citet{1987A&A...175..277Z} who found a reasonable agreement with observed, uncalibrated (\ie in relative units) line profiles. In order to extend the range of gas pressures used by \citet{1978ApJ...221..677H}, \citet{2002A&A...385..273G} computed a larger grid of models of the GHV type, up to $p=1$~dyn cm$^{-2}$. Their principal result is that at pressures higher than 0.1~dyn cm$^{-2}$, the ratio $r$ is strongly dependent on the temperature: $r$ starts to increase for temperatures below 8000~K, and decreases at higher temperatures.  This means that there exists a temperature bifurcation of the ratio $r$ for high pressures. 
Contrary to the results of \citet{1978ApJ...221..677H}, this can explain
higher measured values of $r$, say up to 0.8
\citep{2000SoPh..196..357S} which do indicate lower temperatures and
gas pressures higher than 0.1 dyn cm$^{-2}$. This has important
consequences for prominence momentum balance since higher gas pressures
mean larger values of the plasma-$\beta$ parameter.

Concerning the non-LTE modelling, the Ca~II UV resonance continua are
partially driven by hydrogen Lyman lines \citep{2002A&A...385..273G}. This presents a significant complication for the calcium
transfer modelling. First, the local radiation fields in the Lyman lines
must be carefully computed and then used to evaluate the
photo-ionization rates of calcium UV continua. Note that
\citet{2002A&A...385..273G} used PRD for H and K lines, but did not
treat details of the cross-redistribution with IR triplet lines.
PCTR and multi-threads were also treated later on \citep{2002ESASP.505..421G,2002ESASP.506..451L,2007ASPC..368..337L}. 
Finally, it is worth mentioning that the measurements of \citet{2000SoPh..196..357S}  led to a pretty constant value of $r$ over the observed prominences. This indicated a rather constant value of the gas pressure, assuming a uniform temperature in the cool parts.

{\cite{2001A&A...366..686T} used a two-step method to interpret Ca~II (8542~\AA) observations obtained with the MSDP at the VTT telescope. The first step involved the computation of a large grid of models by a multi-level non-LTE transfer code which yields the Ca~II line depth-dependent mean intensity inside an isolated, isothermal cloud lying above the chromosphere. The second step involves the inversion of the observed profiles with the grid of computed Ca~II profiles.
Maps of the temperature, LOS velocity, and microturbulence were built inside a quiescent filament. The temperature was found to peak at 8500~K, the velocity indicated an excess of blueshift (material moving upwards), and the microturbulence peaked at 5~\kmps. This method has not been applied to SOHO data because it does not work if the medium is too optically thick in the observed line (then the incident radiation component of the cloud model is not visible).}

\subsection{Moving Prominences: Doppler Dimming and Doppler Brightening}
\label{sec:9.4}

It is still not clear how solar prominences reach a state of equilibrium. However it is known that during the formation stage, significant large-scale plasma motions are taking place {(see also the discussion in Paper~II, Sect.~3)}. Similarly, mass motions are important during the disappearance of prominences. Between these two stages of the life of prominences, one may also observe periods of activity with internal fine-structure plasma motions (see Sect.~\ref{sec:5}).
Hydrogen and helium resonance lines are difficult to interpret as the prominence plasma is optically thick in these transitions. When the prominence is moving radially outwards, the incident radiation coming from the solar disk and illuminating the structure is shifted to lower frequencies. Therefore, if mass motions are taking place, the diagnostic of the prominence is even more complex.
In order to improve our understanding of  dynamic structures, it is necessary to study  the effect of  motions in prominences on the relevant lines. {Most of this modelling has been done in 1D \citep[apart from][]{1982ApJ...254..780V}, and we therefore describe it here.}

Let us start with  a simple description of the effect of radial motions on the emitted spectrum. Consider a simple two-level atom whose upper level is excited by the radiation coming from the Sun: the absorption profile of the radiative transition between the two levels gets out of resonance with the incident radiation when this atom is moving away from (or towards) the Sun, due to the Doppler effect. We namely have a Doppler dimming effect when the incident line is in emission, or a Doppler brightening effect if the incident line is in absorption \citep[see also][]{1970SoPh...14..147H}.
In a more realistic situation, an atom has more than two energy levels. Consequently, coupling effects take place between the atomic levels. For that reason, a combination of Doppler dimming and brightening can occur, just as it happens when the coupling between the first two excited levels of hydrogen is taken into account \citep{1987SoPh..110..171H,1997SoPh..172..189G}.
Note that this is the same Doppler effect which is also used for the diagnostic of the radial component of the solar wind \citep[see \eg ][]{1982ApJ...256..263K}.

The main factors determining the effects of the radial motions on the emitted prominence spectrum are the line formation mechanisms, including the relative contributions of collisional and radiative excitation, and the characteristics of the incident radiation (\eg\ whether it is in absorption or in emission, or the strength of the line). The maximum effect is achieved when radiative excitation dominates the collisional one. More discussion can be found in \citet{1987SoPh..110..171H,1997A&A...325..803G,2006IAUJD...3E..47L,2007A&A...463.1171L,2007ASPC..368..337L}.

\cite{1987SoPh..110..171H} and \cite{1997A&A...325..803G,1997SoPh..172..189G}  computed the hydrogen spectrum emitted by an eruptive prominence. The prominence is modelled as a 1D plane-parallel slab standing vertically above the solar limb. Its geometrical thickness along the LOS and its altitude above the limb are free parameters. The other free parameters, which define the prominence atmosphere, are the electron temperature, the gas pressure, and the microturbulent velocity. The incident radiation field is identical on both sides of the prominence.
This velocity-dependent incident radiation illuminating the prominence slab can be represented by the mean intensity $J_\nu$:
\begin{equation}
  J_\nu = \frac{1}{4\pi} \oint{I_0 \left(\nu+\frac{\nu_0}{c} \vec{V} \cdot \vec{n^\prime}, \vec{n^\prime} \right) \mathrm{d\vec{n^\prime}}} \ ,
\end{equation}
where $I_0(\nu,\vec{n})$ is the specific intensity of the incident radiation, $\vec{n^\prime}$ is the direction of the incident photon, and $\vec{V}$ is the prominence velocity. $J_\nu$ is calculated at a given height, taking into account the centre-to-limb variations (if any) of the incident radiation.

\begin{figure}
	\center
	\includegraphics[width=0.49\textwidth]{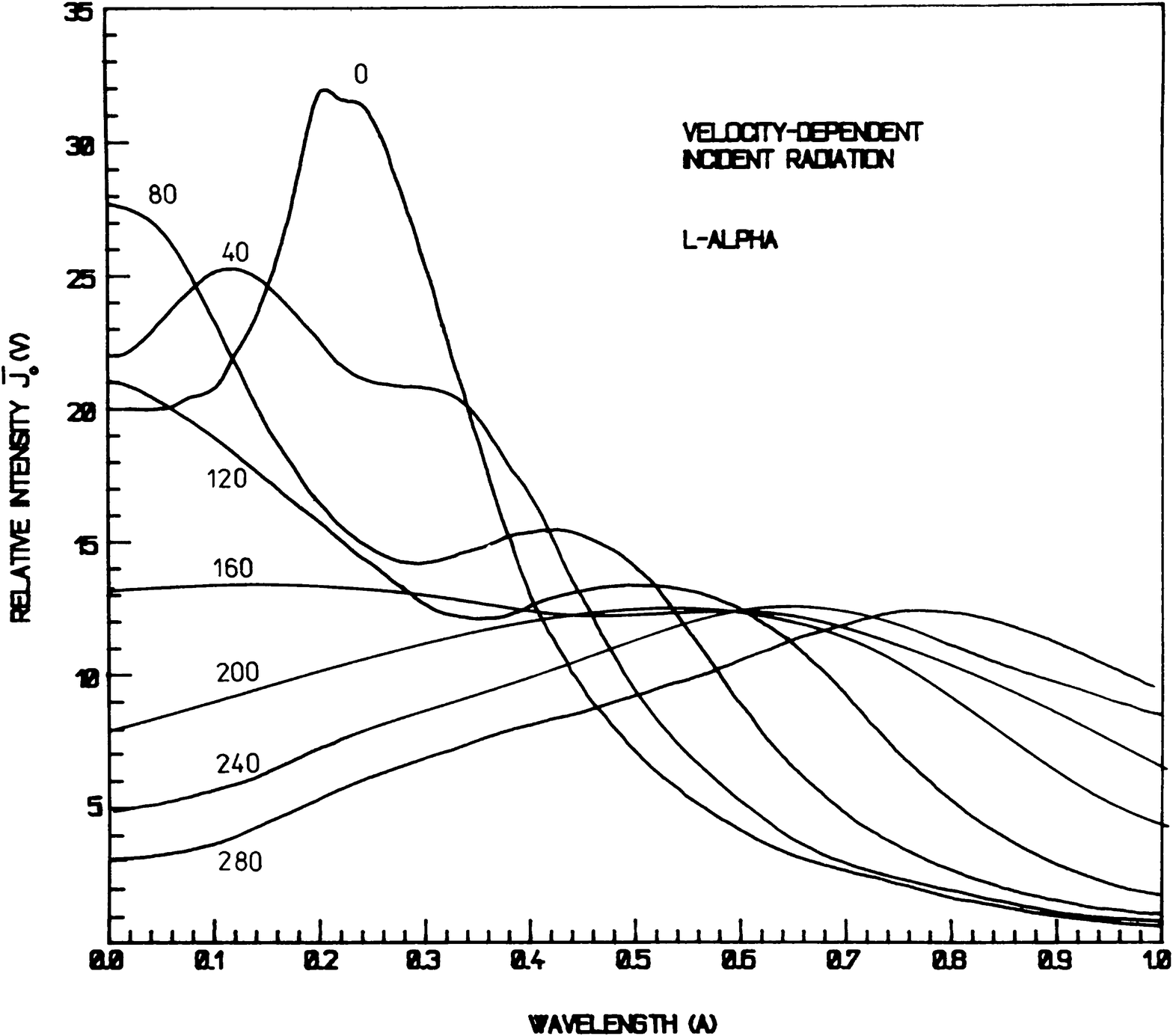}
	\includegraphics[width=0.49\textwidth]{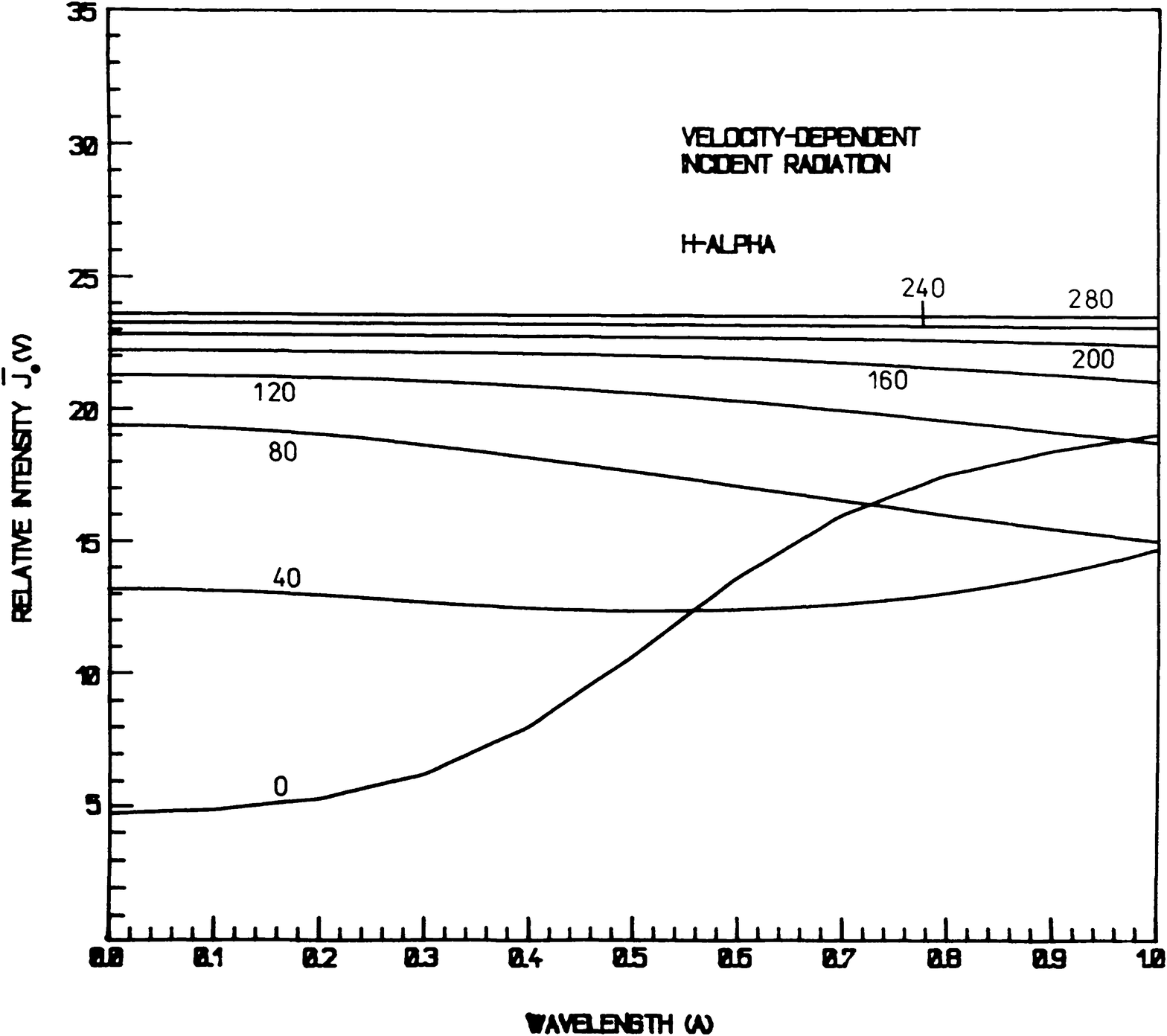}
	\includegraphics[width=0.6\textwidth]{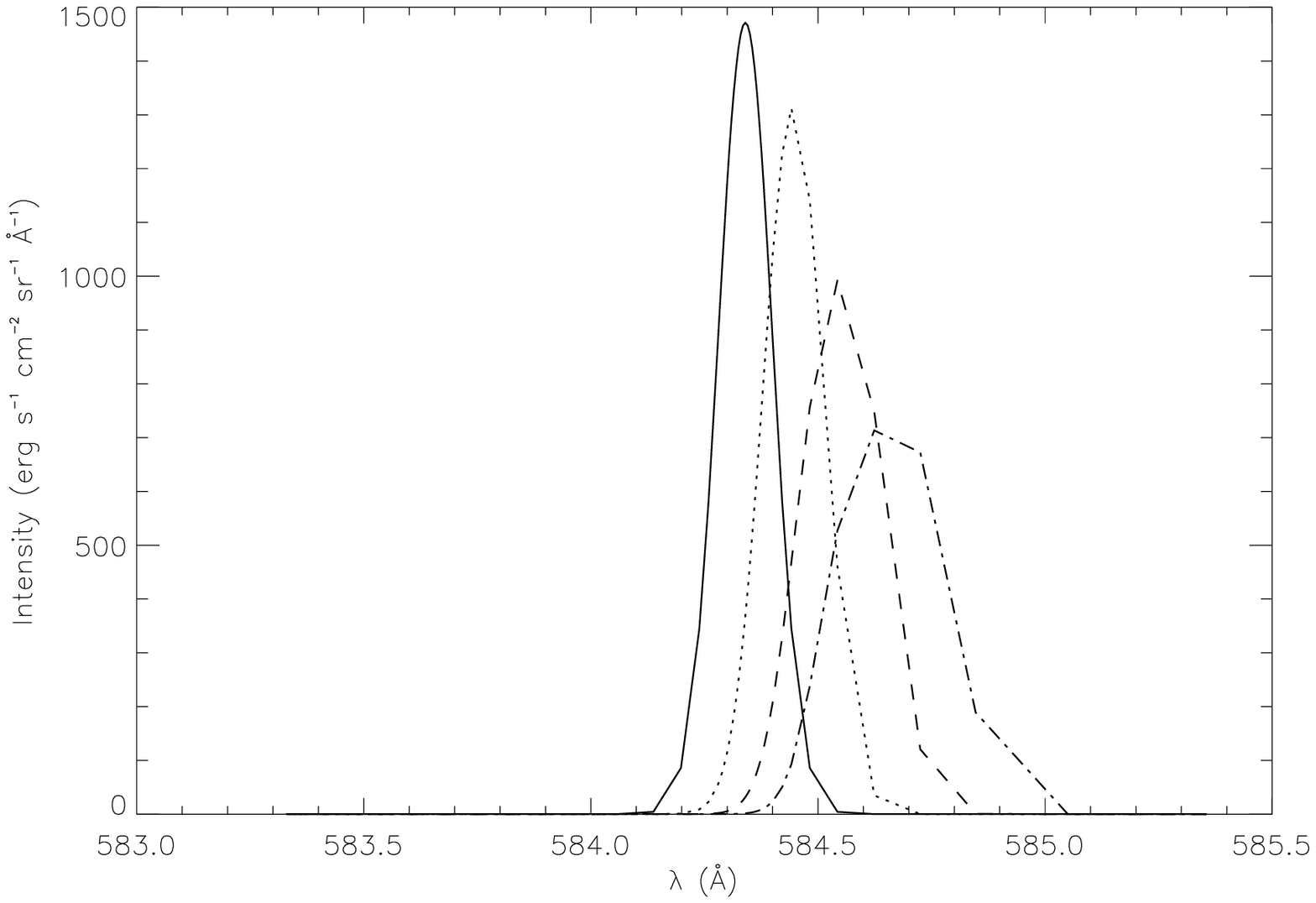}
	\caption{Mean intensities of the incident \La\ (top left), \Ha\ (top right), and He~I 584~\AA\ (bottom) emission profiles as seen by a moving prominence at the height of 50000~km. Top panels: the profiles are displayed for the velocities ranging from $0-280$~\kmps \citep{1987SoPh..110..171H}. Bottom panel: velocities range from 0 (solid line) to 240~\kmps with a step of 80~\kmps \citep{2007A&A...463.1171L}.}
	\label{fig:incidentprof}
\end{figure}

As illustrated in \cite{1987SoPh..110..171H}, \cite{1997A&A...325..803G}, and \cite{2007A&A...463.1171L}, the Doppler effect induces a shift in frequency of the incident line profile, and the variation of the Doppler shift with direction induces a distortion of the incident profile. Figure~\ref{fig:incidentprof} shows this effect for the hydrogen \La\ and \Ha\ lines, and the helium line at 584~\AA.
As the outward velocity increases, the central peak of an incident emission resonance line profile is less intense, it is moved towards the red, and the line width becomes larger.
In order to study this Doppler effect, it is of course necessary to use a realistic incident radiation that varies with frequency in the calculations.

\cite{1987SoPh..110..171H} did their calculations in CRD and were able to show the main effects of the velocity-dependent radiation field on the integrated intensities of the hydrogen Lyman and Balmer lines.
However, it is important to compute the line spectrum in PRD if we want to compare the calculated line profiles with observations, and infer the plasma parameters in moving prominences. PRD differs  from CRD mainly in the line wings (in the line core, CRD dominates). The line wings themselves are greatly affected by the frequency shift of the incident radiation, and this yields potentially large line profile asymmetries.
\cite{1997SoPh..172..189G,1997A&A...325..803G} were the first to include PRD in their modelling of the hydrogen lines emitted by moving prominences.

\begin{figure}
	\center
	\includegraphics[width=\textwidth]{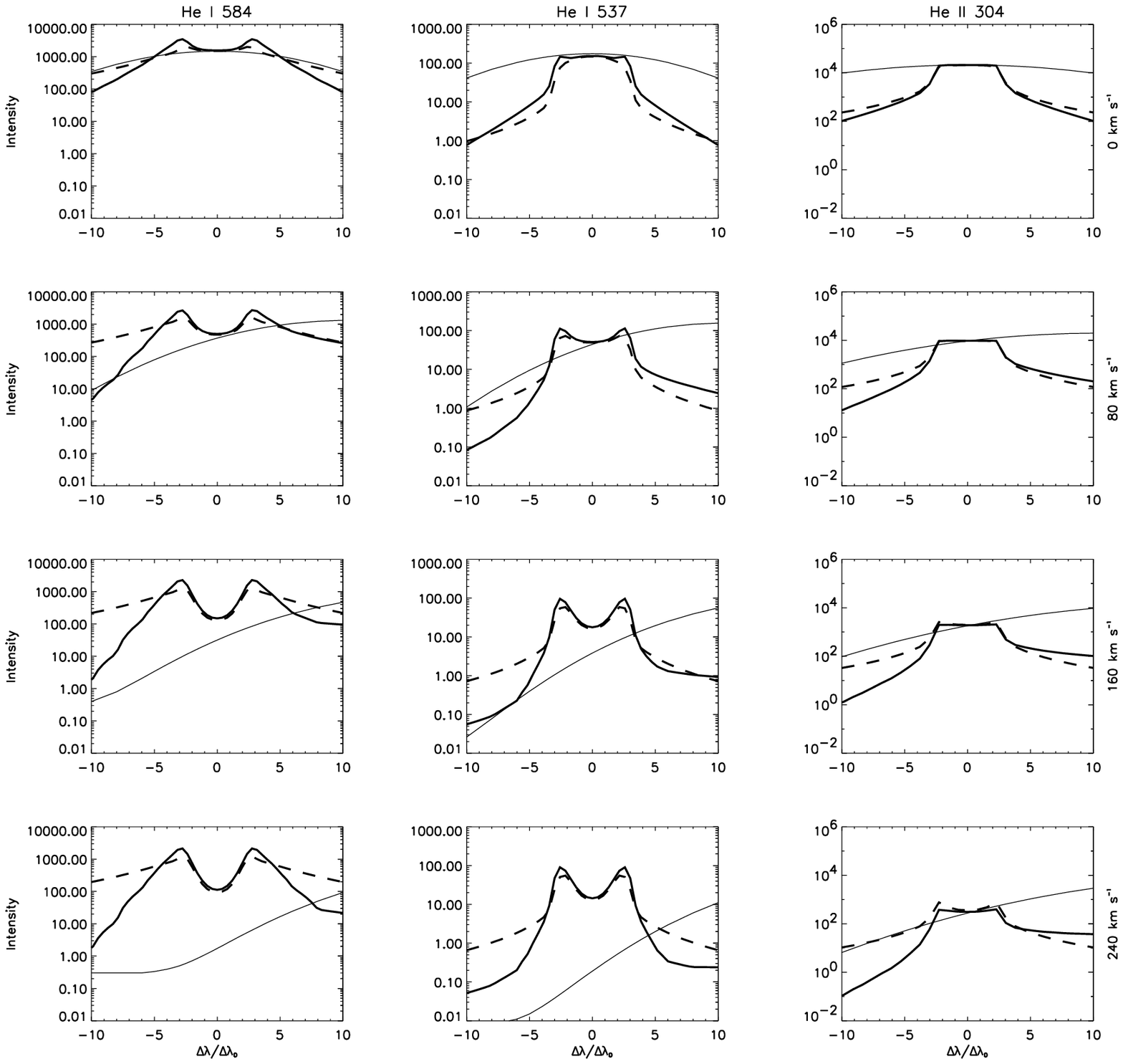}
	\caption{Differences between CRD ({dashed} line) and PRD (solid line) for the profiles of  three helium lines at different velocities: 0, 80, 160, and 240~\kmps from top to bottom. Abscissa: $\Delta\lambda$ (distance to line centre) in  Doppler units and limited to 10 Doppler widths around line centre; vertical axis: specific intensity in erg\,s$^{-1}$\,cm$^{-2}$\,sr$^{-1}$\,\AA$^{-1}$ on a log scale. {Model parameters are altitude $H=50\,000$~km, width $L=650$~km, temperature $T=6500$~K, and pressure $p=0.1$~dyn cm$^{-2}$. The thin solid line shows the incident line profile, gradually shifting towards the red as the prominence is moving radially at increasing velocities. From \cite{2007A&A...463.1171L}.}}
	\label{fig:lgv07-4}
\end{figure}

\cite{2007A&A...463.1171L} presented the first computations of the helium line profiles emitted by a moving prominence. 
The authors focused on the line profile properties and the resulting integrated intensities, and their sensitivities to the prominence plasma parameters. They also studied the effect of frequency redistribution in the line formation mechanisms. Figure~\ref{fig:lgv07-4} shows the differences between CRD and PRD for the profiles of  three helium lines (584, 537, and 304~\AA) at different velocities.
Note the wing asymmetries at higher velocities which are due to PRD effects.
These authors  only considered isothermal and isobaric prominence models similar to those presented in \cite{2001A&A...380..323L}. 

\begin{figure}
	\centering
	\includegraphics[width=0.6\textwidth]{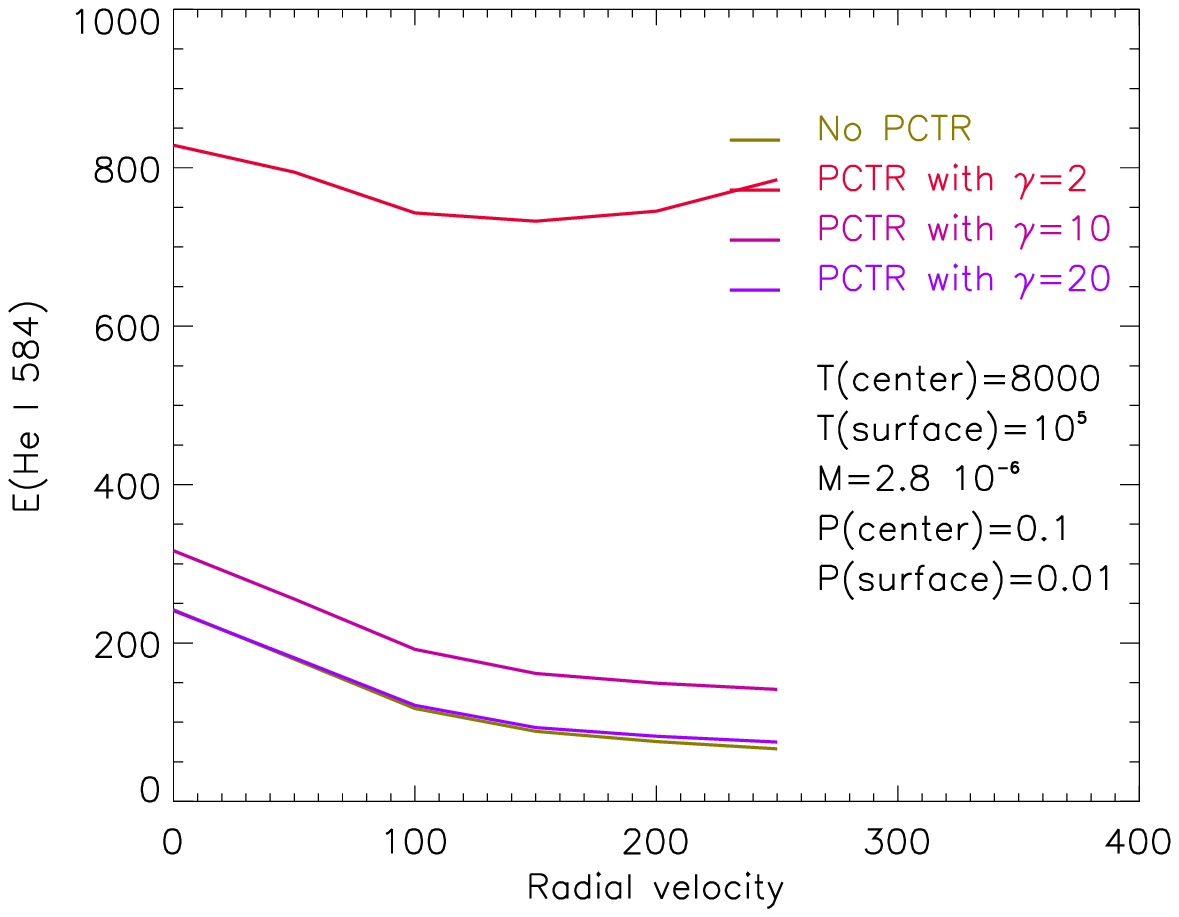}
	\includegraphics[width=0.6\textwidth]{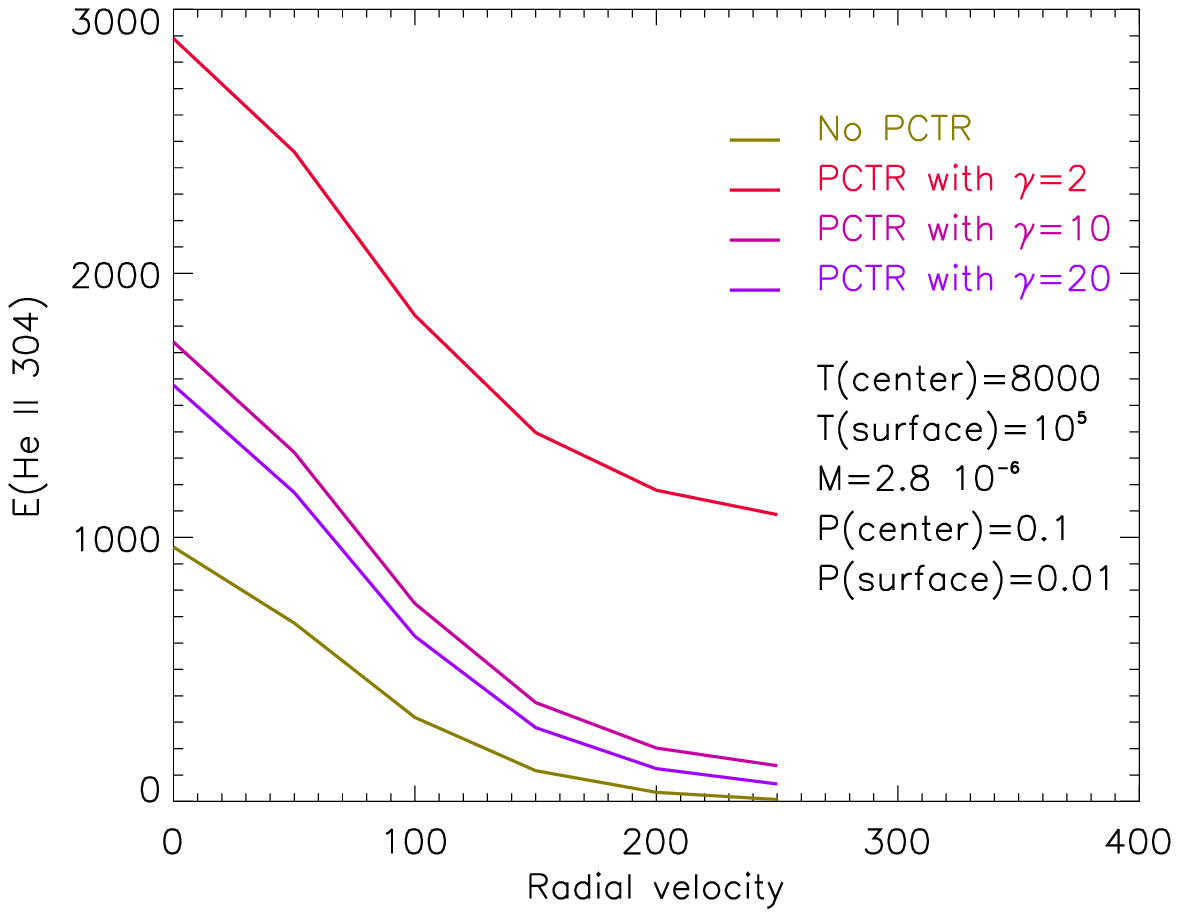}
	\includegraphics[width=0.6\textwidth]{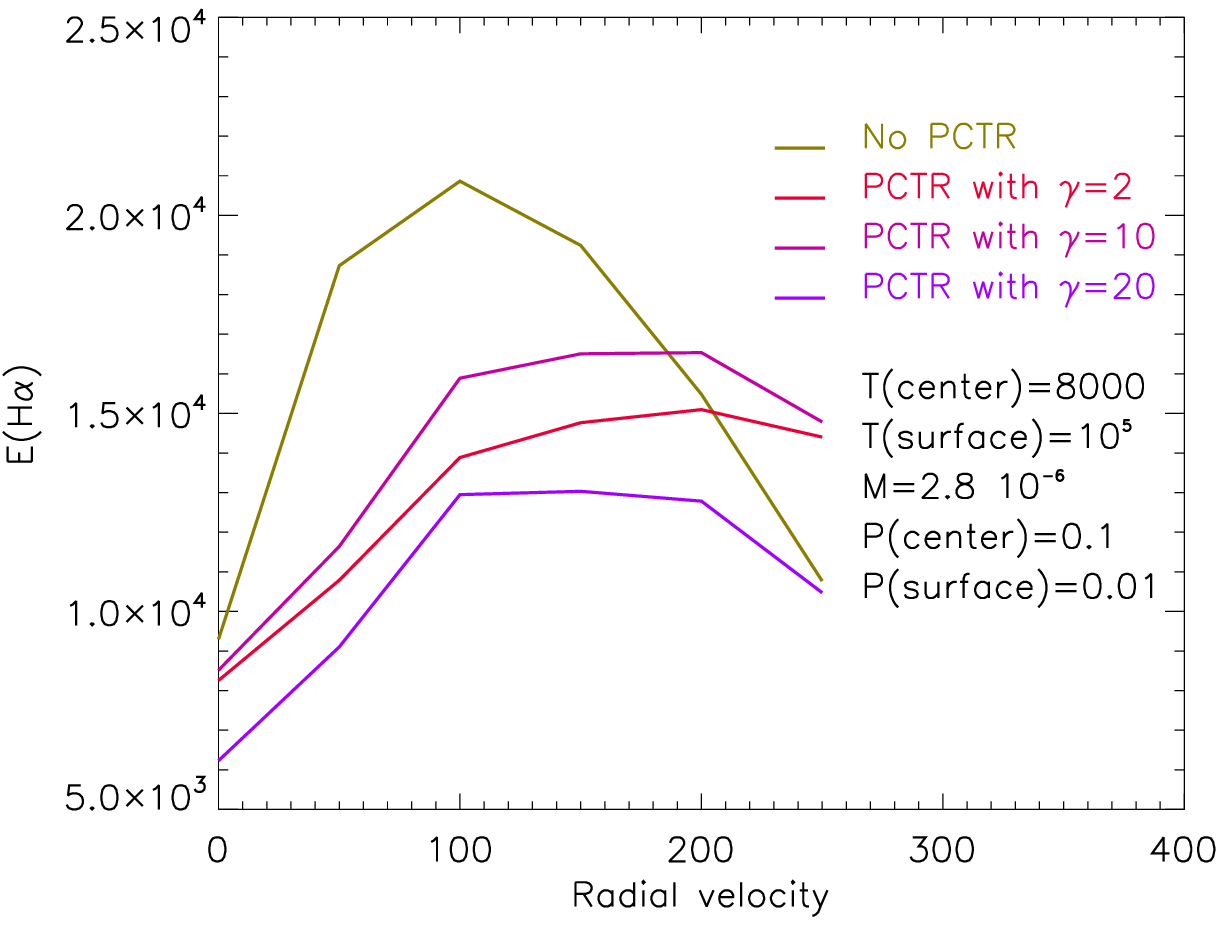}
	\caption{Variation of the He~I 584~\AA\ (top), He~II 304~\AA\ (middle), and hydrogen H$\alpha$ (bottom) line intensities (in erg s$^{-1}$ cm$^{-2}$ sr$^{-1}$) as a function of the radial velocity of the prominence plasma (in \kmps) for 4 different types of prominence atmospheres. The parameter $\gamma$ refers to the pressure and temperature profiles derived by \cite{1999A&A...349..974A} and is related to the temperature gradient in the PCTR: the higher $\gamma$, the higher the gradient, and the narrower the PCTR. From \cite{2008AnGeo..26.2961L}.}
	\label{fig:angeo}
\end{figure}

The effect of the PCTR on the H and He lines in active and eruptive prominences was presented for the first time in \cite{2008AnGeo..26.2961L}. 
The pressure and temperature profiles were taken from \cite{1999A&A...349..974A}. \cite{2008AnGeo..26.2961L} showed that the He~II 304~\AA\ line is strongly dependent on the radial velocity, which can be explained by the dominant role of scattering in the line formation.
	It is thus necessary to carefully consider the role of the radial motion of the plasma in the modelling of this line.
	Other lines such as He I 584~\AA\ and the hydrogen H$\alpha$ line are also sensitive to the Doppler dimming / brightening due to the radial motion of the plasma at velocities up to $\sim$ 100~\kmps (Fig.~\ref{fig:angeo}). 
	At higher velocities, the absorption profile of the transition gets significantly out of resonance with the incident radiation, and the resulting variation of the emergent intensity with the radial velocity then only depends on  parameters such as the strength of collisional excitations, or the coupling with other transitions.

\subsubsection{Summary}
	
The study of the variation of the resonance line intensities with the radial velocity of the plasma indicates that the Doppler dimming effects are essentially present when the relative contribution of the thermal emission compared to the scattering of incident radiation in the lines studied is low. 

Most of the information on the radial velocity can be obtained through the variation of the {relative} integrated intensity, \ie\ by measuring the amount of Doppler dimming on the EUV resonance lines of hydrogen and helium. 
Although this measurement is difficult to perform (mainly because one has to be observing at the right place and at the right time), space-based instruments such as EIT, SUMER, and CDS on SOHO, EUVI {\citep[the Extreme Ultraviolet Imager,][]{2008SSRv..136...67H}} on STEREO, SOT and EIS on Hinode, as well as TRACE, are able to obtain a temporal series of intensity measurements covering an entire prominence eruption, starting before the radial motions take place. In this case, it is possible to measure the dimming in intensity, and to compare it with  predicted relative intensity curves.

Detectors with a spectral pixel size around 0.05 \AA\ or less (\eg\ SUMER) are ideal as they allow line asymmetries to be observed. A lower spectral resolution still allows good comparison with theoretical line profiles, however it makes the task of finding a unique model that fits the observed optically thick profiles more difficult. 
In combination with the study of the apparent motion of the prominence material brought by {imagers on SOHO, Hinode, and STEREO,} the full velocity vector can be inferred.

Further modelling is necessary to take into account the fact that the prominence plasma is heated during an eruption. In this situation, Doppler dimming will be affected by an enhanced contribution of collisional excitation.
	This is one of the major challenges in this type of study: it is essential to make a distinction between a change in the thermodynamic state of the plasma (\eg\ an increase in temperature and density) and a change of the radial velocity. Does a decrease in the intensity of a line formed at low temperatures (say below 10000~K) correspond to the Doppler dimming effect due to the radial motion of the plasma, to variations of the thermodynamic plasma parameters, or to a combination of both?
	To answer this, other lines are needed to perform an independent diagnostic of the prominence plasma. For instance, the intensities of the He I 10830~\AA\ and D3 lines are not very sensitive to the radial velocity \citep{2007ASPC..368..337L}, essentially because of a flat incident spectrum. Therefore these lines can be used to determine the plasma parameters.
	Another mean of distinguishing between variations in intensity due to a change of the plasma state or to the radial velocity is to use the full line profiles. The radial motion of the prominence induces asymmetries in the line profile which cannot be attributed to temperature or pressure effects \citep{1997A&A...325..803G,2007A&A...463.1171L}.

Future studies will need to address active and eruptive filaments. As these structures are seen on the disk, the radial motion of the plasma cannot be inferred from imaging measurements, and Doppler shifts cannot always be interpreted in a straightforward manner. 
The case of eruptive disk filaments is particularly relevant as they can trigger Earth-directed CMEs.
Again, the question of the contribution of density variations or heating of the plasma versus the Doppler dimming effect has yet to be resolved. This is a complex issue, since many properties of the prominence plasma will vary during the eruption. As we  progress towards more sophisticated modelling, we will be able to disentangle this. Efforts should also be directed towards 2D modelling, as discussed in the following section.

\section{Results from 2D Non-LTE Modelling}
\label{sec:10}

The previous section is focused on 1D modelling of solar prominences. Although these 1D models {still} represent a useful and computationally efficient approach for a number of situations, they do not allow us to consistently study the variation of the radiation and the plasma parameters in two dimensions, \eg\ along and across the magnetic field lines. However, prominence observations carried out both from space and from ground reveal the structure of prominences at various angles with respect to the magnetic field orientation. Proper interpretation of such observations requires the use of more general 2D or 3D prominence models.

\subsection{2D Models of the Whole Prominence}

\begin{figure}
	\center
\includegraphics[width=\textwidth]{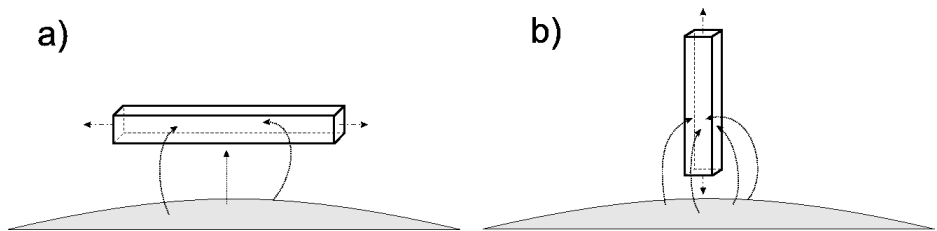}
\caption{Sketch of the geometry of 2D prominence models. An infinite dimension is oriented tangentially (a), or perpendicularly (b), to the solar surface. {Arrows oriented towards the slab indicate the incident radiation.}} 
\label{fig:2D_proms}
\end{figure}

The geometry of 2D prominence models is characterized by two finite and one infinite dimension which can be oriented tangentially, or perpendicularly, to the solar surface (Fig.~\ref{fig:2D_proms}). 
The variation of all quantities takes place within the plane of the cross-section perpendicular to the infinite dimension along which the medium is assumed to be homogeneous.

A robust method for solving the radiative transfer in 2D planar geometries was first developed by \citet{1978ApJ...220.1001M}. It makes use of a Hermitian integration formula on ray segments through grid points (short characteristics) and of CRD, and is also well suited for solving the velocity-dependent problems. This method was used by \citet{1982ApJ...254..780V} to construct a 2D horizontally infinite vertical slab model representing the whole prominence structure. {In this model}, the cross-section dimension along the axis perpendicular to the solar surface (height) dominates over the other cross-section dimension (thickness). Synthetic profiles of H~{I} \La, Ca~{II} H and K, and Mg~{II} h and k lines, obtained using the two-level atom approximation, were compared with OSO-8 observations and a good agreement was found for \La\ and Ca~{II} lines.
However, the method of \citet{1978ApJ...220.1001M} is computationally
prohibitive and was designed to treat CRD and the two-level atom approximation only.
Therefore, \citet{1994A&A...285..675A} developed a new method for the
solution of the non-LTE radiative transfer problem in 2D prominence geometry which can treat also
PRD, important for strong resonance lines
\citep[see][]{1987A&A...183..351H}. 
The method of \citet{1994A&A...285..675A} is based on the ALI technique (see Sect.~\ref{sec:8}), with use of the OAB (Olson-Auer-Buchler) approximate lambda operator \citep{1986JQSRT..35..431O} and the short characteristics method \citep{1988JQSRT..39...67K} to obtain the formal solution along individual ray segments. This method was used by \citet{1993A&A...274..571P} to compute the synthetic profiles of H~{I} (\La), Ca~{II} (H \& K), and Mg~{II} (h \& k) lines in a 2D horizontally infinite slab model using the two-level atom approximation. The authors confirmed a strong influence of PRD on synthetic \La\ profiles.

The models described above treat the 2D radiative transfer problem with the simplified two-level atom approximation. An ALI method for multi-level radiative transfer was further developed by \citet{1991A&A...245..171R,1992A&A...262..209R} (the so called MALI method -- Multilevel Accelerated Lambda Iterations -- see Sect.~\ref{sec:8}). It is based on coupled iterative solution of radiative transfer using ALI and preconditioned equations of statistical equilibrium for the level populations. This method was successfully implemented into 2D horizontally infinite slab geometry by \citet{1994A&A...292..599A} and by \citet{1995A&A...302..587P}, both for CRD and PRD.
Recently, \cite{2009A&A...498..869L} computed emergent neutral helium line profiles (including the He~I 10830~\AA\ and D3 multiplets) using 2D horizontal prominence fine structure models in both single-thread and multi-thread configurations. They considered isothermal and isobaric horizontal slabs irradiated by the solar surface on the sides and on the bottom, and also by the corona on the sides and on the top.
Radiative transfer computations were performed for hydrogen and He~I transitions.  The authors obtained a good agreement with the results of previous models, and confirmed the importance of the multi-thread approach for the modelling of the prominence fine structures.

In the case of the horizontally infinite slab geometry, the incident radiation from the solar surface irradiates the bottom and both side surfaces of the prominence slab. The top of the slab is not irradiated if one does not consider radiation coming from the solar corona (Fig.~\ref{fig:2D_proms}a). The situation is somewhat different if the prominence model assumes a vertical geometry, with the infinite axis pointing perpendicularly to the solar surface. Then the incident radiation irradiates the prominence structure from all sides (see Fig.~\ref{fig:2D_proms}b). \citet{2005A&A...434.1165G} developed 2D vertical models with cylindrical cross-section. In this case, the plasma parameters and the radiation field vary with the distance to the cylinder axis and with azimuth. The ALI method used by \citet{2005A&A...434.1165G} for the solution of the non-LTE radiative transfer in cylindrical geometry with assumption of the two-level atom and CRD was successfully modified to treat also the multi-level atom with anisotropic incident radiation by \citet{2006A&A...448..367G}. \citet{2007A&A...465.1041G} studied the temperature relaxation of 2D cylindrical threads to the radiative equilibrium. \citet{2008A&A...487..805G} developed 2D cylindrical models which take into account the Doppler effects produced by 3D velocity fields. The 2D prominence models with cylindrical geometry can be used also for other solar features such as coronal loops \citep[see \eg ][]{2007ApJ...664.1214P,2008A&A...487..805G}.
The radiative transfer calculations performed under cylindrical geometry could be used to provide additional constraints to the  models described in Sect.~4.3.3 of Paper~II to study the effect of partial ionization on the damping of MHD waves.

\subsection{2D Models of Prominence Fine-Structure Threads}

High-resolution prominence observations reveal that solar prominences
are not homogeneous bodies, but exhibit a variety of fine
structures, in particular elongated thread-like features and knots of
plasma. A considerable number of prominence
fine-structure threads have nearly vertical orientation, although
observations of quiescent prominences composed mainly of horizontal
threads also exist \citep[see Sect.~\ref{sec:5}, and also the review by][]{2007ASPC..370...46H}. Vertical threads which are perpendicular
to the mean magnetic field in the prominence could be formed as plasma
condensations propagating vertically due to the interaction of the
neighbouring magnetic field dips, as suggested by
\citet{1988dssp.conf..133P}.

\begin{figure}
	\center
\includegraphics[width=\textwidth]{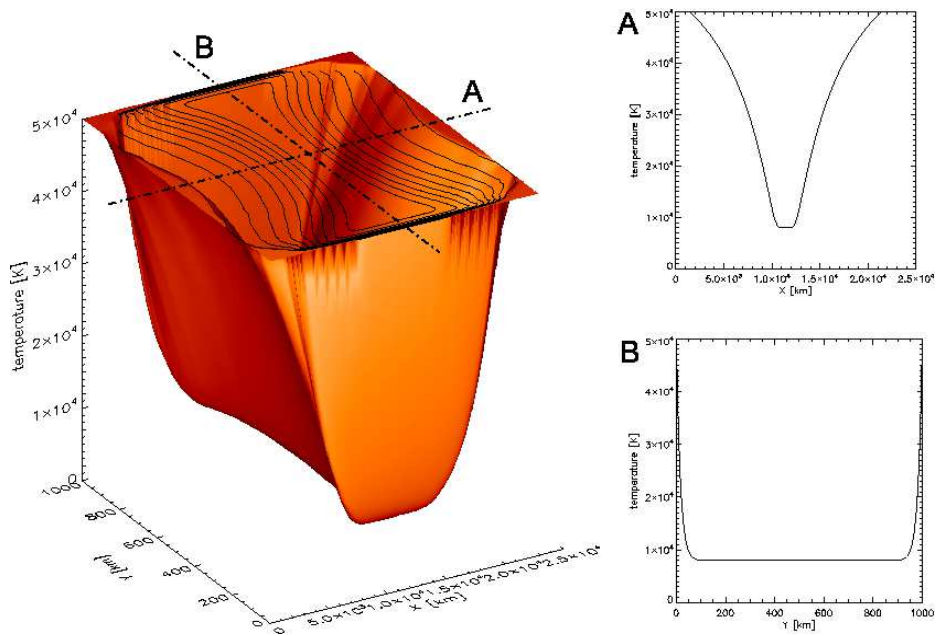}
\caption{Semi-empirical variation of the temperature within the cross-section of the 2D vertical prominence fine-structure thread. The $x$ and $y$ axis, representing geometrical dimensions of the thread, are not drawn to the same scale. Iso-contours of the temperature are shown. This plot clearly shows two different shapes of the PCTR, with A) a gentle rise of the temperature along the $x$-axis (along the field lines), and B) a steep gradient of the temperature along the $y$-axis (across the magnetic field lines).} \label{fig:gun_fig1}
\end{figure}

\citet{2001A&A...375.1082H} described MHS equilibrium of such vertical 2D fine-structure threads hanging in magnetic dips of the horizontal magnetic field. This was obtained by a generalization of the 1D prominence models of \citet{1999A&A...349..974A} in MHS equilibrium of Kippenhahn-Schl\"uter type.
The multi-level non-LTE radiative transfer is solved in this 2D structure using MALI technique \citep{1995A&A...299..563H,1995A&A...302..587P} with short characteristics method to obtain the formal solution along individual ray segments \citep{1994A&A...285..675A}. The 2D temperature variation (Fig.~\ref{fig:gun_fig1}), prescribed semi-empirically, accounts for two different PCTRs. 
The PCTR across the magnetic field lines is very narrow (Fig.~\ref{fig:gun_fig1}B), with steep gradient of the temperature from the cool central part of the thread towards its boundaries. The PCTR along the magnetic field lines is much more extended (Fig.~\ref{fig:gun_fig1}A), with a shallow rise of the temperature. Synthetic profiles of the hydrogen Lyman lines obtained along and across the magnetic field lines show significant differences with considerably reversed profiles obtained across the magnetic field and unreversed profiles obtained along the field. Such a behaviour is in agreement with findings based on the SUMER observations presented by \citet{2001A&A...370..281H}. A modified version of the 2D prominence fine-structure thread model of \citet{2001A&A...375.1082H} was used by \citet{2005A&A...442..331H} to compute a grid of 18 2D models. The 12-level plus continuum hydrogen model atom and adaptive geometrical grid for 2D MHS equilibrium computations \citep{2003ASPC..288..441H} was used. This study confirmed previous findings about the  behaviour of profiles of the hydrogen Lyman series -- see also \citet{2007SoPh..241...53S}. \citet{2007A&A...463..737G} supplemented the previous study with the investigation of the Lyman continuum behaviour and showed that a comparison between synthetic and observed Lyman continuum profiles in the wavelength range $800-911$~\AA\ produces new useful constraints on the temperature variation of the prominence fine structures (see the discussion on the determination of the electron temperature from the slope of the hydrogen Lyman continuum in Sect.~\ref{sec:lycont1D}).

The same 2D thread model was used by \citet{2007A&A...472..929G} to
derive the properties of the prominence fine-structure threads. A
trial-and-error method was used to find a model with the best agreement
between synthetic profiles of the hydrogen Lyman series and the
observed ones. The observed profiles were obtained by SUMER on May
25, 2005 and contain the full hydrogen Lyman line series, including unique
\La\ line observations outside the attenuator. The prominence
fine-structure properties obtained in this way are in a good agreement
with values given in the textbooks, such as
\citet{1995nsp..book.....T}. However, the authors concluded that the
observed Lyman line profiles can be better reproduced by using
multi-thread fine-structure models consisting of a set of identical 2D
threads hanging on separate magnetic field lines
with the LOS perpendicular to the magnetic field.

\begin{figure}
	\center
\includegraphics[width=0.8\textwidth]{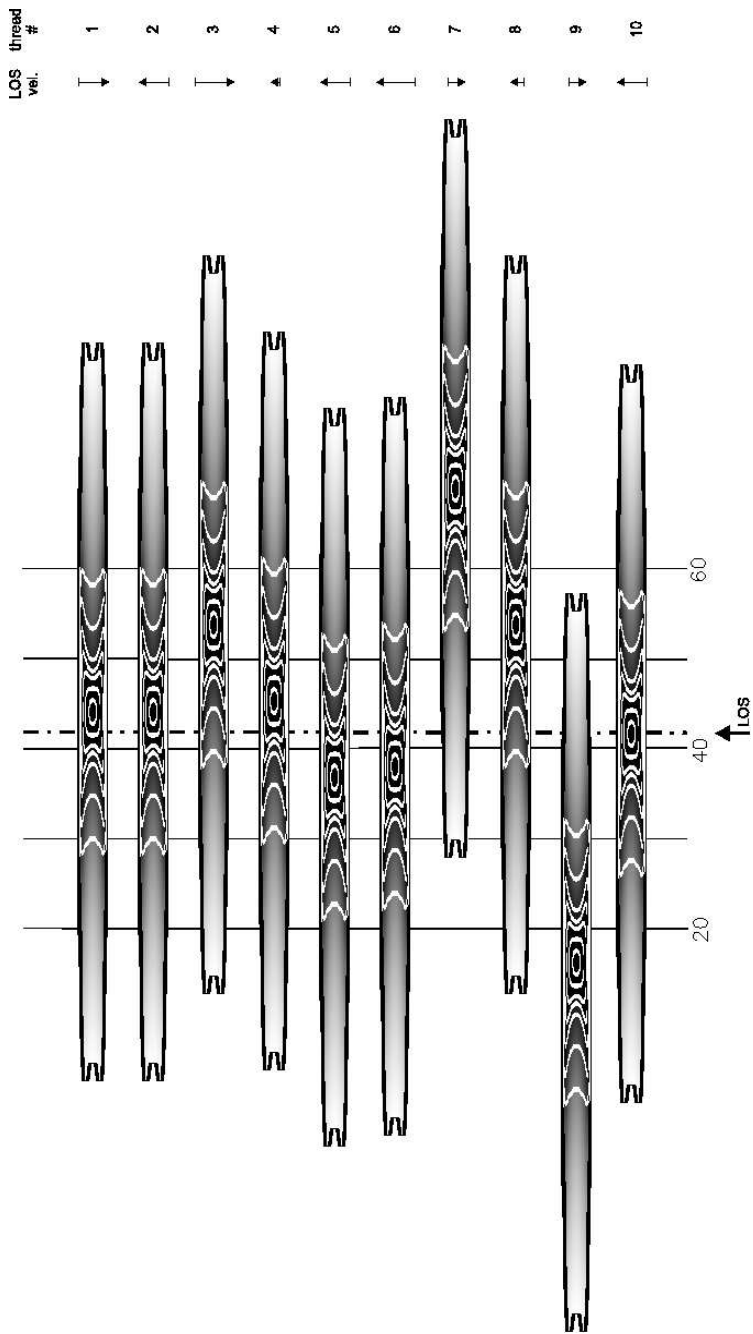}
\caption{Multi-thread 2D fine-structure model with randomly assigned LOS velocities, indicated by the arrows at the top. The grey-scale represents the variation of the temperature from the minimum central temperature (black) to the maximum boundary temperature (white). White contours represent the density from central $5.0 \times 10^{-13}$~g cm$^{-3}$ to external $0.05 \times 10^{-13}$~g cm$^{-3}$ values. From \cite{2008A&A...490..307G}.} \label{fig:gun_fig2}
\end{figure}

Such multi-thread models were used also by \citet{2008A&A...490..307G} to study asymmetries of the hydrogen Lyman line profiles. For this reason each thread of the multi-thread model has a randomly assigned LOS velocity. A set-up of such multi-thread models is shown in Fig.~\ref{fig:gun_fig2}. 
This model consists of 10 identical 2D fine-structure threads randomly shifted with respect to the foremost thread (thread \#10).
LOS velocities are also generated randomly, in this case from the interval $\left<-10,10\right>$~\kmps, and their orientation and magnitudes are indicated by arrows. 
Synthetic Lyman line profiles obtained by such multi-thread modelling exhibit substantial asymmetries, even with the LOS velocities being only of the order of 10~\kmps (note that no mutual radiative interaction is taken into account). The \La\ profiles can also exhibit an opposite asymmetry to that of the higher Lyman lines. This is in agreement with the (so far) rather puzzling behaviour of observed \La\ and L$\beta$ line profiles noticed already by \citet{1982ApJ...253..330V}, and confirmed by \citet{2007A&A...472..929G} and \citet{2007SoPh..246..327V}.

\section{Energy Balance Considerations}
\label{sec:11}

After the work of \citet{1976ApJ...205..273H}, who demonstrated that
the {\em radiative equilibrium} temperature inside 1D prominence slabs is
much lower than what is typically deduced from spectral observations
(see temperature diagnostics in Sects.~\ref{sec:3} and \ref{sec:4}), various authors have
attempted to investigate possible sources of prominence heating. 

The radiative equilibrium means that the radiative flux $F_{\rm r}$ integrated over all frequencies is conserved inside the prominence slab:
\begin{displaymath}
\frac{\mathrm{d}F_{\rm r}}{\mathrm{d}x} = 0 \ ,
\end{displaymath}
\begin{displaymath}
F_{\rm r} = \int_0^\infty F_{\nu} \, \mathrm{d}\nu \ ,
\end{displaymath}
\begin{displaymath}
F_{\nu} = \frac{1}{2} \int_{-1}^{1} I_{\nu,\mu} \mu \, \mathrm{d}\mu \ .
\end{displaymath}
This also means that the net radiation energy emitted at a given point
must be exactly equal to that absorbed:
\begin{equation}
L = 4\pi \int_0^\infty (\eta_{\nu} - \chi_{\nu} J_{\nu}) \mathrm{d}\nu =
4\pi \int_0^\infty \chi_{\nu}(S_{\nu} - J_{\nu}) \mathrm{d}\nu = 0 \ .
\label{eq:radloss}
\end{equation}
In a general case there must exist
sources of heating and cooling which will establish the energy equilibrium.
The energy balance equation then reads:
\begin{equation}
	\label{eq:enbal}
\frac{\mathrm{d}F}{\mathrm{d}x} = L - H \ ,
\end{equation}
where
$F = F_{\rm c} + F_{\rm v}$ represents the conductive flux and the flux associated
with the mass flows (enthalpy flux), 
and
$L$ are the so-called radiative losses defined in (\ref{eq:radloss}). 
Other possible sources of the heating or cooling are included in the term $H$ (\eg\ wave heating, magnetic and mechanical heating, shock dissipation).  This term is usually considered to be proportional to gas density.
This energy balance equation constrains the temperature
structure of the prominence (1D slab in the above case).

The evaluation of the radiative losses inside the cool prominence represents a difficult task because as we see from (\ref{eq:radloss}), $L$ depends on the source function and on the radiation field. Moreover, the integration over frequencies includes various lines and continua of species which are supposed to be important coolants of the prominence plasmas. In the chromosphere, where the thermodynamic conditions are similar to prominences, the most important coolants are hydrogen, calcium and magnesium \citep[see][]{1981ApJS...45..635V}. In the case of prominences, no comprehensive study was devoted to radiative losses of elements other than hydrogen. \citet{1971SoPh...19..401P} made a crude estimate of Ca~II H and K line losses and claimed their relative importance. However, more detailed transfer modelling of \citet{1987A&A...175..277Z} has shown that calcium losses are negligible for prominence slabs and a similar result was also obtained by \citet{1999A&A...349..974A}. On the other hand, some authors have considered the optically thin losses in addition to hydrogen ones computed in detail from non-LTE level populations and radiation intensities \citep{1996ApJ...466..496F,1999A&A...349..974A}, or added also the helium component \citep{1991ApJ...370..763K}.

Concerning the heating, the classical conductive-heating term in (\ref{eq:enbal}) can
be generalised by assuming that the PCTR layers are also heated due to
the {\em ambipolar diffusion}. This was studied in detail by
\citet{1996ApJ...466..496F} who have demonstrated that inclusion of the
ambipolar diffusion dramatically increases \eg\ the \Lb\ line
intensity in multi-slab models. However, such large intensities are
inconsistent with SUMER observations of hydrogen Lyman lines \citep[\eg ][]{2007SoPh..246..327V}. It is not clear whether the problem lies in the method used to compute the Lyman lines with PRD under the energy balance with ambipolar diffusion, or whether the ambipolar diffusion itself is improperly treated in case of prominences \citep[see the discussion in][]{2005A&A...442..331H}.
This certainly deserves further attention.

Prominence heating associated with the mass inflow was estimated for various 1D models by \citet{2000A&A...358L..75A} who considered the inflow of enthalpy and ionization energy. The resulting energy gains were compared with integrated radiative losses obtained for such slabs by \citet{1999A&A...349..974A}. For reasonable inflow velocities, many of the considered models can be in energy equilibrium.
{On the observational side, we have a very large range of velocity values and orientations (see Tables~\ref{t:velocities-ha} and \ref{t:velocities-euv}). The magnitude of velocities is in the bracket $2-35$~\kmps\ with some indication of increased magnitude for increased temperature (compare Table~\ref{t:velocities-ha} with Table~\ref{t:velocities-euv}). But the orientation of these velocities with respect to the prominence axis is unclear. Moreover, one would suspect that in the frame of a \textit{symmetrical} prominence where most UV lines are formed in the PCTR, the inflow velocities, which would be of opposite signs, cancel each other. In fact, there is no direct evidence of inflow of hot material which supports the enthalpy contribution. We also have no values of the radiative losses corresponding to the observed prominences. Consequently, it is difficult to draw any conclusion concerning the role of enthalpy in the energy budget. However, having in mind the values of radiative losses in the modelling of \citet{2000A&A...358L..75A} (which represent some observations) on one hand, and noting that the velocity magnitude is relatively high in transition region lines (more than 10~\kmps\ in Table~\ref{t:velocities-euv}, compared to 1.7~\kmps\ in the work of \citeauthor{2000A&A...358L..75A}) on the other hand, one can derive that the enthalpy flux can match the radiative losses of most of the \citet{2000A&A...358L..75A} models. The issue of comparing enthalpy (derived from observed mass flows in their whole complexity) with radiative losses also derived from observations is certainly an important one to be addressed.}

Returning back to radiative equilibrium models, \citet{1976ApJ...205..273H} have also demonstrated that the penetration of the hydrogen Lyman continuum radiation into a filamentary prominence causes an increase of the central temperature by about $10^3$~K. The reason is that the radiation temperature of the incident Lyman continuum is higher as compared to that of Balmer continuum. The latter one is optically thin in all prominence models, and thus fully penetrates into the central parts. This result suggests that for geometrically thin fine-structure threads, the central radiative equilibrium temperature will be higher compared to more compact and thick slabs where it can reach 4600~K according to \citet{1976ApJ...205..273H}. Such a behaviour was demonstrated by \citet{2007A&A...465.1041G} for vertical cylinders.
By increasing the gas pressure, the equilibrium temperature decreases
because the radiative losses are higher for higher density. However,
contrary to the results of \citet{1976ApJ...205..273H}, the radiative
equilibrium temperatures obtained by \citet{2007A&A...465.1041G} are
much higher, reaching more than 9000 K for low-pressure models. A
similar result was also obtained by \citet{1999A&A...349..974A}, see
their Figs.~7 and 8. By adding losses due to other elements \citep[as \eg\ in][]{1991ApJ...370..763K}, the central temperatures may again
decrease unless some non-radiative heating is considered. This complex
behaviour deserves further study based on detailed non-LTE treatment of
radiative losses.

Energy balance in hotter parts of the prominence was studied by various
authors using the DEM (see Sect.~\ref{sec_dem}). Older work was summarized by
\citet{1989ASSL..150...47E}. Among recent studies let us mention those
of \citet{2007A&A...469.1109P} or \citet{2008A&A...480..537A}. Within
the hotter prominence plasmas, for temperatures above 30000~K, the
energy balance was treated in dynamical models of prominence
condensations -- see Paper~II {(Sect.~3)}.

\section{Open Issues}
\label{sec:12}

This review focuses on results on the thermodynamic parameters brought by spectral
inversion techniques and non-LTE modelling of the prominence plasma. While most
of the thermodynamic parameters are found to be in overall agreement with the Hvar
Reference Model Atmosphere of \cite{1990LNP...363..294E} -- see Table~\ref{tab:hvar}, we still notice
a large range of values concerning densities, ionization degree, filling factor, etc.
Such a large range implies uncertainties on the thermodynamic parameters,
a variety of conditions within an observed prominence and also from prominence to
prominence. As we have seen with the case of the Lyman series, it may also imply that
the properties which are derived depend on the viewing angle, the spatial resolution, or other factors.
It is true that we are now starting
to obtain high resolution ground-based and space-based observations, and to be able
to perform sophisticated non-LTE modelling of prominences with fine structures and
complex geometries. As a result, today's picture of a prominence is that of a dynamic
bundle of small-scale structures in constant evolution. The density, temperature, and
velocity of the plasma all change with time and location, and can vary greatly. This
is vitally connected to the key questions of the magnetic structure and the formation
and {evolution} of prominences, topics that are discussed in Paper~II.
It is tempting to ascribe specific properties to any observed prominence and perform modelling 
adapted to each observational situation.
{However, we must remember that there are some constraints that all prominences are subjected to, \eg\ mechanical support, energetic balance, or long lifetime.}
In order to be able to impose these constraints, it is necessary to constrain the diagnostic
and to obtain typical (if not canonical) values of quiescent and active prominences.

With these considerations in mind, below we attempt to describe some of the outstanding issues that need to be
addressed in the future, both observationally and theoretically.
\begin{enumerate}
\item The issue of the determination of the thermodynamic parameters is completely interrelated with the issue of the \textbf{{fine structure}}. 
\begin{enumerate}
\item \textit{Geometry.} What can imaging observations from space tell us about the 3D structure of prominences?
It is difficult to pick out small features against the disk in the He~II 304~\AA\ channel of EIT and EUVI. 
{Nevertheless, the EUVI observations from STEREO A and B, when A and B are in quadrature, should certainly help to derive the 3D fine-scale structure of filaments and prominences.}
{Meanwhile}, non-LTE modelling should progress towards 3D radiative transfer {including radiative interactions} between  threads. Important steps have been taken recently \citep[\eg][]{2008A&A...490..307G,2008A&A...487..805G}. Non-LTE modelling should also be coupled with MHD. 
\item \textit{Filling factors.} Even if future instruments lack the spatial resolution necessary for identifying the prominence fine structure (cool core and PCTR -- prominence-to-corona transition region), efficient and multi-wavelength spectrometers should allow the derivation of local densities and average squared densities (from DEM studies), from which sound filling factor values could be derived.
\item \textit{Ionization degree.} It is a critical issue because of its impact on the derivation of the mass of the prominence. It also is critical if one tries to understand how material motion can take place perpendicularly to the magnetic field lines,, or what the effects of partial ionization on MHD waves and oscillations are. There is now a need for numerical coupling between oscillations modelling (as described in Paper~II, Sect.~4) and non-LTE ionization balance modelling.
\end{enumerate}
\item The \textbf{{prominence-to-corona transition region}} is a critical region for the transfer of conductive energy (or its absence) from the corona to the prominence.
\begin{enumerate}
\item \textit{General properties derived from EM and DEM analyses.}
In spite of many efforts aimed at a careful analysis of the data and intensive modelling
development, the properties of the PCTR remain uncertain. The sampling of
the thermal structure through the PCTR is still not detailed enough. For example,
the different results concerning the slope of the temperature gradient for $\log T <5$
need to be understood to engage a correct investigation on the heating mechanisms
in prominences. This also includes identifying the location of the temperature where
the DEM has its minimum. How to invert this information to constrain modelling?
Some attempts have been made, but a more systematic interaction between theory
and observations is needed.
\item \textit{The PCTR velocities.}
The reduced spatial resolution of the UV-EUV compared to optical observations
maintains some open questions on PCTR small-scale structure and magnetic configuration.
The NTV (and DEM) may give access to unresolved spatial scales properties.
It is still not clear if the NTV at a given temperature changes with the location
inside the prominence. This can indicate the location of more intense activity inside
the prominence (including heating deposition). Then, much effort needs to be put
in the interpretation of this component of the plasma velocity, either in terms of
waves and micro-turbulence, or flows in a small-scale geometry.
Here again, the role of the spatial resolution should be reminded. \cite{1980A&A....85..326E} found
an anti-correlation between the resolution and the Doppler velocities, and a
correlation between the spatial resolution and the derived NTV.
\item \textit{PCTR  mass flows compared to cool mass flows.}
A pattern has become apparent, especially in filaments observed against the disk, in which roughly vertical motions appear in barbs, and horizontal ones appear in the prominence spine. How does this observed pattern connect with observations of prominences at the limb? What is the connection between observed \Ha\  and EUV motions? Are they from separate elements of the prominence, or aspects of the same motions? Current observations suggest that such motions tend to be faster {in the EUV than in \Ha}. Is this a correct picture, or is it simply due to limitations in the EUV observations? How far can individual moving features be traced? What ramifications are there in these comparisons for our understanding of the PCTR?
\end{enumerate}
\item The long-standing issue of \textbf{{energy balance}} still needs careful investigation.
\begin{enumerate}
\item \textit{Spatial and temporal considerations.}
On which spatial and time scales does the prominence heating take place? Energy balance issues including the role of radiative equilibrium, the importance of ambipolar diffusion, and the heating by waves or shocks, need to be carefully studied {using sophisticated non-LTE models}. 
\item \textit{Energy losses due to radiation emitted by the prominence plasma.}
What would be the modelled time-dependent evolution of prominence condensations (see Paper~II Sect.~3) if these simulations were considering temperatures lower than 30000~K at which hydrogen lines form? Efforts should be made to couple the non-LTE radiative transfer calculations to other types of prominence plasma numerical modelling. This modelling will then lead to radiative signatures and yield new observational scenario.
\item {\textit{Enthalpy.} As we have seen in Sect.~\ref{sec:11}, we do not have precise information on the directions of the observed flows with respect to the cool core of prominence structures. A determination of the direction and magnitude of velocities, along with a precise measurement of densities in a large range of temperatures, is necessary in order to evaluate the enthalpy contribution.}
\end{enumerate}
\item Modelling of \textbf{{filaments}.} \\
What do observations of filaments on the disk in EUV optically thick lines, in \Ha, and in He~I 10830~\AA\ tell us about their physical state? How can we relate the observations to the filaments thermal and magnetic structure? What would be a monochromatic appearance of fine-structures corresponding to various magnetic models (\eg\ dips)? 
In the past decade, much effort has been devoted to the modelling of the radiation emitted by prominences on the limb. {There should be a similar focus on filaments.}  The fine-scale structure and motions of {filaments} should be consistently taken into account in the modelling.  
\item \textbf{{Modelling}} \textit{vs} observations.\\
{How can we explain the multi-wavelength and multi-temperature appearance of prominences and filaments and of their surrounding cavities?} How would synthetic spectra from non-LTE prominence models compare with a full
reference spectrum obtained by an EUV spectrometer? To answer this, we need to
extend the non-LTE modelling to other species. This will give an unprecedented
insight on the thermodynamic structure of prominences, from their cool core to
the boundary with the corona. Much effort needs to be put in deriving prominence
abundances, an important parameter for the choice of the best scenario of prominence
formation. Such a multi-species non-LTE modelling would add constraints and
help interpreting observations.
\item Some \textbf{{basic physical processes}} may need to be revisited.\\
Are prominences in statistical equilibrium? Is ionization equilibrium always a valid
assumption? The recombination time scales seem to be comparable to the fine structure
lifetime, as shown by the SOT movies. It is necessary to investigate
the consequences {of these new observations} for modern non-LTE prominence models, and to develop time-dependent
numerical codes to properly evaluate ionization and recombination times,
and to bridge the gap between simplified models and high-resolution observations.
\end{enumerate}

Large multi-instrument observing campaigns have highlighted the need to reconcile observations performed by various instruments operating at different wavelengths and at a large range of spatial and temporal resolutions. The physical processes responsible for the emitted radiation in these various wavelength regimes are different. This has to be kept in mind when interpreting such observations. 
{A mission offering two simultaneous vantage points, equipped with instruments of high spectral resolution capable of simultaneous observations over a large set of lines, would likely help us to answer central questions related to the prominence-to-corona transition region.} 
Ambitious missions (\eg\ in the frame of Cosmic Vision) should aim at sub-one tenth of arcsecond resolution. 
Imaging spectroscopy is an important aspect of many solar observations and indeed of prominences and filaments. Having imaging and spectrometry capabilities on board future spacecraft with a temperature coverage in the range $10^4-10^6$~K (including key {chromospheric and coronal} lines {in addition to} \La, \Ha, He~II 304~\AA), together with density sensitive lines in the range $10^9-10^{11}$~\cc, would be a major asset, as this provides the crucial plasma parameters, along with the necessary spatial information, to address the questions outlined above.

\begin{acknowledgements}
The authors gratefully acknowledge the support of the International Space Science Institute through its International Team program, and thank the other colleagues from the \textit{Spectroscopy and Imaging of Quiescent and Eruptive Solar Prominences from Space} team for fruitful discussions.
SP acknowledges the support from the Belgian Federal Science Policy Office through the ESA-PRODEX programme.
JCV thanks the Programme National Soleil-Terre for financial support.
SG and PH acknowledge the support from the ESA-PECS project No. 98030.
SG acknowledges the support from the institutional project AV0Z10030501 and from grant 205/07/1100 of the Grant Agency of the Czech Republic.
TK thanks the NASA Heliophysics Guest Investigator Program for financial support.

The authors thank Yong Lin, Tom Berger, and Jongchul Chae, for providing original versions of some of the figures used in this review{, and the careful reading and helpful comments by two anonymous referees}.

The SUMER project is financially supported by DLR, CNES, NASA, and the ESA-PRODEX Programme (Swiss contribution).
SOHO is a mission operated by ESA and NASA.
Hinode is a Japanese mission developed and launched by ISAS/JAXA, collaborating with NAOJ as a domestic partner, NASA and STFC (UK) as international partners. 
{The VAULT instrument development work has been supported by the ONR task area SP033-02-43 and by NASA defense procurement request S-84002F.}

This research has made use of NASA's Astrophysics Data System.
\end{acknowledgements}


\begin{thebibliography}{269}
\ifx \bisbn   \undefined \def \bisbn  #1{ISBN #1}\fi
\ifx \binits  \undefined \def \binits#1{#1} \fi
\ifx \bauthor  \undefined \def \bauthor#1{#1} \fi
\ifx \bjtitle  \undefined \def \bjtitle#1{\textrm{#1}}\fi
\ifx \batitle  \undefined \def \batitle#1{#1} \fi
\ifx \bctitle  \undefined \def \bctitle#1{#1} \fi
\ifx \bvolume  \undefined \def \bvolume#1{\textbf{#1}}\fi
\ifx \byear  \undefined \def \byear#1{#1} \fi
\ifx \bissue  \undefined \def \bissue#1{#1} \fi
\ifx \bfpage  \undefined \def \bfpage#1{#1} \fi
\ifx \blpage  \undefined \def \blpage #1{#1} \fi
\ifx \burl  \undefined \def \burl#1{#1} \fi
\ifx \doiurl  \undefined \def \doiurl#1{#1} \fi
\ifx \betal  \undefined \def \betal{et al.} \fi
\ifx \binstitute  \undefined \def \binstitute#1{#1} \fi
\ifx \beditor  \undefined \def \beditor#1{#1} \fi
\ifx \bpublisher  \undefined \def \bpublisher#1{#1} \fi
\ifx \bbtitle  \undefined \def \bbtitle#1{\textit{#1}} \fi
\ifx \bedition  \undefined \def \bedition#1{#1} \fi
\ifx \bseriesno  \undefined \def \bseriesno#1{#1} \fi
\ifx \blocation  \undefined \def \blocation#1{#1} \fi
\ifx \bsertitle  \undefined \def \bsertitle#1{#1} \fi
\ifx \bsnm \undefined \def \bsnm#1{#1} \fi
\ifx \bsuffix \undefined \def \bsuffix#1{#1} \fi
\ifx \bparticle \undefined \def \bparticle#1{#1} \fi
\ifx \barticle \undefined \def \barticle#1{#1} \fi
\ifx \botherref \undefined \def \botherref #1{#1} \fi
\ifx \url \undefined \def \url#1{#1} \fi
\ifx \bchapter \undefined \def \bchapter#1{#1} \fi
\ifx \bbook \undefined \def \bbook#1{#1} \fi
\ifx \bcomment \undefined \def \bcomment#1{#1} \fi
\ifx \oauthor \undefined \def \oauthor#1{#1} \fi
\ifx \citeauthoryear \undefined \def \citeauthoryear#1{#1} \fi
\ifx \texttildelow  \undefined \def \texttildelow{\symbol{126}} \fi
\def \endbibitem {}

\bibitem[\protect\citeauthoryear{{Anzer} and
  {Heinzel}}{1998}]{1998SoPh..179...75A}
\begin{barticle}
\bauthor{\binits{U.} \bsnm{{Anzer}}}, \bauthor{\binits{P.} \bsnm{{Heinzel}}},
\batitle{{Prominence Parameters Derived from Magnetic-Field Measurements and
  NLTE Diagnostics}}.
\bjtitle{\solphys}
\bvolume{179},
\bfpage{75}--\blpage{87}
(\byear{1998})
\end{barticle}
\endbibitem

\bibitem[\protect\citeauthoryear{{Anzer} and
  {Heinzel}}{1999}]{1999A&A...349..974A}
\begin{barticle}
\bauthor{\binits{U.} \bsnm{{Anzer}}}, \bauthor{\binits{P.} \bsnm{{Heinzel}}},
\batitle{{The energy balance in solar prominences}}.
\bjtitle{\aap}
\bvolume{349},
\bfpage{974}--\blpage{984}
(\byear{1999})
\end{barticle}
\endbibitem

\bibitem[\protect\citeauthoryear{{Anzer} and
  {Heinzel}}{2000}]{2000A&A...358L..75A}
\begin{barticle}
\bauthor{\binits{U.} \bsnm{{Anzer}}}, \bauthor{\binits{P.} \bsnm{{Heinzel}}},
\batitle{{Energy considerations for solar prominences with mass inflow}}.
\bjtitle{\aap}
\bvolume{358},
\bfpage{75}--\blpage{78}
(\byear{2000})
\end{barticle}
\endbibitem

\bibitem[\protect\citeauthoryear{{Anzer} and
  {Heinzel}}{2005}]{2005ApJ...622..714A}
\begin{barticle}
\bauthor{\binits{U.} \bsnm{{Anzer}}}, \bauthor{\binits{P.} \bsnm{{Heinzel}}},
\batitle{{On the Nature of Dark Extreme Ultraviolet Structures Seen by SOHO/EIT
  and TRACE}}.
\bjtitle{\apj}
\bvolume{622},
\bfpage{714}--\blpage{721}
(\byear{2005}).
doi:\doiurl{10.1086/427817}
\end{barticle}
\endbibitem

\bibitem[\protect\citeauthoryear{{Anzer} and
  {Heinzel}}{2008}]{2008A&A...480..537A}
\begin{barticle}
\bauthor{\binits{U.} \bsnm{{Anzer}}}, \bauthor{\binits{P.} \bsnm{{Heinzel}}},
\batitle{{Prominence modelling: from observed emission measures to temperature
  profiles}}.
\bjtitle{\aap}
\bvolume{480},
\bfpage{537}--\blpage{542}
(\byear{2008}).
doi:\doiurl{10.1051/0004-6361:20078832}
\end{barticle}
\endbibitem

\bibitem[\protect\citeauthoryear{{Anzer} et~al.}{2007}]{2007SoPh..242...43A}
\begin{barticle}
\bauthor{\binits{U.} \bsnm{{Anzer}}}, \bauthor{\binits{P.} \bsnm{{Heinzel}}},
  \bauthor{\binits{F.} \bsnm{{F{\'a}rnik}}},
\batitle{{Prominences on the Limb: Diagnostics with UV EUV Lines and the Soft
  X-Ray Continuum}}.
\bjtitle{\solphys}
\bvolume{242},
\bfpage{43}--\blpage{52}
(\byear{2007}).
doi:\doiurl{10.1007/s11207-007-0344-1}
\end{barticle}
\endbibitem

\bibitem[\protect\citeauthoryear{{Auer} et~al.}{1994}]{1994A&A...292..599A}
\begin{barticle}
\bauthor{\binits{L.} \bsnm{{Auer}}}, \bauthor{\binits{P.F.} \bsnm{{Bendicho}}},
  \bauthor{\binits{J.} \bsnm{{Trujillo Bueno}}},
\batitle{{Multidimensional radiative transfer with multilevel atoms. 1: ALI
  method with preconditioning of the rate equations}}.
\bjtitle{\aap}
\bvolume{292},
\bfpage{599}--\blpage{615}
(\byear{1994})
\end{barticle}
\endbibitem

\bibitem[\protect\citeauthoryear{{Auer} and
  {Mihalas}}{1969}]{1969ApJ...156..681A}
\begin{barticle}
\bauthor{\binits{L.H.} \bsnm{{Auer}}}, \bauthor{\binits{D.} \bsnm{{Mihalas}}},
\batitle{{Non-Lte Model Atmospheres. II. Effects of Balmer {$\alpha$}}}.
\bjtitle{\apj}
\bvolume{156},
\bfpage{681}
(\byear{1969}).
doi:\doiurl{10.1086/149998}
\end{barticle}
\endbibitem

\bibitem[\protect\citeauthoryear{{Auer} and
  {Paletou}}{1994}]{1994A&A...285..675A}
\begin{barticle}
\bauthor{\binits{L.H.} \bsnm{{Auer}}}, \bauthor{\binits{F.} \bsnm{{Paletou}}},
\batitle{{Two-dimensional radiative transfer with partial frequency
  redistribution I. General method}}.
\bjtitle{\aap}
\bvolume{285},
\bfpage{675}--\blpage{686}
(\byear{1994})
\end{barticle}
\endbibitem

\bibitem[\protect\citeauthoryear{{Avrett}}{2007}]{2007ASPC..368...81A}
\begin{botherref}
\oauthor{\binits{E.H.} \bsnm{{Avrett}}},
{New Models of the Solar Chromosphere and Transition Region from SUMER
  Observations},
in \textit{The Physics of Chromospheric Plasmas},
ed. by P. {Heinzel}, I. {Dorotovi{\v c}}, R.J. {Rutten}.
Astronomical Society of the Pacific Conference Series,
vol. 368,
2007,
p. 81
\end{botherref}
\endbibitem

\bibitem[\protect\citeauthoryear{{Avrett} and
  {Loeser}}{1987}]{1987nrt..book..135A}
\begin{botherref}
\oauthor{\binits{E.H.} \bsnm{{Avrett}}}, \oauthor{\binits{R.} \bsnm{{Loeser}}},
{Iterative Solution of Multilevel Transfer Problems},
ed. by W. {Kalkofen}
1987,
p. 135
\end{botherref}
\endbibitem

\bibitem[\protect\citeauthoryear{{Avrett} and
  {Loeser}}{1992}]{1992ASPC...26..489A}
\begin{botherref}
\oauthor{\binits{E.H.} \bsnm{{Avrett}}}, \oauthor{\binits{R.} \bsnm{{Loeser}}},
{The PANDORA Atmosphere Program (Invited Review)},
in \textit{Cool Stars, Stellar Systems, and the Sun},
ed. by M.S. {Giampapa}, J.A. {Bookbinder}.
Astronomical Society of the Pacific Conference Series,
vol. 26,
1992,
p. 489
\end{botherref}
\endbibitem

\bibitem[\protect\citeauthoryear{{Bartoe} and
  {Brueckner}}{1975}]{1975JOSA...65...13B}
\begin{barticle}
\bauthor{\binits{J.D.F.} \bsnm{{Bartoe}}}, \bauthor{\binits{G.E.}
  \bsnm{{Brueckner}}},
\batitle{{New stigmatic, coma-free, concave-grating spectrograph.}}
\bjtitle{Journal of the Optical Society of America (1917-1983)}
\bvolume{65},
\bfpage{13}--\blpage{21}
(\byear{1975})
\end{barticle}
\endbibitem

\bibitem[\protect\citeauthoryear{{Bastian} et~al.}{1993}]{1993ApJ...418..510B}
\begin{barticle}
\bauthor{\binits{T.S.} \bsnm{{Bastian}}}, \bauthor{\binits{M.W.} \bsnm{{Ewell}}
  \bsuffix{Jr.}}, \bauthor{\binits{H.} \bsnm{{Zirin}}},
\batitle{{A Study of Solar Prominences near lambda = 1 Millimeter}}.
\bjtitle{\apj}
\bvolume{418},
\bfpage{510}
(\byear{1993}).
doi:\doiurl{10.1086/173413}
\end{barticle}
\endbibitem

\bibitem[\protect\citeauthoryear{{Baudin} et~al.}{2007}]{2007SoPh..241...39B}
\begin{barticle}
\bauthor{\binits{F.} \bsnm{{Baudin}}}, \bauthor{\binits{E.} \bsnm{{Ibarra}}},
  \bauthor{\binits{E.H.} \bsnm{{Avrett}}}, \bauthor{\binits{J.C.}
  \bsnm{{Vial}}}, \bauthor{\binits{K.} \bsnm{{Bocchialini}}},
  \bauthor{\binits{A.} \bsnm{{Costa}}}, \bauthor{\binits{P.} \bsnm{{Lemaire}}},
  \bauthor{\binits{M.} \bsnm{{Rovira}}},
\batitle{{A Contribution to the Understanding of Chromospheric Oscillations}}.
\bjtitle{\solphys}
\bvolume{241},
\bfpage{39}--\blpage{51}
(\byear{2007}).
doi:\doiurl{10.1007/s11207-007-0006-3}
\end{barticle}
\endbibitem

\bibitem[\protect\citeauthoryear{{Beckers}}{1964}]{1964PhDT........83B}
\begin{botherref}
\oauthor{\binits{J.M.} \bsnm{{Beckers}}},
A study of the fine structures in the solar chromosphere,
PhD thesis,
, University of Utrecht (AFCRL-Environmental Research Paper, No.49), (1964),
1964
\end{botherref}
\endbibitem

\bibitem[\protect\citeauthoryear{{Bendlin} et~al.}{1988}]{1988A&A...197..274B}
\begin{barticle}
\bauthor{\binits{C.} \bsnm{{Bendlin}}}, \bauthor{\binits{E.} \bsnm{{Wiehr}}},
  \bauthor{\binits{G.} \bsnm{{Stellmacher}}},
\batitle{{Spectroscopic analysis of prominence emissions}}.
\bjtitle{\aap}
\bvolume{197},
\bfpage{274}--\blpage{280}
(\byear{1988})
\end{barticle}
\endbibitem

\bibitem[\protect\citeauthoryear{{Berger} et~al.}{2008}]{2008ApJ...676L..89B}
\begin{barticle}
\bauthor{\binits{T.E.} \bsnm{{Berger}}}, \bauthor{\binits{R.A.}
  \bsnm{{Shine}}}, \bauthor{\binits{G.L.} \bsnm{{Slater}}},
  \bauthor{\binits{T.D.} \bsnm{{Tarbell}}}, \bauthor{\binits{A.M.}
  \bsnm{{Title}}}, \bauthor{\binits{T.J.} \bsnm{{Okamoto}}},
  \bauthor{\binits{K.} \bsnm{{Ichimoto}}}, \bauthor{\binits{Y.}
  \bsnm{{Katsukawa}}}, \bauthor{\binits{Y.} \bsnm{{Suematsu}}},
  \bauthor{\binits{S.} \bsnm{{Tsuneta}}}, \bauthor{\binits{B.W.}
  \bsnm{{Lites}}}, \bauthor{\binits{T.} \bsnm{{Shimizu}}},
\batitle{{Hinode SOT Observations of Solar Quiescent Prominence Dynamics}}.
\bjtitle{\apjl}
\bvolume{676},
\bfpage{89}--\blpage{92}
(\byear{2008}).
doi:\doiurl{10.1086/587171}
\end{barticle}
\endbibitem

\bibitem[\protect\citeauthoryear{{Bommier} et~al.}{1986b}]{1986A&A...156...90B}
\begin{barticle}
\bauthor{\binits{V.} \bsnm{{Bommier}}}, \bauthor{\binits{J.L.} \bsnm{{Leroy}}},
  \bauthor{\binits{S.} \bsnm{{Sahal-Brechot}}},
\batitle{{The Linear Polarization of Hydrogen H-Beta Radiation and the Joint
  Diagnostic of Magnetic Field Vector and Electron Density in Quiescent
  Prominences. II - The Electron Density}}.
\bjtitle{\aap}
\bvolume{156},
\bfpage{90}--\blpage{94}
(\byear{1986b})
\end{barticle}
\endbibitem

\bibitem[\protect\citeauthoryear{{Bommier} et~al.}{1986a}]{1986A&A...156...79B}
\begin{barticle}
\bauthor{\binits{V.} \bsnm{{Bommier}}}, \bauthor{\binits{S.}
  \bsnm{{Sahal-Brechot}}}, \bauthor{\binits{J.L.} \bsnm{{Leroy}}},
\batitle{{The linear polarization of hydrogen H-beta radiation and the joint
  diagnostic of magnetic field vector and electron density in quiescent
  prominences. I - The magnetic field.}}
\bjtitle{\aap}
\bvolume{156},
\bfpage{79}--\blpage{89}
(\byear{1986a})
\end{barticle}
\endbibitem

\bibitem[\protect\citeauthoryear{{Bommier} et~al.}{1994}]{1994SoPh..154..231B}
\begin{barticle}
\bauthor{\binits{V.} \bsnm{{Bommier}}}, \bauthor{\binits{E.} \bsnm{{Landi
  Degl'Innocenti}}}, \bauthor{\binits{J.L.} \bsnm{{Leroy}}},
  \bauthor{\binits{S.} \bsnm{{Sahal-Brechot}}},
\batitle{{Complete determination of the magnetic field vector and of the
  electron density in 14 prominences from linear polarization measurements in
  the HeI D3 and H-alpha lines}}.
\bjtitle{\solphys}
\bvolume{154},
\bfpage{231}--\blpage{260}
(\byear{1994}).
doi:\doiurl{10.1007/BF00681098}
\end{barticle}
\endbibitem

\bibitem[\protect\citeauthoryear{{Bonnet} et~al.}{1978}]{1978ApJ...221.1032B}
\begin{barticle}
\bauthor{\binits{R.M.} \bsnm{{Bonnet}}}, \bauthor{\binits{P.}
  \bsnm{{Lemaire}}}, \bauthor{\binits{J.C.} \bsnm{{Vial}}},
  \bauthor{\binits{G.} \bsnm{{Artzner}}}, \bauthor{\binits{P.}
  \bsnm{{Gouttebroze}}}, \bauthor{\binits{A.} \bsnm{{Jouchoux}}},
  \bauthor{\binits{A.} \bsnm{{Vidal-Madjar}}}, \bauthor{\binits{J.W.}
  \bsnm{{Leibacher}}}, \bauthor{\binits{A.} \bsnm{{Skumanich}}},
\batitle{{The LPSP instrument on OSO 8. II - In-flight performance and
  preliminary results}}.
\bjtitle{\apj}
\bvolume{221},
\bfpage{1032}--\blpage{1053}
(\byear{1978}).
doi:\doiurl{10.1086/156109}
\end{barticle}
\endbibitem

\bibitem[\protect\citeauthoryear{{Bonnet} et~al.}{1980}]{1980ApJ...237L..47B}
\begin{barticle}
\bauthor{\binits{R.M.} \bsnm{{Bonnet}}}, \bauthor{\binits{M.}
  \bsnm{{Decaudin}}}, \bauthor{\binits{E.C.} \bsnm{{Bruner}} \bsuffix{Jr.}},
  \bauthor{\binits{L.W.} \bsnm{{Acton}}}, \bauthor{\binits{W.A.}
  \bsnm{{Brown}}},
\batitle{{High-resolution Lyman-alpha filtergrams of the sun}}.
\bjtitle{\apjl}
\bvolume{237},
\bfpage{47}--\blpage{50}
(\byear{1980}).
doi:\doiurl{10.1086/183232}
\end{barticle}
\endbibitem

\bibitem[\protect\citeauthoryear{{Chae}}{2003}]{2003ApJ...584.1084C}
\begin{barticle}
\bauthor{\binits{J.} \bsnm{{Chae}}},
\batitle{{The Formation of a Prominence in NOAA Active Region 8668. II. Trace
  Observations of Jets and Eruptions Associated with Canceling Magnetic
  Features}}.
\bjtitle{\apj}
\bvolume{584},
\bfpage{1084}--\blpage{1094}
(\byear{2003}).
doi:\doiurl{10.1086/345739}
\end{barticle}
\endbibitem

\bibitem[\protect\citeauthoryear{{Chae} et~al.}{2007}]{2007JKAS...40...67C}
\begin{barticle}
\bauthor{\binits{J.} \bsnm{{Chae}}}, \bauthor{\binits{H.M.} \bsnm{{Park}}},
  \bauthor{\binits{Y.D.} \bsnm{{Park}}},
\batitle{{H{$\alpha$} Spectral Properties of Velocity Threads Constituting a
  Quiescent Solar Filament}}.
\bjtitle{Journal of Korean Astronomical Society}
\bvolume{40},
\bfpage{67}--\blpage{82}
(\byear{2007})
\end{barticle}
\endbibitem

\bibitem[\protect\citeauthoryear{{Chae} et~al.}{2006}]{2006SoPh..234..115C}
\begin{barticle}
\bauthor{\binits{J.} \bsnm{{Chae}}}, \bauthor{\binits{Y.D.} \bsnm{{Park}}},
  \bauthor{\binits{H.M.} \bsnm{{Park}}},
\batitle{{Imaging Spectroscopy of a Solar Filament Using a Tunable H{$\alpha$}
  Filter}}.
\bjtitle{\solphys}
\bvolume{234},
\bfpage{115}--\blpage{134}
(\byear{2006}).
doi:\doiurl{10.1007/s11207-006-0047-z}
\end{barticle}
\endbibitem

\bibitem[\protect\citeauthoryear{{Chae} et~al.}{1998}]{1998ApJ...505..957C}
\begin{barticle}
\bauthor{\binits{J.} \bsnm{{Chae}}}, \bauthor{\binits{U.}
  \bsnm{{Sch{\"u}hle}}}, \bauthor{\binits{P.} \bsnm{{Lemaire}}},
\batitle{{SUMER Measurements of Nonthermal Motions: Constraints on Coronal
  Heating Mechanisms}}.
\bjtitle{\apj}
\bvolume{505},
\bfpage{957}--\blpage{973}
(\byear{1998}).
doi:\doiurl{10.1086/306179}
\end{barticle}
\endbibitem

\bibitem[\protect\citeauthoryear{{Chae} et~al.}{2000}]{2000SoPh..195..333C}
\begin{barticle}
\bauthor{\binits{J.} \bsnm{{Chae}}}, \bauthor{\binits{C.} \bsnm{{Denker}}},
  \bauthor{\binits{T.J.} \bsnm{{Spirock}}}, \bauthor{\binits{H.}
  \bsnm{{Wang}}}, \bauthor{\binits{P.R.} \bsnm{{Goode}}},
\batitle{{High-Resolution H{$\alpha$} Observations of Proper Motion in NOAA
  8668: Evidence for Filament Mass Injection by Chromospheric Reconnection}}.
\bjtitle{\solphys}
\bvolume{195},
\bfpage{333}--\blpage{346}
(\byear{2000})
\end{barticle}
\endbibitem

\bibitem[\protect\citeauthoryear{{Chae} et~al.}{2008}]{2008ApJ...689L..73C}
\begin{barticle}
\bauthor{\binits{J.} \bsnm{{Chae}}}, \bauthor{\binits{K.} \bsnm{{Ahn}}},
  \bauthor{\binits{E.K.} \bsnm{{Lim}}}, \bauthor{\binits{G.S.} \bsnm{{Choe}}},
  \bauthor{\binits{T.} \bsnm{{Sakurai}}},
\batitle{{Persistent Horizontal Flows and Magnetic Support of Vertical Threads
  in a Quiescent Prominence}}.
\bjtitle{\apjl}
\bvolume{689},
\bfpage{73}--\blpage{76}
(\byear{2008}).
doi:\doiurl{10.1086/595785}
\end{barticle}
\endbibitem

\bibitem[\protect\citeauthoryear{{Chang} and
  {Deming}}{1998}]{1998SoPh..179...89C}
\begin{barticle}
\bauthor{\binits{E.S.} \bsnm{{Chang}}}, \bauthor{\binits{D.} \bsnm{{Deming}}},
\batitle{{Accurate Determination of Electron Densities in Active and Quiescent
  Prominences: the Mid-Infrared Advantage}}.
\bjtitle{\solphys}
\bvolume{179},
\bfpage{89}--\blpage{124}
(\byear{1998})
\end{barticle}
\endbibitem

\bibitem[\protect\citeauthoryear{{Chiuderi} and
  {Chiuderi-Drago}}{1991}]{1991SoPh..132...81C}
\begin{barticle}
\bauthor{\binits{C.} \bsnm{{Chiuderi}}}, \bauthor{\binits{F.}
  \bsnm{{Chiuderi-Drago}}},
\batitle{{Energy balance in the prominence-corona transition region}}.
\bjtitle{\solphys}
\bvolume{132},
\bfpage{81}--\blpage{94}
(\byear{1991}).
doi:\doiurl{10.1007/BF00159131}
\end{barticle}
\endbibitem

\bibitem[\protect\citeauthoryear{{Chiuderi-Drago}}{2005}]{2005A&A...443.1055C}
\begin{barticle}
\bauthor{\binits{F.} \bsnm{{Chiuderi-Drago}}},
\batitle{{The HeI abundance in Solar filaments}}.
\bjtitle{\aap}
\bvolume{443},
\bfpage{1055}--\blpage{1059}
(\byear{2005}).
doi:\doiurl{10.1051/0004-6361:20053341}
\end{barticle}
\endbibitem

\bibitem[\protect\citeauthoryear{{Chiuderi-Drago}
  et~al.}{2001}]{2001SoPh..199..115C}
\begin{barticle}
\bauthor{\binits{F.} \bsnm{{Chiuderi-Drago}}}, \bauthor{\binits{C.E.}
  \bsnm{{Alissandrakis}}}, \bauthor{\binits{T.} \bsnm{{Bastian}}},
  \bauthor{\binits{K.} \bsnm{{Bocchialini}}}, \bauthor{\binits{R.A.}
  \bsnm{{Harrison}}},
\batitle{{Joint EUV/Radio Observations of a Solar Filament}}.
\bjtitle{\solphys}
\bvolume{199},
\bfpage{115}--\blpage{132}
(\byear{2001})
\end{barticle}
\endbibitem

\bibitem[\protect\citeauthoryear{{Ciaravella}
  et~al.}{2000}]{2000ApJ...529..575C}
\begin{barticle}
\bauthor{\binits{A.} \bsnm{{Ciaravella}}}, \bauthor{\binits{J.C.}
  \bsnm{{Raymond}}}, \bauthor{\binits{B.J.} \bsnm{{Thompson}}},
  \bauthor{\binits{A.} \bsnm{{van Ballegooijen}}}, \bauthor{\binits{L.}
  \bsnm{{Strachan}}}, \bauthor{\binits{J.} \bsnm{{Li}}}, \bauthor{\binits{L.}
  \bsnm{{Gardner}}}, \bauthor{\binits{R.} \bsnm{{O'Neal}}},
  \bauthor{\binits{E.} \bsnm{{Antonucci}}}, \bauthor{\binits{J.}
  \bsnm{{Kohl}}}, \bauthor{\binits{G.} \bsnm{{Noci}}},
\batitle{{Solar and Heliospheric Observatory Observations of a Helical Coronal
  Mass Ejection}}.
\bjtitle{\apj}
\bvolume{529},
\bfpage{575}--\blpage{591}
(\byear{2000}).
doi:\doiurl{10.1086/308260}
\end{barticle}
\endbibitem

\bibitem[\protect\citeauthoryear{{Ciaravella}
  et~al.}{2003}]{2003ApJ...597.1118C}
\begin{barticle}
\bauthor{\binits{A.} \bsnm{{Ciaravella}}}, \bauthor{\binits{J.C.}
  \bsnm{{Raymond}}}, \bauthor{\binits{A.} \bsnm{{van Ballegooijen}}},
  \bauthor{\binits{L.} \bsnm{{Strachan}}}, \bauthor{\binits{A.}
  \bsnm{{Vourlidas}}}, \bauthor{\binits{J.} \bsnm{{Li}}}, \bauthor{\binits{J.}
  \bsnm{{Chen}}}, \bauthor{\binits{A.} \bsnm{{Panasyuk}}},
\batitle{{Physical Parameters of the 2000 February 11 Coronal Mass Ejection:
  Ultraviolet Spectra versus White-Light Images}}.
\bjtitle{\apj}
\bvolume{597},
\bfpage{1118}--\blpage{1134}
(\byear{2003}).
doi:\doiurl{10.1086/381220}
\end{barticle}
\endbibitem

\bibitem[\protect\citeauthoryear{{Cirigliano}
  et~al.}{2004}]{2004SoPh..223...95C}
\begin{barticle}
\bauthor{\binits{D.} \bsnm{{Cirigliano}}}, \bauthor{\binits{J.C.}
  \bsnm{{Vial}}}, \bauthor{\binits{M.} \bsnm{{Rovira}}},
\batitle{{Prominence corona transition region plasma diagnostics from SOHO
  observations}}.
\bjtitle{\solphys}
\bvolume{223},
\bfpage{95}--\blpage{118}
(\byear{2004}).
doi:\doiurl{10.1007/s11207-004-5101-0}
\end{barticle}
\endbibitem

\bibitem[\protect\citeauthoryear{{Cram} and
  {Vardavas}}{1978}]{1978SoPh...57...27C}
\begin{barticle}
\bauthor{\binits{L.E.} \bsnm{{Cram}}}, \bauthor{\binits{I.M.}
  \bsnm{{Vardavas}}},
\batitle{{Resonance line scattering from optically thin structures located
  above the solar limb}}.
\bjtitle{\solphys}
\bvolume{57},
\bfpage{27}--\blpage{36}
(\byear{1978}).
doi:\doiurl{10.1007/BF00152041}
\end{barticle}
\endbibitem

\bibitem[\protect\citeauthoryear{{Culhane} et~al.}{2007}]{2007SoPh..243...19C}
\begin{barticle}
\bauthor{\binits{J.L.} \bsnm{{Culhane}}}, \bauthor{\binits{L.K.}
  \bsnm{{Harra}}}, \bauthor{\binits{A.M.} \bsnm{{James}}}, \bauthor{\binits{K.}
  \bsnm{{Al-Janabi}}}, \bauthor{\binits{L.J.} \bsnm{{Bradley}}},
  \bauthor{\binits{R.A.} \bsnm{{Chaudry}}}, \bauthor{\binits{K.}
  \bsnm{{Rees}}}, \bauthor{\binits{J.A.} \bsnm{{Tandy}}}, \bauthor{\binits{P.}
  \bsnm{{Thomas}}}, \bauthor{\binits{M.C.R.} \bsnm{{Whillock}}},
  \bauthor{\binits{B.} \bsnm{{Winter}}}, \bauthor{\binits{G.A.}
  \bsnm{{Doschek}}}, \bauthor{\binits{C.M.} \bsnm{{Korendyke}}},
  \bauthor{\binits{C.M.} \bsnm{{Brown}}}, \bauthor{\binits{S.} \bsnm{{Myers}}},
  \bauthor{\binits{J.} \bsnm{{Mariska}}}, \bauthor{\binits{J.} \bsnm{{Seely}}},
  \bauthor{\binits{J.} \bsnm{{Lang}}}, \bauthor{\binits{B.J.} \bsnm{{Kent}}},
  \bauthor{\binits{B.M.} \bsnm{{Shaughnessy}}}, \bauthor{\binits{P.R.}
  \bsnm{{Young}}}, \bauthor{\binits{G.M.} \bsnm{{Simnett}}},
  \bauthor{\binits{C.M.} \bsnm{{Castelli}}}, \bauthor{\binits{S.}
  \bsnm{{Mahmoud}}}, \bauthor{\binits{H.} \bsnm{{Mapson-Menard}}},
  \bauthor{\binits{B.J.} \bsnm{{Probyn}}}, \bauthor{\binits{R.J.}
  \bsnm{{Thomas}}}, \bauthor{\binits{J.} \bsnm{{Davila}}}, \bauthor{\binits{K.}
  \bsnm{{Dere}}}, \bauthor{\binits{D.} \bsnm{{Windt}}}, \bauthor{\binits{J.}
  \bsnm{{Shea}}}, \bauthor{\binits{R.} \bsnm{{Hagood}}}, \bauthor{\binits{R.}
  \bsnm{{Moye}}}, \bauthor{\binits{H.} \bsnm{{Hara}}}, \bauthor{\binits{T.}
  \bsnm{{Watanabe}}}, \bauthor{\binits{K.} \bsnm{{Matsuzaki}}},
  \bauthor{\binits{T.} \bsnm{{Kosugi}}}, \bauthor{\binits{V.}
  \bsnm{{Hansteen}}}, \bauthor{\binits{{\O}.} \bsnm{{Wikstol}}},
\batitle{{The EUV Imaging Spectrometer for Hinode}}.
\bjtitle{\solphys}
\bvolume{243},
\bfpage{19}--\blpage{61}
(\byear{2007}).
doi:\doiurl{10.1007/s01007-007-0293-1}
\end{barticle}
\endbibitem

\bibitem[\protect\citeauthoryear{{Dammasch} et~al.}{2003}]{2003AN....324..338D}
\begin{barticle}
\bauthor{\binits{I.E.} \bsnm{{Dammasch}}}, \bauthor{\binits{G.}
  \bsnm{{Stellmacher}}}, \bauthor{\binits{E.} \bsnm{{Wiehr}}},
\batitle{{Spectroscopy of solar prominences from space and ground}}.
\bjtitle{Astronomische Nachrichten}
\bvolume{324},
\bfpage{338}--\blpage{339}
(\byear{2003})
\end{barticle}
\endbibitem

\bibitem[\protect\citeauthoryear{{de Boer} et~al.}{1998}]{1998A&A...334..280D}
\begin{barticle}
\bauthor{\binits{C.R.} \bsnm{{de Boer}}}, \bauthor{\binits{G.}
  \bsnm{{Stellmacher}}}, \bauthor{\binits{E.} \bsnm{{Wiehr}}},
\batitle{{The hot prominence periphery in EUV lines}}.
\bjtitle{\aap}
\bvolume{334},
\bfpage{280}--\blpage{288}
(\byear{1998})
\end{barticle}
\endbibitem

\bibitem[\protect\citeauthoryear{{Del Zanna}
  et~al.}{2004}]{2004A&A...420..307D}
\begin{barticle}
\bauthor{\binits{G.} \bsnm{{Del Zanna}}}, \bauthor{\binits{F.}
  \bsnm{{Chiuderi-Drago}}}, \bauthor{\binits{S.} \bsnm{{Parenti}}},
\batitle{{SOHO CDS and SUMER observations of quiescent filaments and their
  interpretation}}.
\bjtitle{\aap}
\bvolume{420},
\bfpage{307}--\blpage{317}
(\byear{2004}).
doi:\doiurl{10.1051/0004-6361:20034267}
\end{barticle}
\endbibitem

\bibitem[\protect\citeauthoryear{{Delaboudini{\`e}re}
  et~al.}{1995}]{1995SoPh..162..291D}
\begin{barticle}
\bauthor{\binits{J.P.} \bsnm{{Delaboudini{\`e}re}}}, \bauthor{\binits{G.E.}
  \bsnm{{Artzner}}}, \bauthor{\binits{J.} \bsnm{{Brunaud}}},
  \bauthor{\binits{A.H.} \bsnm{{Gabriel}}}, \bauthor{\binits{J.F.}
  \bsnm{{Hochedez}}}, \bauthor{\binits{F.} \bsnm{{Millier}}},
  \bauthor{\binits{X.Y.} \bsnm{{Song}}}, \bauthor{\binits{B.} \bsnm{{Au}}},
  \bauthor{\binits{K.P.} \bsnm{{Dere}}}, \bauthor{\binits{R.A.}
  \bsnm{{Howard}}}, \bauthor{\binits{R.} \bsnm{{Kreplin}}},
  \bauthor{\binits{D.J.} \bsnm{{Michels}}}, \bauthor{\binits{J.D.}
  \bsnm{{Moses}}}, \bauthor{\binits{J.M.} \bsnm{{Defise}}},
  \bauthor{\binits{C.} \bsnm{{Jamar}}}, \bauthor{\binits{P.} \bsnm{{Rochus}}},
  \bauthor{\binits{J.P.} \bsnm{{Chauvineau}}}, \bauthor{\binits{J.P.}
  \bsnm{{Marioge}}}, \bauthor{\binits{R.C.} \bsnm{{Catura}}},
  \bauthor{\binits{J.R.} \bsnm{{Lemen}}}, \bauthor{\binits{L.} \bsnm{{Shing}}},
  \bauthor{\binits{R.A.} \bsnm{{Stern}}}, \bauthor{\binits{J.B.}
  \bsnm{{Gurman}}}, \bauthor{\binits{W.M.} \bsnm{{Neupert}}},
  \bauthor{\binits{A.} \bsnm{{Maucherat}}}, \bauthor{\binits{F.}
  \bsnm{{Clette}}}, \bauthor{\binits{P.} \bsnm{{Cugnon}}},
  \bauthor{\binits{E.L.} \bsnm{{van Dessel}}},
\batitle{{EIT: Extreme-Ultraviolet Imaging Telescope for the SOHO Mission}}.
\bjtitle{\solphys}
\bvolume{162},
\bfpage{291}--\blpage{312}
(\byear{1995}).
doi:\doiurl{10.1007/BF00733432}
\end{barticle}
\endbibitem

\bibitem[\protect\citeauthoryear{{Deng} et~al.}{2002}]{2002SoPh..209..153D}
\begin{barticle}
\bauthor{\binits{Y.} \bsnm{{Deng}}}, \bauthor{\binits{Y.} \bsnm{{Lin}}},
  \bauthor{\binits{B.} \bsnm{{Schmieder}}}, \bauthor{\binits{O.}
  \bsnm{{Engvold}}},
\batitle{{Filament activation and magnetic reconnection}}.
\bjtitle{\solphys}
\bvolume{209},
\bfpage{153}--\blpage{170}
(\byear{2002}).
doi:\doiurl{10.1023/A:1020924406991}
\end{barticle}
\endbibitem

\bibitem[\protect\citeauthoryear{{Dere} et~al.}{1997}]{1997A&AS..125..149D}
\begin{barticle}
\bauthor{\binits{K.P.} \bsnm{{Dere}}}, \bauthor{\binits{E.} \bsnm{{Landi}}},
  \bauthor{\binits{H.E.} \bsnm{{Mason}}}, \bauthor{\binits{B.C.}
  \bsnm{{Monsignori Fossi}}}, \bauthor{\binits{P.R.} \bsnm{{Young}}},
\batitle{{CHIANTI - an atomic database for emission lines}}.
\bjtitle{\aaps}
\bvolume{125},
\bfpage{149}--\blpage{173}
(\byear{1997}).
doi:\doiurl{10.1051/aas:1997368}
\end{barticle}
\endbibitem

\bibitem[\protect\citeauthoryear{{Domingo} et~al.}{1995}]{1995SoPh..162....1D}
\begin{barticle}
\bauthor{\binits{V.} \bsnm{{Domingo}}}, \bauthor{\binits{B.} \bsnm{{Fleck}}},
  \bauthor{\binits{A.I.} \bsnm{{Poland}}},
\batitle{{The SOHO Mission: an Overview}}.
\bjtitle{\solphys}
\bvolume{162},
\bfpage{1}--\blpage{2}
(\byear{1995}).
doi:\doiurl{10.1007/BF00733425}
\end{barticle}
\endbibitem

\bibitem[\protect\citeauthoryear{{Dunn}}{1961}]{1961PhDT.........1D}
\begin{botherref}
\oauthor{\binits{R.B.} \bsnm{{Dunn}}},
{Photometry of the Solar Chromosphere.},
PhD thesis,
AA(HARVARD UNIVERSITY.),
1961
\end{botherref}
\endbibitem

\bibitem[\protect\citeauthoryear{{Ebadi} et~al.}{2009}]{2009SoPh..257...91E}
\begin{barticle}
\bauthor{\binits{H.} \bsnm{{Ebadi}}}, \bauthor{\binits{J.} \bsnm{{Vial}}},
  \bauthor{\binits{A.} \bsnm{{Ajabshirizadeh}}},
\batitle{{The He II Lines in the Lyman Series Profiles of Solar Prominences}}.
\bjtitle{\solphys}
\bvolume{257},
\bfpage{91}--\blpage{98}
(\byear{2009}).
doi:\doiurl{10.1007/s11207-009-9368-z}
\end{barticle}
\endbibitem

\bibitem[\protect\citeauthoryear{{Engvold}}{1976}]{1976SoPh...49..283E}
\begin{barticle}
\bauthor{\binits{O.} \bsnm{{Engvold}}},
\batitle{{The fine structure of prominences. I - Observations - H-alpha
  filtergrams}}.
\bjtitle{\solphys}
\bvolume{49},
\bfpage{283}--\blpage{295}
(\byear{1976}).
doi:\doiurl{10.1007/BF00162453}
\end{barticle}
\endbibitem

\bibitem[\protect\citeauthoryear{{Engvold}}{1980}]{1980SoPh...67..351E}
\begin{barticle}
\bauthor{\binits{O.} \bsnm{{Engvold}}},
\batitle{{Thermodynamic models and fine structure of prominences}}.
\bjtitle{\solphys}
\bvolume{67},
\bfpage{351}--\blpage{355}
(\byear{1980}).
doi:\doiurl{10.1007/BF00149812}
\end{barticle}
\endbibitem

\bibitem[\protect\citeauthoryear{{Engvold}}{1988}]{1988sscd.conf..151E}
\begin{botherref}
\oauthor{\binits{O.} \bsnm{{Engvold}}},
{The prominence-corona transition region},
in \textit{Solar and Stellar Coronal Structure and Dynamics},
ed. by R.C. {Altrock},
1988,
pp. 151--169
\end{botherref}
\endbibitem

\bibitem[\protect\citeauthoryear{{Engvold}}{1989}]{1989ASSL..150...47E}
\begin{botherref}
\oauthor{\binits{O.} \bsnm{{Engvold}}},
{Prominence environment},
in \textit{Dynamics and Structure of Quiescent Solar Prominences},
ed. by E.R. {Priest}.
Astrophysics and Space Science Library,
vol. 150,
1989,
pp. 47--76
\end{botherref}
\endbibitem

\bibitem[\protect\citeauthoryear{{Engvold} et~al.}{1980}]{1980A&A....85..326E}
\begin{barticle}
\bauthor{\binits{O.} \bsnm{{Engvold}}}, \bauthor{\binits{E.} \bsnm{{Wiehr}}},
  \bauthor{\binits{A.} \bsnm{{Wittmann}}},
\batitle{{The influence of spatial resolution on the Ca/+/K line width and
  shift in a quiescent prominence}}.
\bjtitle{\aap}
\bvolume{85},
\bfpage{326}--\blpage{328}
(\byear{1980})
\end{barticle}
\endbibitem

\bibitem[\protect\citeauthoryear{{Engvold} et~al.}{1990}]{1990LNP...363..294E}
\begin{botherref}
\oauthor{\binits{O.} \bsnm{{Engvold}}}, \oauthor{\binits{T.}
  \bsnm{{Hirayama}}}, \oauthor{\binits{J.L.} \bsnm{{Leroy}}},
  \oauthor{\binits{E.R.} \bsnm{{Priest}}}, \oauthor{\binits{E.}
  \bsnm{{Tandberg-Hanssen}}},
{Hvar Reference Atmosphere of Quiescent Prominences},
in \textit{IAU Colloq. 117: Dynamics of Quiescent Prominences},
ed. by V. {Ruzdjak}, E. {Tandberg-Hanssen}.
Lecture Notes in Physics, Berlin Springer Verlag,
vol. 363,
1990,
p. 294.
doi:\doiurl{10.1007/BFb0025640}
\end{botherref}
\endbibitem

\bibitem[\protect\citeauthoryear{{Feldman} et~al.}{1977}]{1977ApJ...215..652F}
\begin{barticle}
\bauthor{\binits{U.} \bsnm{{Feldman}}}, \bauthor{\binits{G.A.}
  \bsnm{{Dorschek}}}, \bauthor{\binits{F.D.} \bsnm{{Rosenberg}}},
\batitle{{XUV spectra of the 1973 June 15 solar flare observed from Skylab. II
  - Intersystem and forbidden transitions in transition zone and coronal
  ions}}.
\bjtitle{\apj}
\bvolume{215},
\bfpage{652}--\blpage{665}
(\byear{1977}).
doi:\doiurl{10.1086/155399}
\end{barticle}
\endbibitem

\bibitem[\protect\citeauthoryear{{Fontenla} and
  {Rovira}}{1983}]{1983SoPh...85..141F}
\begin{barticle}
\bauthor{\binits{J.M.} \bsnm{{Fontenla}}}, \bauthor{\binits{M.}
  \bsnm{{Rovira}}},
\batitle{{The Lyman alpha line in solar prominences}}.
\bjtitle{\solphys}
\bvolume{85},
\bfpage{141}--\blpage{156}
(\byear{1983}).
doi:\doiurl{10.1007/BF00148265}
\end{barticle}
\endbibitem

\bibitem[\protect\citeauthoryear{{Fontenla} and
  {Rovira}}{1985}]{1985SoPh...96...53F}
\begin{barticle}
\bauthor{\binits{J.M.} \bsnm{{Fontenla}}}, \bauthor{\binits{M.}
  \bsnm{{Rovira}}},
\batitle{{Quiescent prominence threads models}}.
\bjtitle{\solphys}
\bvolume{96},
\bfpage{53}--\blpage{92}
(\byear{1985}).
doi:\doiurl{10.1007/BF00239794}
\end{barticle}
\endbibitem

\bibitem[\protect\citeauthoryear{{Fontenla} et~al.}{1988}]{1988ApJ...329..464F}
\begin{barticle}
\bauthor{\binits{J.M.} \bsnm{{Fontenla}}}, \bauthor{\binits{E.J.}
  \bsnm{{Reichmann}}}, \bauthor{\binits{E.} \bsnm{{Tandberg-Hanssen}}},
\batitle{{The Lyman-alpha line in various solar features. I - Observations}}.
\bjtitle{\apj}
\bvolume{329},
\bfpage{464}--\blpage{481}
(\byear{1988}).
doi:\doiurl{10.1086/166392}
\end{barticle}
\endbibitem

\bibitem[\protect\citeauthoryear{{Fontenla} et~al.}{1996}]{1996ApJ...466..496F}
\begin{barticle}
\bauthor{\binits{J.M.} \bsnm{{Fontenla}}}, \bauthor{\binits{M.}
  \bsnm{{Rovira}}}, \bauthor{\binits{J.C.} \bsnm{{Vial}}}, \bauthor{\binits{P.}
  \bsnm{{Gouttebroze}}},
\batitle{{Prominence Thread Models Including Ambipolar Diffusion}}.
\bjtitle{\apj}
\bvolume{466},
\bfpage{496}
(\byear{1996}).
doi:\doiurl{10.1086/177527}
\end{barticle}
\endbibitem

\bibitem[\protect\citeauthoryear{{Gilbert} et~al.}{2007}]{2007ApJ...671..978G}
\begin{barticle}
\bauthor{\binits{H.} \bsnm{{Gilbert}}}, \bauthor{\binits{G.} \bsnm{{Kilper}}},
  \bauthor{\binits{D.} \bsnm{{Alexander}}},
\batitle{{Observational Evidence Supporting Cross-field Diffusion of Neutral
  Material in Solar Filaments}}.
\bjtitle{\apj}
\bvolume{671},
\bfpage{978}--\blpage{989}
(\byear{2007}).
doi:\doiurl{10.1086/522884}
\end{barticle}
\endbibitem

\bibitem[\protect\citeauthoryear{{Gilbert} et~al.}{2002}]{2002ApJ...577..464G}
\begin{barticle}
\bauthor{\binits{H.R.} \bsnm{{Gilbert}}}, \bauthor{\binits{V.H.}
  \bsnm{{Hansteen}}}, \bauthor{\binits{T.E.} \bsnm{{Holzer}}},
\batitle{{Neutral Atom Diffusion in a Partially Ionized Prominence Plasma}}.
\bjtitle{\apj}
\bvolume{577},
\bfpage{464}--\blpage{474}
(\byear{2002}).
doi:\doiurl{10.1086/342165}
\end{barticle}
\endbibitem

\bibitem[\protect\citeauthoryear{{Gilbert} et~al.}{2005}]{2005ApJ...618..524G}
\begin{barticle}
\bauthor{\binits{H.R.} \bsnm{{Gilbert}}}, \bauthor{\binits{T.E.}
  \bsnm{{Holzer}}}, \bauthor{\binits{R.M.} \bsnm{{MacQueen}}},
\batitle{{A New Technique for Deriving Prominence Mass from SOHO/EIT Fe XII
  (19.5 Nanometers) Absorption Features}}.
\bjtitle{\apj}
\bvolume{618},
\bfpage{524}--\blpage{536}
(\byear{2005}).
doi:\doiurl{10.1086/425975}
\end{barticle}
\endbibitem

\bibitem[\protect\citeauthoryear{{Gilbert} et~al.}{2006}]{2006ApJ...641..606G}
\begin{barticle}
\bauthor{\binits{H.R.} \bsnm{{Gilbert}}}, \bauthor{\binits{L.E.}
  \bsnm{{Falco}}}, \bauthor{\binits{T.E.} \bsnm{{Holzer}}},
  \bauthor{\binits{R.M.} \bsnm{{MacQueen}}},
\batitle{{Application of a New Technique for Deriving Prominence Mass from SOHO
  EIT Fe XII (19.5 nm) Absorption Features}}.
\bjtitle{\apj}
\bvolume{641},
\bfpage{606}--\blpage{610}
(\byear{2006}).
doi:\doiurl{10.1086/500354}
\end{barticle}
\endbibitem

\bibitem[\protect\citeauthoryear{{Gilbert} et~al.}{2010}]{2009Gilbert}
\begin{botherref}
\oauthor{\binits{H.R.} \bsnm{{Gilbert}}}, \oauthor{\binits{G.}
  \bsnm{{Kilper}}}, \oauthor{\binits{T.A.} \bsnm{{Kucera}}},
  \oauthor{\binits{D.} \bsnm{{Alexander}}}, \oauthor{\binits{J.B.}
  \bsnm{{Gurman}}},
{Comparing Prominence Absorption in Different Coronal Lines}.
\textrm{in preparation}
(2010).
in preparation
\end{botherref}
\endbibitem

\bibitem[\protect\citeauthoryear{{Golub} et~al.}{1999}]{1999PhPl....6.2205G}
\begin{barticle}
\bauthor{\binits{L.} \bsnm{{Golub}}}, \bauthor{\binits{J.}
  \bsnm{{Bookbinder}}}, \bauthor{\binits{E.} \bsnm{{Deluca}}},
  \bauthor{\binits{M.} \bsnm{{Karovska}}}, \bauthor{\binits{H.}
  \bsnm{{Warren}}}, \bauthor{\binits{C.J.} \bsnm{{Schrijver}}},
  \bauthor{\binits{R.} \bsnm{{Shine}}}, \bauthor{\binits{T.} \bsnm{{Tarbell}}},
  \bauthor{\binits{A.} \bsnm{{Title}}}, \bauthor{\binits{J.} \bsnm{{Wolfson}}},
  \bauthor{\binits{B.} \bsnm{{Handy}}}, \bauthor{\binits{C.}
  \bsnm{{Kankelborg}}},
\batitle{{A new view of the solar corona from the Transition Region and Coronal
  Explorer (TRACE).}}
\bjtitle{Physics of Plasmas}
\bvolume{6},
\bfpage{2205}--\blpage{2216}
(\byear{1999}).
doi:\doiurl{10.1063/1.873473}
\end{barticle}
\endbibitem

\bibitem[\protect\citeauthoryear{{Golub} et~al.}{2007}]{2007SoPh..243...63G}
\begin{barticle}
\bauthor{\binits{L.} \bsnm{{Golub}}}, \bauthor{\binits{E.} \bsnm{{Deluca}}},
  \bauthor{\binits{G.} \bsnm{{Austin}}}, \bauthor{\binits{J.}
  \bsnm{{Bookbinder}}}, \bauthor{\binits{D.} \bsnm{{Caldwell}}},
  \bauthor{\binits{P.} \bsnm{{Cheimets}}}, \bauthor{\binits{J.}
  \bsnm{{Cirtain}}}, \bauthor{\binits{M.} \bsnm{{Cosmo}}}, \bauthor{\binits{P.}
  \bsnm{{Reid}}}, \bauthor{\binits{A.} \bsnm{{Sette}}}, \bauthor{\binits{M.}
  \bsnm{{Weber}}}, \bauthor{\binits{T.} \bsnm{{Sakao}}}, \bauthor{\binits{R.}
  \bsnm{{Kano}}}, \bauthor{\binits{K.} \bsnm{{Shibasaki}}},
  \bauthor{\binits{H.} \bsnm{{Hara}}}, \bauthor{\binits{S.} \bsnm{{Tsuneta}}},
  \bauthor{\binits{K.} \bsnm{{Kumagai}}}, \bauthor{\binits{T.}
  \bsnm{{Tamura}}}, \bauthor{\binits{M.} \bsnm{{Shimojo}}},
  \bauthor{\binits{J.} \bsnm{{McCracken}}}, \bauthor{\binits{J.}
  \bsnm{{Carpenter}}}, \bauthor{\binits{H.} \bsnm{{Haight}}},
  \bauthor{\binits{R.} \bsnm{{Siler}}}, \bauthor{\binits{E.} \bsnm{{Wright}}},
  \bauthor{\binits{J.} \bsnm{{Tucker}}}, \bauthor{\binits{H.}
  \bsnm{{Rutledge}}}, \bauthor{\binits{M.} \bsnm{{Barbera}}},
  \bauthor{\binits{G.} \bsnm{{Peres}}}, \bauthor{\binits{S.} \bsnm{{Varisco}}},
\batitle{{The X-Ray Telescope (XRT) for the Hinode Mission}}.
\bjtitle{\solphys}
\bvolume{243},
\bfpage{63}--\blpage{86}
(\byear{2007}).
doi:\doiurl{10.1007/s11207-007-0182-1}
\end{barticle}
\endbibitem

\bibitem[\protect\citeauthoryear{{Gontikakis}
  et~al.}{1997a}]{1997A&A...325..803G}
\begin{barticle}
\bauthor{\binits{C.} \bsnm{{Gontikakis}}}, \bauthor{\binits{J.C.}
  \bsnm{{Vial}}}, \bauthor{\binits{P.} \bsnm{{Gouttebroze}}},
\batitle{{Emission of hydrogen lines by moving solar prominences.}}
\bjtitle{\aap}
\bvolume{325},
\bfpage{803}--\blpage{812}
(\byear{1997a})
\end{barticle}
\endbibitem

\bibitem[\protect\citeauthoryear{{Gontikakis}
  et~al.}{1997b}]{1997SoPh..172..189G}
\begin{barticle}
\bauthor{\binits{C.} \bsnm{{Gontikakis}}}, \bauthor{\binits{J.C.}
  \bsnm{{Vial}}}, \bauthor{\binits{P.} \bsnm{{Gouttebroze}}},
\batitle{{Spectral diagnostics for eruptive prominences}}.
\bjtitle{\solphys}
\bvolume{172},
\bfpage{189}--\blpage{197}
(\byear{1997b})
\end{barticle}
\endbibitem

\bibitem[\protect\citeauthoryear{{Gouttebroze}}{2005}]{2005A&A...434.1165G}
\begin{barticle}
\bauthor{\binits{P.} \bsnm{{Gouttebroze}}},
\batitle{{Radiative transfer in cylindrical threads with incident radiation.
  II. 2D azimuth-dependent case}}.
\bjtitle{\aap}
\bvolume{434},
\bfpage{1165}--\blpage{1171}
(\byear{2005}).
doi:\doiurl{10.1051/0004-6361:20042309}
\end{barticle}
\endbibitem

\bibitem[\protect\citeauthoryear{{Gouttebroze}}{2006}]{2006A&A...448..367G}
\begin{barticle}
\bauthor{\binits{P.} \bsnm{{Gouttebroze}}},
\batitle{{Radiative transfer in cylindrical threads with incident radiation.
  III. Hydrogen spectrum}}.
\bjtitle{\aap}
\bvolume{448},
\bfpage{367}--\blpage{374}
(\byear{2006}).
doi:\doiurl{10.1051/0004-6361:20054139}
\end{barticle}
\endbibitem

\bibitem[\protect\citeauthoryear{{Gouttebroze}}{2007}]{2007A&A...465.1041G}
\begin{barticle}
\bauthor{\binits{P.} \bsnm{{Gouttebroze}}},
\batitle{{Radiative transfer in cylindrical threads with incident radiation.
  IV. Time-dependent and thermal equilibrium models}}.
\bjtitle{\aap}
\bvolume{465},
\bfpage{1041}--\blpage{1049}
(\byear{2007}).
doi:\doiurl{10.1051/0004-6361:20066636}
\end{barticle}
\endbibitem

\bibitem[\protect\citeauthoryear{{Gouttebroze}}{2008}]{2008A&A...487..805G}
\begin{barticle}
\bauthor{\binits{P.} \bsnm{{Gouttebroze}}},
\batitle{{Radiative transfer in cylindrical threads with incident radiation. V.
  2D transfer with 3D velocity fields}}.
\bjtitle{\aap}
\bvolume{487},
\bfpage{805}--\blpage{813}
(\byear{2008}).
doi:\doiurl{10.1051/0004-6361:20079272}
\end{barticle}
\endbibitem

\bibitem[\protect\citeauthoryear{{Gouttebroze} and
  {Heinzel}}{2002}]{2002A&A...385..273G}
\begin{barticle}
\bauthor{\binits{P.} \bsnm{{Gouttebroze}}}, \bauthor{\binits{P.}
  \bsnm{{Heinzel}}},
\batitle{{Calcium to hydrogen line ratios in solar prominences}}.
\bjtitle{\aap}
\bvolume{385},
\bfpage{273}--\blpage{280}
(\byear{2002}).
doi:\doiurl{10.1051/0004-6361:20020142}
\end{barticle}
\endbibitem

\bibitem[\protect\citeauthoryear{{Gouttebroze} and
  {Labrosse}}{2000}]{2000SoPh..196..349G}
\begin{barticle}
\bauthor{\binits{P.} \bsnm{{Gouttebroze}}}, \bauthor{\binits{N.}
  \bsnm{{Labrosse}}},
\batitle{{A ready-made code for the computation of prominence NLTE models}}.
\bjtitle{\solphys}
\bvolume{196},
\bfpage{349}--\blpage{355}
(\byear{2000})
\end{barticle}
\endbibitem

\bibitem[\protect\citeauthoryear{{Gouttebroze} and
  {Labrosse}}{2009}]{2009A&A...503..663G}
\begin{barticle}
\bauthor{\binits{P.} \bsnm{{Gouttebroze}}}, \bauthor{\binits{N.}
  \bsnm{{Labrosse}}},
\batitle{{Radiative transfer in cylindrical threads with incident radiation.
  VI. A hydrogen plus helium system}}.
\bjtitle{\aap}
\bvolume{503},
\bfpage{663}--\blpage{671}
(\byear{2009}).
doi:\doiurl{10.1051/0004-6361/200811483}
\end{barticle}
\endbibitem

\bibitem[\protect\citeauthoryear{{Gouttebroze}
  et~al.}{1993}]{1993A&AS...99..513G}
\begin{barticle}
\bauthor{\binits{P.} \bsnm{{Gouttebroze}}}, \bauthor{\binits{P.}
  \bsnm{{Heinzel}}}, \bauthor{\binits{J.C.} \bsnm{{Vial}}},
\batitle{{The hydrogen spectrum of model prominences}}.
\bjtitle{\aaps}
\bvolume{99},
\bfpage{513}--\blpage{543}
(\byear{1993})
\end{barticle}
\endbibitem

\bibitem[\protect\citeauthoryear{{Gouttebroze}
  et~al.}{1978}]{1978ApJ...225..655G}
\begin{barticle}
\bauthor{\binits{P.} \bsnm{{Gouttebroze}}}, \bauthor{\binits{P.}
  \bsnm{{Lemaire}}}, \bauthor{\binits{J.C.} \bsnm{{Vial}}},
  \bauthor{\binits{G.} \bsnm{{Artzner}}},
\batitle{{The solar hydrogen Lyman-beta and Lyman-alpha lines - Disk center
  observations from OSO 8 compared with theoretical profiles}}.
\bjtitle{\apj}
\bvolume{225},
\bfpage{655}--\blpage{664}
(\byear{1978}).
doi:\doiurl{10.1086/156526}
\end{barticle}
\endbibitem

\bibitem[\protect\citeauthoryear{{Gouttebroze}
  et~al.}{2002}]{2002ESASP.505..421G}
\begin{botherref}
\oauthor{\binits{P.} \bsnm{{Gouttebroze}}}, \oauthor{\binits{N.}
  \bsnm{{Labrosse}}}, \oauthor{\binits{P.} \bsnm{{Heinzel}}},
  \oauthor{\binits{J.C.} \bsnm{{Vial}}},
{Prediction of line intensity ratios in solar prominences},
in \textit{SOLMAG 2002. Proceedings of the Magnetic Coupling of the Solar
  Atmosphere Euroconference},
ed. by H. {Sawaya-Lacoste}.
ESA Special Publication,
vol. 505,
2002,
pp. 421--424
\end{botherref}
\endbibitem

\bibitem[\protect\citeauthoryear{{Gun{\'a}r}
  et~al.}{2007a}]{2007A&A...463..737G}
\begin{barticle}
\bauthor{\binits{S.} \bsnm{{Gun{\'a}r}}}, \bauthor{\binits{P.}
  \bsnm{{Heinzel}}}, \bauthor{\binits{U.} \bsnm{{Anzer}}},
\batitle{{Prominence fine structures in a magnetic equilibrium. III. Lyman
  continuum in 2D configurations}}.
\bjtitle{\aap}
\bvolume{463},
\bfpage{737}--\blpage{743}
(\byear{2007a}).
doi:\doiurl{10.1051/0004-6361:20066142}
\end{barticle}
\endbibitem

\bibitem[\protect\citeauthoryear{{Gun{\'a}r}
  et~al.}{2007b}]{2007A&A...472..929G}
\begin{barticle}
\bauthor{\binits{S.} \bsnm{{Gun{\'a}r}}}, \bauthor{\binits{P.}
  \bsnm{{Heinzel}}}, \bauthor{\binits{B.} \bsnm{{Schmieder}}},
  \bauthor{\binits{P.} \bsnm{{Schwartz}}}, \bauthor{\binits{U.}
  \bsnm{{Anzer}}},
\batitle{{Properties of prominence fine-structure threads derived from
  SOHO/SUMER hydrogen Lyman lines}}.
\bjtitle{\aap}
\bvolume{472},
\bfpage{929}--\blpage{936}
(\byear{2007b}).
doi:\doiurl{10.1051/0004-6361:20077785}
\end{barticle}
\endbibitem

\bibitem[\protect\citeauthoryear{{Gun{\'a}r}
  et~al.}{2008}]{2008A&A...490..307G}
\begin{barticle}
\bauthor{\binits{S.} \bsnm{{Gun{\'a}r}}}, \bauthor{\binits{P.}
  \bsnm{{Heinzel}}}, \bauthor{\binits{U.} \bsnm{{Anzer}}}, \bauthor{\binits{B.}
  \bsnm{{Schmieder}}},
\batitle{{On Lyman-line asymmetries in quiescent prominences}}.
\bjtitle{\aap}
\bvolume{490},
\bfpage{307}--\blpage{313}
(\byear{2008}).
doi:\doiurl{10.1051/0004-6361:200810127}
\end{barticle}
\endbibitem

\bibitem[\protect\citeauthoryear{{Handy} et~al.}{1999}]{1999SoPh..187..229H}
\begin{barticle}
\bauthor{\binits{B.N.} \bsnm{{Handy}}}, \bauthor{\binits{L.W.} \bsnm{{Acton}}},
  \bauthor{\binits{C.C.} \bsnm{{Kankelborg}}}, \bauthor{\binits{C.J.}
  \bsnm{{Wolfson}}}, \bauthor{\binits{D.J.} \bsnm{{Akin}}},
  \bauthor{\binits{M.E.} \bsnm{{Bruner}}}, \bauthor{\binits{R.}
  \bsnm{{Caravalho}}}, \bauthor{\binits{R.C.} \bsnm{{Catura}}},
  \bauthor{\binits{R.} \bsnm{{Chevalier}}}, \bauthor{\binits{D.W.}
  \bsnm{{Duncan}}}, \bauthor{\binits{C.G.} \bsnm{{Edwards}}},
  \bauthor{\binits{C.N.} \bsnm{{Feinstein}}}, \bauthor{\binits{S.L.}
  \bsnm{{Freeland}}}, \bauthor{\binits{F.M.} \bsnm{{Friedlaender}}},
  \bauthor{\binits{C.H.} \bsnm{{Hoffmann}}}, \bauthor{\binits{N.E.}
  \bsnm{{Hurlburt}}}, \bauthor{\binits{B.K.} \bsnm{{Jurcevich}}},
  \bauthor{\binits{N.L.} \bsnm{{Katz}}}, \bauthor{\binits{G.A.}
  \bsnm{{Kelly}}}, \bauthor{\binits{J.R.} \bsnm{{Lemen}}}, \bauthor{\binits{M.}
  \bsnm{{Levay}}}, \bauthor{\binits{R.W.} \bsnm{{Lindgren}}},
  \bauthor{\binits{D.P.} \bsnm{{Mathur}}}, \bauthor{\binits{S.B.}
  \bsnm{{Meyer}}}, \bauthor{\binits{S.J.} \bsnm{{Morrison}}},
  \bauthor{\binits{M.D.} \bsnm{{Morrison}}}, \bauthor{\binits{R.W.}
  \bsnm{{Nightingale}}}, \bauthor{\binits{T.P.} \bsnm{{Pope}}},
  \bauthor{\binits{R.A.} \bsnm{{Rehse}}}, \bauthor{\binits{C.J.}
  \bsnm{{Schrijver}}}, \bauthor{\binits{R.A.} \bsnm{{Shine}}},
  \bauthor{\binits{L.} \bsnm{{Shing}}}, \bauthor{\binits{K.T.}
  \bsnm{{Strong}}}, \bauthor{\binits{T.D.} \bsnm{{Tarbell}}},
  \bauthor{\binits{A.M.} \bsnm{{Title}}}, \bauthor{\binits{D.D.}
  \bsnm{{Torgerson}}}, \bauthor{\binits{L.} \bsnm{{Golub}}},
  \bauthor{\binits{J.A.} \bsnm{{Bookbinder}}}, \bauthor{\binits{D.}
  \bsnm{{Caldwell}}}, \bauthor{\binits{P.N.} \bsnm{{Cheimets}}},
  \bauthor{\binits{W.N.} \bsnm{{Davis}}}, \bauthor{\binits{E.E.}
  \bsnm{{Deluca}}}, \bauthor{\binits{R.A.} \bsnm{{McMullen}}},
  \bauthor{\binits{H.P.} \bsnm{{Warren}}}, \bauthor{\binits{D.}
  \bsnm{{Amato}}}, \bauthor{\binits{R.} \bsnm{{Fisher}}}, \bauthor{\binits{H.}
  \bsnm{{Maldonado}}}, \bauthor{\binits{C.} \bsnm{{Parkinson}}},
\batitle{{The transition region and coronal explorer}}.
\bjtitle{\solphys}
\bvolume{187},
\bfpage{229}--\blpage{260}
(\byear{1999}).
doi:\doiurl{10.1023/A:1005166902804}
\end{barticle}
\endbibitem

\bibitem[\protect\citeauthoryear{{Harrison} et~al.}{1993}]{1993A&A...274L...9H}
\begin{barticle}
\bauthor{\binits{R.A.} \bsnm{{Harrison}}}, \bauthor{\binits{M.K.}
  \bsnm{{Carter}}}, \bauthor{\binits{T.A.} \bsnm{{Clark}}},
  \bauthor{\binits{C.} \bsnm{{Lindsey}}}, \bauthor{\binits{J.T.}
  \bsnm{{Jefferies}}}, \bauthor{\binits{D.G.} \bsnm{{Sime}}},
  \bauthor{\binits{G.} \bsnm{{Watt}}}, \bauthor{\binits{T.L.}
  \bsnm{{Roellig}}}, \bauthor{\binits{E.E.} \bsnm{{Becklin}}},
  \bauthor{\binits{D.A.} \bsnm{{Naylor}}}, \bauthor{\binits{G.J.}
  \bsnm{{Tompkins}}}, \bauthor{\binits{D.} \bsnm{{Braun}}},
\batitle{{An active solar prominence in 1.3 MM radiation}}.
\bjtitle{\aap}
\bvolume{274},
\bfpage{9}
(\byear{1993})
\end{barticle}
\endbibitem

\bibitem[\protect\citeauthoryear{{Harrison} et~al.}{1995}]{1995SoPh..162..233H}
\begin{barticle}
\bauthor{\binits{R.A.} \bsnm{{Harrison}}}, \bauthor{\binits{E.C.}
  \bsnm{{Sawyer}}}, \bauthor{\binits{M.K.} \bsnm{{Carter}}},
  \bauthor{\binits{A.M.} \bsnm{{Cruise}}}, \bauthor{\binits{R.M.}
  \bsnm{{Cutler}}}, \bauthor{\binits{A.} \bsnm{{Fludra}}},
  \bauthor{\binits{R.W.} \bsnm{{Hayes}}}, \bauthor{\binits{B.J.}
  \bsnm{{Kent}}}, \bauthor{\binits{J.} \bsnm{{Lang}}}, \bauthor{\binits{D.J.}
  \bsnm{{Parker}}}, \bauthor{\binits{J.} \bsnm{{Payne}}},
  \bauthor{\binits{C.D.} \bsnm{{Pike}}}, \bauthor{\binits{S.C.}
  \bsnm{{Peskett}}}, \bauthor{\binits{A.G.} \bsnm{{Richards}}},
  \bauthor{\binits{J.L.} \bsnm{{Gulhane}}}, \bauthor{\binits{K.}
  \bsnm{{Norman}}}, \bauthor{\binits{A.A.} \bsnm{{Breeveld}}},
  \bauthor{\binits{E.R.} \bsnm{{Breeveld}}}, \bauthor{\binits{K.F.} \bsnm{{Al
  Janabi}}}, \bauthor{\binits{A.J.} \bsnm{{McCalden}}}, \bauthor{\binits{J.H.}
  \bsnm{{Parkinson}}}, \bauthor{\binits{D.G.} \bsnm{{Self}}},
  \bauthor{\binits{P.D.} \bsnm{{Thomas}}}, \bauthor{\binits{A.I.}
  \bsnm{{Poland}}}, \bauthor{\binits{R.J.} \bsnm{{Thomas}}},
  \bauthor{\binits{W.T.} \bsnm{{Thompson}}}, \bauthor{\binits{O.}
  \bsnm{{Kjeldseth-Moe}}}, \bauthor{\binits{P.} \bsnm{{Brekke}}},
  \bauthor{\binits{J.} \bsnm{{Karud}}}, \bauthor{\binits{P.} \bsnm{{Maltby}}},
  \bauthor{\binits{B.} \bsnm{{Aschenbach}}}, \bauthor{\binits{H.}
  \bsnm{{Br{\"a}uninger}}}, \bauthor{\binits{M.} \bsnm{{K{\"u}hne}}},
  \bauthor{\binits{J.} \bsnm{{Hollandt}}}, \bauthor{\binits{O.H.W.}
  \bsnm{{Siegmund}}}, \bauthor{\binits{M.C.E.} \bsnm{{Huber}}},
  \bauthor{\binits{A.H.} \bsnm{{Gabriel}}}, \bauthor{\binits{H.E.}
  \bsnm{{Mason}}}, \bauthor{\binits{B.J.I.} \bsnm{{Bromage}}},
\batitle{{The Coronal Diagnostic Spectrometer for the Solar and Heliospheric
  Observatory}}.
\bjtitle{\solphys}
\bvolume{162},
\bfpage{233}--\blpage{290}
(\byear{1995}).
doi:\doiurl{10.1007/BF00733431}
\end{barticle}
\endbibitem

\bibitem[\protect\citeauthoryear{{Heasley} and
  {Mihalas}}{1976}]{1976ApJ...205..273H}
\begin{barticle}
\bauthor{\binits{J.N.} \bsnm{{Heasley}}}, \bauthor{\binits{D.}
  \bsnm{{Mihalas}}},
\batitle{{Structure and spectrum of quiescent prominences - Energy balance and
  hydrogen spectrum}}.
\bjtitle{\apj}
\bvolume{205},
\bfpage{273}--\blpage{285}
(\byear{1976}).
doi:\doiurl{10.1086/154273}
\end{barticle}
\endbibitem

\bibitem[\protect\citeauthoryear{{Heasley} and
  {Milkey}}{1976}]{1976ApJ...210..827H}
\begin{barticle}
\bauthor{\binits{J.N.} \bsnm{{Heasley}}}, \bauthor{\binits{R.W.}
  \bsnm{{Milkey}}},
\batitle{{Structure and spectrum of quiescent prominences. II - Hydrogen and
  helium spectra}}.
\bjtitle{\apj}
\bvolume{210},
\bfpage{827}--\blpage{835}
(\byear{1976}).
doi:\doiurl{10.1086/154892}
\end{barticle}
\endbibitem

\bibitem[\protect\citeauthoryear{{Heasley} and
  {Milkey}}{1978}]{1978ApJ...221..677H}
\begin{barticle}
\bauthor{\binits{J.N.} \bsnm{{Heasley}}}, \bauthor{\binits{R.W.}
  \bsnm{{Milkey}}},
\batitle{{Structure and spectrum of quiescent prominences. III - Application of
  theoretical models in helium abundance determinations}}.
\bjtitle{\apj}
\bvolume{221},
\bfpage{677}--\blpage{688}
(\byear{1978}).
doi:\doiurl{10.1086/156072}
\end{barticle}
\endbibitem

\bibitem[\protect\citeauthoryear{{Heasley} and
  {Milkey}}{1983}]{1983ApJ...268..398H}
\begin{barticle}
\bauthor{\binits{J.N.} \bsnm{{Heasley}}}, \bauthor{\binits{R.W.}
  \bsnm{{Milkey}}},
\batitle{{Structure and spectrum of quiescent prominences. IV - The ultraviolet
  ionization continua of hydrogen and helium}}.
\bjtitle{\apj}
\bvolume{268},
\bfpage{398}--\blpage{402}
(\byear{1983}).
doi:\doiurl{10.1086/160965}
\end{barticle}
\endbibitem

\bibitem[\protect\citeauthoryear{{Heasley} et~al.}{1974}]{1974ApJ...192..181H}
\begin{barticle}
\bauthor{\binits{J.N.} \bsnm{{Heasley}}}, \bauthor{\binits{D.}
  \bsnm{{Mihalas}}}, \bauthor{\binits{A.I.} \bsnm{{Poland}}},
\batitle{{Theoretical Helium I Emission-Line Intensities for Quiescent
  Prominences}}.
\bjtitle{\apj}
\bvolume{192},
\bfpage{181}--\blpage{192}
(\byear{1974}).
doi:\doiurl{10.1086/153049}
\end{barticle}
\endbibitem

\bibitem[\protect\citeauthoryear{{Heinzel}}{1983}]{1983BAICz..34....1H}
\begin{barticle}
\bauthor{\binits{P.} \bsnm{{Heinzel}}},
\batitle{{Resonance scattering of radiation in solar prominences. I Partial
  redistribution in optically thin subordinate lines}}.
\bjtitle{Bulletin of the Astronomical Institutes of Czechoslovakia}
\bvolume{34},
\bfpage{1}--\blpage{17}
(\byear{1983})
\end{barticle}
\endbibitem

\bibitem[\protect\citeauthoryear{{Heinzel}}{1989}]{1989HvaOB..13..317H}
\begin{barticle}
\bauthor{\binits{P.} \bsnm{{Heinzel}}},
\batitle{{Hydrogen Lines Formation in Filamentary Prominences}}.
\bjtitle{Hvar Observatory Bulletin}
\bvolume{13},
\bfpage{317}
(\byear{1989})
\end{barticle}
\endbibitem

\bibitem[\protect\citeauthoryear{{Heinzel}}{1990}]{1990LNP...363..279H}
\begin{botherref}
\oauthor{\binits{P.} \bsnm{{Heinzel}}},
{Hydrogen Line Formation in Filamentary Prominences},
in \textit{IAU Colloq. 117: Dynamics of Quiescent Prominences},
ed. by V. {Ruzdjak}, E. {Tandberg-Hanssen}.
Lecture Notes in Physics, Berlin Springer Verlag,
vol. 363,
1990,
p. 279.
doi:\doiurl{10.1007/BFb0025640}
\end{botherref}
\endbibitem

\bibitem[\protect\citeauthoryear{{Heinzel}}{1995}]{1995A&A...299..563H}
\begin{barticle}
\bauthor{\binits{P.} \bsnm{{Heinzel}}},
\batitle{{Multilevel NLTE radiative transfer in isolated atmospheric
  structures: implementation of the MALI-technique.}}
\bjtitle{\aap}
\bvolume{299},
\bfpage{563}
(\byear{1995})
\end{barticle}
\endbibitem

\bibitem[\protect\citeauthoryear{{Heinzel}}{2007}]{2007ASPC..370...46H}
\begin{botherref}
\oauthor{\binits{P.} \bsnm{{Heinzel}}},
{Multiwavelength Observations of Solar Prominences},
in \textit{Solar and Stellar Physics Through Eclipses},
ed. by O. {Demircan}, S.O. {Selam}, B. {Albayrak}.
Astronomical Society of the Pacific Conference Series,
vol. 370,
2007,
p. 46
\end{botherref}
\endbibitem

\bibitem[\protect\citeauthoryear{{Heinzel} and
  {Anzer}}{1999}]{1999SoPh..184..103H}
\begin{barticle}
\bauthor{\binits{P.} \bsnm{{Heinzel}}}, \bauthor{\binits{U.} \bsnm{{Anzer}}},
\batitle{{Magnetic Dips in Prominences}}.
\bjtitle{\solphys}
\bvolume{184},
\bfpage{103}--\blpage{111}
(\byear{1999})
\end{barticle}
\endbibitem

\bibitem[\protect\citeauthoryear{{Heinzel} and
  {Anzer}}{2001}]{2001A&A...375.1082H}
\begin{barticle}
\bauthor{\binits{P.} \bsnm{{Heinzel}}}, \bauthor{\binits{U.} \bsnm{{Anzer}}},
\batitle{{Prominence fine structures in a magnetic equilibrium: Two-dimensional
  models with multilevel radiative transfer}}.
\bjtitle{\aap}
\bvolume{375},
\bfpage{1082}--\blpage{1090}
(\byear{2001}).
doi:\doiurl{10.1051/0004-6361:20010926}
\end{barticle}
\endbibitem

\bibitem[\protect\citeauthoryear{{Heinzel} and
  {Anzer}}{2003}]{2003ASPC..288..441H}
\begin{botherref}
\oauthor{\binits{P.} \bsnm{{Heinzel}}}, \oauthor{\binits{U.} \bsnm{{Anzer}}},
{2D Radiative Transfer in Magnetically Confined Structures},
in \textit{Stellar Atmosphere Modeling},
ed. by I. {Hubeny}, D. {Mihalas}, K. {Werner}.
Astronomical Society of the Pacific Conference Series,
vol. 288,
2003,
p. 441
\end{botherref}
\endbibitem

\bibitem[\protect\citeauthoryear{{Heinzel} and
  {Rompolt}}{1987}]{1987SoPh..110..171H}
\begin{barticle}
\bauthor{\binits{P.} \bsnm{{Heinzel}}}, \bauthor{\binits{B.} \bsnm{{Rompolt}}},
\batitle{{Hydrogen emission from moving solar prominences}}.
\bjtitle{\solphys}
\bvolume{110},
\bfpage{171}--\blpage{189}
(\byear{1987}).
doi:\doiurl{10.1007/BF00148210}
\end{barticle}
\endbibitem

\bibitem[\protect\citeauthoryear{{Heinzel} and
  {Vial}}{1983}]{1983A&A...121..155H}
\begin{barticle}
\bauthor{\binits{P.} \bsnm{{Heinzel}}}, \bauthor{\binits{J.C.} \bsnm{{Vial}}},
\batitle{{OSO-8 observations of a quiescent prominence - A comparison of
  Lyman-alpha with theoretical intensities}}.
\bjtitle{\aap}
\bvolume{121},
\bfpage{155}--\blpage{157}
(\byear{1983})
\end{barticle}
\endbibitem

\bibitem[\protect\citeauthoryear{{Heinzel} et~al.}{2005}]{2005A&A...442..331H}
\begin{barticle}
\bauthor{\binits{P.} \bsnm{{Heinzel}}}, \bauthor{\binits{U.} \bsnm{{Anzer}}},
  \bauthor{\binits{S.} \bsnm{{Gun{\'a}r}}},
\batitle{{Prominence fine structures in a magnetic equilibrium. II. A grid of
  two-dimensional models}}.
\bjtitle{\aap}
\bvolume{442},
\bfpage{331}--\blpage{343}
(\byear{2005}).
doi:\doiurl{10.1051/0004-6361:20053360}
\end{barticle}
\endbibitem

\bibitem[\protect\citeauthoryear{{Heinzel} et~al.}{2003a}]{2003SoPh..216..159H}
\begin{barticle}
\bauthor{\binits{P.} \bsnm{{Heinzel}}}, \bauthor{\binits{U.} \bsnm{{Anzer}}},
  \bauthor{\binits{B.} \bsnm{{Schmieder}}},
\batitle{{A Spectroscopic Model of euv Filaments}}.
\bjtitle{\solphys}
\bvolume{216},
\bfpage{159}--\blpage{171}
(\byear{2003a}).
doi:\doiurl{10.1023/A:1026130028966}
\end{barticle}
\endbibitem

\bibitem[\protect\citeauthoryear{{Heinzel} et~al.}{1996}]{1996SoPh..164..211H}
\begin{barticle}
\bauthor{\binits{P.} \bsnm{{Heinzel}}}, \bauthor{\binits{V.} \bsnm{{Bommier}}},
  \bauthor{\binits{J.C.} \bsnm{{Vial}}},
\batitle{{A Complex Diagnostic of Solar Prominences}}.
\bjtitle{\solphys}
\bvolume{164},
\bfpage{211}--\blpage{222}
(\byear{1996}).
doi:\doiurl{10.1007/BF00146635}
\end{barticle}
\endbibitem

\bibitem[\protect\citeauthoryear{{Heinzel} et~al.}{1987}]{1987A&A...183..351H}
\begin{barticle}
\bauthor{\binits{P.} \bsnm{{Heinzel}}}, \bauthor{\binits{P.}
  \bsnm{{Gouttebroze}}}, \bauthor{\binits{J.C.} \bsnm{{Vial}}},
\batitle{{Formation of the hydrogen spectrum in quiescent prominences -
  One-dimensional models with standard partial redistribution}}.
\bjtitle{\aap}
\bvolume{183},
\bfpage{351}--\blpage{362}
(\byear{1987})
\end{barticle}
\endbibitem

\bibitem[\protect\citeauthoryear{{Heinzel} et~al.}{1988}]{1988dssp.conf...71H}
\begin{botherref}
\oauthor{\binits{P.} \bsnm{{Heinzel}}}, \oauthor{\binits{P.}
  \bsnm{{Gouttebroze}}}, \oauthor{\binits{J.C.} \bsnm{{Vial}}},
{Non LTE modelling of prominences.},
in \textit{Universitat de les Illes Balears, Palma de Mallorca (Spain)},
1988,
p. 71
\end{botherref}
\endbibitem

\bibitem[\protect\citeauthoryear{{Heinzel} et~al.}{1994}]{1994A&A...292..656H}
\begin{barticle}
\bauthor{\binits{P.} \bsnm{{Heinzel}}}, \bauthor{\binits{P.}
  \bsnm{{Gouttebroze}}}, \bauthor{\binits{J.C.} \bsnm{{Vial}}},
\batitle{{Theoretical correlations between prominence plasma parameters and the
  emitted radiation}}.
\bjtitle{\aap}
\bvolume{292},
\bfpage{656}--\blpage{668}
(\byear{1994})
\end{barticle}
\endbibitem

\bibitem[\protect\citeauthoryear{{Heinzel} et~al.}{1999}]{1999A&A...346..322H}
\begin{barticle}
\bauthor{\binits{P.} \bsnm{{Heinzel}}}, \bauthor{\binits{N.} \bsnm{{Mein}}},
  \bauthor{\binits{P.} \bsnm{{Mein}}},
\batitle{{Cloud model with variable source function for solar H{$\alpha$}
  structures. II. Dynamical models}}.
\bjtitle{\aap}
\bvolume{346},
\bfpage{322}--\blpage{328}
(\byear{1999})
\end{barticle}
\endbibitem

\bibitem[\protect\citeauthoryear{{Heinzel} et~al.}{2001b}]{2001ApJ...561L.223H}
\begin{barticle}
\bauthor{\binits{P.} \bsnm{{Heinzel}}}, \bauthor{\binits{B.}
  \bsnm{{Schmieder}}}, \bauthor{\binits{K.} \bsnm{{Tziotziou}}},
\batitle{{Why Are Solar Filaments More Extended in Extreme-Ultraviolet Lines
  than in H{$\alpha$}?}}
\bjtitle{\apjl}
\bvolume{561},
\bfpage{223}--\blpage{227}
(\byear{2001b}).
doi:\doiurl{10.1086/324755}
\end{barticle}
\endbibitem

\bibitem[\protect\citeauthoryear{{Heinzel} et~al.}{2001a}]{2001A&A...370..281H}
\begin{barticle}
\bauthor{\binits{P.} \bsnm{{Heinzel}}}, \bauthor{\binits{B.}
  \bsnm{{Schmieder}}}, \bauthor{\binits{J.C.} \bsnm{{Vial}}},
  \bauthor{\binits{P.} \bsnm{{Kotr{\v c}}}},
\batitle{{SOHO/SUMER observations and analysis of the hydrogen Lyman spectrum
  in solar prominences}}.
\bjtitle{\aap}
\bvolume{370},
\bfpage{281}--\blpage{297}
(\byear{2001a}).
doi:\doiurl{10.1051/0004-6361:20010265}
\end{barticle}
\endbibitem

\bibitem[\protect\citeauthoryear{{Heinzel} et~al.}{2003b}]{2003ESASP.535..447H}
\begin{botherref}
\oauthor{\binits{P.} \bsnm{{Heinzel}}}, \oauthor{\binits{U.} \bsnm{{Anzer}}},
  \oauthor{\binits{B.} \bsnm{{Schmieder}}}, \oauthor{\binits{P.}
  \bsnm{{Schwartz}}},
{EUV-filaments and their mass loading},
in \textit{Solar Variability as an Input to the Earth's Environment},
ed. by A. {Wilson}.
ESA Special Publication,
vol. 535,
2003b,
pp. 447--457
\end{botherref}
\endbibitem

\bibitem[\protect\citeauthoryear{{Heinzel} et~al.}{2008}]{2008ApJ...686.1383H}
\begin{barticle}
\bauthor{\binits{P.} \bsnm{{Heinzel}}}, \bauthor{\binits{B.}
  \bsnm{{Schmieder}}}, \bauthor{\binits{F.} \bsnm{{F{\'a}rn{\'{\i}}k}}},
  \bauthor{\binits{P.} \bsnm{{Schwartz}}}, \bauthor{\binits{N.}
  \bsnm{{Labrosse}}}, \bauthor{\binits{P.} \bsnm{{Kotr{\v c}}}},
  \bauthor{\binits{U.} \bsnm{{Anzer}}}, \bauthor{\binits{G.} \bsnm{{Molodij}}},
  \bauthor{\binits{A.} \bsnm{{Berlicki}}}, \bauthor{\binits{E.E.}
  \bsnm{{DeLuca}}}, \bauthor{\binits{L.} \bsnm{{Golub}}}, \bauthor{\binits{T.}
  \bsnm{{Watanabe}}}, \bauthor{\binits{T.} \bsnm{{Berger}}},
\batitle{{Hinode, TRACE, SOHO, and Ground-based Observations of a Quiescent
  Prominence}}.
\bjtitle{\apj}
\bvolume{686},
\bfpage{1383}--\blpage{1396}
(\byear{2008}).
doi:\doiurl{10.1086/591018}
\end{barticle}
\endbibitem

\bibitem[\protect\citeauthoryear{{Hirayama}}{1963}]{1963PASJ...15..122H}
\begin{barticle}
\bauthor{\binits{T.} \bsnm{{Hirayama}}},
\batitle{{On the Model of the Solar Quiescent Prominence and the Effect of the
  Solar UV Radiation on the Prominence}}.
\bjtitle{\pasj}
\bvolume{15},
\bfpage{122}
(\byear{1963})
\end{barticle}
\endbibitem

\bibitem[\protect\citeauthoryear{{Hirayama}}{1971}]{1971SoPh...17...50H}
\begin{barticle}
\bauthor{\binits{T.} \bsnm{{Hirayama}}},
\batitle{{Spectral Analysis of Four Quiescent Prominences Observed at the
  Peruvian Eclipse}}.
\bjtitle{\solphys}
\bvolume{17},
\bfpage{50}--\blpage{75}
(\byear{1971}).
doi:\doiurl{10.1007/BF00152861}
\end{barticle}
\endbibitem

\bibitem[\protect\citeauthoryear{{Hirayama}}{1972}]{1972SoPh...24..310H}
\begin{barticle}
\bauthor{\binits{T.} \bsnm{{Hirayama}}},
\batitle{{Ionized Helium in Prominences and in the Chromosphere}}.
\bjtitle{\solphys}
\bvolume{24},
\bfpage{310}--\blpage{323}
(\byear{1972}).
doi:\doiurl{10.1007/BF00153371}
\end{barticle}
\endbibitem

\bibitem[\protect\citeauthoryear{{Hirayama}}{1985}]{1985SoPh..100..415H}
\begin{barticle}
\bauthor{\binits{T.} \bsnm{{Hirayama}}},
\batitle{{Modern observations of solar prominences}}.
\bjtitle{\solphys}
\bvolume{100},
\bfpage{415}--\blpage{434}
(\byear{1985}).
doi:\doiurl{10.1007/BF00158439}
\end{barticle}
\endbibitem

\bibitem[\protect\citeauthoryear{{Hirayama}}{1990}]{1990LNP...363..187H}
\begin{botherref}
\oauthor{\binits{T.} \bsnm{{Hirayama}}},
{Physical conditions in prominences},
in \textit{IAU Colloq. 117: Dynamics of Quiescent Prominences},
ed. by V. {Ruzdjak}, E. {Tandberg-Hanssen}.
Lecture Notes in Physics, Berlin Springer Verlag,
vol. 363,
1990,
pp. 187--203
\end{botherref}
\endbibitem

\bibitem[\protect\citeauthoryear{{Howard} et~al.}{2008}]{2008SSRv..136...67H}
\begin{barticle}
\bauthor{\binits{R.A.} \bsnm{{Howard}}}, \bauthor{\binits{J.D.}
  \bsnm{{Moses}}}, \bauthor{\binits{A.} \bsnm{{Vourlidas}}},
  \bauthor{\binits{J.S.} \bsnm{{Newmark}}}, \bauthor{\binits{D.G.}
  \bsnm{{Socker}}}, \bauthor{\binits{S.P.} \bsnm{{Plunkett}}},
  \bauthor{\binits{C.M.} \bsnm{{Korendyke}}}, \bauthor{\binits{J.W.}
  \bsnm{{Cook}}}, \bauthor{\binits{A.} \bsnm{{Hurley}}}, \bauthor{\binits{J.M.}
  \bsnm{{Davila}}}, \bauthor{\binits{W.T.} \bsnm{{Thompson}}},
  \bauthor{\binits{O.C.} \bsnm{{St Cyr}}}, \bauthor{\binits{E.}
  \bsnm{{Mentzell}}}, \bauthor{\binits{K.} \bsnm{{Mehalick}}},
  \bauthor{\binits{J.R.} \bsnm{{Lemen}}}, \bauthor{\binits{J.P.}
  \bsnm{{Wuelser}}}, \bauthor{\binits{D.W.} \bsnm{{Duncan}}},
  \bauthor{\binits{T.D.} \bsnm{{Tarbell}}}, \bauthor{\binits{C.J.}
  \bsnm{{Wolfson}}}, \bauthor{\binits{A.} \bsnm{{Moore}}},
  \bauthor{\binits{R.A.} \bsnm{{Harrison}}}, \bauthor{\binits{N.R.}
  \bsnm{{Waltham}}}, \bauthor{\binits{J.} \bsnm{{Lang}}},
  \bauthor{\binits{C.J.} \bsnm{{Davis}}}, \bauthor{\binits{C.J.}
  \bsnm{{Eyles}}}, \bauthor{\binits{H.} \bsnm{{Mapson-Menard}}},
  \bauthor{\binits{G.M.} \bsnm{{Simnett}}}, \bauthor{\binits{J.P.}
  \bsnm{{Halain}}}, \bauthor{\binits{J.M.} \bsnm{{Defise}}},
  \bauthor{\binits{E.} \bsnm{{Mazy}}}, \bauthor{\binits{P.} \bsnm{{Rochus}}},
  \bauthor{\binits{R.} \bsnm{{Mercier}}}, \bauthor{\binits{M.F.}
  \bsnm{{Ravet}}}, \bauthor{\binits{F.} \bsnm{{Delmotte}}},
  \bauthor{\binits{F.} \bsnm{{Auchere}}}, \bauthor{\binits{J.P.}
  \bsnm{{Delaboudiniere}}}, \bauthor{\binits{V.} \bsnm{{Bothmer}}},
  \bauthor{\binits{W.} \bsnm{{Deutsch}}}, \bauthor{\binits{D.} \bsnm{{Wang}}},
  \bauthor{\binits{N.} \bsnm{{Rich}}}, \bauthor{\binits{S.} \bsnm{{Cooper}}},
  \bauthor{\binits{V.} \bsnm{{Stephens}}}, \bauthor{\binits{G.}
  \bsnm{{Maahs}}}, \bauthor{\binits{R.} \bsnm{{Baugh}}}, \bauthor{\binits{D.}
  \bsnm{{McMullin}}}, \bauthor{\binits{T.} \bsnm{{Carter}}},
\batitle{{Sun Earth Connection Coronal and Heliospheric Investigation
  (SECCHI)}}.
\bjtitle{Space Science Reviews}
\bvolume{136},
\bfpage{67}--\blpage{115}
(\byear{2008}).
doi:\doiurl{10.1007/s11214-008-9341-4}
\end{barticle}
\endbibitem

\bibitem[\protect\citeauthoryear{{Hubeny}}{1985}]{1985A&A...145..461H}
\begin{barticle}
\bauthor{\binits{I.} \bsnm{{Hubeny}}},
\batitle{{A modified Rybicki method and the partial coherent scattering
  approximation}}.
\bjtitle{\aap}
\bvolume{145},
\bfpage{461}--\blpage{474}
(\byear{1985})
\end{barticle}
\endbibitem

\bibitem[\protect\citeauthoryear{{Hubeny}}{2003}]{2003ASPC..288...17H}
\begin{botherref}
\oauthor{\binits{I.} \bsnm{{Hubeny}}},
{Accelerated Lambda Iteration: An Overview},
in \textit{Stellar Atmosphere Modeling},
ed. by I. {Hubeny}, D. {Mihalas}, K. {Werner}.
Astronomical Society of the Pacific Conference Series,
vol. 288,
2003,
p. 17
\end{botherref}
\endbibitem

\bibitem[\protect\citeauthoryear{{Hubeny} and
  {Lanz}}{1995}]{1995ApJ...439..875H}
\begin{barticle}
\bauthor{\binits{I.} \bsnm{{Hubeny}}}, \bauthor{\binits{T.} \bsnm{{Lanz}}},
\batitle{{Non-LTE line-blanketed model atmospheres of hot stars. 1: Hybrid
  complete linearization/accelerated lambda iteration method}}.
\bjtitle{\apj}
\bvolume{439},
\bfpage{875}--\blpage{904}
(\byear{1995}).
doi:\doiurl{10.1086/175226}
\end{barticle}
\endbibitem

\bibitem[\protect\citeauthoryear{{Hubeny} and
  {Lites}}{1995}]{1995ApJ...455..376H}
\begin{barticle}
\bauthor{\binits{I.} \bsnm{{Hubeny}}}, \bauthor{\binits{B.W.} \bsnm{{Lites}}},
\batitle{{Partial Redistribution in Multilevel Atoms. I. Method and Application
  to the Solar Hydrogen Line Formation}}.
\bjtitle{\apj}
\bvolume{455},
\bfpage{376}
(\byear{1995}).
doi:\doiurl{10.1086/176584}
\end{barticle}
\endbibitem

\bibitem[\protect\citeauthoryear{{Hummer}}{1962}]{1962MNRAS.125...21H}
\begin{barticle}
\bauthor{\binits{D.G.} \bsnm{{Hummer}}},
\batitle{{Non-coherent scattering: I. The redistribution function with Doppler
  broadening}}.
\bjtitle{\mnras}
\bvolume{125},
\bfpage{21}--\blpage{37}
(\byear{1962})
\end{barticle}
\endbibitem

\bibitem[\protect\citeauthoryear{{Hummer}}{1969}]{1969MNRAS.145...95H}
\begin{barticle}
\bauthor{\binits{D.G.} \bsnm{{Hummer}}},
\batitle{{Non-coherent scattering-VI. Solutions of the transfer problem with a
  frequency-dependent source function}}.
\bjtitle{\mnras}
\bvolume{145},
\bfpage{95}
(\byear{1969})
\end{barticle}
\endbibitem

\bibitem[\protect\citeauthoryear{{Hyder} and
  {Lites}}{1970}]{1970SoPh...14..147H}
\begin{barticle}
\bauthor{\binits{C.L.} \bsnm{{Hyder}}}, \bauthor{\binits{B.W.} \bsnm{{Lites}}},
\batitle{{H{$\alpha$} Doppler Brightening and Lyman-{$\alpha$} Doppler Dimming
  in Moving H{$\alpha$} Prominences}}.
\bjtitle{\solphys}
\bvolume{14},
\bfpage{147}--\blpage{156}
(\byear{1970}).
doi:\doiurl{10.1007/BF00240170}
\end{barticle}
\endbibitem

\bibitem[\protect\citeauthoryear{{Inglis} and
  {Teller}}{1939}]{1939ApJ....90..439I}
\begin{barticle}
\bauthor{\binits{D.R.} \bsnm{{Inglis}}}, \bauthor{\binits{E.} \bsnm{{Teller}}},
\batitle{{Ionic Depression of Series Limits in Cne-Electron Spectra.}}
\bjtitle{\apj}
\bvolume{90},
\bfpage{439}
(\byear{1939}).
doi:\doiurl{10.1086/144118}
\end{barticle}
\endbibitem

\bibitem[\protect\citeauthoryear{{Irimajiri}
  et~al.}{1995}]{1995SoPh..156..363I}
\begin{barticle}
\bauthor{\binits{Y.} \bsnm{{Irimajiri}}}, \bauthor{\binits{T.}
  \bsnm{{Takano}}}, \bauthor{\binits{H.} \bsnm{{Nakajima}}},
  \bauthor{\binits{K.} \bsnm{{Shibasaki}}}, \bauthor{\binits{Y.}
  \bsnm{{Hanaoka}}}, \bauthor{\binits{K.} \bsnm{{Ichimoto}}},
\batitle{{Simultaneous multifrequency observations of an eruptive prominence at
  millimeter wavelengths}}.
\bjtitle{\solphys}
\bvolume{156},
\bfpage{363}--\blpage{375}
(\byear{1995}).
doi:\doiurl{10.1007/BF00670232}
\end{barticle}
\endbibitem

\bibitem[\protect\citeauthoryear{{Ivanov-Kholodnyi} and
  {Nikol'skii}}{1961}]{1961AZh....38...45I}
\begin{barticle}
\bauthor{\binits{G.S.} \bsnm{{Ivanov-Kholodnyi}}}, \bauthor{\binits{G.M.}
  \bsnm{{Nikol'skii}}},
\batitle{{Ultraviolet Solar Radiation and the Transition Layer Between the
  Chromosphere and the Corona}}.
\bjtitle{\azh}
\bvolume{38},
\bfpage{45}
(\byear{1961})
\end{barticle}
\endbibitem

\bibitem[\protect\citeauthoryear{{Jej{\v c}i{\v c}} and
  {Heinzel}}{2009}]{2009SoPh..254...89J}
\begin{barticle}
\bauthor{\binits{S.} \bsnm{{Jej{\v c}i{\v c}}}}, \bauthor{\binits{P.}
  \bsnm{{Heinzel}}},
\batitle{{Electron Densities in Quiescent Prominences Derived from Eclipse
  Observations}}.
\bjtitle{\solphys}
\bvolume{254},
\bfpage{89}--\blpage{100}
(\byear{2009}).
doi:\doiurl{10.1007/s11207-008-9289-2}
\end{barticle}
\endbibitem

\bibitem[\protect\citeauthoryear{{Judge}}{2000}]{2000ApJ...531..585J}
\begin{barticle}
\bauthor{\binits{P.G.} \bsnm{{Judge}}},
\batitle{{On Spectroscopic Filling Factors and the Solar Transition Region}}.
\bjtitle{\apj}
\bvolume{531},
\bfpage{585}--\blpage{590}
(\byear{2000}).
doi:\doiurl{10.1086/308458}
\end{barticle}
\endbibitem

\bibitem[\protect\citeauthoryear{{Kanno} et~al.}{1981}]{1981SoPh...69..313K}
\begin{barticle}
\bauthor{\binits{M.} \bsnm{{Kanno}}}, \bauthor{\binits{G.L.}
  \bsnm{{Withbroe}}}, \bauthor{\binits{R.W.} \bsnm{{Noyes}}},
\batitle{{Analysis of extreme-ultraviolet spectroheliograms of solar
  prominences}}.
\bjtitle{\solphys}
\bvolume{69},
\bfpage{313}--\blpage{326}
(\byear{1981}).
doi:\doiurl{10.1007/BF00149997}
\end{barticle}
\endbibitem

\bibitem[\protect\citeauthoryear{Keady and Kilcrease}{2000}]{Keady2000}
\begin{bchapter}
\bauthor{\binits{J.} \bsnm{Keady}}, \bauthor{\binits{D.} \bsnm{Kilcrease}},
\bctitle{Allen's astrophysical quantities},
ed. by \beditor{\binits{A.N.} \bsnm{Cox}}
(\bpublisher{Springer},
\blocation{New York}, \byear{2000}),
pp. \bfpage{95}--\blpage{120}.
\bcomment{Chap. Radiation}
\end{bchapter}
\endbibitem

\bibitem[\protect\citeauthoryear{{Kilper} et~al.}{2009}]{2009ApJ...704..522K}
\begin{barticle}
\bauthor{\binits{G.} \bsnm{{Kilper}}}, \bauthor{\binits{H.} \bsnm{{Gilbert}}},
  \bauthor{\binits{D.} \bsnm{{Alexander}}},
\batitle{{Mass Composition in Pre-eruption Quiet Sun Filaments}}.
\bjtitle{\apj}
\bvolume{704},
\bfpage{522}--\blpage{530}
(\byear{2009}).
doi:\doiurl{10.1088/0004-637X/704/1/522}
\end{barticle}
\endbibitem

\bibitem[\protect\citeauthoryear{{Kjeldseth-Moe}
  et~al.}{1979}]{1979SoPh...61..319M}
\begin{barticle}
\bauthor{\binits{O.} \bsnm{{Kjeldseth-Moe}}}, \bauthor{\binits{J.W.}
  \bsnm{{Cook}}}, \bauthor{\binits{S.A.} \bsnm{{Mango}}},
\batitle{{EUV observations of quiescent prominences from SKYLAB}}.
\bjtitle{\solphys}
\bvolume{61},
\bfpage{319}--\blpage{334}
(\byear{1979}).
doi:\doiurl{10.1007/BF00150417}
\end{barticle}
\endbibitem

\bibitem[\protect\citeauthoryear{{Kohl} and
  {Withbroe}}{1982}]{1982ApJ...256..263K}
\begin{barticle}
\bauthor{\binits{J.L.} \bsnm{{Kohl}}}, \bauthor{\binits{G.L.}
  \bsnm{{Withbroe}}},
\batitle{{EUV spectroscopic plasma diagnostics for the solar wind acceleration
  region}}.
\bjtitle{\apj}
\bvolume{256},
\bfpage{263}--\blpage{270}
(\byear{1982}).
doi:\doiurl{10.1086/159904}
\end{barticle}
\endbibitem

\bibitem[\protect\citeauthoryear{{Kohl} et~al.}{1995}]{1995SoPh..162..313K}
\begin{barticle}
\bauthor{\binits{J.L.} \bsnm{{Kohl}}}, \bauthor{\binits{R.} \bsnm{{Esser}}},
  \bauthor{\binits{L.D.} \bsnm{{Gardner}}}, \bauthor{\binits{S.}
  \bsnm{{Habbal}}}, \bauthor{\binits{P.S.} \bsnm{{Daigneau}}},
  \bauthor{\binits{E.F.} \bsnm{{Dennis}}}, \bauthor{\binits{G.U.}
  \bsnm{{Nystrom}}}, \bauthor{\binits{A.} \bsnm{{Panasyuk}}},
  \bauthor{\binits{J.C.} \bsnm{{Raymond}}}, \bauthor{\binits{P.L.}
  \bsnm{{Smith}}}, \bauthor{\binits{L.} \bsnm{{Strachan}}},
  \bauthor{\binits{A.A.} \bsnm{{van Ballegooijen}}}, \bauthor{\binits{G.}
  \bsnm{{Noci}}}, \bauthor{\binits{S.} \bsnm{{Fineschi}}}, \bauthor{\binits{M.}
  \bsnm{{Romoli}}}, \bauthor{\binits{A.} \bsnm{{Ciaravella}}},
  \bauthor{\binits{A.} \bsnm{{Modigliani}}}, \bauthor{\binits{M.C.E.}
  \bsnm{{Huber}}}, \bauthor{\binits{E.} \bsnm{{Antonucci}}},
  \bauthor{\binits{C.} \bsnm{{Benna}}}, \bauthor{\binits{S.}
  \bsnm{{Giordano}}}, \bauthor{\binits{G.} \bsnm{{Tondello}}},
  \bauthor{\binits{P.} \bsnm{{Nicolosi}}}, \bauthor{\binits{G.}
  \bsnm{{Naletto}}}, \bauthor{\binits{C.} \bsnm{{Pernechele}}},
  \bauthor{\binits{D.} \bsnm{{Spadaro}}}, \bauthor{\binits{G.}
  \bsnm{{Poletto}}}, \bauthor{\binits{S.} \bsnm{{Livi}}}, \bauthor{\binits{O.}
  \bsnm{{von der L{\"u}he}}}, \bauthor{\binits{J.} \bsnm{{Geiss}}},
  \bauthor{\binits{J.G.} \bsnm{{Timothy}}}, \bauthor{\binits{G.}
  \bsnm{{Gloeckler}}}, \bauthor{\binits{A.} \bsnm{{Allegra}}},
  \bauthor{\binits{G.} \bsnm{{Basile}}}, \bauthor{\binits{R.} \bsnm{{Brusa}}},
  \bauthor{\binits{B.} \bsnm{{Wood}}}, \bauthor{\binits{O.H.W.}
  \bsnm{{Siegmund}}}, \bauthor{\binits{W.} \bsnm{{Fowler}}},
  \bauthor{\binits{R.} \bsnm{{Fisher}}}, \bauthor{\binits{M.}
  \bsnm{{Jhabvala}}},
\batitle{{The Ultraviolet Coronagraph Spectrometer for the Solar and
  Heliospheric Observatory}}.
\bjtitle{\solphys}
\bvolume{162},
\bfpage{313}--\blpage{356}
(\byear{1995}).
doi:\doiurl{10.1007/BF00733433}
\end{barticle}
\endbibitem

\bibitem[\protect\citeauthoryear{{Korendyke}
  et~al.}{2001}]{2001SoPh..200...63K}
\begin{barticle}
\bauthor{\binits{C.M.} \bsnm{{Korendyke}}}, \bauthor{\binits{A.}
  \bsnm{{Vourlidas}}}, \bauthor{\binits{J.W.} \bsnm{{Cook}}},
  \bauthor{\binits{K.P.} \bsnm{{Dere}}}, \bauthor{\binits{R.A.}
  \bsnm{{Howard}}}, \bauthor{\binits{J.S.} \bsnm{{Morrill}}},
  \bauthor{\binits{J.D.} \bsnm{{Moses}}}, \bauthor{\binits{N.E.}
  \bsnm{{Moulton}}}, \bauthor{\binits{D.G.} \bsnm{{Socker}}},
\batitle{{High-resolution Imaging of the Upper Solar Chromosphere: First Light
  Performance of the Very-high-Resolution Advanced ULtraviolet Telescope}}.
\bjtitle{\solphys}
\bvolume{200},
\bfpage{63}--\blpage{73}
(\byear{2001})
\end{barticle}
\endbibitem

\bibitem[\protect\citeauthoryear{{Kosugi} et~al.}{2007}]{2007SoPh..243....3K}
\begin{barticle}
\bauthor{\binits{T.} \bsnm{{Kosugi}}}, \bauthor{\binits{K.}
  \bsnm{{Matsuzaki}}}, \bauthor{\binits{T.} \bsnm{{Sakao}}},
  \bauthor{\binits{T.} \bsnm{{Shimizu}}}, \bauthor{\binits{Y.} \bsnm{{Sone}}},
  \bauthor{\binits{S.} \bsnm{{Tachikawa}}}, \bauthor{\binits{T.}
  \bsnm{{Hashimoto}}}, \bauthor{\binits{K.} \bsnm{{Minesugi}}},
  \bauthor{\binits{A.} \bsnm{{Ohnishi}}}, \bauthor{\binits{T.}
  \bsnm{{Yamada}}}, \bauthor{\binits{S.} \bsnm{{Tsuneta}}},
  \bauthor{\binits{H.} \bsnm{{Hara}}}, \bauthor{\binits{K.} \bsnm{{Ichimoto}}},
  \bauthor{\binits{Y.} \bsnm{{Suematsu}}}, \bauthor{\binits{M.}
  \bsnm{{Shimojo}}}, \bauthor{\binits{T.} \bsnm{{Watanabe}}},
  \bauthor{\binits{S.} \bsnm{{Shimada}}}, \bauthor{\binits{J.M.}
  \bsnm{{Davis}}}, \bauthor{\binits{L.D.} \bsnm{{Hill}}},
  \bauthor{\binits{J.K.} \bsnm{{Owens}}}, \bauthor{\binits{A.M.}
  \bsnm{{Title}}}, \bauthor{\binits{J.L.} \bsnm{{Culhane}}},
  \bauthor{\binits{L.K.} \bsnm{{Harra}}}, \bauthor{\binits{G.A.}
  \bsnm{{Doschek}}}, \bauthor{\binits{L.} \bsnm{{Golub}}},
\batitle{{The Hinode (Solar-B) Mission: An Overview}}.
\bjtitle{\solphys}
\bvolume{243},
\bfpage{3}--\blpage{17}
(\byear{2007}).
doi:\doiurl{10.1007/s11207-007-9014-6}
\end{barticle}
\endbibitem

\bibitem[\protect\citeauthoryear{{Koutchmy} et~al.}{1983}]{1983A&A...119..261K}
\begin{barticle}
\bauthor{\binits{S.} \bsnm{{Koutchmy}}}, \bauthor{\binits{C.} \bsnm{{Lebecq}}},
  \bauthor{\binits{G.} \bsnm{{Stellmacher}}},
\batitle{{The electron density of faint prominences observed during the solar
  eclipse of July 31, 1981}}.
\bjtitle{\aap}
\bvolume{119},
\bfpage{261}--\blpage{264}
(\byear{1983})
\end{barticle}
\endbibitem

\bibitem[\protect\citeauthoryear{{Kucera} and
  {Landi}}{2006}]{2006ApJ...645.1525K}
\begin{barticle}
\bauthor{\binits{T.A.} \bsnm{{Kucera}}}, \bauthor{\binits{E.} \bsnm{{Landi}}},
\batitle{{Ultraviolet Observations of Prominence Activation and Cool Loop
  Dynamics}}.
\bjtitle{\apj}
\bvolume{645},
\bfpage{1525}--\blpage{1536}
(\byear{2006}).
doi:\doiurl{10.1086/504398}
\end{barticle}
\endbibitem

\bibitem[\protect\citeauthoryear{{Kucera} and
  {Landi}}{2008}]{2008ApJ...673..611K}
\begin{barticle}
\bauthor{\binits{T.A.} \bsnm{{Kucera}}}, \bauthor{\binits{E.} \bsnm{{Landi}}},
\batitle{{An Observation of Low-Level Heating in an Erupting Prominence}}.
\bjtitle{\apj}
\bvolume{673},
\bfpage{611}--\blpage{620}
(\byear{2008}).
doi:\doiurl{10.1086/523694}
\end{barticle}
\endbibitem

\bibitem[\protect\citeauthoryear{{Kucera} et~al.}{1998}]{1998SoPh..183..107K}
\begin{barticle}
\bauthor{\binits{T.A.} \bsnm{{Kucera}}}, \bauthor{\binits{V.}
  \bsnm{{Andretta}}}, \bauthor{\binits{A.I.} \bsnm{{Poland}}},
\batitle{{Neutral Hydrogen Column Depths in Prominences Using EUV Absorption
  Features}}.
\bjtitle{\solphys}
\bvolume{183},
\bfpage{107}--\blpage{121}
(\byear{1998})
\end{barticle}
\endbibitem

\bibitem[\protect\citeauthoryear{{Kucera} et~al.}{2003}]{2003SoPh..212...81K}
\begin{barticle}
\bauthor{\binits{T.A.} \bsnm{{Kucera}}}, \bauthor{\binits{M.} \bsnm{{Tovar}}},
  \bauthor{\binits{B.} \bsnm{{de Pontieu}}},
\batitle{{Prominence Motions Observed at High Cadences in Temperatures from 10
  000 to 250 000 K}}.
\bjtitle{\solphys}
\bvolume{212},
\bfpage{81}--\blpage{97}
(\byear{2003}).
doi:\doiurl{10.1023/A:1022900604972}
\end{barticle}
\endbibitem

\bibitem[\protect\citeauthoryear{{Kucera} et~al.}{1993}]{1993ApJ...412..853K}
\begin{barticle}
\bauthor{\binits{T.A.} \bsnm{{Kucera}}}, \bauthor{\binits{G.A.} \bsnm{{Dulk}}},
  \bauthor{\binits{A.L.} \bsnm{{Kiplinger}}}, \bauthor{\binits{R.M.}
  \bsnm{{Winglee}}}, \bauthor{\binits{T.S.} \bsnm{{Bastian}}},
  \bauthor{\binits{M.} \bsnm{{Graeter}}},
\batitle{{Multiple wavelength observations of an off-limb eruptive solar
  flare}}.
\bjtitle{\apj}
\bvolume{412},
\bfpage{853}--\blpage{864}
(\byear{1993}).
doi:\doiurl{10.1086/172967}
\end{barticle}
\endbibitem

\bibitem[\protect\citeauthoryear{{Kucera} et~al.}{1999}]{1999SoPh..186..259K}
\begin{barticle}
\bauthor{\binits{T.A.} \bsnm{{Kucera}}}, \bauthor{\binits{G.}
  \bsnm{{Aulanier}}}, \bauthor{\binits{B.} \bsnm{{Schmieder}}},
  \bauthor{\binits{N.} \bsnm{{Mein}}}, \bauthor{\binits{J.C.} \bsnm{{Vial}}},
\batitle{{Filament channel structures in a SI IV line related to a 3d magnetic
  model}}.
\bjtitle{\solphys}
\bvolume{186},
\bfpage{259}--\blpage{280}
(\byear{1999})
\end{barticle}
\endbibitem

\bibitem[\protect\citeauthoryear{{Kuin} and
  {Poland}}{1991}]{1991ApJ...370..763K}
\begin{barticle}
\bauthor{\binits{N.P.M.} \bsnm{{Kuin}}}, \bauthor{\binits{A.I.}
  \bsnm{{Poland}}},
\batitle{{Opacity effects on the radiative losses of coronal loops}}.
\bjtitle{\apj}
\bvolume{370},
\bfpage{763}--\blpage{774}
(\byear{1991}).
doi:\doiurl{10.1086/169859}
\end{barticle}
\endbibitem

\bibitem[\protect\citeauthoryear{{Kunasz} and
  {Auer}}{1988}]{1988JQSRT..39...67K}
\begin{barticle}
\bauthor{\binits{P.} \bsnm{{Kunasz}}}, \bauthor{\binits{L.H.} \bsnm{{Auer}}},
\batitle{{Short characteristic integration of radiative transfer problems:
  formal solution in two-dimensional slabs.}}
\bjtitle{Journal of Quantitative Spectroscopy and Radiative Transfer}
\bvolume{39},
\bfpage{67}--\blpage{79}
(\byear{1988}).
doi:\doiurl{10.1016/0022-4073(88)90021-0}
\end{barticle}
\endbibitem

\bibitem[\protect\citeauthoryear{{Labrosse} and
  {Gouttebroze}}{2001}]{2001A&A...380..323L}
\begin{barticle}
\bauthor{\binits{N.} \bsnm{{Labrosse}}}, \bauthor{\binits{P.}
  \bsnm{{Gouttebroze}}},
\batitle{{Formation of helium spectrum in solar quiescent prominences}}.
\bjtitle{\aap}
\bvolume{380},
\bfpage{323}--\blpage{340}
(\byear{2001}).
doi:\doiurl{10.1051/0004-6361:20011395}
\end{barticle}
\endbibitem

\bibitem[\protect\citeauthoryear{{Labrosse} and
  {Gouttebroze}}{2004}]{2004ApJ...617..614L}
\begin{barticle}
\bauthor{\binits{N.} \bsnm{{Labrosse}}}, \bauthor{\binits{P.}
  \bsnm{{Gouttebroze}}},
\batitle{{Non-LTE Radiative Transfer in Model Prominences. I. Integrated
  Intensities of He I Triplet Lines}}.
\bjtitle{\apj}
\bvolume{617},
\bfpage{614}--\blpage{622}
(\byear{2004}).
doi:\doiurl{10.1086/425168}
\end{barticle}
\endbibitem

\bibitem[\protect\citeauthoryear{{Labrosse}
  et~al.}{2007a}]{2007A&A...463.1171L}
\begin{barticle}
\bauthor{\binits{N.} \bsnm{{Labrosse}}}, \bauthor{\binits{P.}
  \bsnm{{Gouttebroze}}}, \bauthor{\binits{J.C.} \bsnm{{Vial}}},
\batitle{{Effect of motions in prominences on the helium resonance lines in the
  extreme ultraviolet}}.
\bjtitle{\aap}
\bvolume{463},
\bfpage{1171}--\blpage{1179}
(\byear{2007a}).
doi:\doiurl{10.1051/0004-6361:20065775}
\end{barticle}
\endbibitem

\bibitem[\protect\citeauthoryear{{Labrosse}
  et~al.}{2007b}]{2007ASPC..368..337L}
\begin{botherref}
\oauthor{\binits{N.} \bsnm{{Labrosse}}}, \oauthor{\binits{P.}
  \bsnm{{Gouttebroze}}}, \oauthor{\binits{J.C.} \bsnm{{Vial}}},
{Spectral Diagnostics of Active Prominences},
in \textit{The Physics of Chromospheric Plasmas},
ed. by P. {Heinzel}, I. {Dorotovi{\v c}}, R.J. {Rutten}.
Astronomical Society of the Pacific Conference Series,
vol. 368,
2007b,
p. 337
\end{botherref}
\endbibitem

\bibitem[\protect\citeauthoryear{{Labrosse} et~al.}{2006}]{2006sf2a.conf..549L}
\begin{botherref}
\oauthor{\binits{N.} \bsnm{{Labrosse}}}, \oauthor{\binits{J.C.} \bsnm{{Vial}}},
  \oauthor{\binits{P.} \bsnm{{Gouttebroze}}},
{Plasma diagnostic of a solar prominence from hydrogen and helium resonance
  lines},
in \textit{SF2A-2006: Semaine de l'Astrophysique Francaise},
ed. by D. {Barret}, F. {Casoli}, G. {Lagache}, A. {Lecavelier}, L. {Pagani},
2006,
p. 549
\end{botherref}
\endbibitem

\bibitem[\protect\citeauthoryear{{Labrosse}
  et~al.}{2006b}]{2006IAUJD...3E..47L}
\begin{botherref}
\oauthor{\binits{N.} \bsnm{{Labrosse}}}, \oauthor{\binits{J.C.} \bsnm{{Vial}}},
  \oauthor{\binits{P.} \bsnm{{Gouttebroze}}},
{The Helium Spectrum in Erupting Solar Prominences}.
\textrm{Solar Active Regions and 3D Magnetic Structure, 26th meeting of the
  IAU, Joint Discussion 3, 16-17 August, 2006, Prague, Czech Republic, JD03,
  \#47}
\textbf{3}
(2006b)
\end{botherref}
\endbibitem

\bibitem[\protect\citeauthoryear{{Labrosse} et~al.}{2008}]{2008AnGeo..26.2961L}
\begin{barticle}
\bauthor{\binits{N.} \bsnm{{Labrosse}}}, \bauthor{\binits{J.C.} \bsnm{{Vial}}},
  \bauthor{\binits{P.} \bsnm{{Gouttebroze}}},
\batitle{{Diagnostics of active and eruptive prominences through hydrogen and
  helium lines modelling}}.
\bjtitle{Annales Geophysicae}
\bvolume{26},
\bfpage{2961}--\blpage{2965}
(\byear{2008})
\end{barticle}
\endbibitem

\bibitem[\protect\citeauthoryear{{Labrosse} et~al.}{2002}]{2002ESASP.506..451L}
\begin{botherref}
\oauthor{\binits{N.} \bsnm{{Labrosse}}}, \oauthor{\binits{P.}
  \bsnm{{Gouttebroze}}}, \oauthor{\binits{P.} \bsnm{{Heinzel}}},
  \oauthor{\binits{J.C.} \bsnm{{Vial}}},
{Line profiles and intensity ratios in prominence models with a prominence to
  corona interface},
in \textit{Solar Variability: From Core to Outer Frontiers},
ed. by J. {Kuijpers}.
ESA Special Publication,
vol. 506,
2002,
pp. 451--454
\end{botherref}
\endbibitem

\bibitem[\protect\citeauthoryear{{Landi} and
  {Landini}}{1997}]{1997A&A...327.1230L}
\begin{barticle}
\bauthor{\binits{E.} \bsnm{{Landi}}}, \bauthor{\binits{M.} \bsnm{{Landini}}},
\batitle{{Simultaneous temperature and density diagnostics of optically thin
  plasmas}}.
\bjtitle{\aap}
\bvolume{327},
\bfpage{1230}--\blpage{1241}
(\byear{1997})
\end{barticle}
\endbibitem

\bibitem[\protect\citeauthoryear{{Landman}}{1984}]{1984ApJ...279..438L}
\begin{barticle}
\bauthor{\binits{D.A.} \bsnm{{Landman}}},
\batitle{{Physical conditions in the cool parts of prominences. II - The MG
  triplet lines}}.
\bjtitle{\apj}
\bvolume{279},
\bfpage{438}--\blpage{445}
(\byear{1984}).
doi:\doiurl{10.1086/161906}
\end{barticle}
\endbibitem

\bibitem[\protect\citeauthoryear{{Landman}}{1986}]{1986ApJ...305..546L}
\begin{barticle}
\bauthor{\binits{D.A.} \bsnm{{Landman}}},
\batitle{{Physical conditions in the cool parts of prominences and spicules -
  The effects of model atom level truncation on the derived plasma
  parameters}}.
\bjtitle{\apj}
\bvolume{305},
\bfpage{546}--\blpage{552}
(\byear{1986}).
doi:\doiurl{10.1086/164267}
\end{barticle}
\endbibitem

\bibitem[\protect\citeauthoryear{{L{\'e}ger} and
  {Paletou}}{2009}]{2009A&A...498..869L}
\begin{barticle}
\bauthor{\binits{L.} \bsnm{{L{\'e}ger}}}, \bauthor{\binits{F.}
  \bsnm{{Paletou}}},
\batitle{{2D non-LTE radiative modelling of He I spectral lines formed in solar
  prominences}}.
\bjtitle{\aap}
\bvolume{498},
\bfpage{869}--\blpage{875}
(\byear{2009}).
doi:\doiurl{10.1051/0004-6361/200810296}
\end{barticle}
\endbibitem

\bibitem[\protect\citeauthoryear{{Li} et~al.}{2000}]{2000MNRAS.313..761L}
\begin{barticle}
\bauthor{\binits{K.} \bsnm{{Li}}}, \bauthor{\binits{X.} \bsnm{{Gu}}},
  \bauthor{\binits{X.} \bsnm{{Chen}}},
\batitle{{Calculations and physical properties of the D3 emission lines of a
  prominence}}.
\bjtitle{\mnras}
\bvolume{313},
\bfpage{761}--\blpage{766}
(\byear{2000}).
doi:\doiurl{10.1046/j.1365-8711.2000.03336.x}
\end{barticle}
\endbibitem

\bibitem[\protect\citeauthoryear{{Li} et~al.}{1998}]{1998SoPh..183..323L}
\begin{barticle}
\bauthor{\binits{K.} \bsnm{{Li}}}, \bauthor{\binits{B.} \bsnm{{Schmieder}}},
  \bauthor{\binits{J.M.} \bsnm{{Malherbe}}}, \bauthor{\binits{T.}
  \bsnm{{Roudier}}}, \bauthor{\binits{J.E.} \bsnm{{Wiik}}},
\batitle{{Physical properties of the quiescent prominence of 5June 1996, from
  H{$\alpha$} observations}}.
\bjtitle{\solphys}
\bvolume{183},
\bfpage{323}--\blpage{338}
(\byear{1998})
\end{barticle}
\endbibitem

\bibitem[\protect\citeauthoryear{{Lin} et~al.}{2003}]{2003SoPh..216..109L}
\begin{barticle}
\bauthor{\binits{Y.} \bsnm{{Lin}}}, \bauthor{\binits{O.R.} \bsnm{{Engvold}}},
  \bauthor{\binits{J.E.} \bsnm{{Wiik}}},
\batitle{{Counterstreaming in a Large Polar Crown Filament}}.
\bjtitle{\solphys}
\bvolume{216},
\bfpage{109}--\blpage{120}
(\byear{2003}).
doi:\doiurl{10.1023/A:1026150809598}
\end{barticle}
\endbibitem

\bibitem[\protect\citeauthoryear{{Lin} et~al.}{2005}]{2005SoPh..226..239L}
\begin{barticle}
\bauthor{\binits{Y.} \bsnm{{Lin}}}, \bauthor{\binits{O.} \bsnm{{Engvold}}},
  \bauthor{\binits{L.} \bsnm{{Rouppe van der Voort}}}, \bauthor{\binits{J.E.}
  \bsnm{{Wiik}}}, \bauthor{\binits{T.E.} \bsnm{{Berger}}},
\batitle{{Thin Threads of Solar Filaments}}.
\bjtitle{\solphys}
\bvolume{226},
\bfpage{239}--\blpage{254}
(\byear{2005}).
doi:\doiurl{10.1007/s11207-005-6876-3}
\end{barticle}
\endbibitem

\bibitem[\protect\citeauthoryear{{Lin} et~al.}{2007}]{2007SoPh..246...65L}
\begin{barticle}
\bauthor{\binits{Y.} \bsnm{{Lin}}}, \bauthor{\binits{O.} \bsnm{{Engvold}}},
  \bauthor{\binits{L.H.M.} \bsnm{{Rouppe van der Voort}}}, \bauthor{\binits{M.}
  \bsnm{{van Noort}}},
\batitle{{Evidence of Traveling Waves in Filament Threads}}.
\bjtitle{\solphys}
\bvolume{246},
\bfpage{65}--\blpage{72}
(\byear{2007}).
doi:\doiurl{10.1007/s11207-007-0402-8}
\end{barticle}
\endbibitem

\bibitem[\protect\citeauthoryear{{Lin} et~al.}{2008}]{2008AdSpR..42..803L}
\begin{barticle}
\bauthor{\binits{Y.} \bsnm{{Lin}}}, \bauthor{\binits{S.F.} \bsnm{{Martin}}},
  \bauthor{\binits{O.} \bsnm{{Engvold}}}, \bauthor{\binits{L.H.M.}
  \bsnm{{Rouppe van der Voort}}}, \bauthor{\binits{M.} \bsnm{{van Noort}}},
\batitle{{On small active region filaments, fibrils and surges}}.
\bjtitle{Advances in Space Research}
\bvolume{42},
\bfpage{803}--\blpage{811}
(\byear{2008}).
doi:\doiurl{10.1016/j.asr.2007.05.052}
\end{barticle}
\endbibitem

\bibitem[\protect\citeauthoryear{{Litvinenko} and
  {Martin}}{1999}]{1999SoPh..190...45L}
\begin{barticle}
\bauthor{\binits{Y.E.} \bsnm{{Litvinenko}}}, \bauthor{\binits{S.F.}
  \bsnm{{Martin}}},
\batitle{{Magnetic reconnection as the cause of a photospheric canceling
  feature and mass flows in a filament}}.
\bjtitle{\solphys}
\bvolume{190},
\bfpage{45}--\blpage{58}
(\byear{1999}).
doi:\doiurl{10.1023/A:1005284116353}
\end{barticle}
\endbibitem

\bibitem[\protect\citeauthoryear{{Liu} et~al.}{2005}]{2005ApJ...631L..93L}
\begin{barticle}
\bauthor{\binits{Y.} \bsnm{{Liu}}}, \bauthor{\binits{H.} \bsnm{{Kurokawa}}},
  \bauthor{\binits{K.} \bsnm{{Shibata}}},
\batitle{{Production of Filaments by Surges}}.
\bjtitle{\apjl}
\bvolume{631},
\bfpage{93}--\blpage{96}
(\byear{2005}).
doi:\doiurl{10.1086/496919}
\end{barticle}
\endbibitem

\bibitem[\protect\citeauthoryear{{Mackay} and
  {Galsgaard}}{2001}]{2001SoPh..198..289M}
\begin{barticle}
\bauthor{\binits{D.H.} \bsnm{{Mackay}}}, \bauthor{\binits{K.}
  \bsnm{{Galsgaard}}},
\batitle{{Evolution of a Density Enhancement in a Stratified Atmosphere With
  Uniform Vertical Magnetic Field}}.
\bjtitle{\solphys}
\bvolume{198},
\bfpage{289}--\blpage{312}
(\byear{2001}).
doi:\doiurl{10.1023/A:1005266330720}
\end{barticle}
\endbibitem

\bibitem[\protect\citeauthoryear{{Mackay} et~al.}{2010}]{2010SSRv..tmp...32M}
\begin{botherref}
\oauthor{\binits{D.H.} \bsnm{{Mackay}}}, \oauthor{\binits{J.T.}
  \bsnm{{Karpen}}}, \oauthor{\binits{J.L.} \bsnm{{Ballester}}},
  \oauthor{\binits{B.} \bsnm{{Schmieder}}}, \oauthor{\binits{G.}
  \bsnm{{Aulanier}}},
{Physics of Solar Prominences: II -- Magnetic Structure and Dynamics}.
\textrm{Space Science Reviews},
32
(2010).
doi:\doiurl{10.1007/s11214-010-9628-0}
\end{botherref}
\endbibitem

\bibitem[\protect\citeauthoryear{{Madjarska}
  et~al.}{1999}]{1999ESASP.446..467M}
\begin{botherref}
\oauthor{\binits{M.S.} \bsnm{{Madjarska}}}, \oauthor{\binits{J.C.}
  \bsnm{{Vial}}}, \oauthor{\binits{K.} \bsnm{{Bocchialini}}},
  \oauthor{\binits{V.N.} \bsnm{{Dermendjiev}}},
{Plasma Diagnostics Of A Solar Prominence Observed On 12 June 1997 by EIT,
  Sumer And CDS},
in \textit{8th SOHO Workshop: Plasma Dynamics and Diagnostics in the Solar
  Transition Region and Corona},
ed. by J.C. {Vial}, B. {Kaldeich-Sch{\"u}}.
ESA Special Publication,
vol. 446,
1999,
p. 467
\end{botherref}
\endbibitem

\bibitem[\protect\citeauthoryear{{Mariska}}{1980}]{1980ApJ...235..268M}
\begin{barticle}
\bauthor{\binits{J.T.} \bsnm{{Mariska}}},
\batitle{{Relative chemical abundances in different solar regions}}.
\bjtitle{\apj}
\bvolume{235},
\bfpage{268}--\blpage{273}
(\byear{1980}).
doi:\doiurl{10.1086/157630}
\end{barticle}
\endbibitem

\bibitem[\protect\citeauthoryear{{Mariska}}{1992}]{1992str..book.....M}
\begin{bbook}
\bauthor{\binits{J.T.} \bsnm{{Mariska}}},
\bbtitle{{The solar transition region}}
(\bpublisher{Cambridge University Press},
\blocation{Cambridge}, \byear{1992})
\end{bbook}
\endbibitem

\bibitem[\protect\citeauthoryear{{Mariska} et~al.}{1979}]{1979ApJ...232..929M}
\begin{barticle}
\bauthor{\binits{J.T.} \bsnm{{Mariska}}}, \bauthor{\binits{G.A.}
  \bsnm{{Doschek}}}, \bauthor{\binits{U.} \bsnm{{Feldman}}},
\batitle{{Extreme-ultraviolet limb spectra of a prominence observed from
  SKYLAB}}.
\bjtitle{\apj}
\bvolume{232},
\bfpage{929}--\blpage{939}
(\byear{1979}).
doi:\doiurl{10.1086/157356}
\end{barticle}
\endbibitem

\bibitem[\protect\citeauthoryear{{Mason} and {Monsignori
  Fossi}}{1994}]{1994A&ARv...6..123M}
\begin{barticle}
\bauthor{\binits{H.E.} \bsnm{{Mason}}}, \bauthor{\binits{B.C.}
  \bsnm{{Monsignori Fossi}}},
\batitle{{Spectroscopic diagnostics in the VUV for solar and stellar plasmas}}.
\bjtitle{\aapr}
\bvolume{6},
\bfpage{123}--\blpage{179}
(\byear{1994}).
doi:\doiurl{10.1007/BF01208253}
\end{barticle}
\endbibitem

\bibitem[\protect\citeauthoryear{{Mein} et~al.}{1996}]{1996A&A...309..275M}
\begin{barticle}
\bauthor{\binits{N.} \bsnm{{Mein}}}, \bauthor{\binits{P.} \bsnm{{Mein}}},
  \bauthor{\binits{P.} \bsnm{{Heinzel}}}, \bauthor{\binits{J.C.}
  \bsnm{{Vial}}}, \bauthor{\binits{J.M.} \bsnm{{Malherbe}}},
  \bauthor{\binits{J.} \bsnm{{Staiger}}},
\batitle{{Cloud model with variable source function for solar H{$\alpha$}
  structures.}}
\bjtitle{\aap}
\bvolume{309},
\bfpage{275}--\blpage{283}
(\byear{1996})
\end{barticle}
\endbibitem

\bibitem[\protect\citeauthoryear{{Mein} and {Mein}}{1991}]{1991SoPh..136..317M}
\begin{barticle}
\bauthor{\binits{P.} \bsnm{{Mein}}}, \bauthor{\binits{N.} \bsnm{{Mein}}},
\batitle{{Dynamical fine structure of a quiescent prominence}}.
\bjtitle{\solphys}
\bvolume{136},
\bfpage{317}--\blpage{333}
(\byear{1991}).
doi:\doiurl{10.1007/BF00146539}
\end{barticle}
\endbibitem

\bibitem[\protect\citeauthoryear{{Mihalas}}{1978}]{1978stat.book.....M}
\begin{bbook}
\bauthor{\binits{D.} \bsnm{{Mihalas}}},
\bbtitle{{Stellar atmospheres /2nd edition/}}
(\bpublisher{W.~H.~Freeman and Co.},
\blocation{San Francisco}, \byear{1978})
\end{bbook}
\endbibitem

\bibitem[\protect\citeauthoryear{{Mihalas} et~al.}{1978}]{1978ApJ...220.1001M}
\begin{barticle}
\bauthor{\binits{D.} \bsnm{{Mihalas}}}, \bauthor{\binits{L.H.} \bsnm{{Auer}}},
  \bauthor{\binits{B.R.} \bsnm{{Mihalas}}},
\batitle{{Two-dimensional radiative transfer. I - Planar geometry}}.
\bjtitle{\apj}
\bvolume{220},
\bfpage{1001}--\blpage{1023}
(\byear{1978}).
doi:\doiurl{10.1086/155988}
\end{barticle}
\endbibitem

\bibitem[\protect\citeauthoryear{{Milkey} and
  {Mihalas}}{1973a}]{1973ApJ...185..709M}
\begin{barticle}
\bauthor{\binits{R.W.} \bsnm{{Milkey}}}, \bauthor{\binits{D.}
  \bsnm{{Mihalas}}},
\batitle{{Resonance-Line Transfer with Partial Redistribution: a Preliminary
  Study of Lyman~$\alpha$ in the Solar Chromosphere}}.
\bjtitle{\apj}
\bvolume{185},
\bfpage{709}--\blpage{726}
(\byear{1973a}).
doi:\doiurl{10.1086/152448}
\end{barticle}
\endbibitem

\bibitem[\protect\citeauthoryear{{Milkey} et~al.}{1973bp}]{1973ApJ...186.1043M}
\begin{barticle}
\bauthor{\binits{R.W.} \bsnm{{Milkey}}}, \bauthor{\binits{J.N.}
  \bsnm{{Heasley}}}, \bauthor{\binits{H.A.} \bsnm{{Beebe}}},
\batitle{{Helium Excitation in the Solar Chromosphere: he i in a Homogeneous
  Chromosphere}}.
\bjtitle{\apj}
\bvolume{186},
\bfpage{1043}--\blpage{1052}
(\byear{1973bp}).
doi:\doiurl{10.1086/152568}
\end{barticle}
\endbibitem

\bibitem[\protect\citeauthoryear{{Milkey} et~al.}{1979}]{1979phsp.coll...53M}
\begin{botherref}
\oauthor{\binits{R.W.} \bsnm{{Milkey}}}, \oauthor{\binits{J.N.}
  \bsnm{{Heasley}}}, \oauthor{\binits{E.J.} \bsnm{{Schmahl}}},
  \oauthor{\binits{O.} \bsnm{{Engvold}}},
{Frequency redistribution effects in the formation of Lyman alpha in
  prominences and their influence on the ratio of H-alpha to L-alpha},
in \textit{IAU Colloq. 44: Physics of Solar Prominences},
ed. by E. {Jensen}, P. {Maltby}, F.Q. {Orrall},
1979,
pp. 53--55
\end{botherref}
\endbibitem

\bibitem[\protect\citeauthoryear{{Monsignori Fossi} and
  {Landini}}{1991}]{1991AdSpR..11..281M}
\begin{barticle}
\bauthor{\binits{B.C.} \bsnm{{Monsignori Fossi}}}, \bauthor{\binits{M.}
  \bsnm{{Landini}}},
\batitle{{Models for inner corona parameters}}.
\bjtitle{Advances in Space Research}
\bvolume{11},
\bfpage{281}--\blpage{292}
(\byear{1991}).
doi:\doiurl{10.1016/0273-1177(91)90121-Y}
\end{barticle}
\endbibitem

\bibitem[\protect\citeauthoryear{{Morozhenko}}{1970}]{1970asas.book..176M}
\begin{botherref}
\oauthor{\binits{N.N.} \bsnm{{Morozhenko}}},
{Helium Excitation and Structure of Quiescent Solar Prominences},
ed. by V.I. {Voroshilov}
1970,
p. 176
\end{botherref}
\endbibitem

\bibitem[\protect\citeauthoryear{{Morozhenko}}{1978}]{1978SoPh...58...47M}
\begin{barticle}
\bauthor{\binits{N.N.} \bsnm{{Morozhenko}}},
\batitle{{Radiation transfer in prominences with filamentary structure}}.
\bjtitle{\solphys}
\bvolume{58},
\bfpage{47}--\blpage{56}
(\byear{1978}).
doi:\doiurl{10.1007/BF00152554}
\end{barticle}
\endbibitem

\bibitem[\protect\citeauthoryear{{Morozhenko}}{1984}]{1984SoPh...92..153M}
\begin{barticle}
\bauthor{\binits{N.N.} \bsnm{{Morozhenko}}},
\batitle{{On the excitation of lower levels of singlet helium in quiescent
  prominences}}.
\bjtitle{\solphys}
\bvolume{92},
\bfpage{153}--\blpage{160}
(\byear{1984}).
doi:\doiurl{10.1007/BF00157242}
\end{barticle}
\endbibitem

\bibitem[\protect\citeauthoryear{{Morozhenko} and
  {Zharkova}}{1982}]{1982AAfz...47...34M}
\begin{barticle}
\bauthor{\binits{N.N.} \bsnm{{Morozhenko}}}, \bauthor{\binits{V.V.}
  \bsnm{{Zharkova}}},
\batitle{{The spectral properties of filamentary, physically inhomogeneous
  prominences. II - Hydrogen (second level excitation, ionization)}}.
\bjtitle{Astrometriia i Astrofizika}
\bvolume{47},
\bfpage{34}--\blpage{41}
(\byear{1982})
\end{barticle}
\endbibitem

\bibitem[\protect\citeauthoryear{{Morton} and
  {Widing}}{1961}]{1961ApJ...133..596M}
\begin{barticle}
\bauthor{\binits{D.C.} \bsnm{{Morton}}}, \bauthor{\binits{K.G.}
  \bsnm{{Widing}}},
\batitle{{The Solar Lyman-Alpha Emission Line.}}
\bjtitle{\apj}
\bvolume{133},
\bfpage{596}
(\byear{1961}).
doi:\doiurl{10.1086/147062}
\end{barticle}
\endbibitem

\bibitem[\protect\citeauthoryear{{Noyes} and
  {Kalkofen}}{1970}]{1970SoPh...15..120N}
\begin{barticle}
\bauthor{\binits{R.W.} \bsnm{{Noyes}}}, \bauthor{\binits{W.}
  \bsnm{{Kalkofen}}},
\batitle{{The Solar Lyman Continuum and the Structure of the Solar
  Chromosphere}}.
\bjtitle{\solphys}
\bvolume{15},
\bfpage{120}--\blpage{138}
(\byear{1970}).
doi:\doiurl{10.1007/BF00149479}
\end{barticle}
\endbibitem

\bibitem[\protect\citeauthoryear{{Noyes} et~al.}{1972}]{1972ApJ...178..515N}
\begin{barticle}
\bauthor{\binits{R.W.} \bsnm{{Noyes}}}, \bauthor{\binits{A.K.}
  \bsnm{{Dupree}}}, \bauthor{\binits{M.C.E.} \bsnm{{Huber}}},
  \bauthor{\binits{W.H.} \bsnm{{Parkinson}}}, \bauthor{\binits{E.M.}
  \bsnm{{Reeves}}}, \bauthor{\binits{G.L.} \bsnm{{Withbroe}}},
\batitle{{Extreme-Ultraviolet Emission from Solar Prominences}}.
\bjtitle{\apj}
\bvolume{178},
\bfpage{515}--\blpage{526}
(\byear{1972}).
doi:\doiurl{10.1086/151812}
\end{barticle}
\endbibitem

\bibitem[\protect\citeauthoryear{{Ofman} et~al.}{1998}]{1998SoPh..183...97O}
\begin{barticle}
\bauthor{\binits{L.} \bsnm{{Ofman}}}, \bauthor{\binits{T.A.} \bsnm{{Kucera}}},
  \bauthor{\binits{Z.} \bsnm{{Mouradian}}}, \bauthor{\binits{A.I.}
  \bsnm{{Poland}}},
\batitle{{SUMER Observations of the Evolution and the Disappearance of a Solar
  Prominence}}.
\bjtitle{\solphys}
\bvolume{183},
\bfpage{97}--\blpage{106}
(\byear{1998})
\end{barticle}
\endbibitem

\bibitem[\protect\citeauthoryear{{Okamoto} et~al.}{2007}]{2007Sci...318.1577O}
\begin{barticle}
\bauthor{\binits{T.J.} \bsnm{{Okamoto}}}, \bauthor{\binits{S.}
  \bsnm{{Tsuneta}}}, \bauthor{\binits{T.E.} \bsnm{{Berger}}},
  \bauthor{\binits{K.} \bsnm{{Ichimoto}}}, \bauthor{\binits{Y.}
  \bsnm{{Katsukawa}}}, \bauthor{\binits{B.W.} \bsnm{{Lites}}},
  \bauthor{\binits{S.} \bsnm{{Nagata}}}, \bauthor{\binits{K.}
  \bsnm{{Shibata}}}, \bauthor{\binits{T.} \bsnm{{Shimizu}}},
  \bauthor{\binits{R.A.} \bsnm{{Shine}}}, \bauthor{\binits{Y.}
  \bsnm{{Suematsu}}}, \bauthor{\binits{T.D.} \bsnm{{Tarbell}}},
  \bauthor{\binits{A.M.} \bsnm{{Title}}},
\batitle{{Coronal Transverse Magnetohydrodynamic Waves in a Solar Prominence}}.
\bjtitle{Science}
\bvolume{318},
\bfpage{1577}
(\byear{2007}).
doi:\doiurl{10.1126/science.1145447}
\end{barticle}
\endbibitem

\bibitem[\protect\citeauthoryear{{Oliver}}{2009}]{2009SSRv..tmp...39O}
\begin{botherref}
\oauthor{\binits{R.} \bsnm{{Oliver}}},
{Prominence Seismology Using Small Amplitude Oscillations}.
\textrm{Space Science Reviews},
39
(2009).
doi:\doiurl{10.1007/s11214-009-9527-4}
\end{botherref}
\endbibitem

\bibitem[\protect\citeauthoryear{{Olson} et~al.}{1986}]{1986JQSRT..35..431O}
\begin{barticle}
\bauthor{\binits{G.L.} \bsnm{{Olson}}}, \bauthor{\binits{L.H.} \bsnm{{Auer}}},
  \bauthor{\binits{J.R.} \bsnm{{Buchler}}},
\batitle{{A rapidly convergent iterative solution of the non-LTE radiation
  transfer problem.}}
\bjtitle{Journal of Quantitative Spectroscopy and Radiative Transfer}
\bvolume{35},
\bfpage{431}--\blpage{442}
(\byear{1986}).
doi:\doiurl{10.1016/0022-4073(86)90030-0}
\end{barticle}
\endbibitem

\bibitem[\protect\citeauthoryear{{Omont} et~al.}{1972}]{1972ApJ...175..185O}
\begin{barticle}
\bauthor{\binits{A.} \bsnm{{Omont}}}, \bauthor{\binits{E.W.} \bsnm{{Smith}}},
  \bauthor{\binits{J.} \bsnm{{Cooper}}},
\batitle{{Redistribution of Resonance Radiation. I. The Effect of Collisions}}.
\bjtitle{\apj}
\bvolume{175},
\bfpage{185}
(\byear{1972}).
doi:\doiurl{10.1086/151548}
\end{barticle}
\endbibitem

\bibitem[\protect\citeauthoryear{{Orrall} and
  {Schmahl}}{1976}]{1976SoPh...50..365O}
\begin{barticle}
\bauthor{\binits{F.Q.} \bsnm{{Orrall}}}, \bauthor{\binits{E.J.}
  \bsnm{{Schmahl}}},
\batitle{{The prominence-corona interface compared with the chromosphere-corona
  transition region}}.
\bjtitle{\solphys}
\bvolume{50},
\bfpage{365}--\blpage{381}
(\byear{1976}).
doi:\doiurl{10.1007/BF00155299}
\end{barticle}
\endbibitem

\bibitem[\protect\citeauthoryear{{Orrall} and
  {Schmahl}}{1980}]{1980ApJ...240..908O}
\begin{barticle}
\bauthor{\binits{F.Q.} \bsnm{{Orrall}}}, \bauthor{\binits{E.J.}
  \bsnm{{Schmahl}}},
\batitle{{The H I Lyman continuum in solar prominences and its interpretation
  in the presence of inhomogeneities}}.
\bjtitle{\apj}
\bvolume{240},
\bfpage{908}--\blpage{922}
(\byear{1980}).
doi:\doiurl{10.1086/158304}
\end{barticle}
\endbibitem

\bibitem[\protect\citeauthoryear{{Paletou}}{1995}]{1995A&A...302..587P}
\begin{barticle}
\bauthor{\binits{F.} \bsnm{{Paletou}}},
\batitle{{Two-dimensional multilevel radiative transfer with standard partial
  frequency redistribution in isolated solar atmospheric structures.}}
\bjtitle{\aap}
\bvolume{302},
\bfpage{587}
(\byear{1995})
\end{barticle}
\endbibitem

\bibitem[\protect\citeauthoryear{{Paletou} et~al.}{1993}]{1993A&A...274..571P}
\begin{barticle}
\bauthor{\binits{F.} \bsnm{{Paletou}}}, \bauthor{\binits{J.C.} \bsnm{{Vial}}},
  \bauthor{\binits{L.H.} \bsnm{{Auer}}},
\batitle{{Two-dimensional radiative transfer with partial frequency
  redistribution. II. Application to resonance lines in quiescent
  prominences}}.
\bjtitle{\aap}
\bvolume{274},
\bfpage{571}
(\byear{1993})
\end{barticle}
\endbibitem

\bibitem[\protect\citeauthoryear{{Parenti} and
  {Vial}}{2007}]{2007A&A...469.1109P}
\begin{barticle}
\bauthor{\binits{S.} \bsnm{{Parenti}}}, \bauthor{\binits{J.C.} \bsnm{{Vial}}},
\batitle{{Prominence and quiet-Sun plasma parameters derived from FUV spectral
  emission}}.
\bjtitle{\aap}
\bvolume{469},
\bfpage{1109}--\blpage{1115}
(\byear{2007}).
doi:\doiurl{10.1051/0004-6361:20077196}
\end{barticle}
\endbibitem

\bibitem[\protect\citeauthoryear{{Parenti} et~al.}{2005a}]{2005A&A...443..685P}
\begin{barticle}
\bauthor{\binits{S.} \bsnm{{Parenti}}}, \bauthor{\binits{P.} \bsnm{{Lemaire}}},
  \bauthor{\binits{J.C.} \bsnm{{Vial}}},
\batitle{{Solar hydrogen-Lyman continuum observations with SOHO/SUMER}}.
\bjtitle{\aap}
\bvolume{443},
\bfpage{685}--\blpage{689}
(\byear{2005a}).
doi:\doiurl{10.1051/0004-6361:20053431}
\end{barticle}
\endbibitem

\bibitem[\protect\citeauthoryear{{Parenti} et~al.}{2004}]{2004SoPh..220...61P}
\begin{barticle}
\bauthor{\binits{S.} \bsnm{{Parenti}}}, \bauthor{\binits{J.C.} \bsnm{{Vial}}},
  \bauthor{\binits{P.} \bsnm{{Lemaire}}},
\batitle{{Prominence atlas in the SUMER range 800 1250 {\AA}: I. Observations,
  data reduction and preliminary results}}.
\bjtitle{\solphys}
\bvolume{220},
\bfpage{61}--\blpage{80}
(\byear{2004}).
doi:\doiurl{10.1023/B:sola.0000023444.58697.e7}
\end{barticle}
\endbibitem

\bibitem[\protect\citeauthoryear{{Parenti} et~al.}{2005b}]{2005A&A...443..679P}
\begin{barticle}
\bauthor{\binits{S.} \bsnm{{Parenti}}}, \bauthor{\binits{J.C.} \bsnm{{Vial}}},
  \bauthor{\binits{P.} \bsnm{{Lemaire}}},
\batitle{{Prominence atlas in the SUMER range 800-1250 {\AA}. II. Line profile
  properties and ions identifications}}.
\bjtitle{\aap}
\bvolume{443},
\bfpage{679}--\blpage{684}
(\byear{2005b}).
doi:\doiurl{10.1051/0004-6361:20053122}
\end{barticle}
\endbibitem

\bibitem[\protect\citeauthoryear{{Patsourakos} and
  {Vial}}{2002}]{2002SoPh..208..253P}
\begin{barticle}
\bauthor{\binits{S.} \bsnm{{Patsourakos}}}, \bauthor{\binits{J.C.}
  \bsnm{{Vial}}},
\batitle{{Soho Contribution to Prominence Science}}.
\bjtitle{\solphys}
\bvolume{208},
\bfpage{253}--\blpage{281}
(\byear{2002})
\end{barticle}
\endbibitem

\bibitem[\protect\citeauthoryear{{Patsourakos}
  et~al.}{2007}]{2007ApJ...664.1214P}
\begin{barticle}
\bauthor{\binits{S.} \bsnm{{Patsourakos}}}, \bauthor{\binits{P.}
  \bsnm{{Gouttebroze}}}, \bauthor{\binits{A.} \bsnm{{Vourlidas}}},
\batitle{{The Quiet Sun Network at Subarcsecond Resolution: VAULT Observations
  and Radiative Transfer Modeling of Cool Loops}}.
\bjtitle{\apj}
\bvolume{664},
\bfpage{1214}--\blpage{1220}
(\byear{2007}).
doi:\doiurl{10.1086/518645}
\end{barticle}
\endbibitem

\bibitem[\protect\citeauthoryear{{Penn}}{2000}]{2000SoPh..197..313P}
\begin{barticle}
\bauthor{\binits{M.J.} \bsnm{{Penn}}},
\batitle{{An Erupting Active Region Filament: Three-Dimensional Trajectory and
  Hydrogen Column Density}}.
\bjtitle{\solphys}
\bvolume{197},
\bfpage{313}--\blpage{335}
(\byear{2000})
\end{barticle}
\endbibitem

\bibitem[\protect\citeauthoryear{{Phillips} et~al.}{2008}]{2008Phillips}
\begin{bbook}
\bauthor{\binits{K.J.H.} \bsnm{{Phillips}}}, \bauthor{\binits{U.}
  \bsnm{{Feldman}}}, \bauthor{\binits{E.} \bsnm{{Landi}}},
\bbtitle{{Ultraviolet and X-ray Spectroscopy of the Solar Atmosphere}}
(\bpublisher{Cambridge University Press},
\blocation{Cambridge}, \byear{2008})
\end{bbook}
\endbibitem

\bibitem[\protect\citeauthoryear{{Pojoga}}{1994}]{1994scs..conf..357P}
\begin{botherref}
\oauthor{\binits{S.} \bsnm{{Pojoga}}},
{Emission Measure of Prominence-Corona Transition Region},
in \textit{IAU Colloq. 144: Solar Coronal Structures},
1994,
p. 357
\end{botherref}
\endbibitem

\bibitem[\protect\citeauthoryear{{Pojoga} and
  {Molowny-Horas}}{1999}]{1999SoPh..185..113P}
\begin{barticle}
\bauthor{\binits{S.} \bsnm{{Pojoga}}}, \bauthor{\binits{R.}
  \bsnm{{Molowny-Horas}}},
\batitle{{The Transverse Velocity Field Of an EUV SolarProminence}}.
\bjtitle{\solphys}
\bvolume{185},
\bfpage{113}--\blpage{125}
(\byear{1999})
\end{barticle}
\endbibitem

\bibitem[\protect\citeauthoryear{{Pojoga} et~al.}{1998}]{1998A&A...332..325P}
\begin{barticle}
\bauthor{\binits{S.} \bsnm{{Pojoga}}}, \bauthor{\binits{A.G.}
  \bsnm{{Nikoghossian}}}, \bauthor{\binits{Z.} \bsnm{{Mouradian}}},
\batitle{{A statistical approach to the investigation of fine structure of
  solar prominences}}.
\bjtitle{\aap}
\bvolume{332},
\bfpage{325}--\blpage{338}
(\byear{1998})
\end{barticle}
\endbibitem

\bibitem[\protect\citeauthoryear{{Poland} and
  {Anzer}}{1971}]{1971SoPh...19..401P}
\begin{barticle}
\bauthor{\binits{A.} \bsnm{{Poland}}}, \bauthor{\binits{U.} \bsnm{{Anzer}}},
\batitle{{Energy Balance in Cool Quiescent Prominences}}.
\bjtitle{\solphys}
\bvolume{19},
\bfpage{401}--\blpage{413}
(\byear{1971}).
doi:\doiurl{10.1007/BF00146067}
\end{barticle}
\endbibitem

\bibitem[\protect\citeauthoryear{{Poland} and
  {Mariska}}{1988}]{1988dssp.conf..133P}
\begin{botherref}
\oauthor{\binits{A.I.} \bsnm{{Poland}}}, \oauthor{\binits{J.T.}
  \bsnm{{Mariska}}},
{A model for the structure and formation of prominences.},
in \textit{Universitat de les Illes Balears, Palma de Mallorca (Spain)},
1988,
p. 133
\end{botherref}
\endbibitem

\bibitem[\protect\citeauthoryear{{Poland} and
  {Tandberg-Hanssen}}{1983}]{1983SoPh...84...63P}
\begin{barticle}
\bauthor{\binits{A.I.} \bsnm{{Poland}}}, \bauthor{\binits{E.}
  \bsnm{{Tandberg-Hanssen}}},
\batitle{{Physical conditions in a quiescent prominence derived from UV spectra
  obtained with the UVSP instrument on the SMM}}.
\bjtitle{\solphys}
\bvolume{84},
\bfpage{63}--\blpage{70}
(\byear{1983}).
doi:\doiurl{10.1007/BF00157443}
\end{barticle}
\endbibitem

\bibitem[\protect\citeauthoryear{{Pottasch}}{1963}]{1963ApJ...137..945P}
\begin{barticle}
\bauthor{\binits{S.R.} \bsnm{{Pottasch}}},
\batitle{{The Lower Solar Corona: Interpretation of the Ultraviolet Spectrum.}}
\bjtitle{\apj}
\bvolume{137},
\bfpage{945}
(\byear{1963}).
doi:\doiurl{10.1086/147569}
\end{barticle}
\endbibitem

\bibitem[\protect\citeauthoryear{{Pottasch}}{1964}]{1964SSRv....3..816P}
\begin{barticle}
\bauthor{\binits{S.R.} \bsnm{{Pottasch}}},
\batitle{{On the Interpretation of the Solar Ultraviolet Emission Line
  Spectrum}}.
\bjtitle{Space Science Reviews}
\bvolume{3},
\bfpage{816}--\blpage{855}
(\byear{1964}).
doi:\doiurl{10.1007/BF00177958}
\end{barticle}
\endbibitem

\bibitem[\protect\citeauthoryear{{Purcell} and
  {Tousey}}{1960}]{1960JGR....65..370P}
\begin{barticle}
\bauthor{\binits{J.D.} \bsnm{{Purcell}}}, \bauthor{\binits{R.}
  \bsnm{{Tousey}}},
\batitle{{The Profile of Solar Hydrogen-Lyman-{$\alpha$}}}.
\bjtitle{\jgr}
\bvolume{65},
\bfpage{370}
(\byear{1960}).
doi:\doiurl{10.1029/JZ065i001p00370}
\end{barticle}
\endbibitem

\bibitem[\protect\citeauthoryear{{Rudawy} and
  {Heinzel}}{1992}]{1992SoPh..138..123R}
\begin{barticle}
\bauthor{\binits{P.} \bsnm{{Rudawy}}}, \bauthor{\binits{P.} \bsnm{{Heinzel}}},
\batitle{{Hydrogen photoionization rates for chromospheric and prominence
  plasmas}}.
\bjtitle{\solphys}
\bvolume{138},
\bfpage{123}--\blpage{131}
(\byear{1992}).
doi:\doiurl{10.1007/BF00146200}
\end{barticle}
\endbibitem

\bibitem[\protect\citeauthoryear{{Rybicki} and
  {Hummer}}{1991}]{1991A&A...245..171R}
\begin{barticle}
\bauthor{\binits{G.B.} \bsnm{{Rybicki}}}, \bauthor{\binits{D.G.}
  \bsnm{{Hummer}}},
\batitle{{An accelerated lambda iteration method for multilevel radiative
  transfer. I - Non-overlapping lines with background continuum}}.
\bjtitle{\aap}
\bvolume{245},
\bfpage{171}--\blpage{181}
(\byear{1991})
\end{barticle}
\endbibitem

\bibitem[\protect\citeauthoryear{{Rybicki} and
  {Hummer}}{1992}]{1992A&A...262..209R}
\begin{barticle}
\bauthor{\binits{G.B.} \bsnm{{Rybicki}}}, \bauthor{\binits{D.G.}
  \bsnm{{Hummer}}},
\batitle{{An accelerated lambda iteration method for multilevel radiative
  transfer. II - Overlapping transitions with full continuum}}.
\bjtitle{\aap}
\bvolume{262},
\bfpage{209}--\blpage{215}
(\byear{1992})
\end{barticle}
\endbibitem

\bibitem[\protect\citeauthoryear{{Schmahl} and
  {Orrall}}{1986}]{1986NASCP2442..127S}
\begin{barticle}
\bauthor{\binits{E.J.} \bsnm{{Schmahl}}}, \bauthor{\binits{F.Q.}
  \bsnm{{Orrall}}},
\batitle{{Interpretation of the prominence differential emissions measure for 3
  geometries}}.
\bjtitle{NASA Conference Publication}
\bvolume{2442},
\bfpage{127}--\blpage{133}
(\byear{1986})
\end{barticle}
\endbibitem

\bibitem[\protect\citeauthoryear{{Schmieder}}{1988}]{1988dsqs.work...15S}
\begin{botherref}
\oauthor{\binits{B.} \bsnm{{Schmieder}}},
{Overall properties and steady flows.},
in \textit{Dynamics and structure of quiescent solar prominences, p. 15 - 46},
1988,
pp. 15--46
\end{botherref}
\endbibitem

\bibitem[\protect\citeauthoryear{{Schmieder}
  et~al.}{1991}]{1991A&A...252..353S}
\begin{barticle}
\bauthor{\binits{B.} \bsnm{{Schmieder}}}, \bauthor{\binits{M.A.}
  \bsnm{{Raadu}}}, \bauthor{\binits{J.E.} \bsnm{{Wiik}}},
\batitle{{Fine structure of solar filaments. II - Dynamics of threads and
  footpoints}}.
\bjtitle{\aap}
\bvolume{252},
\bfpage{353}--\blpage{365}
(\byear{1991})
\end{barticle}
\endbibitem

\bibitem[\protect\citeauthoryear{{Schmieder}
  et~al.}{2003}]{2003A&A...401..361S}
\begin{barticle}
\bauthor{\binits{B.} \bsnm{{Schmieder}}}, \bauthor{\binits{K.}
  \bsnm{{Tziotziou}}}, \bauthor{\binits{P.} \bsnm{{Heinzel}}},
\batitle{{Spectroscopic diagnostics of an H{$\alpha$} and EUV filament observed
  with THEMIS and SOHO}}.
\bjtitle{\aap}
\bvolume{401},
\bfpage{361}--\blpage{375}
(\byear{2003}).
doi:\doiurl{10.1051/0004-6361:20030126}
\end{barticle}
\endbibitem

\bibitem[\protect\citeauthoryear{{Schmieder}
  et~al.}{1998}]{1998SoPh..181..309S}
\begin{barticle}
\bauthor{\binits{B.} \bsnm{{Schmieder}}}, \bauthor{\binits{P.}
  \bsnm{{Heinzel}}}, \bauthor{\binits{T.} \bsnm{{Kucera}}},
  \bauthor{\binits{J.C.} \bsnm{{Vial}}},
\batitle{{Filament Observations with SOHO Sumer/cds: The Behaviour of Hydrogen
  Lyman Lines}}.
\bjtitle{\solphys}
\bvolume{181},
\bfpage{309}--\blpage{326}
(\byear{1998})
\end{barticle}
\endbibitem

\bibitem[\protect\citeauthoryear{{Schmieder}
  et~al.}{1999}]{1999SoPh..189..109S}
\begin{barticle}
\bauthor{\binits{B.} \bsnm{{Schmieder}}}, \bauthor{\binits{P.}
  \bsnm{{Heinzel}}}, \bauthor{\binits{J.C.} \bsnm{{Vial}}},
  \bauthor{\binits{P.} \bsnm{{Rudawy}}},
\batitle{{SOHO/SUMER observations and analysis of hydrogen Lyman lines in a
  quiescent prominence}}.
\bjtitle{\solphys}
\bvolume{189},
\bfpage{109}--\blpage{127}
(\byear{1999})
\end{barticle}
\endbibitem

\bibitem[\protect\citeauthoryear{{Schmieder}
  et~al.}{2000}]{2000A&A...358..728S}
\begin{barticle}
\bauthor{\binits{B.} \bsnm{{Schmieder}}}, \bauthor{\binits{C.}
  \bsnm{{Delann{\'e}e}}}, \bauthor{\binits{D.Y.} \bsnm{{Yong}}},
  \bauthor{\binits{J.C.} \bsnm{{Vial}}}, \bauthor{\binits{M.}
  \bsnm{{Madjarska}}},
\batitle{{Multi-wavelength study of the slow ``disparition brusque'' of a
  filament observed with SOHO}}.
\bjtitle{\aap}
\bvolume{358},
\bfpage{728}--\blpage{740}
(\byear{2000})
\end{barticle}
\endbibitem

\bibitem[\protect\citeauthoryear{{Schmieder}
  et~al.}{2004a}]{2004SoPh..221..297S}
\begin{barticle}
\bauthor{\binits{B.} \bsnm{{Schmieder}}}, \bauthor{\binits{Y.} \bsnm{{Lin}}},
  \bauthor{\binits{P.} \bsnm{{Heinzel}}}, \bauthor{\binits{P.}
  \bsnm{{Schwartz}}},
\batitle{{Multi-wavelength study of a high-latitude EUV filament}}.
\bjtitle{\solphys}
\bvolume{221},
\bfpage{297}--\blpage{323}
(\byear{2004a}).
doi:\doiurl{10.1023/B:SOLA.0000035059.50427.68}
\end{barticle}
\endbibitem

\bibitem[\protect\citeauthoryear{{Schmieder}
  et~al.}{2004b}]{2004SoPh..223..119S}
\begin{barticle}
\bauthor{\binits{B.} \bsnm{{Schmieder}}}, \bauthor{\binits{N.} \bsnm{{Mein}}},
  \bauthor{\binits{Y.} \bsnm{{Deng}}}, \bauthor{\binits{C.}
  \bsnm{{Dumitrache}}}, \bauthor{\binits{J.M.} \bsnm{{Malherbe}}},
  \bauthor{\binits{J.} \bsnm{{Staiger}}}, \bauthor{\binits{E.E.}
  \bsnm{{Deluca}}},
\batitle{{Magnetic changes observed in the formation of two filaments in a
  complex active region: TRACE and MSDP observations}}.
\bjtitle{\solphys}
\bvolume{223},
\bfpage{119}--\blpage{141}
(\byear{2004b}).
doi:\doiurl{10.1007/s11207-004-1107-x}
\end{barticle}
\endbibitem

\bibitem[\protect\citeauthoryear{{Schmieder}
  et~al.}{2007}]{2007SoPh..241...53S}
\begin{barticle}
\bauthor{\binits{B.} \bsnm{{Schmieder}}}, \bauthor{\binits{S.}
  \bsnm{{Gun{\'a}r}}}, \bauthor{\binits{P.} \bsnm{{Heinzel}}},
  \bauthor{\binits{U.} \bsnm{{Anzer}}},
\batitle{{Spectral Diagnostics of the Magnetic Field Orientation in a
  Prominence Observed with SOHO/SUMER}}.
\bjtitle{\solphys}
\bvolume{241},
\bfpage{53}--\blpage{66}
(\byear{2007}).
doi:\doiurl{10.1007/s11207-007-0251-5}
\end{barticle}
\endbibitem

\bibitem[\protect\citeauthoryear{{Schmieder}
  et~al.}{2008}]{2008SoPh..247..321S}
\begin{barticle}
\bauthor{\binits{B.} \bsnm{{Schmieder}}}, \bauthor{\binits{V.}
  \bsnm{{Bommier}}}, \bauthor{\binits{R.} \bsnm{{Kitai}}}, \bauthor{\binits{T.}
  \bsnm{{Matsumoto}}}, \bauthor{\binits{T.T.} \bsnm{{Ishii}}},
  \bauthor{\binits{M.} \bsnm{{Hagino}}}, \bauthor{\binits{H.} \bsnm{{Li}}},
  \bauthor{\binits{L.} \bsnm{{Golub}}},
\batitle{{Magnetic Causes of the Eruption of a Quiescent Filament}}.
\bjtitle{\solphys}
\bvolume{247},
\bfpage{321}--\blpage{333}
(\byear{2008}).
doi:\doiurl{10.1007/s11207-007-9100-9}
\end{barticle}
\endbibitem

\bibitem[\protect\citeauthoryear{{Schmieder}
  et~al.}{2010}]{2009arXiv0911.5091S}
\begin{botherref}
\oauthor{\binits{B.} \bsnm{{Schmieder}}}, \oauthor{\binits{R.}
  \bsnm{{Chandra}}}, \oauthor{\binits{A.} \bsnm{{Berlicki}}},
  \oauthor{\binits{P.} \bsnm{{Mein}}},
{Velocity vectors of a quiescent prominence observed by Hinode/SOT and the MSDP
  (Meudon)}.
\textrm{ArXiv e-prints}
(2010)
\end{botherref}
\endbibitem

\bibitem[\protect\citeauthoryear{{Schwartz} et~al.}{2004}]{2004A&A...421..323S}
\begin{barticle}
\bauthor{\binits{P.} \bsnm{{Schwartz}}}, \bauthor{\binits{P.}
  \bsnm{{Heinzel}}}, \bauthor{\binits{U.} \bsnm{{Anzer}}}, \bauthor{\binits{B.}
  \bsnm{{Schmieder}}},
\batitle{{Determination of the 3D structure of an EUV-filament observed by
  SoHO/CDS, SoHO/SUMER and VTT/MSDP}}.
\bjtitle{\aap}
\bvolume{421},
\bfpage{323}--\blpage{338}
(\byear{2004}).
doi:\doiurl{10.1051/0004-6361:20034199}
\end{barticle}
\endbibitem

\bibitem[\protect\citeauthoryear{{Schwartz} et~al.}{2006}]{2006A&A...459..651S}
\begin{barticle}
\bauthor{\binits{P.} \bsnm{{Schwartz}}}, \bauthor{\binits{P.}
  \bsnm{{Heinzel}}}, \bauthor{\binits{B.} \bsnm{{Schmieder}}},
  \bauthor{\binits{U.} \bsnm{{Anzer}}},
\batitle{{Study of an extended EUV filament using SoHO/SUMER observations of
  the hydrogen Lyman lines}}.
\bjtitle{\aap}
\bvolume{459},
\bfpage{651}--\blpage{661}
(\byear{2006}).
doi:\doiurl{10.1051/0004-6361:20065619}
\end{barticle}
\endbibitem

\bibitem[\protect\citeauthoryear{{Sheeley}}{1995}]{1995ApJ...440..884S}
\begin{barticle}
\bauthor{\binits{N.R.} \bsnm{{Sheeley}} \bsuffix{Jr.}},
\batitle{{A Volcanic Origin for High-FIP Material in the Solar Atmosphere}}.
\bjtitle{\apj}
\bvolume{440},
\bfpage{884}
(\byear{1995}).
doi:\doiurl{10.1086/175326}
\end{barticle}
\endbibitem

\bibitem[\protect\citeauthoryear{{Smith} et~al.}{2001}]{2001ApJ...556L..91S}
\begin{barticle}
\bauthor{\binits{R.K.} \bsnm{{Smith}}}, \bauthor{\binits{N.S.}
  \bsnm{{Brickhouse}}}, \bauthor{\binits{D.A.} \bsnm{{Liedahl}}},
  \bauthor{\binits{J.C.} \bsnm{{Raymond}}},
\batitle{{Collisional Plasma Models with APEC/APED: Emission-Line Diagnostics
  of Hydrogen-like and Helium-like Ions}}.
\bjtitle{\apjl}
\bvolume{556},
\bfpage{91}--\blpage{95}
(\byear{2001}).
doi:\doiurl{10.1086/322992}
\end{barticle}
\endbibitem

\bibitem[\protect\citeauthoryear{{Spicer} et~al.}{1998}]{1998ApJ...494..450S}
\begin{barticle}
\bauthor{\binits{D.S.} \bsnm{{Spicer}}}, \bauthor{\binits{U.}
  \bsnm{{Feldman}}}, \bauthor{\binits{K.G.} \bsnm{{Widing}}},
  \bauthor{\binits{M.} \bsnm{{Rilee}}},
\batitle{{The Neon-to-Magnesium Abundance Ratio as a Tracer of the Source
  Region of Prominence Material}}.
\bjtitle{\apj}
\bvolume{494},
\bfpage{450}
(\byear{1998}).
doi:\doiurl{10.1086/305203}
\end{barticle}
\endbibitem

\bibitem[\protect\citeauthoryear{{Stellmacher} and
  {Wiehr}}{1997}]{1997A&A...319..669S}
\begin{barticle}
\bauthor{\binits{G.} \bsnm{{Stellmacher}}}, \bauthor{\binits{E.}
  \bsnm{{Wiehr}}},
\batitle{{The helium singlet-to-triplet line ratio in solar prominences.}}
\bjtitle{\aap}
\bvolume{319},
\bfpage{669}--\blpage{672}
(\byear{1997})
\end{barticle}
\endbibitem

\bibitem[\protect\citeauthoryear{{Stellmacher} and
  {Wiehr}}{2000}]{2000SoPh..196..357S}
\begin{barticle}
\bauthor{\binits{G.} \bsnm{{Stellmacher}}}, \bauthor{\binits{E.}
  \bsnm{{Wiehr}}},
\batitle{{Two-dimensional photometric analysis of emission lines in quiescent
  prominences}}.
\bjtitle{\solphys}
\bvolume{196},
\bfpage{357}--\blpage{367}
(\byear{2000})
\end{barticle}
\endbibitem

\bibitem[\protect\citeauthoryear{{Stellmacher}
  et~al.}{1986}]{1986A&A...162..307S}
\begin{barticle}
\bauthor{\binits{G.} \bsnm{{Stellmacher}}}, \bauthor{\binits{S.}
  \bsnm{{Koutchmy}}}, \bauthor{\binits{C.} \bsnm{{Lebecq}}},
\batitle{{The 1981 total solar eclipse. III - Photometric study of the
  prominence remnant in the reversing south polar field}}.
\bjtitle{\aap}
\bvolume{162},
\bfpage{307}--\blpage{311}
(\byear{1986})
\end{barticle}
\endbibitem

\bibitem[\protect\citeauthoryear{{Stellmacher}
  et~al.}{2003}]{2003SoPh..217..133S}
\begin{barticle}
\bauthor{\binits{G.} \bsnm{{Stellmacher}}}, \bauthor{\binits{E.}
  \bsnm{{Wiehr}}}, \bauthor{\binits{I.E.} \bsnm{{Dammasch}}},
\batitle{{Spectroscopy of Solar Prominences Simultaneously From Space and
  Ground}}.
\bjtitle{\solphys}
\bvolume{217},
\bfpage{133}--\blpage{155}
(\byear{2003})
\end{barticle}
\endbibitem

\bibitem[\protect\citeauthoryear{{Tandberg-Hanssen}}{1995}]{1995nsp..book.....%
T}
\begin{bbook}
\bauthor{\binits{E.} \bsnm{{Tandberg-Hanssen}}},
\bbtitle{{The nature of solar prominences}}
(\bpublisher{Dordrecht ; Boston : Kluwer, c1995.}, \blocation{???},
  \byear{1995})
\end{bbook}
\endbibitem

\bibitem[\protect\citeauthoryear{{Tousey}}{1977}]{1977ApOpt..16..825T}
\begin{barticle}
\bauthor{\binits{R.} \bsnm{{Tousey}}},
\batitle{{Apollo Telescope Mount of SKYLAB - an overview}}.
\bjtitle{\ao}
\bvolume{16},
\bfpage{825}--\blpage{836}
(\byear{1977})
\end{barticle}
\endbibitem

\bibitem[\protect\citeauthoryear{{Tousey} et~al.}{1965}]{1965AnAp...28..755T}
\begin{barticle}
\bauthor{\binits{R.} \bsnm{{Tousey}}}, \bauthor{\binits{W.E.} \bsnm{{Austin}}},
  \bauthor{\binits{J.D.} \bsnm{{Purcell}}}, \bauthor{\binits{K.G.}
  \bsnm{{Widing}}},
\batitle{{The extreme ultraviolet emission from the Sun between the Lyman-alpha
  lines of H I and C VI}}.
\bjtitle{Annales d'Astrophysique}
\bvolume{28},
\bfpage{755}
(\byear{1965})
\end{barticle}
\endbibitem

\bibitem[\protect\citeauthoryear{{Tsuneta} et~al.}{2008}]{2008SoPh..249..167T}
\begin{barticle}
\bauthor{\binits{S.} \bsnm{{Tsuneta}}}, \bauthor{\binits{K.}
  \bsnm{{Ichimoto}}}, \bauthor{\binits{Y.} \bsnm{{Katsukawa}}},
  \bauthor{\binits{S.} \bsnm{{Nagata}}}, \bauthor{\binits{M.} \bsnm{{Otsubo}}},
  \bauthor{\binits{T.} \bsnm{{Shimizu}}}, \bauthor{\binits{Y.}
  \bsnm{{Suematsu}}}, \bauthor{\binits{M.} \bsnm{{Nakagiri}}},
  \bauthor{\binits{M.} \bsnm{{Noguchi}}}, \bauthor{\binits{T.}
  \bsnm{{Tarbell}}}, \bauthor{\binits{A.} \bsnm{{Title}}}, \bauthor{\binits{R.}
  \bsnm{{Shine}}}, \bauthor{\binits{W.} \bsnm{{Rosenberg}}},
  \bauthor{\binits{C.} \bsnm{{Hoffmann}}}, \bauthor{\binits{B.}
  \bsnm{{Jurcevich}}}, \bauthor{\binits{G.} \bsnm{{Kushner}}},
  \bauthor{\binits{M.} \bsnm{{Levay}}}, \bauthor{\binits{B.} \bsnm{{Lites}}},
  \bauthor{\binits{D.} \bsnm{{Elmore}}}, \bauthor{\binits{T.}
  \bsnm{{Matsushita}}}, \bauthor{\binits{N.} \bsnm{{Kawaguchi}}},
  \bauthor{\binits{H.} \bsnm{{Saito}}}, \bauthor{\binits{I.} \bsnm{{Mikami}}},
  \bauthor{\binits{L.D.} \bsnm{{Hill}}}, \bauthor{\binits{J.K.}
  \bsnm{{Owens}}},
\batitle{{The Solar Optical Telescope for the Hinode Mission: An Overview}}.
\bjtitle{\solphys}
\bvolume{249},
\bfpage{167}--\blpage{196}
(\byear{2008}).
doi:\doiurl{10.1007/s11207-008-9174-z}
\end{barticle}
\endbibitem

\bibitem[\protect\citeauthoryear{{Tziotziou}}{2007}]{2007ASPC..368..217T}
\begin{botherref}
\oauthor{\binits{K.} \bsnm{{Tziotziou}}},
{Chromospheric Cloud-Model Inversion Techniques},
in \textit{The Physics of Chromospheric Plasmas},
ed. by P. {Heinzel}, I. {Dorotovi{\v c}}, R.J. {Rutten}.
Astronomical Society of the Pacific Conference Series,
vol. 368,
2007,
p. 217
\end{botherref}
\endbibitem

\bibitem[\protect\citeauthoryear{{Tziotziou}
  et~al.}{2001}]{2001A&A...366..686T}
\begin{barticle}
\bauthor{\binits{K.} \bsnm{{Tziotziou}}}, \bauthor{\binits{P.}
  \bsnm{{Heinzel}}}, \bauthor{\binits{P.} \bsnm{{Mein}}}, \bauthor{\binits{N.}
  \bsnm{{Mein}}},
\batitle{{Non-LTE inversion of chromospheric $\{$$\backslash$Ca Ii$\}$
  cloud-like features}}.
\bjtitle{\aap}
\bvolume{366},
\bfpage{686}--\blpage{698}
(\byear{2001}).
doi:\doiurl{10.1051/0004-6361:20000257}
\end{barticle}
\endbibitem

\bibitem[\protect\citeauthoryear{{van
  Ballegooijen}}{2004}]{2004ApJ...612..519V}
\begin{barticle}
\bauthor{\binits{A.A.} \bsnm{{van Ballegooijen}}},
\batitle{{Observations and Modeling of a Filament on the Sun}}.
\bjtitle{\apj}
\bvolume{612},
\bfpage{519}--\blpage{529}
(\byear{2004}).
doi:\doiurl{10.1086/422512}
\end{barticle}
\endbibitem

\bibitem[\protect\citeauthoryear{{Vernazza} and
  {Noyes}}{1972}]{1972SoPh...22..358V}
\begin{barticle}
\bauthor{\binits{J.E.} \bsnm{{Vernazza}}}, \bauthor{\binits{R.W.}
  \bsnm{{Noyes}}},
\batitle{{Inhomogeneous Structure of the Solar Chromosphere from
  Lyman-Continuum Data}}.
\bjtitle{\solphys}
\bvolume{22},
\bfpage{358}--\blpage{374}
(\byear{1972}).
doi:\doiurl{10.1007/BF00148702}
\end{barticle}
\endbibitem

\bibitem[\protect\citeauthoryear{{Vernazza} et~al.}{1981}]{1981ApJS...45..635V}
\begin{barticle}
\bauthor{\binits{J.E.} \bsnm{{Vernazza}}}, \bauthor{\binits{E.H.}
  \bsnm{{Avrett}}}, \bauthor{\binits{R.} \bsnm{{Loeser}}},
\batitle{{Structure of the solar chromosphere. III - Models of the EUV
  brightness components of the quiet-sun}}.
\bjtitle{\apjs}
\bvolume{45},
\bfpage{635}--\blpage{725}
(\byear{1981}).
doi:\doiurl{10.1086/190731}
\end{barticle}
\endbibitem

\bibitem[\protect\citeauthoryear{{Vial}}{1982a}]{1982ApJ...253..330V}
\begin{barticle}
\bauthor{\binits{J.C.} \bsnm{{Vial}}},
\batitle{{Optically thick lines in a quiescent prominence - Profiles of
  Lyman-alpha, Lyman-beta /H I/, K and H /Mg II/, and K and H /Ca II/ lines
  with the OSO 8 LPSP instrument}}.
\bjtitle{\apj}
\bvolume{253},
\bfpage{330}--\blpage{352}
(\byear{1982a}).
doi:\doiurl{10.1086/159639}
\end{barticle}
\endbibitem

\bibitem[\protect\citeauthoryear{{Vial}}{1982b}]{1982ApJ...254..780V}
\begin{barticle}
\bauthor{\binits{J.C.} \bsnm{{Vial}}},
\batitle{{Two-dimensional nonlocal thermodynamic equilibrium transfer
  computations of resonance lines in quiescent prominences}}.
\bjtitle{\apj}
\bvolume{254},
\bfpage{780}--\blpage{795}
(\byear{1982b}).
doi:\doiurl{10.1086/159789}
\end{barticle}
\endbibitem

\bibitem[\protect\citeauthoryear{{Vial}}{1990}]{1990LNP...363..106V}
\begin{botherref}
\oauthor{\binits{J.C.} \bsnm{{Vial}}},
{The prominence-corona interface},
in \textit{IAU Colloq. 117: Dynamics of Quiescent Prominences},
ed. by V. {Ruzdjak}, E. {Tandberg-Hanssen}.
Lecture Notes in Physics, Berlin Springer Verlag,
vol. 363,
1990,
pp. 106--119
\end{botherref}
\endbibitem

\bibitem[\protect\citeauthoryear{{Vial} et~al.}{2007}]{2007SoPh..246..327V}
\begin{barticle}
\bauthor{\binits{J.C.} \bsnm{{Vial}}}, \bauthor{\binits{H.} \bsnm{{Ebadi}}},
  \bauthor{\binits{A.} \bsnm{{Ajabshirizadeh}}},
\batitle{{The Ly {$\alpha$} and Ly {$\beta$} Profiles in Solar Prominences and
  Prominence Fine Structure}}.
\bjtitle{\solphys}
\bvolume{246},
\bfpage{327}--\blpage{338}
(\byear{2007}).
doi:\doiurl{10.1007/s11207-007-9080-9}
\end{barticle}
\endbibitem

\bibitem[\protect\citeauthoryear{{Vial} et~al.}{1980}]{1980SoPh...68..187V}
\begin{barticle}
\bauthor{\binits{J.C.} \bsnm{{Vial}}}, \bauthor{\binits{P.} \bsnm{{Lemaire}}},
  \bauthor{\binits{G.} \bsnm{{Artzner}}}, \bauthor{\binits{P.}
  \bsnm{{Gouttebroze}}},
\batitle{{O VI /lambda equals 1032 A/ profiles in and above an active region
  prominence, compared to quiet sun center and limb profiles}}.
\bjtitle{\solphys}
\bvolume{68},
\bfpage{187}--\blpage{206}
(\byear{1980})
\end{barticle}
\endbibitem

\bibitem[\protect\citeauthoryear{{Vial} et~al.}{1990}]{1990LNP...363..282V}
\begin{botherref}
\oauthor{\binits{J.C.} \bsnm{{Vial}}}, \oauthor{\binits{M.} \bsnm{{Rovira}}},
  \oauthor{\binits{J.M.} \bsnm{{Fontenla}}}, \oauthor{\binits{P.}
  \bsnm{{Gouttebroze}}},
{Multi-Thread Structure as a Possible Solution for the L-Beta Problem in Solar
  Prominences},
in \textit{IAU Colloq. 117: Dynamics of Quiescent Prominences},
ed. by V. {Ruzdjak}, E. {Tandberg-Hanssen}.
Lecture Notes in Physics, Berlin Springer Verlag,
vol. 363,
1990,
p. 282.
doi:\doiurl{10.1007/BFb0025640}
\end{botherref}
\endbibitem

\bibitem[\protect\citeauthoryear{{Wang}}{1999}]{1999ApJ...520L..71W}
\begin{barticle}
\bauthor{\binits{Y.M.} \bsnm{{Wang}}},
\batitle{{The Jetlike Nature of HE II lambda304 Prominences}}.
\bjtitle{\apjl}
\bvolume{520},
\bfpage{71}--\blpage{74}
(\byear{1999}).
doi:\doiurl{10.1086/312149}
\end{barticle}
\endbibitem

\bibitem[\protect\citeauthoryear{{Wang}}{2001}]{2001ApJ...560..456W}
\begin{barticle}
\bauthor{\binits{Y.M.} \bsnm{{Wang}}},
\batitle{{On the Relationship between He II {$\lambda$}304 Prominences and the
  Photospheric Magnetic Field}}.
\bjtitle{\apj}
\bvolume{560},
\bfpage{456}--\blpage{465}
(\byear{2001}).
doi:\doiurl{10.1086/322495}
\end{barticle}
\endbibitem

\bibitem[\protect\citeauthoryear{{Widing} et~al.}{1986}]{1986ApJ...308..982W}
\begin{barticle}
\bauthor{\binits{K.G.} \bsnm{{Widing}}}, \bauthor{\binits{U.}
  \bsnm{{Feldman}}}, \bauthor{\binits{A.K.} \bsnm{{Bhatia}}},
\batitle{{The extreme-ultraviolet spectrum (300-630 A) of an erupting
  prominence observed from SKYLAB}}.
\bjtitle{\apj}
\bvolume{308},
\bfpage{982}--\blpage{992}
(\byear{1986}).
doi:\doiurl{10.1086/164566}
\end{barticle}
\endbibitem

\bibitem[\protect\citeauthoryear{{Wiik} et~al.}{1993}]{1993A&A...273..267W}
\begin{barticle}
\bauthor{\binits{J.E.} \bsnm{{Wiik}}}, \bauthor{\binits{K.} \bsnm{{Dere}}},
  \bauthor{\binits{B.} \bsnm{{Schmieder}}},
\batitle{{UV prominences observed with the HRTS: structure and physical
  properties}}.
\bjtitle{\aap}
\bvolume{273},
\bfpage{267}
(\byear{1993})
\end{barticle}
\endbibitem

\bibitem[\protect\citeauthoryear{{Wiik} et~al.}{1992}]{1992A&A...260..419W}
\begin{barticle}
\bauthor{\binits{J.E.} \bsnm{{Wiik}}}, \bauthor{\binits{P.} \bsnm{{Heinzel}}},
  \bauthor{\binits{B.} \bsnm{{Schmieder}}},
\batitle{{Determination of plasma parameters in a quiescent prominence}}.
\bjtitle{\aap}
\bvolume{260},
\bfpage{419}--\blpage{430}
(\byear{1992})
\end{barticle}
\endbibitem

\bibitem[\protect\citeauthoryear{{Wiik} et~al.}{1997}]{1997SoPh..175..411W}
\begin{barticle}
\bauthor{\binits{J.E.} \bsnm{{Wiik}}}, \bauthor{\binits{B.}
  \bsnm{{Schmieder}}}, \bauthor{\binits{T.} \bsnm{{Kucera}}},
  \bauthor{\binits{A.} \bsnm{{Poland}}}, \bauthor{\binits{P.} \bsnm{{Brekke}}},
  \bauthor{\binits{G.} \bsnm{{Simnett}}},
\batitle{{Eruptive prominence and associated CME observed with SUMER, CDS and
  LASCO (SOHO)}}.
\bjtitle{\solphys}
\bvolume{175},
\bfpage{411}--\blpage{436}
(\byear{1997}).
doi:\doiurl{10.1023/A:1004925024794}
\end{barticle}
\endbibitem

\bibitem[\protect\citeauthoryear{{Wiik} et~al.}{1999}]{1999SoPh..187..405W}
\begin{barticle}
\bauthor{\binits{J.E.} \bsnm{{Wiik}}}, \bauthor{\binits{I.E.}
  \bsnm{{Dammasch}}}, \bauthor{\binits{B.} \bsnm{{Schmieder}}},
  \bauthor{\binits{K.} \bsnm{{Wilhelm}}},
\batitle{{Multiple-Thread Model of a Prominence Observed by SUMER and EIT on
  SOHO}}.
\bjtitle{\solphys}
\bvolume{187},
\bfpage{405}--\blpage{426}
(\byear{1999}).
doi:\doiurl{10.1023/A:1005151015043}
\end{barticle}
\endbibitem

\bibitem[\protect\citeauthoryear{{Wilhelm} et~al.}{1995}]{1995SoPh..162..189W}
\begin{barticle}
\bauthor{\binits{K.} \bsnm{{Wilhelm}}}, \bauthor{\binits{W.} \bsnm{{Curdt}}},
  \bauthor{\binits{E.} \bsnm{{Marsch}}}, \bauthor{\binits{U.}
  \bsnm{{Sch{\"u}hle}}}, \bauthor{\binits{P.} \bsnm{{Lemaire}}},
  \bauthor{\binits{A.} \bsnm{{Gabriel}}}, \bauthor{\binits{J.C.}
  \bsnm{{Vial}}}, \bauthor{\binits{M.} \bsnm{{Grewing}}},
  \bauthor{\binits{M.C.E.} \bsnm{{Huber}}}, \bauthor{\binits{S.D.}
  \bsnm{{Jordan}}}, \bauthor{\binits{A.I.} \bsnm{{Poland}}},
  \bauthor{\binits{R.J.} \bsnm{{Thomas}}}, \bauthor{\binits{M.}
  \bsnm{{K{\"u}hne}}}, \bauthor{\binits{J.G.} \bsnm{{Timothy}}},
  \bauthor{\binits{D.M.} \bsnm{{Hassler}}}, \bauthor{\binits{O.H.W.}
  \bsnm{{Siegmund}}},
\batitle{{SUMER - Solar Ultraviolet Measurements of Emitted Radiation}}.
\bjtitle{\solphys}
\bvolume{162},
\bfpage{189}--\blpage{231}
(\byear{1995}).
doi:\doiurl{10.1007/BF00733430}
\end{barticle}
\endbibitem

\bibitem[\protect\citeauthoryear{{Woodgate} et~al.}{1980}]{1980SoPh...65...73W}
\begin{barticle}
\bauthor{\binits{B.E.} \bsnm{{Woodgate}}}, \bauthor{\binits{J.C.}
  \bsnm{{Brandt}}}, \bauthor{\binits{M.W.} \bsnm{{Kalet}}},
  \bauthor{\binits{P.J.} \bsnm{{Kenny}}}, \bauthor{\binits{E.A.}
  \bsnm{{Tandberg-Hanssen}}}, \bauthor{\binits{E.C.} \bsnm{{Bruner}}},
  \bauthor{\binits{J.M.} \bsnm{{Beckers}}}, \bauthor{\binits{W.}
  \bsnm{{Henze}}}, \bauthor{\binits{E.D.} \bsnm{{Knox}}},
  \bauthor{\binits{C.L.} \bsnm{{Hyder}}},
\batitle{{The Ultraviolet Spectrometer and Polarimeter on the Solar Maximum
  Mission}}.
\bjtitle{\solphys}
\bvolume{65},
\bfpage{73}--\blpage{90}
(\byear{1980}).
doi:\doiurl{10.1007/BF00151385}
\end{barticle}
\endbibitem

\bibitem[\protect\citeauthoryear{{Yakovkin} and
  {Zel'dina}}{1964}]{1964SvA.....8..262Y}
\begin{barticle}
\bauthor{\binits{N.A.} \bsnm{{Yakovkin}}}, \bauthor{\binits{M.Y.}
  \bsnm{{Zel'dina}}},
\batitle{{Excitation and Ionization of Hydrogen in Prominences.}}
\bjtitle{Soviet Astronomy}
\bvolume{8},
\bfpage{262}
(\byear{1964})
\end{barticle}
\endbibitem

\bibitem[\protect\citeauthoryear{{Yakovkin} and
  {Zel'dina}}{1968}]{1968SvA....12...40Y}
\begin{barticle}
\bauthor{\binits{N.A.} \bsnm{{Yakovkin}}}, \bauthor{\binits{M.Y.}
  \bsnm{{Zel'dina}}},
\batitle{{The Lyman-{$\alpha$} Radiation Field in a Chromospheric Filament.}}
\bjtitle{Soviet Astronomy}
\bvolume{12},
\bfpage{40}
(\byear{1968})
\end{barticle}
\endbibitem

\bibitem[\protect\citeauthoryear{{Yakovkin} et~al.}{1982}]{1982SoPh...81..339I}
\begin{barticle}
\bauthor{\binits{N.A.} \bsnm{{Yakovkin}}}, \bauthor{\binits{M.Y.}
  \bsnm{{Zeldina}}}, \bauthor{\binits{C.} \bsnm{{Lhagvazhav}}},
\batitle{Helium radiation diffusion in prominences.}
\bjtitle{\solphys}
\bvolume{81},
\bfpage{339}--\blpage{354}
(\byear{1982})
\end{barticle}
\endbibitem

\bibitem[\protect\citeauthoryear{{Zhang} and
  {Fang}}{1987}]{1987A&A...175..277Z}
\begin{barticle}
\bauthor{\binits{Q.Z.} \bsnm{{Zhang}}}, \bauthor{\binits{C.} \bsnm{{Fang}}},
\batitle{{Semi-empirical models of a quiescent prominence}}.
\bjtitle{\aap}
\bvolume{175},
\bfpage{277}--\blpage{281}
(\byear{1987})
\end{barticle}
\endbibitem

\bibitem[\protect\citeauthoryear{{Zharkova}}{1989}]{1989HvaOB..13..331Z}
\begin{barticle}
\bauthor{\binits{V.V.} \bsnm{{Zharkova}}},
\batitle{{Toward Hydrogen Emission in Filamentary Quiescent Prominences}}.
\bjtitle{Hvar Observatory Bulletin}
\bvolume{13},
\bfpage{331}
(\byear{1989})
\end{barticle}
\endbibitem

\bibitem[\protect\citeauthoryear{{Zirin}}{1976}]{1976SoPh...50..399Z}
\begin{barticle}
\bauthor{\binits{H.} \bsnm{{Zirin}}},
\batitle{{Production of a short-lived filament by a surge}}.
\bjtitle{\solphys}
\bvolume{50},
\bfpage{399}--\blpage{404}
(\byear{1976}).
doi:\doiurl{10.1007/BF00155302}
\end{barticle}
\endbibitem

\bibitem[\protect\citeauthoryear{{Zirker}}{1985}]{1985SoPh..102...33Z}
\begin{barticle}
\bauthor{\binits{J.B.} \bsnm{{Zirker}}},
\batitle{{Prominence hydrogen lines at 10-20 microns}}.
\bjtitle{\solphys}
\bvolume{102},
\bfpage{33}--\blpage{40}
(\byear{1985}).
doi:\doiurl{10.1007/BF00154035}
\end{barticle}
\endbibitem

\bibitem[\protect\citeauthoryear{{Zirker} and
  {Koutchmy}}{1991}]{1991SoPh..131..107Z}
\begin{barticle}
\bauthor{\binits{J.B.} \bsnm{{Zirker}}}, \bauthor{\binits{S.}
  \bsnm{{Koutchmy}}},
\batitle{{Prominence fine structure. II - Diagnostics}}.
\bjtitle{\solphys}
\bvolume{131},
\bfpage{107}--\blpage{118}
(\byear{1991}).
doi:\doiurl{10.1007/BF00151747}
\end{barticle}
\endbibitem

\bibitem[\protect\citeauthoryear{{Zirker} et~al.}{1998}]{1998Natur.396..440Z}
\begin{barticle}
\bauthor{\binits{J.B.} \bsnm{{Zirker}}}, \bauthor{\binits{O.}
  \bsnm{{Engvold}}}, \bauthor{\binits{S.F.} \bsnm{{Martin}}},
\batitle{{Counter-streaming gas flows in solar prominences as evidence for
  vertical magnetic fields}}.
\bjtitle{\nat}
\bvolume{396},
\bfpage{440}
(\byear{1998}).
doi:\doiurl{10.1038/24798}
\end{barticle}
\endbibitem

\end{thebibliography}

\end{document}